\documentclass{article}

\usepackage[margin=0.75in]{geometry}
\usepackage[utf8]{inputenc}
\usepackage[english]{babel}
\usepackage{times}

\usepackage{amsmath, amssymb, amsthm, mathtools, bm, bbm, extpfeil}

\usepackage{graphicx}
\usepackage[width=\textwidth]{caption}
\usepackage{booktabs, array, multirow, color}

\usepackage[authoryear]{natbib}
\usepackage{enumitem}
\usepackage{hyperref}
\usepackage{url}

\graphicspath{{figures/}}

\newtheorem{theorem}{Theorem}
\newtheorem{proposition}{Proposition}
\newtheorem{remark}{Remark}
\newtheorem{definition}{Definition}
\newtheorem{assumption}{Assumption}

\newtheorem{lemma}{Lemma}
\newtheorem{model}{Model}
\numberwithin{equation}{section}

\newcommand{\E}{\mathbb{E}}

\newcommand{\cov}{\text{Cov}}
\newcommand{\corr}{\text{Corr}}

\newcommand{\past}{\text{past}}

\newcommand{\PP}{\mathbb{P}}
\newcommand{\Z}{\mathbb{Z}}
\newcommand{\N}{\mathbb{N}}
\newcommand{\R}{\mathbb{R}}
\newcommand{\C}{\mathbb{C}}

\newcommand{\F}{\mathcal{F}}

\newcommand{\J}{\mathcal{J}}

\newcommand{\X}{{X}}
\newcommand{\Y}{{Y}}
\newcommand{\ZZ}{{Z}}
\newcommand{\vw}{\bm{w}}
\newcommand{\vv}{\bm{v}}

\newcommand{\1}{\mathbbm{1}}

\begin{document}
\title{Causal tail coefficient for compound extremes in multivariate time series}
\author{Cathy Yin\textsuperscript{1,}\textsuperscript{*}, Adam M. Sykulski\textsuperscript{1} and Almut E.D. Veraart\textsuperscript{1}}
\date{}
\maketitle

\begin{abstract}
Extreme events are often multivariate in nature. A compound extreme occurs when a combination of variables jointly produces a significant impact, even if individual components are not necessarily marginally extreme. 
Compound extremes have been observed across a wide range of domains, including space weather, climate, and environmental science. 
For example, a single episode of heavy rainfall can trigger a compound hydrological response involving elevated river levels, increased soil saturation, and substantial surface runoff, which together may result in a sustained flooding event.
However, most existing methods for detecting extremal causality focus primarily on individual extreme values and lack the flexibility to capture causal relationships between compound extremes.
This work introduces a novel framework for detecting causal dependencies between extreme events, including compound extremes. 
We introduce the compound causal tail coefficient for stationary time series, which quantifies the extremal dependence between an extreme in a candidate cause series and a subsequent compound response jointly defined by multiple future values of one or more effect series.
Based on a consistent estimator of this coefficient, we develop a novel bootstrap hypothesis test to evaluate the presence and direction of causal relationships. Our method can accommodate nonlinearity and latent confounding variables. We demonstrate the effectiveness of our method by establishing theoretic properties, and through simulation studies and an application to space-weather data.
\let\thefootnote\relax\footnote{\textsuperscript{1} Department of Mathematics, Imperial College London, 180 Queen's Gate, SW7 2AZ, United Kingdom}
\let\thefootnote\relax\footnote{\textsuperscript{*} Corresponding author. Email: \texttt{c.yin24@imperial.ac.uk}}
\end{abstract}

\section{Introduction}
Statistical modeling of extreme values has drawn increasing attention due to the rising prevalence of rare but high-impact events, such as heatwaves, floods, financial crises, and infrastructure failures. At a more local scale, extreme value methods are also used to monitor acute medical conditions, including epileptic seizures and cardiac arrests. Across these domains, a central challenge is to understand the dependencies between extremes. Identifying how one extreme event influences another is critical not only for predicting rare events but also for managing systemic risks and mitigating their impacts.

Multivariate time series provide a natural framework for studying such dependence, as they capture both cross-sectional and temporal relationships. Within this framework, causality represents the strongest form of dependence, where one process directly influences another. 
The most widely used notion is Granger causality \citep{Granger1969,Granger1980}, which evaluates whether a potential cause series has predictive power for future values of an effect series through regression-based tests. While effective for the bulk of the distribution, such general causality measures are inadequate for extremes, since extreme values often behave differently from regular values and many causal mechanisms may manifest only in the tails.
Several approaches have sought to adapt causality concepts to extreme value theory. Granger causality in risks \citep{HLW2009} restricts Granger’s framework to tail events but remains limited by its reliance on Value-at-Risk exceedances. More recently, causal tail coefficients have been introduced to quantify extremal causal influence, first in structural causal models \citep{GMPE2021} and later in time series \citep{BPP2024}, with conditional versions developed to account for confounding \citep{PCDD2023}.
By exploiting temporal ordering, these coefficients assess whether an extreme in one process predicts a subsequent extreme response, providing an observational framework analogous to a natural experiment. Such a predictive notion of causality is particularly appropriate for extreme events, for which controlled interventions are generally infeasible and causal information must be inferred from naturally occurring variation in time series.

However, existing causal tail coefficients are restrictive, as they define extremes only in marginal terms and fail to capture the compounding nature of many real-world events. 
For example, an extreme rainfall event may be followed by a compound hydrological response involving elevated river levels, increased soil saturation, and substantial surface runoff, which together can overwhelm drainage systems and lead to subsequent sustained flooding. In health, extreme heat exposure may induce concurrent changes in body temperature, heart rate, and blood pressure, which can collectively pose a serious cardiovascular risk. Similar patterns arise in other domains: an extreme disturbance to a power grid may produce simultaneous disruptions to electricity supply, communication networks, and transportation systems, thereby amplifying societal and economic impacts. Likewise, an extreme financial shock may generate a compound response involving falling asset prices, increased volatility, and reduced market liquidity across successive time periods, threatening broader financial stability.
These types of multivariate responses, in which several variables over successive time periods jointly contribute to an adverse impact, are referred to as \textit{compound extremes} \citep{BMHWV2017}. 
Their societal impacts have received growing attention in the contexts of disaster risk management, climate adaptation, and emergency preparedness \citep{Zscheischler2018,Zscheischler2020,BMM2021}. 
This literature emphasizes the importance of accounting for interactions among drivers to obtain realistic risk estimates and inform early warning, resource allocation, and other risk reduction measures.
Compound extremes are also increasingly recognized in studies of causal mechanisms in climate and environmental science \citep[e.g.,][]{HN2018,KN2020,CV2021}.
Nevertheless, no formal statistical measure has been developed to incorporate compound extremes into causal analysis.

This paper addresses this gap by introducing the compound causal tail coefficient, a flexible extension of the causal tail coefficient that accommodates compounding effects across multiple variables in time series. Building on this new coefficient, we develop a block bootstrap hypothesis test with a novel time-shifting step for causal inference in extremes. Our method demonstrates high accuracy and computational efficiency across diverse time series models, and mitigates confounding from feedback structures in the data-generating process. We illustrate its use with a space-weather application, where the causal drivers of geomagnetic storms remain debated, and highlight broader relevance to climate science, finance, epidemiology, and infrastructure risk.

The remainder of this paper is organized as follows. Section~\ref{sec:prelim} introduces notation and reviews relevant concepts. Section~\ref{sec:background} summarizes existing causal tail coefficients. Section~\ref{sec:cctc_main} develops the compound causal tail coefficient for time series and extends it to settings with confounders and higher dimensions. Section~\ref{sec:estimation} outlines estimation and optimization procedures and introduces a block bootstrap test with a time-shifting technique for causal inference. 
Section~\ref{sec:simulation} presents Monte Carlo results benchmarking our method against existing pairwise causal discovery approaches. 
Section~\ref{sec:spacew} applies the methodology to space-weather data, corroborating prior findings on the relationships among geomagnetic storms, substorms, and the interplanetary magnetic field. 
Section~\ref{sec:conclusion} concludes with remarks on practical implications and directions for future research. 
The supplementary material provides additional Monte Carlo results, sensitivity analyses, and details of the space-weather application, along with statements and proofs of the theoretical results developed herein.
All code used in this paper is publicly available at 
\href{https://github.com/icxmas/compound_causal_tail_coefficient}{https://github.com/icxmas/compound\_causal\_tail\_coefficient}.

\section{Preliminaries}\label{sec:prelim}
We consider a time series as a sequence of random variables defined on a common probability space $(\Omega, \F, \{\F_t, t\in \Z \}, \PP)$, written $\X = \{\X_t, t\in \Z\}$, with integer time indices. We use the term ``stationary time series" to refer to strict stationarity and write i.i.d.\ to mean independent and identically distributed.
For a random variable $X$, let $F(x)$ be its cumulative distribution function (CDF), $\bar{F}(x) = 1 - F(x) = \PP(X>x)$ its survival function, and $Q(u) = \inf \{x\in \R: F(x)\geq u \}$, $u\in[0,1]$, its quantile function. For functions $f$ and $g$, $f \sim g$ means $\lim_{x\to\infty} f(x)/g(x) = 1$. We use $x_{+} := \max\{x,0\}$, bold capitals for vectors, and $\mathbf{1}$ for a vector of ones.  

We work within the framework of regular variation \citep{Resnick1987, EKM1997}, a central tool in extreme value theory. A random variable $X$ has \textit{regularly varying tails} with index $\alpha>0$ if  
$
\bar{F}(x) \sim \ell(x) x^{-\alpha}
$
as $x \to \infty$, where $\ell$ is slowly varying, i.e.\ $\lim_{x \to \infty} \ell(tx) / \ell(x) = 1$ for all $t > 0$. The tail index $\alpha$ determines the rate of decay: smaller $\alpha$ corresponds to heavier tails.  
Regular variation arises in many time series models, including autoregressive moving average (ARMA) and stochastic volatility models with i.i.d.\ regularly varying noise, generalized autoregressive conditional heteroskedasticity (GARCH) processes with infinite-support noise (e.g., normal or Student-$t$), and stable processes with infinite variance. Furthermore, any strictly stationary series whose finite-dimensional distributions lie in the maximum domain of attraction of a multivariate extreme value distribution can be transformed into a regularly varying process by a monotone transformation of its marginal distribution \citep{Resnick1987}. This means that results for regularly varying processes presented in our work have a wider applicability within extreme value theory.

We follow the notion of causality proposed by \citet{Granger1969}, where past values of a series $X$ improve the prediction of future values of another series $Y$ beyond what is possible using $Y$’s own past. While predictability alone reflects statistical dependence, imposing a temporal ordering allows such dependence to be interpreted causally \citep{Kuersteiner2010}.  
Suppose $(X_t, Y_t, Z_t)$ contains all relevant information at time $t$. Let $X_{\past(t)} = \{X_s: s \leq t\}$, with analogous definitions for $Y_{\past(t)}$ and $Z_{\past(t)}$ (more precisely, these can be interpreted as the $\sigma$-fields generated by the respective histories).
The process $\X$ is said to \textit{Granger cause} $\Y$ if $Y_{t+1}$ is not conditionally independent of past values of $X$ given all other relevant information up to time $t$; that is,
$ Y_{t+1} \not\!\perp\!\!\!\perp X_{\past(t)} \mid Y_{\past(t)}, Z_{\past(t)} $ for all $t$.
Note that Granger causality is always defined relative to the available information set $\sigma( \X_{\past(t)},\Y_{\past(t)},\ZZ_{\past(t)} )$; adding variables to this set may change the inferred relationship.

\section{Existing work on causal tail coefficients}\label{sec:background}
The causal tail coefficient (CTC) is an extremal causality measure introduced by \citet{GMPE2021} in the context of structural causal models (SCMs). Unlike Granger causality, the causal tail coefficient captures the asymmetric dependencies in the tails of distributions rather than over the bulk of the data. These dependencies manifest primarily during extreme events, where conventional methods often fail. For random variables $X$ and $Y$ with CDFs $F_X$ and $F_Y$, the \textit{causal tail coefficient} from $X$ to $Y$ is defined as $\Gamma_{X\to Y} := \lim_{u \to 1^-} \E \left[ F_Y(Y) \mid F_X(X) > u \right]$, if the limit exists. This coefficient lies in $[0,1]$ and is invariant to strictly increasing transformations. If large values of $X$ cause large values of $Y$ but not vice versa, then $\Gamma_{\X \rightarrow \Y} = 1$, whereas $\Gamma_{\Y \rightarrow \X} < 1$.

\citet{GMPE2021} proposed a nonparametric estimator using the empirical distribution functions of $X$ and $Y$. Given $n$ i.i.d.\ samples $(x_i,y_i)$,  
$
    \hat{\Gamma}_{X \rightarrow Y} = \tfrac{1}{k} \sum_{i=1}^n \hat{F}_Y(y_i) \cdot
    \1_{ \{ x_i \geq x_{(n-k+1)} \} },
$
where $x_{(n-k+1)}$ is the $k$-th largest $x$-value, and $\hat{F}_Y(y) = \frac{1}{n}\sum_{j=1}^n \1_{ \{y_j \leq y\} }$ is the empirical CDF of $Y$. Consistency holds in heavy-tailed linear SCMs under mild conditions. The threshold parameter $k=k_n$ must satisfy $k_n\to\infty$ and $k_n/n\to 0$; in practice $k=\sqrt{n}$ is common.

\citet{PCDD2023} proposed a parametric version by fitting a generalized Pareto distribution (GPD) to the upper tails of $F_X$ and $F_Y$, motivated by the Second Extreme Value Theorem \citep{Pickands1975}. For a high threshold $u$, exceedances $X-u$ have an approximate GPD distribution with scale $\sigma>0$ and shape $\xi\in\R$. Let $\hat{F}_X(\cdot; \hat{\sigma}_x, \hat{\xi}_x)$ be a hybrid estimator using the empirical CDF below $u$ and a fitted GPD above $u$. The parametric CTC estimator is
$
    \hat{\Gamma}^{GPD}_{X \rightarrow Y} = \frac{1}{k_g} \sum_{i=1}^n \hat{F}_Y (y_i;\hat{\sigma}_y, \hat{\xi}_y) \cdot
    \1_{ \left\{\hat{F}_X (x_i;\hat{\sigma}_x, \hat{\xi}_x) \geq 1 - \frac{k}{n} \right\} },
$
where $k_g := \left|\{i\in \{1,\dots,n\}: \hat{F}_X (x_i;\hat{\sigma}_x, \hat{\xi}_x) \geq 1 - \frac{k}{n} \right|$ may differ from $k$, since it now depends on the fitted $\hat{F}_X (x;\hat{\sigma}_x, \hat{\xi}_x)$.
\cite{PCDD2023} argued that this parametric estimator is better suited to reduce confounding from covariates.

To account for lagged dependence in extreme events, \citet{BPP2024} extended the CTC to stationary time series $\X = \{X_t\}$ and $\Y = \{Y_t\}$. The \textit{causal tail coefficient for time series} from $\X$ to $\Y$ is defined as follows:
\begin{align}\label{CTC1}
    \Gamma_{\X \rightarrow \Y}(p) :=
    \lim_{u\to 1^-} \E \left[ \max_{1 \le j \le p} F_Y(Y_j) \mid F_X(X_0)>u \right],
\end{align}
where $p\in\N$ is the \textit{extremal delay}, representing the hypothesized maximal lag between cause and effect. This formulation allows for delayed causal influence in extremes.
A corresponding nonparametric estimator is
\begin{align}\label{npestimator}
    \hat{\Gamma}_{\X \rightarrow \Y} (p) = \frac{1}{k} \sum_{i=1}^{n-p} \max \left(\hat{F}_Y(y_{i+1}), \ldots, \hat{F}_Y(y_{i+p}) \right) 
    \cdot \1_{\{ x_i \geq x_{(n-k+1)} \} }.
\end{align}
Here, $p$ is chosen via domain knowledge or heuristics. When $X\to Y$ is the true causal direction, $\hat{\Gamma}_{\X \rightarrow \Y}(p)$ increases much faster than $\hat{\Gamma}_{\Y \rightarrow \X}(p)$ as $p$ increases before reaching the ``correct" extremal delay; for $p$ sufficiently large, both coefficients converge to $1$ \citep{BPP2024}. Selecting an appropriate extremal delay $p$ therefore remains a key practical challenge.

\section{The causal tail coefficient for compound extremes in time series}\label{sec:cctc_main}

The aforementioned CTCs do not account for compounding effects in multivariate time series. 
As a first step towards addressing this limitation, we consider the bivariate time series setting $(\X,\Y)$, in which the compound response is jointly defined by multiple future values of the single effect series $\Y$. For example, if $X$ is a single rainfall time series and $Y$ is a single river level time series, then a single rainfall event in $X$ might cause a sustained elevation in river levels in $Y$ over several time periods.
We later extend this definition in \eqref{CCTC_mv} to a multivariate compound response jointly defined across multiple effect series and time points. In both settings, flexible modeling of the compound response is achieved through a carefully chosen impact function that combines its components.

\begin{definition}
Let $h:[0,1]^p \rightarrow[0,1]$ be an impact function. The \textit{compound causal tail coefficient for time series} from $X$ to $Y$ with extremal delay $p$ is defined as
    \begin{align}\label{CCTC}
        \Gamma_{{\X} \rightarrow {h(\Y})}(p):=\lim_{u\to 1^-}\E \, \left[  h(F_Y(Y_1), \ldots, F_Y(Y_p)) \mid F_X(X_0)>u \right],
    \end{align}
where $F_X$ and $F_Y$ are the marginal CDFs of $X_t$ and $Y_t$.
\end{definition}

The compound CTC measures the effect of an extreme event in $\X$ on a compound response formed from the subsequent values of $\Y$. Since the impact function $h$ maps $[0,1]^p$ into $[0,1]$, the compound CTC also lies in $[0,1]$, with values close to $1$ indicating a strong upper tail causal influence of $\X$ on $\Y$. As with other CTCs, the measure is directional and need not be symmetric. The theoretical results below in Section~\ref{sec:property} establish conditions under which the asymmetry between $\Gamma_{\X\rightarrow h(\Y)}(p)$ and $\Gamma_{\Y\rightarrow h(\X)}(p)$ can be leveraged to identify the causal direction. Its precise interpretation depends on the choice of impact function $h$ in \eqref{CCTC}. We now introduce a proposed impact function that can flexibly represent different compositions of compound extremes.

\subsection{Impact function for modeling compound extremes}\label{sec:impact}

The CTC for time series of \citet{BPP2024} \eqref{CTC1} implicitly uses the maximum as its impact function:
$
h(v_1,\ldots,v_p) = \max(v_1,\ldots,v_p), \,v_i \in [0,1].
$
This function exhibits a \textit{nil cancellation} effect: if any $v_i$ is close to $1$, then $h$ is close to $1$. This property is worth preserving, as a single large value can also trigger an extreme impact.
A similar cancellation effect can be achieved using the Blalock formula (\citet[p.105]{Blalock1982}; see also \cite{FH2002}),
$
h(v_1,\dots,v_p; \vw) = 1-\prod_{i=1}^p (1-v_i)^{w_i},
$
where $\vw=(w_1,\ldots,w_p)\in\R_{\geq0}^p$ contains nonnegative weights that need not sum to $1$.

More generally, a suitable impact function should satisfy several other properties. It should map inputs on the marginal probability scale from $[0,1]^p$ into $[0,1]$, so that the resulting compound CTC remains comparable with existing CTCs, and it should be nondecreasing in each argument, so that increasing the extremeness of any component cannot reduce the overall impact. In addition to preserving nil cancellation, it should accommodate an additive form of accumulation, which we refer to as a \textit{linear cancellation} effect, since a compound extreme may arise from several moderately large contributions. Finally, its parameters should have clear interpretations in terms of the relative importance of different lags and the type of cancellation.
Neither a weighted sum, the maximum, nor the Blalock impact function can accommodate both linear and nil cancellation within the same functional form. Therefore, representing both mechanisms within a single family requires additional flexibility to control the transition between them. 
\citet{Firth2002} provides such a generalization of the Blalock formula:
$
  \log\!\left(1- h(v_1,\ldots,v_p)(1-e^{-\alpha})\right)
  = \sum_{i=1}^p w_i \log \left( 1-v_i(1-e^{-\alpha}) \right),
$ for $\alpha>0$.
This construction introduces only one additional scalar parameter, $\alpha$, beyond the weights $w_i$, and therefore provides a parsimonious extension capable of representing both mechanisms.
As $\alpha \to \infty$, this reduces to the Blalock formula; as $\alpha \to 0$, it becomes the weighted sum $\sum_i w_i v_i$. Proposition~\ref{prop:3} in the Supplementary Section~\ref{sec:impact_pf} justifies this limiting behavior. Varying $\alpha$ yields a continuum from nil to linear cancellation; such functions have been applied, for example, to construct overall local deprivation indices \citep{NWSD2006}.

We adopt Firth’s generalization as the impact function to define compound extremes:
\begin{align}\label{impactfunction}
h(v_1,\ldots,v_p; \vw,\alpha)
=\left[1-\prod_{i=1}^p \left\{ 1-v_i(1-e^{-\alpha})\right\}^{w_i}\right] \left(1-e^{-\alpha}\right)^{-1},
\end{align}
with non-negative weights normalized so that $\vw \cdot \mathbf{1} = 1$. Normalizing to a unit total weight ensures that $h$ takes values in $[0,1]$, as required by the definition of the compound CTC, and enables direct comparison of individual components and of values across the impact function.
The impact shape parameter $\alpha$ controls the type of cancellation, while the $w_i$ reflects the marginal importance of each lag. Uniform weights imply equal contributions; $w_i=1$ for some $i$ implies that lag $i$ alone determines the impact. Parameters may be set from domain knowledge or chosen to maximize the CTC.
Together, $\vw$ and $\alpha$ allow the impact function to represent a range of different compositions of compound extremes.

To make the distinct roles of these parameters more explicit, we examine the geometry of the upper contour sets of the impact function.
For an impact level $\eta\in(0,1)$, define
\begin{align*}
\mathcal{E}_{\eta}(\vw,\alpha)
=
\left\{
(v_1,\ldots,v_p)\in[0,1]^p:
h(v_1,\ldots,v_p;\vw,\alpha)\geq\eta
\right\}.
\end{align*}
The set $\mathcal{E}_{\eta}(\vw,\alpha)$ contains precisely those combinations of component values that are classified as compound extreme at level $\eta$.
Writing $c_\alpha :=1-e^{-\alpha}$, it can equivalently be expressed as
\begin{align*}
\mathcal{E}_{\eta}(\vw,\alpha)
=
\left\{
(v_1,\ldots,v_p)\in[0,1]^p:
\prod_{i=1}^p(1-c_\alpha v_i)^{w_i}
\leq 1-\eta c_\alpha
\right\}.
\end{align*}
Thus, contours of $h$ characterize the different compositions of marginal inputs $(v_1,\ldots,v_p)$ that produce the same aggregate impact.

\begin{figure}[htb]
    \centering
    \includegraphics[width=0.8\textwidth]
    {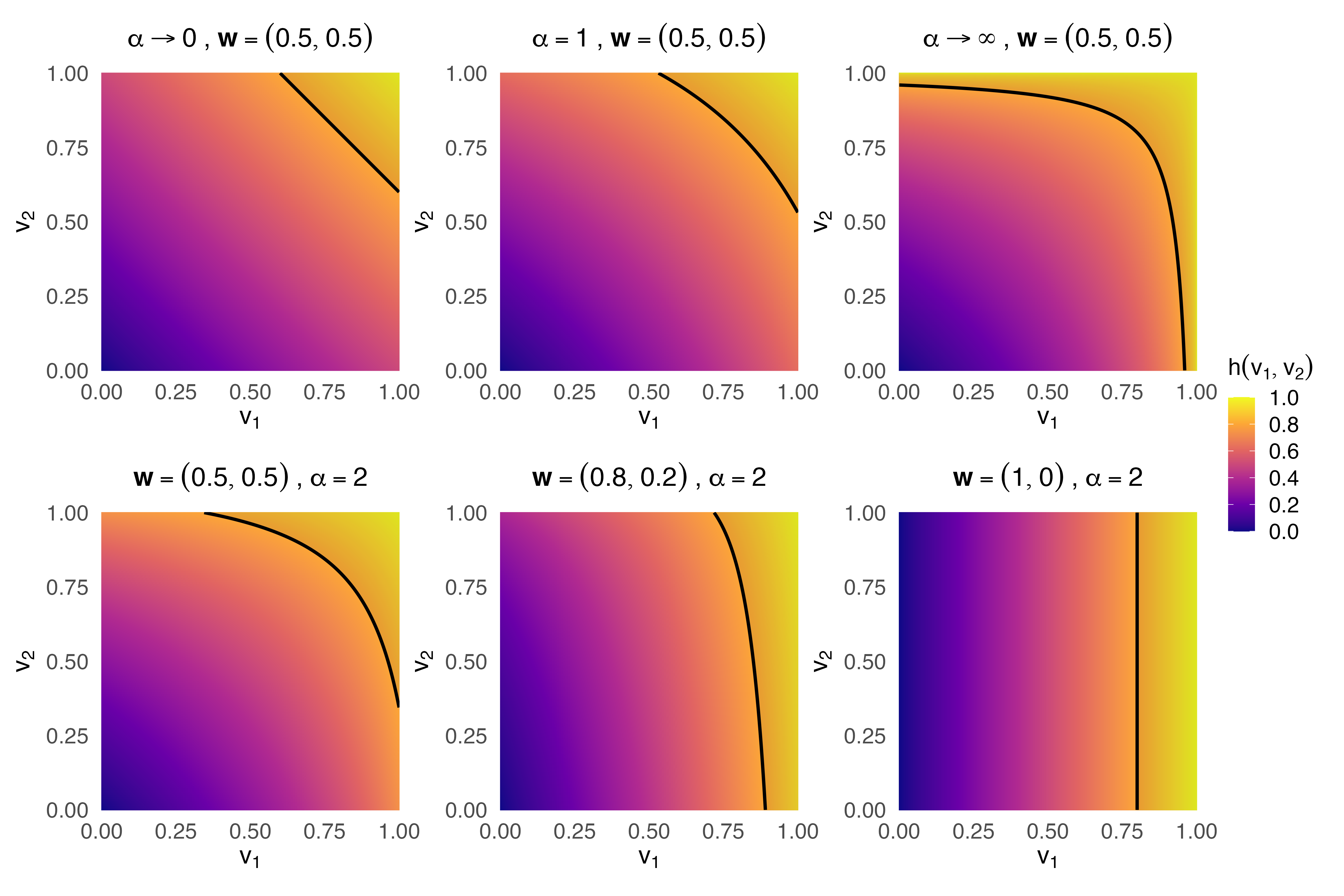}
    \caption{\footnotesize
    Geometry of the two-component impact function in \eqref{impactfunction}. Colors indicate the values of $h(v_1,v_2;\vw,\alpha)$, with black contour marking the boundary of the set $\mathcal{E}_{0.8}(\vw,\alpha)$. 
    The top row fixes $\vw=(1/2,1/2)$ and varies $\alpha$, illustrating the transition from the weighted-sum limit as $\alpha\to0$ to the Blalock limit as $\alpha\to\infty$. The bottom row fixes $\alpha=2$ and varies $\vw$, illustrating equal-weight, unequal-weight, and single-lag specifications.}
    \label{fig:impact_geometry}
\end{figure}

Figure~\ref{fig:impact_geometry} visualizes this geometry by showing how the contours and upper contour sets of the impact function change with $\alpha$ and $\vw$ in the two-dimensional case.
The top row shows that the impact shape parameter $\alpha$ governs how the positively weighted components are aggregated.
As $\alpha\to0$, the impact function converges to
$w_1 v_1 + w_2 v_2$.
In this limit, the boundaries of $\mathcal{E}_{\eta}(\vw,\alpha)$ are linear.
This corresponds to an accumulation-dominated compound extreme, in which several moderately large components can jointly generate a high impact through their weighted total.
As $\alpha$ increases, the contours become increasingly curved and extend further towards the coordinate axes. This indicates greater sensitivity to uneven compositions, where one component is particularly large.
In the limit $\alpha\to\infty$, the impact function converges to the Blalock form
$1- (1-v_1)^{w_1} (1-v_2)^{w_2}$.
This represents a component-dominated compound extreme: if any positively weighted component approaches $1$, the overall impact also approaches $1$, regardless of the remaining components.
Intermediate values of $\alpha$ capture hybrid compositions that combine weighted accumulation with sensitivity to a particularly large component.

The bottom row of Figure~\ref{fig:impact_geometry} distinguishes the role of the weights from that of $\alpha$.
Equal weights produce a symmetric impact region in which both lags contribute equally, whereas, for $\vw=(0.8,0.2)$, the contours reflect the greater importance assigned to the first lag.
At the boundary case $\vw=(1,0)$, the impact
$ h(v_1,v_2;(1,0),\alpha)=v_1 $
for every $\alpha>0$, so the event is determined entirely by the first lag.
An expanded $3\times3$ comparison covering all combinations of $\alpha\to0$, $\alpha=1$, and $\alpha\to\infty$ with $\vw=(0.5,0.5)$, $\vw=(0.8,0.2)$, and $\vw=(1,0)$ is provided in Figure~\ref{fig:impact_geometry_full} of the Supplementary material.


Figure~\ref{fig:impact_fixed_average} further isolates the effect of $\alpha$ by holding the average magnitude of the inputs $(v_1, v_2)$ fixed.
Under equal weights, let $v_1=m+d$ and $v_2=m-d$,
where $|d|$ measures the extent to which the input concentrates in one component.
In the linear limit,
\(
    \lim_{\alpha\to0}
    h\left( m+d,m-d; (0.5, 0.5),\alpha \right)
    = m,
\)
so all compositions with the same average input produce the same impact.
For $\alpha>0$, the impact increases with $|d|$: a more uneven composition produces a larger value of $h$, even though its average magnitude is unchanged.
Moreover, the normalization of the weights implies the calibration property
\(
    h(m,\ldots,m; \vw,\alpha) = m
\)
for every $\alpha>0$,
and thus the divergence of the curves away from the dotted line $h=m$ illustrates the increasing sensitivity to component-dominated extremes as $\alpha$ grows.

\begin{figure}[htb]
    \centering
    \includegraphics[width=0.6\textwidth]
    {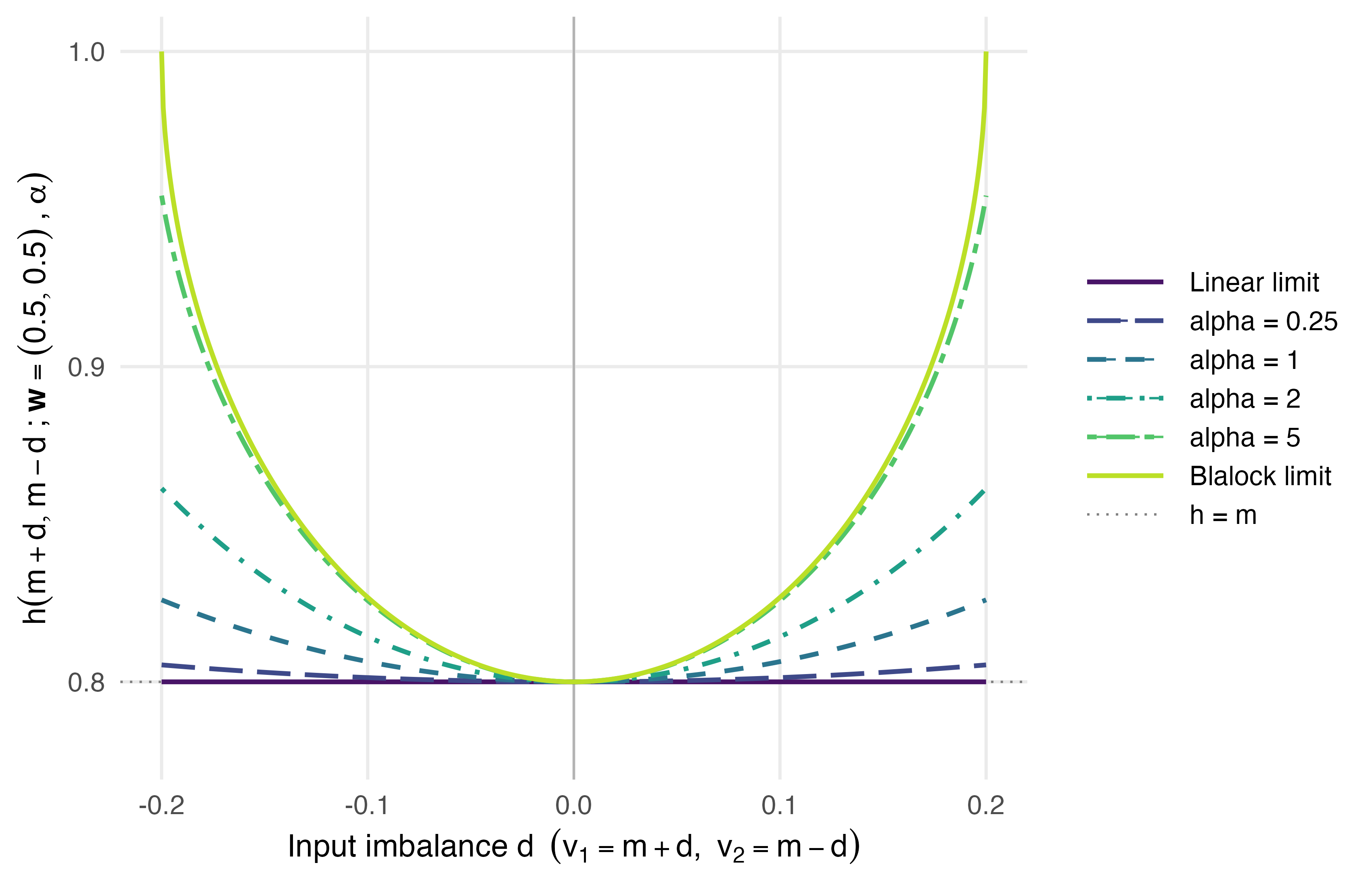}
    \caption{\footnotesize 
    Impact under different compositions with a fixed average input.
    We set $\vw=(0.5, 0.5)$, $v_1=m+d$, $v_2=m-d$, and $m=0.8$, so that $(v_1+v_2)/2=m$ for every $d$. The quantity $|d|$ measures the imbalance between the two components.
    In the linear limit, the impact remains equal to $m$, whereas increasing $\alpha$ assigns increasingly greater impact to more uneven compositions.}
    \label{fig:impact_fixed_average}
\end{figure}

In summary, the weights $\vw$ determine which components participate in the compound event and their relative importance, while the impact shape parameter $\alpha$ controls how the positively weighted components are combined. 
The resulting regimes also admit natural real-world interpretations. For example, a high cumulative impact may arise either from moderately elevated stress sustained over consecutive periods, corresponding to an accumulation-dominated event, or from a single exceptionally severe shock at one relevant lag, corresponding to a component-dominated event. Intermediate values of $\alpha$, together with non-degenerate weight specifications, allow both sustained contributions across several lags and dominant shocks at individual lags to be represented within the same model.

\subsection{Properties of the CCTC}\label{sec:property}
We examine the theoretical properties of the compound causal tail coefficient within a flexible class of autoregressive models formally defined in Supplementary Section~\ref{sec:models}, which includes the heavy-tailed vector autoregressive (hVAR) and the heavy-tailed nonlinear additive autoregressive (hNAAR) models. 
The corresponding proofs are provided in Supplementary Section~\ref{sec:cctc_pf}.

We first formalize \textit{correctly specified} weight parameters in Assumption~\ref{weight assumption}. To make this assumption more mild, we only require zero weights for non-causal lags, without constraining how weight is distributed among causal lags; for example, a single causal lag may receive full weight.

\begin{assumption}\label{weight assumption}
Let $(\X, \Y)$ be a bivariate hVAR($p$) or hNAAR($p$) time series (Definitions~\ref{def:hVAR} and \ref{def:hNAAR} respectively in Section~\ref{sec:models}).  
We say that the weight vector $\vw$ is correctly specified for the compound CTC $\Gamma_{\X \rightarrow h(\Y)}(p)$ if $w_i = 0$ for all $i \in \{1, \dots, p\}$ such that $\gamma^X_i = 0$ in the hVAR case or $g_i$ is constant on the support of $X_{t-i}$ in the hNAAR case.
We assume correct specification for both $\Gamma_{\X \rightarrow h(\Y)}(p)$ and $\Gamma_{\Y \rightarrow h(\X)}(p)$.
\end{assumption}

The next result establishes that the compound CTC equals one in the true causal direction and is strictly less than one otherwise. This asymmetry provides a theoretical basis for identifying the correct causal direction.

\begin{theorem}\label{thm:1}
Let $(\X, \Y)$ follow an hVAR($p$) or hNAAR($p$) model, and suppose that Assumption~\ref{weight assumption} holds.
If $\X$ causes $\Y$, then $\Gamma_{\X \rightarrow h(\Y)}(p) = 1$.
If $\X$ does not cause $\Y$, then $\Gamma_{\X \rightarrow h(\Y)}(p) < 1$ for all $p \in \N$.
\end{theorem}

\noindent This result can be extended to extremal behavior in both the upper and lower tails.

\begin{theorem}\label{thm:3}
Let $(\X,\Y)$ follow an hVAR($p$) model with possibly negative coefficients, satisfying the extremal causal condition, and suppose that Assumption~\ref{weight assumption} holds for the weight vector defining $\Gamma_{|\X|\rightarrow h(|\Y|)}(p)$. 
Suppose further that $\varepsilon_t^X$ and $\varepsilon_t^Y$ have full support on $\R$, and that $|\varepsilon_t^X|$ and $|\varepsilon_t^Y|$ are regularly varying with index $\theta$.
If $\X$ causes $\Y$, then $\Gamma_{|\X| \rightarrow h(|\Y|)}(p) = 1$.
If $\X$ does not cause $\Y$, then $\Gamma_{|\X| \rightarrow h(|\Y|)}(p) < 1$ for all $p \in \N$.
\end{theorem}

\subsection{Extension to confounders}
A key extension of the CTC is the handling of covariates or confounding variables.
A confounder $Z$ for $(X,Y)$ is a common cause that influences both variables and can thereby either induce spurious dependence or mask a true causal relationship. Graphically, $Z$ is a parent node of both $X$ and $Y$, while a directed edge between $X$ and $Y$ may or may not also be present.
Consistent with Reichenbach's common cause principle \citep{Reichenbach1956}, any correlation between $X$ and $Y$ may arise from a direct causal relationship or from a common cause $Z$. 
The key challenge is to distinguish true causality from spurious dependence induced by a confounder.

CTCs have been shown to be asymptotically robust to hidden confounders in a linear SCM, provided that confounding variables have regularly varying tails no heavier than those of the variables of interest \citep{GMPE2021}. 
A similar property holds for our compound CTC in the time series setting. The proof of the first claim in Theorem~\ref{thm:1} implies that adding a common cause $\ZZ$ of both $\X$ and $\Y$ does not change the coefficient when $\X$ causes $\Y$. In the reverse direction, Theorem~\ref{thm:4} below provides the analogue in the presence of confounding.

\begin{theorem}\label{thm:4}
Let $(\X,\Y,\ZZ)$ follow a trivariate hVAR($p$) model,\footnote{ Here, a trivariate hVAR($p$) model is understood as the direct extension of Definition~\ref{def:hVAR} in Section~\ref{sec:models}: the process is stable, its innovations are i.i.d.\ regularly varying with a common tail index, its VAR coefficients are non-negative, and all coefficient sequences in its causal representation satisfy condition~(c) of Definition~\ref{def:hVAR}.}
and suppose that Assumption~\ref{weight assumption} holds for the weight vector defining $\Gamma_{\X\rightarrow h(\Y)}(p)$.
Suppose $\ZZ$ is a common cause of $\X$ and $\Y$, and that neither $\X$ nor $\Y$ causes $\ZZ$. If $\X$ does not cause $\Y$, then $\Gamma_{\X\rightarrow h(\Y)}(p)<1$
for all $p\in\N$.
\end{theorem}

\noindent Theorem~\ref{thm:4} assumes equal tail indices for $\X$, $\Y$, and $\ZZ$, but extends easily to cases where $\ZZ$ has lighter tails. However, the assumption that confounders do not have heavier tails is often unverifiable and unrealistic. When confounders are observed, conditioning on them can offer more accurate estimates. We therefore extend the compound CTC to incorporate the effect of a confounding series $\ZZ$.

\begin{definition}
The \textit{conditional compound causal tail coefficient for time series} from $\X$ to $\Y$ given a confounder $\ZZ$ is
\begin{align}\label{CCTC_C}
\Gamma_{\X \rightarrow h(\Y)\mid \ZZ}(p)
:= \lim_{u\to 1^-} \E\!\left[ h(F_Y(Y_1), \ldots, F_Y(Y_p)) \,\middle|\, F_X(X_0)>u, \ZZ_{\past(p)} \right],
\end{align}
where $\ZZ_{\past(p)} = \{Z_1, \dots,Z_{p-1}, Z_p\}$ denotes all past values of the confounding series $\ZZ$ up to time $p$.
This formulation generalizes naturally to multiple confounders $\ZZ^1, \dots, \ZZ^k$, $k \in \N$, by adding $\ZZ^j_{\past(p)}$ for $j = 1, \dots, k$ into the conditioning set.
\end{definition}

An alternative is to replace $F_X$ and $F_Y$ in \eqref{CCTC} with the conditional CDFs $F_{X\mid Z}$ and $F_{Y\mid Z}$. This approach, however, adjusts only for contemporaneous confounding from $Z_t$ and ignores lagged effects within the extremal delay, hence we recommend the use of \eqref{CCTC_C}.

\subsection{Extension to higher dimensions}\label{sec:mv}
The compound events considered thus far focus on temporal compounding, where the impact function $h$ aggregates lagged values from a univariate effect series $\Y$ within the extremal delay. The flexible multivariate form of the proposed impact function also allows the framework to extend readily to higher dimensions, accommodating compounding across both time and multiple effect variables.

Let $\bm{Y} := (Y^1,\dots,Y^k)$, $k\in\N$, denote a $k$-dimensional effect process with $\bm{Y}_t \in \R^k$.  
We extend \eqref{CCTC} to incorporate both temporal and multivariate compounding as follows: 
\begin{align} \label{CCTC_mv}
\Gamma_{\X \rightarrow h(\bm{Y})}(p)
:= \lim_{u\to 1^-} \E \!\left[
h \!\left(F_{Y^1}(Y^1_1),\ldots,F_{Y^1}(Y^1_p),\ldots,F_{Y^k}(Y^k_1),\ldots,F_{Y^k}(Y^k_p) \right)
\,\middle|\, F_X(X_0) > u
\right],
\end{align} 
where $h$ is given by
\(
    h(v_1,\ldots,v_{kp})
    :=
    \left[
    1-\prod_{i=1}^{kp}
    \left\{1-v_i(1-e^{-\alpha})\right\}^{w_i}
    \right](1-e^{-\alpha})^{-1},
\)
with its inputs corresponding to the cross-sectional and temporal components
\(
    \{
    F_{Y^j}(Y^j_i):
    j=1,\ldots,k, i=1,\ldots,p
    \}.
\)
Each of the $kp$ components represents an effect variable and lag combination, with the weights determining its relative importance and $\alpha$ controlling how the contributions are combined.
The same impact function accommodates temporal compounding within individual series, cross-sectional compounding across effect variables, and combinations of both.
Thus, the resulting coefficient directly measures the joint response of $\bm Y$ following an extreme value of $X$, without aggregating pairwise coefficients.
Observed confounders may also be incorporated by combining this multivariate construction with the conditional coefficient in \eqref{CCTC_C}.

The corresponding nonparametric estimator and bootstrap hypothesis test of Section~\ref{sec:estimation} extend directly by evaluating $h$ using the componentwise empirical distribution transforms of all $kp$ components. Full details of the estimator, together with a Monte Carlo study covering six three-dimensional time series models, are provided in Supplementary Section~\ref{sec:mv_ap}.

\section{Estimation and hypothesis testing}\label{sec:estimation}

\subsection{Nonparametric estimator}

Given a finite sample $(x_1,y_1),\ldots,(x_n,y_n)$, $n \in \N$, we estimate the compound CTC of \eqref{CCTC} by
\begin{align}\label{estimator}
\hat{\Gamma}_{\X \rightarrow h(\Y)}(p)
= \frac{1}{k} \sum_{i=1}^{n-p}
h \left(\hat{F}_Y(y_{i+1}), \ldots, \hat{F}_Y(y_{i+p}) \right)
\cdot \1_{\{x_i \ge x_{(n-k+1)}\}},
\end{align}
where $x_{(n-k+1)}$ is the $k$-th largest $x_t$, and  
$\hat{F}_Y(y) = \frac{1}{n} \sum_{j=1}^n \1_{\{y_j \le y\}}$ is the empirical CDF of $Y$.  
A corresponding nonparametric estimator of the conditional compound CTC of \eqref{CCTC_C} is provided in Supplementary Section~\ref{sec:sensi_confounding}.
We follow the standard practice and set $k = \sqrt{n}$.
By construction, $\hat{\Gamma}_{\X \rightarrow h(\Y)}(p) \in [0,1]$.  
The next result establishes consistency under mild regularity conditions.

\begin{theorem}\label{thm:5}
    Let (X,Y) be a stationary and ergodic bivariate time series with absolutely continuous marginal distributions supported on some neighborhood of infinity. Suppose that $\Gamma_{X\to h(Y)}(p)$ is well-defined for some $p$, and that $k=k_n$ satisfies $k \to\infty$ and $\frac{k}{n} \to 0$, as $ n\to\infty$.
    Further, assume that for some $\xi>0$,
    $ \frac{n}{k^{1-\xi}} \sup_{x \in\R} \left|\hat{F}_X(x) - F_X(x) \right| \xlongrightarrow{\PP}0$ and $ \frac{n}{k} \sup_{y \in\R} \left|\hat{F}_Y(y) - F_Y(y) \right| \xlongrightarrow{\PP}0$.
    Then $\hat{\Gamma}_{X\to h(Y)}(p)  \xlongrightarrow{\PP}  {\Gamma}_{X\to h(Y)}(p)$ as $n\to\infty$.
\end{theorem}

The different rate assumptions on $\hat F_X$ and $\hat F_Y$ reflect their distinct roles in the estimator. Since $\hat F_X$ determines the extreme conditioning set through $\1_{{\hat F_X(X_i)>1-k/n}}$, estimation errors near the threshold may alter the selected observations and therefore require stronger control. By contrast, $\hat F_Y$ enters through the bounded Lipschitz impact function and perturbs summand values continuously without changing the tail sample. This asymmetry concerns tail selection versus impact evaluation, not the intrinsic estimability of the two series. The roles are reversed for $\Gamma_{\Y\rightarrow h(\X)}(p)$. Hence, when estimating both directions, imposing the stricter rate condition on both marginal empirical distribution functions is natural.

\subsubsection{Estimating the extremal delay}\label{sec:pccf}
In applied settings, the extremal delay $p$ may be selected based on domain knowledge or estimated from data.  
We consider two data-driven approaches, starting with the extremogram of \citet{DM2009}.
The cross-extremogram quantifies the extremal dependence between two time series across varying lags $\tau$. Using a peak-over-threshold (POT) formulation \citep{Pickands1975}, it is defined as
$ \rho_{XY}(\tau) = \lim_{u \to \infty} \PP(Y_{t+\tau} > u \mid X_t > u)$. 
We estimate $\rho_{XY}(\tau)$ as the proportion of times $Y_{t+\tau}$ exceeds its 95th percentile given that $X_t$ exceeds its 95th percentile.
Given a threshold $\bar{c}$, we set $p = \sup \{\tau: \hat{\rho}_{XY}(\tau)\geq \bar{c} \}$. Based on simulation results, we suggest setting $\bar{c}=0.1$. 

The second approach is based on the partial cross-correlation function (PCCF). PCCF generalizes the partial autocorrelation function to measure the linear association between $X_t$ and $Y_{t+\tau}$, after adjusting for intermediate lags.
We propose an \textit{asymmetric} PCCF that removes only the linear influence of the lagged effect series $Y$ at shorter lags, thus capturing directional dependence from $X$ to $Y$:
$\phi_{XY} (\tau) = \corr(X_t - \tilde{P}_{t+\tau-1}X_t, \, Y_{t+\tau} - P_{t+\tau-1} Y_{t+\tau})$,
where $\tilde{P}_{t+\tau-1} X_t$ and $P_{t+\tau-1} Y_{t+\tau}$ are best linear predictors of $X_t$ and $Y_{t+\tau}$ based on $Y_{t+1}, \dots, Y_{t+\tau-1}$.
Let $\hat{\phi}_{XY}(\tau)$ denote the empirical PCCF and $\bar{c}$ a given threshold. We set $p$ to be the largest lag $\tau$ such that $\hat{\phi}_{XY}(\tau) \geq \bar{c}$.
Simulation results (see the Supplementary material) suggest using $\bar{c} \in \{0.10, 0.15\}$ in practice.

Each method has strengths and limitations.
The cross-extremogram relies on the upper $5\%$ of the data to estimate $\hat{\rho}_{XY}$, which can result in high variance and unstable parameter selection.
The PCCF uses the full dataset and yields more stable estimates but is less sensitive when the causal relationship is present only in the tails rather than across the entire distribution.
An empirical comparison of the two approaches is provided in the Supplementary Section~\ref{sec:sensi_lag}.

\subsubsection{Weight optimization}\label{sec:weight_opt}
The weight parameters $w_i$ control the relative contribution of each lagged variable $Y_{t+i}$ to a compound extreme event. These weights may be specified \textit{a priori} using domain knowledge, for example, when the composition of a compound event is known from physical models. 
Alternatively, they can be treated as optimization parameters to maximize the compound CTC. The motivation is to identify the combination of lagged values in the effect series most likely to produce a compound extreme event in response to an extreme value in the cause series.
Practical details of the optimization procedure are given in the Supplementary Section~\ref{sec:weight_distr}.

The optimized weights should be distinguished from the correct specification condition in Assumption~\ref{weight assumption}, which is a sufficient condition for population-level directional separation. In finite samples, the optimized weights are data adaptive and need not recover the exact zero pattern of the causal lags. They should instead be interpreted as tuning parameters that identify the lagged components of the effect series that are most strongly associated with an extreme in the cause series.

\subsection{Block bootstrapping hypothesis test}\label{sec:mbb}
One major goal of developing any causality measure is to perform causal discovery, that is, inferring about the true underlying causal model from data. 
In our case, this requires a statistical procedure to determine whether the estimated compound CTC $\hat{\Gamma}_{\X \to h(\Y)}(p)$ is statistically significant. 
\citet{BPP2024} proposed a hard threshold of $0.9$ to determine causality. This threshold was based solely on simulation results and lacked theoretical justification. Given the variety of time series models and the complexity of real-world data, a hard threshold seems problematic. Instead, we propose a hypothesis testing framework to formally evaluate the existence, strength, and direction of any causal relation. 

Our causal discovery framework tests the null hypothesis that extremes in the time series $\X$ do not Granger cause extremes in $\Y$, against the alternative that $\X$ does cause $\Y$, using the estimated CTC:
\vspace{-0.5em}
\begin{align*}
    H_0\!: \X \not\to \Y, \quad H_A\!: \X \to Y .
\end{align*}
To generate samples under $H_0$, we must remove any true causal link from $\X$ to $\Y$ while preserving the serial dependence within each series, so that the compound event defined in the CTC remains meaningful.

A widely adopted method for time series data is the moving block bootstrap (MBB) \citep{Kunsch1989,LS1992}. MBB keeps neighboring observations together and resamples blocks of observations at a time, thereby preserving serial dependence at lags shorter than the block length.
In the multivariate setting, resampling both series jointly preserves cross-series dependence, including any causal link, which is undesirable under $H_0$. 
Resampling the series independently removes causality but also eliminates feedback loops, which may arise from the autoregressive structure of the series or from confounding variables. This can cause bootstrapped coefficients to be systematically smaller than the observed coefficient, potentially leading to spurious statistical significance in the non-causal direction and inflating false positives.

\begin{figure}[htbp]
    \centering
    \includegraphics[width=0.4\textwidth]{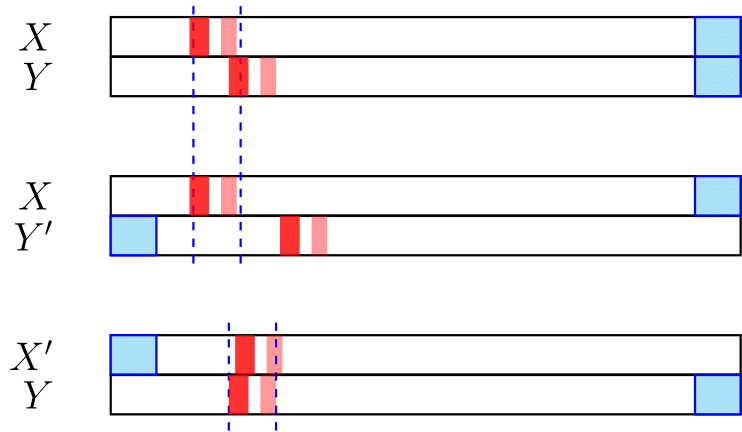}
    \caption{\footnotesize Illustration of time-shifting before moving block bootstrap. Black rectangles represent some observed bivariate series $(\X,\Y)$; red segments denote extreme events and pink segments their aftershocks. Two vertical dashed blue lines mark the extremal delay $p$. Blue rectangles indicate observations wrapped around to start the shifted series $\X',\Y'$, with widths equal to $p$.}
    \label{fig:tikz}
\end{figure}

To address this, we introduce a \emph{time-shifting} step before applying the standard multivariate MBB. 
Given an extremal delay $p$, we shift the hypothesized effect series in each direction forward in time by $p$ observations. The resulting shifted series are denoted $\X'$ and $\Y'$:
\begin{align*}
    X'_t = X_{\,(t-p) \!\! \mod n}, \; Y'_t = Y_{\,(t-p) \!\! \mod n}, \ \forall t\in \Z.
\end{align*}
An illustration of our time-shifting procedure is shown in Figure \ref{fig:tikz}, where the true causal direction is $\X\to\Y$.
The time-shifting step disrupts the temporal alignment between $\X$ and $\Y'$, so extreme events in the two series no longer fall within the extremal delay window. As a result, bootstrap coefficients computed from $(\X, \Y')$ cannot capture causality, and the MBB test will tend to reject the null in favor of $\X \to \Y$.
In the reverse direction, bootstrap coefficients are computed based on $(\X', \Y)$, where any feedback structure is preserved within the extremal delay window. This produces slightly stronger evidence for $\Y \to \X'$ than for $\Y \to \X$, leading the MBB test to correctly conclude that $\Y$ does not cause $\X$.

To achieve the desired time-shifting effect as discussed above, $p$ must be at least as large as the true extremal delay, and the block length $\ell$ must satisfy $\ell > p$ to ensure that feedback structure is preserved in the bootstrap samples. The shifting step can be applied either before or after block resampling, with similar results in simulations.
With the shifting in place, we use the compound CTC estimator of \eqref{estimator} as the test statistic and perform a one-sided bootstrap hypothesis test given the observed data. The $p$-value is computed as the proportion of bootstrapped coefficients greater than or equal to the observed coefficient, and the null is rejected at the $5\%$ level if the $p$-value is below $0.05$.
The same time-shifting procedure can, in principle, be combined with the conditional compound CTC by shifting the hypothesized effect series together with the corresponding confounder history, although explicit conditioning may not be necessary in practice, given the robustness of the unconditional compound CTC to confounding.

Our proposed MBB test with time-shifting is asymptotically valid and consistent under suitable regularity conditions, including stationarity and $\alpha$-mixing of the time series, as well as the standard assumption on the block length, $\ell = O(n^{\varepsilon})$ for some $0 < \varepsilon < \tfrac{1}{2}$ \citep{NR1994,Lahiri1999}. A formal statement and proof are given in the Supplementary  Section~\ref{sec:mbb_pf}.

\section{Simulations}\label{sec:simulation}
\subsection{Setup}
We study the finite sample behavior of $\hat{\Gamma}_{{\X} \rightarrow {h(\Y})}$ and evaluate the performance of our bootstrap hypothesis test through Monte Carlo simulations across nine model specifications for the bivariate series $(\X,\Y)$. These specifications cover the two model classes introduced in Section~\ref{sec:models}. For comparability, all models are stationary and share a true extremal delay of $3$.
We begin with basic VAR(3) models. Model~{1} is a baseline noise model with no causal dependence between $X$ and $Y$. Model~\ref{model:2} specifies $X$ as an AR(1) process that causally impacts $Y$ with a lag of 3, while Model~\ref{model:3} complicates this structure by adding autoregressive dynamics to both series. Model~\ref{model:4} introduces a bidirectional causal loop between $X$ and $Y$. Models~\ref{model:5} and \ref{model:6} extend the single-lag structures of Models~\ref{model:2} and \ref{model:3} to include cross-dependence across multiple lags, capturing compounding effects in extremes. Model~\ref{model:7} adds nonlinearity to Model~\ref{model:3}, where $X$ influences $Y$ only in the upper tail, representing causality specific to extreme events. Finally, to assess robustness under hidden confounding, Models~\ref{model:8} and \ref{model:9} introduce a common cause $Z$ to Models~\ref{model:2} and \ref{model:7}, respectively. We assume $\ZZ$ is an AR(1) process that has the same type of noise distribution as $\X$ and $\Y$, and the tails of $\epsilon^Z_t$ are no heavier than those of $\epsilon^X_t$ and $\epsilon^Y_t$.
Details of the simulation procedures and results are given in Section~\ref{sec:setup} of the Supplementary material.

\subsection{Comparison with state-of-the-art methods}\label{sec:compare}

This section compares our method with three established pairwise causal discovery techniques for time series. First, we applied the baseline Granger causality test with default parameters at significance level $0.05$. Second, we followed \citet{BPP2024}, using the CTC in \eqref{npestimator} with a hard threshold of $0.9$ to infer $X \to Y$. Third, we implemented the permutation test of \citet{PCDD2023}, combined with a time series extension of their parametric GPD causal tail coefficient estimator.
We also considered a hybrid approach that combines the original CTC of \citet{BPP2024} with our block bootstrap procedure, replacing our compound CTC estimator with \eqref{npestimator} while retaining the time-shifting and resampling steps of Section~\ref{sec:mbb}. For our proposed method, we fixed the impact shape parameter at $\alpha=10^4$ for comparability, noting that results are robust to variations in $\alpha$.
We evaluated all five methods through repeated simulations on the nine time series models under three noise distributions: Student’s $t$, Pareto, and Poisson noise (details in the Supplementary material). 
These distributions represent qualitatively different tail regimes and are therefore intended to provide a broad robustness assessment; tail-index sensitivity within the Student’s $t$ family is examined separately in Section~\ref{sec:sensi_tail} of the Supplementary material.
Each simulation used time series of length $n=1000$ with $k=\sqrt{n}$ extremes, block length $\ell=n^{1/3}$, and $b=100$ bootstrap samples, repeated 100 times. The true extremal delay of 3 was used consistently across all methods. Tables~\ref{tab:t}, \ref{tab:pareto}, and \ref{tab:poisson} report the percentage of correct inferences across models with Student’s $t$, Pareto, and Poisson noise, respectively.

\begin{table}[htbp]
\caption{\footnotesize A comparison of five causal inference methods on nine time series models with Student's $t$ noise with $df=10$ for $X$ and $df=2$ for $Y$. The percentage obtained after 100 repetitions represents the proportion of correct inferences, with bold numbers highlighting the method with the lowest overall error rate for each model.}
\label{tab:t}
\centering
\resizebox{\textwidth}{!}{%
\begin{tabular}{c|*{2}{>{\centering\arraybackslash}p{1.2cm}}|*{2}{>{\centering\arraybackslash}p{1.2cm}}|*{2}{>{\centering\arraybackslash}p{1.2cm}}|*{2}{>{\centering\arraybackslash}p{1.2cm}}|*{2}{>{\centering\arraybackslash}p{1.2cm}}}
\toprule
 & \multicolumn{2}{c|}{\shortstack{Compound CTC \\ w. bootstrap}} 
 & \multicolumn{2}{c|}{\shortstack{CTC (Bodik et al.) \\ w. bootstrap}} 
 & \multicolumn{2}{c|}{\shortstack{CTC (Bodik et al.) \\ w. hard threshold}} 
 & \multicolumn{2}{c|}{\shortstack{Granger \\ causality test}} 
 & \multicolumn{2}{c}{\shortstack{Permutation test \\ (Pasche et al.)}} \\
\cmidrule(lr){2-3} \cmidrule(lr){4-5} \cmidrule(lr){6-7} \cmidrule(lr){8-9} \cmidrule(lr){10-11}
Model & $X$→$Y$ & $Y$→$X$ & $X$→$Y$ & $Y$→$X$ & $X$→$Y$ & $Y$→$X$ & $X$→$Y$ & $Y$→$X$ & $X$→$Y$ & $Y$→$X$ \\
\midrule
1 & 96\% & 96\% & 92\% & 95\% & \textbf{100\%} & \textbf{98\%} & 96\% & 93\% & 96\% & 92\% \\
2 & \textbf{98\%} & \textbf{96\%} & 94\% & 97\% & 9\% & 100\% & \textbf{100\%} & \textbf{94\%} & 78\% & 100\% \\
3 & \textbf{99\%} & \textbf{97\%} & 88\% & 97\% & 0\% & 100\% & \textbf{100\%} & \textbf{96\%} & 47\% & 100\% \\
4 & 87\% & 100\% & 80\% & 100\% & 45\% & 100\% & \textbf{100\%} & \textbf{100\%} & 0\% & 22\% \\
5 & \textbf{100\%} & \textbf{99\%} & {99\%} & {99\%} & 47\% & 100\% & 98\% & 97\% & 88\% & 100\% \\
6 & \textbf{100\%} & \textbf{100\%} & {100\%} & {99\%} & 2\% & 100\% & 100\% & 97\% & 80\% & 100\% \\
7 & \textbf{100\%} & \textbf{97\%} & {97\%} & {99\%} & 2\% & 100\% & 61\% & 94\% & 76\% & 100\% \\
8 & \textbf{97\%} & \textbf{99\%} & 90\% & 96\% & 9\% & 100\% & 97\% & 88\% & 87\% & 100\% \\
9 & \textbf{99\%} & \textbf{99\%} & \textbf{100\%} & \textbf{98\%} & 51\% & 100\% & 76\% & 87\% & 94\% & 100\% \\
\bottomrule
\end{tabular}
}
\end{table}

\begin{table}[htbp]
\caption{\footnotesize A comparison of five causal inference methods on nine time series models with Pareto(1,1) noise.}
\label{tab:pareto}
\centering
\resizebox{\textwidth}{!}{%
\begin{tabular}{c|*{2}{>{\centering\arraybackslash}p{1.2cm}}|*{2}{>{\centering\arraybackslash}p{1.2cm}}|*{2}{>{\centering\arraybackslash}p{1.2cm}}|*{2}{>{\centering\arraybackslash}p{1.2cm}}|*{2}{>{\centering\arraybackslash}p{1.2cm}}}
\toprule
 & \multicolumn{2}{c|}{\shortstack{Compound CTC \\ w. bootstrap}} 
 & \multicolumn{2}{c|}{\shortstack{CTC (Bodik et al.) \\ w. bootstrap}} 
 & \multicolumn{2}{c|}{\shortstack{CTC (Bodik et al.) \\ w. hard threshold}} 
 & \multicolumn{2}{c|}{\shortstack{Granger \\ causality test}} 
 & \multicolumn{2}{c}{\shortstack{Permutation test \\ (Pasche et al.)}} \\
\cmidrule(lr){2-3} \cmidrule(lr){4-5} \cmidrule(lr){6-7} \cmidrule(lr){8-9} \cmidrule(lr){10-11}
Model & $X$→$Y$ & $Y$→$X$ & $X$→$Y$ & $Y$→$X$ & $X$→$Y$ & $Y$→$X$ & $X$→$Y$ & $Y$→$X$ & $X$→$Y$ & $Y$→$X$ \\
\midrule
1 & 95\% & 93\% & 93\% & 91\% & \textbf{100\%} & \textbf{100\%} & 95\% & 93\% & 95\% & 94\% \\
2 & \textbf{100\%} & \textbf{100\%} & \textbf{100\%} & \textbf{100\%} & \textbf{100\%} & \textbf{100\%} & 89\% & 97\% & 93\% & 100\% \\
3 & \textbf{100\%} & \textbf{100\%} & {100\%} & {98\%} & \textbf{100\%} & \textbf{100\%} & 93\% & 99\% & 93\% & 100\% \\
4 & \textbf{100\%} & \textbf{100\%} & \textbf{100\%} & \textbf{100\%} & \textbf{100\%} & \textbf{100\%} & \textbf{100\%} & \textbf{100\%} & 0\% & 0\% \\
5 & \textbf{100\%} & \textbf{100\%} & \textbf{100\%} & \textbf{100\%} & 100\% & 92\% & 91\% & 95\% & 78\% & 100\% \\
6 & \textbf{100\%} & \textbf{100\%} & \textbf{100\%} & \textbf{100\%} & 100\% & 91\% & 92\% & 96\% & 74\% & 100\% \\
7 & \textbf{100\%} & \textbf{100\%} & \textbf{100\%} & \textbf{100\%} & 98\% & 100\% & 85\% & 92\% & 99\% & 100\% \\
8 & 100\% & 90\% & \textbf{100\%} & \textbf{96\%} & 100\% & 90\% & 92\% & 59\% & 84\% & 100\% \\
9 & 100\% & 79\% & \textbf{100\%} & \textbf{94\%} & \textbf{100\%} & \textbf{94\%} & 86\% & 62\% & 84\% & 100\% \\
\bottomrule
\end{tabular}
}
\end{table}

\begin{table}[htbp]
\caption{\footnotesize A comparison of five causal inference methods on nine time series models with Poisson(3) noise.}
\label{tab:poisson}
\centering
\resizebox{\textwidth}{!}{%
\begin{tabular}{c|*{2}{>{\centering\arraybackslash}p{1.2cm}}|*{2}{>{\centering\arraybackslash}p{1.2cm}}|*{2}{>{\centering\arraybackslash}p{1.2cm}}|*{2}{>{\centering\arraybackslash}p{1.2cm}}|*{2}{>{\centering\arraybackslash}p{1.2cm}}}
\toprule
 & \multicolumn{2}{c|}{\shortstack{Compound CTC \\ w. bootstrap}} 
 & \multicolumn{2}{c|}{\shortstack{CTC (Bodik et al.) \\ w. bootstrap}} 
 & \multicolumn{2}{c|}{\shortstack{CTC (Bodik et al.) \\ w. hard threshold}} 
 & \multicolumn{2}{c|}{\shortstack{Granger \\ causality test}} 
 & \multicolumn{2}{c}{\shortstack{Permutation test \\ (Pasche et al.)}} \\
\cmidrule(lr){2-3} \cmidrule(lr){4-5} \cmidrule(lr){6-7} \cmidrule(lr){8-9} \cmidrule(lr){10-11}
Model & $X$→$Y$ & $Y$→$X$ & $X$→$Y$ & $Y$→$X$ & $X$→$Y$ & $Y$→$X$ & $X$→$Y$ & $Y$→$X$ & $X$→$Y$ & $Y$→$X$ \\
\midrule
1 & 91\% & 93\% & 92\% & 91\% & \textbf{100\%} & \textbf{100\%} & 95\% & 93\% & 98\% & 95\% \\
2 & \textbf{100\%} & \textbf{100\%} & {100\%} & {99\%} & 44\% & 100\% & 100\% & 96\% & 97\% & 100\% \\
3 & \textbf{100\%} & \textbf{100\%} & {99\%} & {99\%} & 4\% & 100\% & 100\% & 95\% & 77\% & 100\% \\
4 & \textbf{100\%} & \textbf{100\%} & 96\% & 100\% & 25\% & 21\% & \textbf{100\%} & \textbf{100\%} & 3\% & 0\% \\
5 & \textbf{100\%} & \textbf{100\%} & \textbf{100\%} & \textbf{100\%} & 91\% & 100\% & 100\% & 93\% & 91\% & 100\% \\
6 & \textbf{100\%} & \textbf{100\%} & \textbf{100\%} & \textbf{100\%} & 71\% & 100\% & 100\% & 91\% & 83\% & 100\% \\
7 & \textbf{100\%} & \textbf{100\%} & \textbf{100\%} & \textbf{100\%} & 8\% & 100\% & 100\% & 96\% & 70\% & 100\% \\
8 & \textbf{99\%} & \textbf{99\%} & 97\% & 99\% & 45\% & 100\% & 100\% & 52\% & 93\% & 100\% \\
9 & \textbf{100\%} & \textbf{100\%} & {97\%} & {100\%} & 22\% & 100\% & 100\% & 66\% & 68\% & 100\% \\
\bottomrule
\end{tabular}
}
\vspace{-1em}
\end{table}

Overall, our method outperforms existing approaches by achieving the lowest total error rate (Type I + Type II) across nearly all models and noise types. Performance is especially strong in the multi-lag models~\ref{model:5}, \ref{model:6} and the thresholded models~\ref{model:7}, \ref{model:9}, reflecting the design of the compound CTC to address multi-lag compounding and heavy tails.

The second-best method is the original CTC of \citet{BPP2024} combined with our block bootstrap test. Compared with our method, this hybrid approach is most effective under Pareto noise, as the maximum impact function in \eqref{CTC1} easily captures the extreme values generated by Pareto noise. By contrast, our compound CTC in \eqref{CCTC} is tailored to multi-lag dependencies and outperforms the hybrid approach in Models~\ref{model:5} and \ref{model:6} with Student’s $t$ noise, though the advantage narrows in other cases where the hybrid approach also performs well. The strong performance of the hybrid approach further confirms that our block bootstrap with time-shifting provides a valid test for pairwise causality in time series, with potential applications beyond extremes. Both our method and the hybrid approach outperform other approaches in Models~\ref{model:4}, \ref{model:6}, and \ref{model:7} with autoregressive effects, and in Models~\ref{model:8} and \ref{model:9} with hidden confounders. These models share feedback structures, suggesting that our time-shifting step effectively distinguishes true causality from spurious dependencies.

The testing framework of \cite{BPP2024} shows acceptable performance only for models with Pareto noise, but fails to identify the correct causal direction in most models with Student’s $t$ or Poisson noise. Many estimated CTCs fall below the hard threshold of $0.9$, even for the true causal direction $X \to Y$, which also explains the low false positive rate for $Y \to X$. Thus, hard thresholding is inadequate for causal inference, and a formal hypothesis test is preferable.

The classical Granger test performs well for simple VAR models with non-extreme noise. By regressing one series on another, it captures autoregressive structure in Models~\ref{model:3}, \ref{model:4}, and \ref{model:6}, and its performance under Poisson noise is acceptable. However, it deteriorates with heavy-tailed noise and is particularly poor in Models~\ref{model:7} and \ref{model:9}, where causality arises only in the tail. This is expected, as the test is not designed for extremes. It also struggles with more complex settings such as Models~\ref{model:8} and \ref{model:9}, where hidden confounders induce spurious effects misidentified as reverse causality $Y \to X$. 
In contrast, the permutation test underperforms across most settings. Its poor results on Model~\ref{model:4} are unsurprising: by inferring causality from directional asymmetries between $X \to Y$ and $Y \to X$, it fails in the presence of feedback loops. In other models, it often accepts the null of no causality, leading to high Type II but low Type I error rates.

In summary, the results presented in Tables \ref{tab:t}, \ref{tab:pareto}, and \ref{tab:poisson} suggest that our new CTC for compound extremes is particularly well-suited for detecting extremal causality that manifests across multiple time lags. The proposed bootstrap test with time-shifting further distinguishes true causality from spurious associations induced by feedback loops or hidden confounders.

\subsection{Sensitivity analysis}
Both the estimated compound CTCs and the associated bootstrap test results are robust across a wide range of parameter settings. We assessed sensitivity to the impact shape parameter $\alpha$ and the weight vector $\vw$ in the impact function, as well as the extreme threshold $k$ and the extremal delay parameter $p$. We also examined the effect of varying the shift parameter in the block bootstrap when it differs from $p$. Furthermore, we evaluated robustness to potential confounders by comparing compound CTCs with their conditional counterparts in \eqref{CCTC_C}. Detailed sensitivity results and discussion are provided in Section~\ref{sec:sensitivity} of the Supplementary material.

\section{Space-weather application}\label{sec:spacew}
We illustrate our method with a dataset on geomagnetic storms, a key phenomenon in space weather. Space weather examines the impact of solar activity on the near-Earth environment, where extreme events can pose threats to critical infrastructure, such as satellite systems, power grids, communication networks, and navigation services. Identifying the drivers of these events is therefore essential for risk mitigation and infrastructure protection. 
Our analysis focuses on geomagnetic storms and substorms: storms are large-scale events with sustained energy input from the solar wind, whereas substorms are localized, transient disturbances that mainly affect auroral zones. Both phenomena are primarily triggered by changes in the solar wind and the interplanetary magnetic field it carries.
Understanding the interaction between magnetic storms and substorms remains a central challenge in space-weather research. Whether substorms directly cause storms has long been debated \citep{KBDG1998,SBGK2003}. Early studies viewed storms as the accumulation of successive substorms, but \citet{MBCPP2021} challenged this view, suggesting instead that a vertical component of the interplanetary magnetic field acts as a common trigger for both. In this section, we apply our methodology to evaluate this hypothesis by identifying causal relationships among key variables in extreme space-weather scenarios.

The dataset we use is publicly available from NASA (\url{https://cdaweb.gsfc.nasa.gov}). Our dataset consists of three time series: the geomagnetic storm index (SYM), the substorm index (AE), and the vertical component of the interplanetary magnetic field (BZ). Each series contains about 500,000 observations at 5-minute resolution from January 1, 1999, to December 31, 2003, covering the maximum phase of Solar Cycle 23. 
We focus on Cycle 23 rather than the more recent Cycle 24, as it exhibits stronger solar activity and is thus more suitable for analyzing extreme space-weather events. Results over a longer time horizon are provided in Section~\ref{sec:25yr} of the Supplementary material.

\subsection{Results}\label{sec:results}
\begin{figure}[htbp]
  \centering

  \begin{minipage}[b]{0.41\linewidth}
    \centering
    \includegraphics[width=\linewidth]{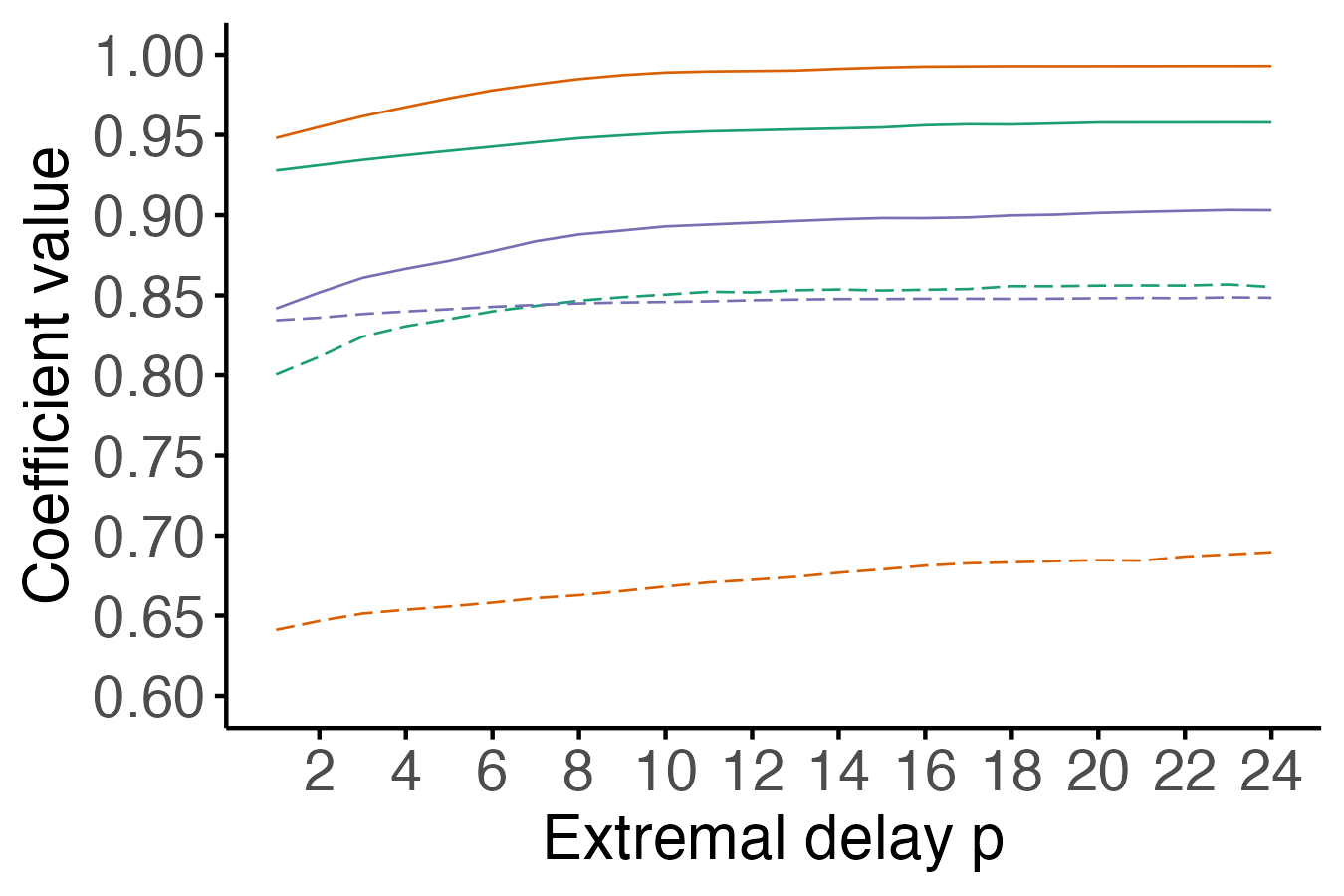}
    \par\vspace{0.3em}
    {\footnotesize (a) Compound causal tail coefficients}
  \end{minipage}\hfill
  \begin{minipage}[b]{0.55\linewidth}
    \centering
    \includegraphics[width=\linewidth]{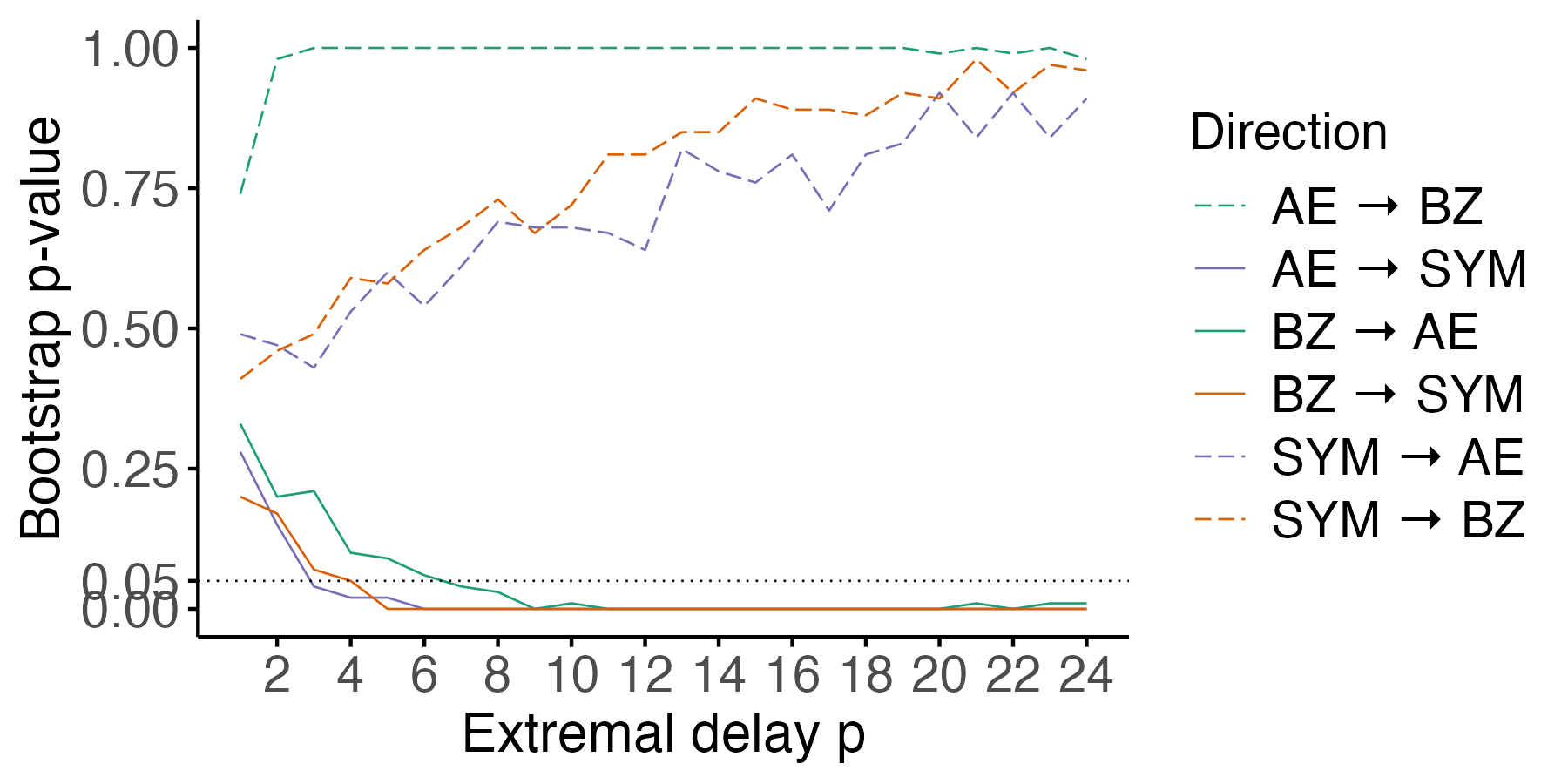}
    \par\vspace{0.3em}
    {\footnotesize (b) Bootstrap $p$-values}
  \end{minipage}

  \caption{\footnotesize Results for all directed variable pairs across extremal delays $p\in[1,24]$. }
  \label{fig:spacew_results}
\end{figure}

We now present empirical results for the estimated causal tail coefficients and bootstrap hypothesis tests between each pair of series in the space-weather dataset. 
Prior studies suggest that geomagnetic responses occur with delays of 15–60 minutes, up to about 2 hours \citep{BM2021,VRBM2020}, which corresponds to $p = 24$ at 5-minute resolution. The threshold for extremes is set to $k = \lfloor\sqrt{n}\rfloor= 710$, and the impact shape parameter is $\alpha = 10^4$.
Figure~\ref{fig:spacew_results} displays compound CTC estimates and bootstrap $p$-values as functions of the extremal delay $p$, with $p$-values below the dotted line at 0.05 indicating statistical evidence of a directional causal effect.

There are several key findings. Estimated CTCs for BZ$\to$SYM (orange) and BZ$\to$AE (green) are substantially larger than those for other pairs, especially compared with their reverse directions. This asymmetry provides preliminary evidence that BZ has a causal influence on both SYM and AE. This is reinforced by the bootstrap tests: 
$p$-values for BZ$\to$SYM and BZ$\to$AE drop below 0.05 after a few lags and remain near zero, whereas those for SYM$\to$BZ and AE$\to$BZ stay large and converge to one.
These results support the hypothesis that BZ is a common driver of AE and SYM.
There is also evidence for AE$\to$SYM (purple): bootstrap $p$-values remain low, and estimated coefficients are slightly larger than for the reverse direction. However, the asymmetry is modest, making the evidence weaker than for BZ$\to$SYM and BZ$\to$AE.

\begin{figure}[htb]
    \centering
    \includegraphics[width=0.6\linewidth]{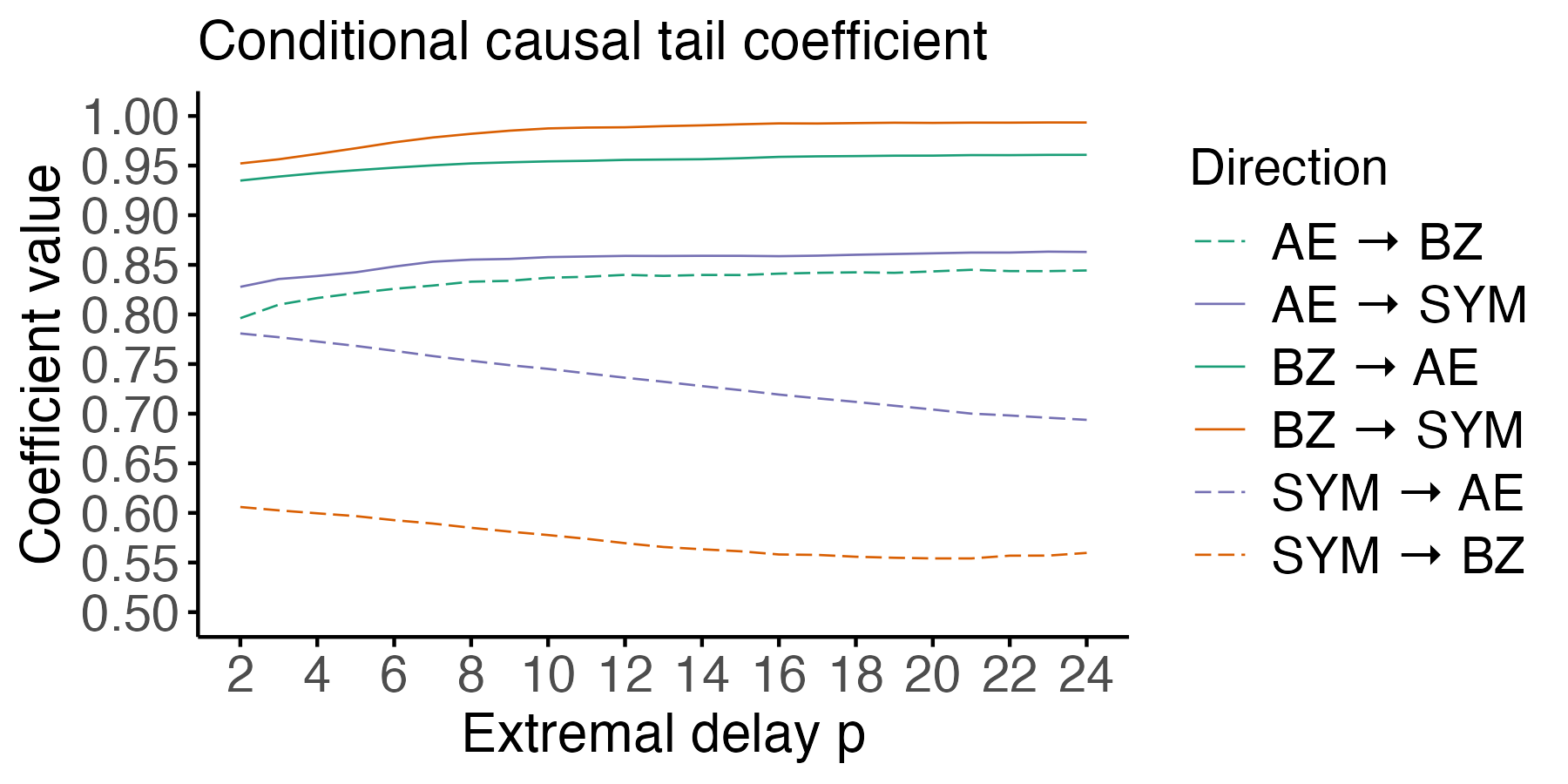}
    \caption{\footnotesize Conditional compound CTC values for all directed variable pairs across extremal delays $p \in [2, 24]$. 
    }
    \label{fig:spacew_coef_wC}
\end{figure}

To assess robustness, and in particular whether the observed causal effect AE$\to$SYM is explained by BZ, we estimated conditional compound CTCs by treating the third variable as a potential confounder for each directed pair. Figure~\ref{fig:spacew_coef_wC} shows the results across extremal delays. Compared with Figure~\ref{fig:spacew_results}(a), CTCs for BZ$\to$SYM and BZ$\to$AE remain largely unchanged after conditioning on AE and SYM, respectively. In contrast, coefficients for AE$\to$SYM decrease by approximately 0.05, and those for SYM$\to$AE decrease even more. These results indicate that the causal influence of AE on SYM is confounded by BZ, and that conditioning on BZ mitigates this spurious effect. The bootstrap $p$-values remain largely unchanged from the unconditional case in Figure~\ref{fig:spacew_results}(b), with all three directions (BZ$\to$SYM, BZ$\to$AE, and AE$\to$SYM) rejected for $p > 6$. This highlights the robustness of our bootstrap test to hidden confounding.

Finally, Table~\ref{tab:spacew} reports compound CTC estimates and bootstrap $p$-values with extremal delay $p$ selected using three PCCF threshold levels, as described in Section~\ref{sec:pccf}. For the pair (AE, SYM), applying the threshold $\bar{c}=0.15$ yields $p=1$ and a bootstrap $p$-value of 0.36 for AE$\to$SYM, suggesting limited statistical evidence for extremal causality in this direction. By contrast, the hypothesis that BZ acts as a common cause of SYM and AE appears more robust to the choice of PCCF threshold. 
Overall, these results support established hypotheses that the interplanetary magnetic field is a common driver of both magnetic storms and substorms, with substorms serving as a possible secondary driver of storms.

\begin{table}[htbp]
    \centering
    \caption{\footnotesize Estimated compound CTCs, with bootstrap $p$-values reported in parentheses, across varying PCCF thresholds used to select extremal delay $p$.
    }
    \label{tab:spacew}
    \begin{tabular}{l|c c c}
    \toprule
    \multicolumn{1}{c|}{} & $\bar{c}=0.15$ & $\bar{c}=0.10$ & $\bar{c}=0.05$ \\
    \midrule
    AE $\rightarrow$ BZ  & 0.824 (1.00) & 0.835 (1.00) & 0.851 (1.00) \\
    AE $\rightarrow$ SYM & 0.842 (0.36) & 0.852 (0.10) & 0.884 (0.01) \\
    BZ $\rightarrow$ AE  & 0.934 (0.09) & 0.940 (0.06) & 0.951 (0.02) \\
    BZ $\rightarrow$ SYM & 0.973 (0.01) & 0.985 (0.00) & 0.990 (0.00) \\
    SYM $\rightarrow$ BZ & 0.656 (0.62) & 0.662 (0.76) & 0.670 (0.74) \\
    SYM $\rightarrow$ AE & 0.834 (0.44) & 0.836 (0.52) & 0.844 (0.51) \\
    \bottomrule
    \end{tabular}
    \vspace{-1em}
\end{table}

\section{Discussion and conclusion}\label{sec:conclusion}
Understanding causal mechanisms behind extreme events is increasingly important, particularly as real-world extremes often arise from the joint occurrence of multiple components. This paper introduced the compound causal tail coefficient, a new measure designed to capture such compound extremes in multivariate time series. We combined this measure with a shifting technique in the moving block bootstrap to construct a hypothesis testing framework for causal discovery. We demonstrated its effectiveness through Monte Carlo simulations and a space-weather application.

Several extensions could broaden the practical scope of our work. The proposed multivariate compound CTC assesses whether an extreme value of a candidate cause is followed by a compound response jointly defined across several effect variables and time lags. While \eqref{CCTC_mv} extends the compound structure on the effect side, the initiating event remains a single extreme value of $\X$. A natural extension would allow the initiating event itself to be compound by applying a separate impact function to multiple variables or lags of the cause process and conditioning the coefficient on the resulting compound extreme event. The weights of this additional impact function could be specified using expert knowledge, allowing the extension to be incorporated without altering the overall mathematical framework.

A further direction for multivariate analysis is to embed the pairwise compound CTC and its associated test within graphical models for extremes \citep[e.g.,][]{EH2020,EI2021}. Recovering a causal graph or topological order would require additional assumptions concerning acyclicity, adjustment for common causes, and extremal faithfulness, together with methods that remain computationally feasible in high-dimensional systems. Such a framework would facilitate applications in which risks propagate through many interconnected components, including power grids, financial markets, and climate networks.

The current framework also has two further limitations that motivate methodological development. First, it targets extremal causality through compound effects at the mean level and does not capture second-order effects that manifest in the variance.
Extending the method to conditional variance models, such as GARCH \citep{HH2008}, would make it applicable in settings where volatility clustering is central, for example, in energy prices or financial contagion.
Second, our analysis assumes stationarity, which is often violated in practice. While preprocessing steps like deseasonalization or detrending can help, they tend to work better for the bulk than for the tails of the distribution. Developing tools that remain valid under non-stationarity would be especially valuable in applications such as climate extremes, economic shocks, and pandemic dynamics, where structural changes are the rule rather than the exception.


\clearpage
\thispagestyle{empty}

\setcounter{section}{0}
\renewcommand\thesection{S\arabic{section}}

\pagenumbering{arabic}
\setcounter{page}{0}

\pagestyle{plain}
\setlength{\footskip}{25pt}

\begin{center}
\Large\bfseries Supplementary material
\end{center}

\vspace{1em}

\noindent This document provides the supplementary material for the paper \textit{Causal tail coefficient for compound extremes in multivariate time series}. The structure of this supplement is as follows. 
Section~\ref{sec:proof} contains the assumptions, statements, and proofs underpinning the theoretical developments in the main paper.
Section~\ref{sec:setup} details the Monte Carlo simulation procedure and results. Section~\ref{sec:sensitivity} presents a sensitivity analysis demonstrating the robustness of our method across a wide range of parameter settings. Section~\ref{sec:mv_ap} reports Monte Carlo results for the multivariate extension of our approach. 
Finally, Section~\ref{sec:25yr} provides supplementary findings for the space-weather application. 

\noindent All code is publicly available at 
\href{https://github.com/icxmas/compound_causal_tail_coefficient}{https://github.com/icxmas/compound\_causal\_tail\_coefficient}.

\vspace{2em}

{\noindent \large\bfseries Contents}

\vspace{1em}

\begingroup
\setlength{\parindent}{0pt}

\newcommand{\suppcontentssection}[2]{%
  \hyperref[#1]{\textbf{\ref*{#1}\quad #2}}%
  \dotfill\pageref{#1}\par
  \vspace{0.2em}
}

\newcommand{\suppcontentssubsection}[2]{%
  \hspace*{1.5em}%
  \hyperref[#1]{\ref*{#1}\quad #2}%
  \dotfill\pageref{#1}\par
}

\suppcontentssection
  {sec:proof}
  {Theoretical results}

\suppcontentssubsection
  {sec:models}
  {Time series models}

\suppcontentssubsection
  {sec:impact_pf}
  {Properties of the impact function}

\suppcontentssubsection
  {sec:cctc_pf}
  {Properties of the compound causal tail coefficient}

\suppcontentssubsection
  {sec:estimator_pf}
  {Asymptotic results of the nonparametric estimator}

\suppcontentssubsection
  {sec:mbb_pf}
  {Asymptotic validity of the shifted MBB test}

\vspace{0.4em}

\suppcontentssection
  {sec:setup}
  {Monte Carlo simulations}

\suppcontentssubsection
  {sec:mc_setup}
  {Setup}

\suppcontentssubsection
  {sec:mc_results}
  {Results}

\vspace{0.4em}

\suppcontentssection
  {sec:sensitivity}
  {Sensitivity analysis}

\suppcontentssubsection
  {sec:shape}
  {Impact shape parameter}

\suppcontentssubsection
  {sec:threshold}
  {Extreme threshold}

\suppcontentssubsection
  {sec:sensi_tail}
  {Tail index}

\suppcontentssubsection
  {sec:sensi_lag}
  {Extremal delay parameter}

\suppcontentssubsection
  {sec:weight_distr}
  {Weight distribution}

\suppcontentssubsection
  {sec:weight_spec}
  {Weight specification}

\suppcontentssubsection
  {sec:sensi_shift}
  {Shift parameter in block bootstrap}

\suppcontentssubsection
  {sec:sensi_confounding}
  {Adjustment for confounding variables}

\vspace{0.4em}

\suppcontentssection
  {sec:mv_ap}
  {Multivariate extension}

\suppcontentssubsection
  {sec:mvsetup}
  {Setup}

\suppcontentssubsection
  {sec:mv_results}
  {Results}

\vspace{0.4em}

\suppcontentssection
  {sec:25yr}
  {Space-weather application}

\suppcontentssubsection
  {sec:extended_horizon}
  {Extended horizon analysis (1995--2019)}

\endgroup

\vspace{1.5em}

\clearpage
\section{Theoretical results}\label{sec:proof}

This section develops the theoretical component of the paper, with detailed model assumptions, statements, and proofs provided as a supplement to Section~\ref{sec:property}.

\subsection{Time series models}\label{sec:models}
We now introduce the model classes underpinning our theoretical results, adapting Section~2.1 in \citet{BPP2024} with modifications as needed.

\subsubsection{VAR models}
\begin{definition}\label{def:VAR}
We say that $(\X, \Y) = \{(X_t, Y_t), t \in \Z\}$ follows a bivariate vector autoregressive model of order $p$, denoted by VAR($p$), if it satisfies the following equations for all $t\in \Z$:
\begin{align*}
    X_t &= \alpha_1^X X_{t-1} + \dots + \alpha_p^X X_{t-p} + \gamma^X_1 Y_{t-1} + \dots + \gamma^X_p Y_{t-p} + \varepsilon^X_t, \\
    Y_t &= \alpha_1^Y Y_{t-1} + \dots + \alpha_p^Y Y_{t-p} + \gamma^Y_1 X_{t-1} + \dots + \gamma^Y_p X_{t-p} + \varepsilon^Y_t,
\end{align*}
where $\alpha^s_i, \gamma^s_i \in\R, \:i\in \{1,\dots,p\}, \,s\in\{X,Y\}$ are real constants with at least one of the leading coefficients $\alpha^s_p, \gamma^s_p$ non-zero, and $\varepsilon^s_t, t\in\Z , s\in\{X,Y\}$ are white noises. 
\end{definition}

With the above formulation, we say that $\X$ (Granger) causes $\Y$, written $\X \to \Y$, if $\gamma^Y_r \neq 0$ for some $r \in \{1,\dots,p\}$, and conversely $\Y \to \X$ if $\gamma^X_t \neq 0$ for some $t \in \{1,\dots,p\}$. Both directions may hold simultaneously, giving a bidirected edge in a cyclic graph, often called a \textit{causal loop}.

The process $(\X, \Y)$ satisfies the \textit{stability condition} if $\det(I_2 - A_1 z - \dots - A_p z^p) \neq 0$ for all $z \in \C$ such that $|z| \leq 1$, where $I_2$ denotes the $2$-dimensional identity matrix and $A_i = \left[ \begin{smallmatrix} \alpha^X_i & \gamma^X_i \\ \alpha^Y_i & \gamma^Y_i \end{smallmatrix} \right]$.
For a VAR($p$) process, stability ensures stationarity and yields the causal representation  
\begin{align}\label{eq:var_causal_rep}
    X_t = \sum_{i=0}^\infty a^X_i \varepsilon_{t-i}^X + \sum_{i=0}^\infty c^X_i \varepsilon_{t-i}^Y, \ \
    Y_t = \sum_{i=0}^\infty a^Y_i \varepsilon_{t-i}^X + \sum_{i=0}^\infty c^Y_i \varepsilon_{t-i}^Y,
\end{align}
for constants $a^X_i, a^Y_i, c^X_i, c^Y_i$. In this framework, $\X$ causes $\Y$ if and only if $a^Y_i \neq 0$ for some $i \in \N$ \citep{Kuersteiner2010}.  
Furthermore, we say that $(\X, \Y)$ satisfies the \textit{extremal causal condition} if there exists $r \leq p$ such that $a^X_i \neq 0$ implies $a^Y_{i+r} \neq 0$ for all $i \in \N$ \citep{BPP2024}.

\begin{definition}\label{def:hVAR}
Given a stable VAR($p$) specification with causal representation given by Eq.~\eqref{eq:var_causal_rep}, we say that $(\X, \Y)$ follows a \textit{heavy-tailed VAR($p$) model} (abbreviated as hVAR($p$)) if the following conditions are satisfied:
\begin{enumerate}[label=(\alph*), labelsep=1em, leftmargin=*]
    \item $\varepsilon_{t}^X, \varepsilon_{t}^Y \overset{\text{i.i.d.}}{\sim}RV_{\theta}
    $ for some $\theta>0$,
    \item $\alpha^X_i, \alpha^Y_i, \gamma^X_i, \gamma^Y_i \geq 0$,
    \item $\exists\, \delta>0$ such that $\sum^{\infty}_{i=0} (a^X_i)^{\theta-\delta},
    \sum^{\infty}_{i=0} (c^X_i)^{\theta-\delta},
    \sum^{\infty}_{i=0} (a^Y_i)^{\theta-\delta},
    \sum^{\infty}_{i=0} (c^Y_i)^{\theta-\delta}<\infty $.
\end{enumerate}
\end{definition}

\noindent Condition (a), together with (c), ensures that the stationary linear process is regularly varying. The stability condition imposed on the underlying VAR model guarantees the existence of its stationary causal representation, while condition (c) ensures that the infinite moving average sums are well defined and permits their tail probabilities to be expressed asymptotically in terms of the tails of the individual innovation terms. Condition (b) prevents cancellation of extreme contributions. In particular, if $\X$ causes $\Y$ through lag $r$, so that $\gamma_r^Y>0$, then any innovation contributing positively to $X_t$ also contributes positively to $Y_{t+r}$ through the term $\gamma_r^Y X_t$. Hence, $a_i^X>0$ implies $a_{i+r}^Y>0$, establishing the extremal causal condition.

\subsubsection{NAAR models}
\begin{definition}\label{def:NAAR}
    We say that $(\X, \Y)$ follows a bivariate nonlinear additive autoregressive model of order $p$, denoted by NAAR($p$), if it satisfies the following equations for all $t\in \Z$:
    \begin{align*}
        X_t = f_0(X_{t-1}) + f_1(Y_{t-1}) + \dots + f_p(Y_{t-p}) + \varepsilon^X_t, \
        Y_t = g_0(Y_{t-1}) + \dots + g_p(X_{t-p}) + \varepsilon^Y_t,
    \end{align*}
    where $f_i, g_i, \:i\in \{0,1,\dots,p\}, $ are continuous, non-negative real-valued functions, with the condition that $f_p, g_p$ are not both constant, and $\varepsilon^s_t, t\in\Z, s\in\{X,Y\}$ are white noises.
\end{definition}

We say that $\X$ (Granger) causes $\Y$ if $g_r$ is non-constant on the support of $X_{t-r}$ for some $r$. 

\begin{definition}\label{def:hNAAR}
Given a NAAR($p$) specification, we say that $(\X, \Y)$ follows a \textit{heavy-tailed NAAR($p$) model} (abbreviated as hNAAR($p$)) if the following conditions are satisfied:
\begin{enumerate}[label=(\alph*), labelsep=1em, leftmargin=*]
    \item $\varepsilon_{t}^X, \varepsilon_{t}^Y \overset{\text{i.i.d.}}{\sim}RV_{\theta}
    $ are non-negative for some $\theta>0$,
    \item $\lim_{s\to\infty}f_i(s)=\infty$ and $\lim_{s\to\infty}g_i(s)=\infty$ for $i\in\{ 1,\dots, p \}$,
    \item $\lim_{s\to\infty} \frac{f_i(s)}{s}<1$ and $\lim_{s\to\infty} \frac{g_i(s)}{s}<1$ for $i\in\{ 1,\dots, p \}$.
\end{enumerate}
\end{definition}

\subsection{Properties of the impact function}\label{sec:impact_pf}

\begin{proposition}\label{prop:1}
Let $h: [0,1]^p \to \R$ be the impact function defined by
\begin{align*}
    h(v_1, \ldots, v_p)=h(v_1,\ldots, v_p; {\vw}, \alpha)
  =\left[ 1-\prod_{i=1} ^p \left\{ 1-v_i(1-e^{-\alpha})\right\} ^{w_i}\right] (1-e^{-\alpha})^{-1},
\end{align*}
where $\alpha > 0$, and the weights $\vw = (w_1, \ldots, w_p)$ are non-negative and satisfy the normalization condition $\vw \cdot \mathbf{1} = 1$. Then the function $h$ takes values in the interval $[0, 1]$.
\end{proposition}

\begin{proof}[Proof of Proposition 1.]
In the following proofs, we set $c := 1 - e^{-\alpha}$, and note that $c\in (0,1)$ for all $\alpha > 0$.
By definition of the impact function,
\[
    h(v_1, \ldots, v_p) = \frac{ 1 - \prod_{i=1}^p (1 - v_i c )^{w_i} }{ c }.
\]
For each $i = 1, \ldots, p$, $v_i \in [0,1]$ implies
\[
    1 - v_i c \in [1 - c, 1].
\]
Since $w_i \geq 0$, raising both sides to the power $w_i$ preserves the inequality:
\[
    (1 - c)^{w_i} \leq (1 - v_i c)^{w_i} \leq 1.
\]
Taking the product over $i = 1, \ldots, p$ and using the normalization $\sum_{i=1}^p w_i = 1$, we obtain
\[
    1 - c =  (1 - c)^{\sum w_i} = \prod_{i=1}^p (1 - c)^{w_i}  \leq \prod_{i=1}^p (1 - v_i c)^{w_i} \leq 1.
\]
It follows that
\[
    0 \leq 1 - \prod_{i=1}^p (1 - v_i c)^{w_i} \leq c.
\]
Multiplying both sides by $1/c > 0$ yields the desired result
\[
    0 \leq h(v_1, \ldots, v_p)  \leq 1.\qedhere
\]
\end{proof}

\begin{proposition}
    Under the set-up of Proposition~\ref{prop:1}, suppose there exists a subset of indices $\J \subseteq \{1, \dots, p\}$ such that $v_{j} = 1$ for all $j \in\J$. If the corresponding weights satisfy
    \[
        \sum_{j\in\J} w_{j} = 1
    \]
    then for all $\alpha>0$, the impact function satisfies
    \[
        h(v_1, \ldots, v_p)=1.
    \]
\end{proposition}

\begin{proof}[Proof of Proposition 2.] 
Suppose there exists a subset of indices $\J \subseteq \{1, \dots, p\}$ such that $v_{j} = 1$ for all $j \in\J $, and $\sum_{j\in \J} w_{j} = 1$. 
For each $j \in\J$, we have
\[
    (1 - v_{j} c)^{w_{j}} = (1 - c)^{w_{j}}.
\]
Moreover, since $\sum_{j\in\J} w_{j} = 1$, it follows that $\sum_{i \notin \J } w_i = 0$, and hence $w_i = 0$ for all $i \notin \J$.
Therefore,
\[
    \prod_{i \notin \J} (1 - v_i c)^{w_i} = 1.
\]
The full product over all indices decomposes as
\begin{align*}
    \prod_{i=1}^p (1 - v_i c)^{w_i} 
    &= \prod_{j\in\J} (1 - v_j c)^{w_{j}} \cdot \prod_{i \notin \J} (1 - v_i c)^{w_i} \\
    &= \prod_{j\in\J} (1 - c)^{w_{j}} \cdot 1 \\ 
    & = (1 - c)^{\sum_{j\in\J} w_{j}} 
    = (1 - c)^1 = 1-c.
\end{align*}
Substituting into the definition of the impact function,
\[
    h(v_1, \ldots, v_p) = \frac{1 - \prod_{i=1}^p (1 - v_i c)^{w_i}}{c} = \frac{1 - (1 - c)}{c} = \frac{c}{c} = 1.\qedhere
\]
\end{proof}

\medskip

The following proposition shows that, in the limiting case as $\alpha \to \infty$, our proposed impact function reduces to the Blalock formula \eqref{impactfunction}, while as $\alpha \to 0$, it converges to a direct weighted linear combination. Hence, varying the impact shape parameter $\alpha$ provides a continuum range of possibilities from nil cancellation to linear cancellation.

\begin{proposition}\label{prop:3}
    Let $h$ be the impact function as defined in Proposition~\ref{prop:1}. Then:
\begin{itemize}
    \item[(i)] As $\alpha \to 0$,  
    \[
    h(v_1, \ldots, v_p) \to \sum_{i=1}^p w_i v_i.
    \]
    
    \item[(ii)] As $\alpha \to \infty$,  
    \[
    h(v_1, \ldots, v_p) \to 1 - \prod_{i=1}^p (1 - v_i)^{w_i}.
    \]
\end{itemize}
\end{proposition}

\begin{proof}[Proof of Proposition 3.]
As $\alpha \to 0$, we have $c := 1 - e^{-\alpha} \to 0$. For each $i$, since $v_i c \to 0$, the first-order Taylor expansion of $\log(1 - v_i c)$ gives
\[
    \log(1 - v_i c) = -v_i c + r_i(c), \quad \text{with } r_i(c) = o(c).
\]
Therefore,
\[
    \sum_{i=1}^p w_i \log(1 - v_i c) = -c \sum_{i=1}^p w_i v_i + \sum_{i=1}^p w_i r_i(c) = -c \sum_{i=1}^p w_i v_i + o(c),
\]
where $o(c)$ denotes a remainder term that vanishes faster than $c$ as $c \to 0$.
Exponentiating both sides yields
\[
    \prod_{i=1}^p (1 - v_i c)^{w_i} = \exp\left( -c \sum_{i=1}^p w_i v_i + o(c) \right) = 1 - c \sum_{i=1}^p w_i v_i + o(c),
\]
where the second equality follows from the Taylor expansion $\exp(-x + o(x)) = 1 - x + o(x)$ as $x \to 0$.
Substituting into the definition of $h$, we obtain
\[
    h(v_1, \ldots, v_p) = \frac{1 - \prod_{i=1}^p (1 - v_i c)^{w_i}}{c} =
    \frac{1 - \left(1 - c \sum_{i=1}^p w_i v_i +o(c) \right)}{c} = \sum_{i=1}^p w_i v_i + o(1),
\]
where $o(1)$ denotes a term that vanishes as $c \to 0$. Thus, $h(v_1, \ldots, v_p) \to \sum_{i=1}^p w_i v_i$.

For the second case, as $\alpha \to \infty$, we have $c \to 1$. By continuity, it follows that
\[
    h(v_1, \ldots, v_p) = \frac{1 - \prod_{i=1}^p (1 - v_i c)^{w_i}}{c} 
    \xrightarrow{c \to 1} 
    1 - \prod_{i=1}^p (1 - v_i)^{w_i}. \qedhere
\]
\end{proof}

An implication of Proposition~\ref{prop:3} concerns the behavior of the optimized weight distribution discussed in Section \ref{sec:weight_distr}. 
As $\alpha \to 0$, the impact function converges to $\sum_{i=1}^p w_i v_i$, which is linear in the weights $w_i$. Therefore, under the constraint $\sum_{i=1}^p w_i = 1$, the maximizer will always assign full weight to the index corresponding to the largest $v_i$.
In contrast, as $\alpha \to \infty$, the product structure of the impact function becomes dominant. 
In this regime, the limiting form $1 - \prod_{i=1}^p (1 - v_i)^{w_i}$ is concave over the probability simplex 
\[
\Delta^{p-1} := \left\{ \vw = (w_1, \ldots, w_p) \in \R^p \, :\, w_i \geq 0 \text{ for all } i,\; \sum_{i=1}^p w_i = 1 \right\}.
\]
As a result, the maximizer tends to distribute mass more evenly across components that have larger values of $v_i$, balancing the contribution of multiple inputs rather than concentrating all mass on a single coordinate.

\medskip

The following lemma provides upper and lower bounds for the impact function $h$, which will help us establish the numerical values of the compound causal tail coefficient in the causal and non-causal directions.
\begin{lemma}\label{lemma:1}
     Let $h$ be the impact function as defined in Proposition~\ref{prop:1}. Then:
     \[
        \sum_{i=1}^p w_i v_i 
        \leq h(v_1, \ldots, v_p) 
        \leq \max(v_1, \ldots, v_p).
     \]
\end{lemma}

\begin{proof}[Proof of Lemma 1.]
We first show the lower bound $ h(v_1, \ldots, v_p) \geq \sum_{i=1}^p w_i v_i $. Note that the function $f:[0,1] \to \R$, defined by $f(x)= \log(1-cx)$, is concave. Then by the finite form of Jensen's inequality,
\[
    \sum_{i=1}^p w_i \log(1-c\, v_i) \leq \log(1 - c \sum_{i=1}^p w_i v_i).
\]
It follows that 
\[
    h(v_1, \ldots, v_p) = \frac{1 - \exp\left( \sum_{i=1}^p w_i \log(1 - c\, v_i) \right)}{c} 
    \geq \frac{1 - \exp\left( \log\left(1 - c \sum_{i=1}^p w_i v_i \right) \right)}{c} 
    = \sum_{i=1}^p w_i v_i. 
\]

We now prove the upper bound $ h(v_1, \ldots, v_p) \leq \max(v_1, \ldots, v_p)$. First, observe that since $w_i \in [0,1]$ and $1-c \, v_i \in (0,1]$, we have
\[
    \frac{\partial h}{\partial v_i} = w_i \cdot (1 - c\, v_i)^{w_i - 1} \cdot \prod_{j \ne i} (1 - c \, v_j)^{w_j} \ \in [0, w_i]  ,
\]
which implies that $h(v_1,\dots,v_p)$ is non-decreasing in each coordinate $v_i$.
Now let $v_{\max} := \max(v_1,\ldots,v_p)$. By monotonicity, we have
\[
    h(v_1, \ldots, v_p) \leq h(v_{\max}, \ldots, v_{\max}).
\]
Evaluating the function at this constant vector $(v_{\max}, \ldots, v_{\max})$ yields
\begin{align*}
    h(v_{\max}, \dots, v_{\max}) 
    = \frac{1 - (1 - c\, v_{\max})^{\sum w_i}}{c} = \frac{1 - (1 - c\, v_{\max})^1}{c} 
    = v_{\max},
\end{align*}
where the second equality follows from \( \sum_{i=1}^p w_i = 1 \). Hence, $h(v_1, \ldots, v_p) \leq v_{\max}$, as claimed.
\end{proof}

\subsection{Properties of the compound causal tail coefficient}\label{sec:cctc_pf}
Throughout this section, we assume that the weight parameters are correctly specified, as stated in Assumption~\ref{weight assumption}. All other definitions and assumptions follow those introduced in Section~\ref{sec:models}.

\setcounter{theorem}{0}
\begin{theorem}
    Let $(\X, \Y)$ follow an hVAR($p$) or hNAAR($p$) model.  
    If $\X$ causes $\Y$, then $\Gamma_{\X \rightarrow h(\Y)}(p) = 1$.
    If $\X$ does not cause $\Y$, then $\Gamma_{\X \rightarrow h(\Y)}(p) < 1$ for all $p \in \N$.
\end{theorem}

\begin{proof}[Proof of Theorem 1.]
\text{Claim 1:} We first establish that $\X$ causing $\Y$ implies $\Gamma_{\X \rightarrow h(\Y)}(p) = 1$.
Using the definition of the compound causal tail coefficient and the lower bound in Lemma~\ref{lemma:1}, we have
\begin{align*}
    \Gamma_{{\X} \rightarrow {h(\Y})}(p)
    &= \lim_{u\to 1^-}\E \left[ h(F_Y(Y_1), \ldots, F_Y(Y_p)) \mid F_X(X_0)>u \right] \\
    &\geq \lim_{u\to 1^-}\E \left[ \sum_{i=1}^p w_i F_Y(Y_i) \mid F_X(X_0)>u \right] \\
    &= \sum_{i=1}^p w_i \lim_{u\to 1^-} \E \left[ F_Y(Y_i) \mid F_X(X_0)>u \right],
\end{align*}
where the final equality follows from the linearity of expectation and the sum being finite.
Let $v_i \in [0,1]$ for all $i \in \{1, \dots, p\}$. Observe that $\sum_{i=1}^p w_i v_i = 1$ if and only if $v_i = 1$ for all $i$ such that $w_i > 0$. The ``if'' direction is immediate. For the ``only if'' direction, note that since $v_i \in [0,1]$ for all $i$, it follows that
\[
    \sum_{i=1}^p w_i v_i \leq \sum_{i=1}^p w_i = 1.
\]
Given that $\sum_{i=1}^p w_i v_i = 1$, equality must hold throughout. In particular, $\sum_{i=1}^p w_i v_i = \sum_{i=1}^p w_i$
implies that $v_i = 1$ for all indices $i$ such that $w_i > 0$.
From this observation, to establish that $\Gamma_{{\X} \rightarrow {h(\Y})}(p) = 1$, it suffices to verify that for all $i$ with $w_i > 0$,
\[
    \lim_{u \to 1^-} \E \left[ F_Y(Y_i) \mid F_X(X_0) > u \right] = 1.
\]
Since the weight vector is assumed to be correctly specified, this is equivalent to verifying the above limit for all causal lags $r \leq p$ in the hVAR($p$) or hNAAR($p$) model.

We claim that $\lim_{u \to 1^-} \E \left[ F_Y(Y_r) \mid F_X(X_0) > u \right] = 1$ if and only if
\begin{align}\label{eq:1}
    \lim_{v \to \infty} \PP(Y_r > \lambda \mid X_0 > v) = 1, \quad \forall \lambda \in \R.
\end{align}
To see this, we first prove the ``if'' direction. Fix $\varepsilon>0$, and choose $\lambda\in\R$ such that $F_Y(\lambda) > 1-\varepsilon$. Then
\begin{align*}
    &\lim_{v\to\infty} \PP(Y_r >\lambda\mid X_0>v) = 1, \\
    \implies & \lim_{v\to\infty} \PP(F_Y (Y_r) > 1-\varepsilon \mid X_0>v) = 1, \\
    \implies & \lim_{v\to\infty} \E[ F_Y (Y_r)  \mid X_0>v] > 1-\varepsilon.
\end{align*}
Letting $\varepsilon\to 0$ yields the result.
For the ``only if'' direction, fix $\lambda \in \R$. Choose $\varepsilon > 0$ such that $F_Y(\lambda) < 1 - \varepsilon$. Then, by assumption, there exists $v$ such that
\[
\E[F_Y(Y_r) \mid X_0 > v] > 1 - \varepsilon^2.
\]
We decompose the conditional expectation:
\begin{align*}
    \E[F_Y(Y_r) \mid X_0 > v] 
    &= \E\left[ F_Y(Y_r) \cdot \1_{\{F_Y(Y_r) \leq 1 - \varepsilon\}} \mid X_0 > v \right]  + \E\left[ F_Y(Y_r) \cdot \1_{\{F_Y(Y_r) > 1 - \varepsilon\}} \mid X_0 > v \right] \\
    &\leq (1 - \varepsilon) \cdot \PP(F_Y(Y_r) \leq 1 - \varepsilon \mid X_0 > v) + 1 \cdot \PP(F_Y(Y_r) > 1 - \varepsilon \mid X_0 > v) \\
    &= 1 - \varepsilon \cdot \left\{1 - \PP(F_Y(Y_r) > 1 - \varepsilon \mid X_0 > v) \right\}.
\end{align*}
Combining this with the lower bound $\E[F_Y(Y_r) \mid X_0 > v] > 1 - \varepsilon^2$, we obtain
\[
    \PP(F_Y(Y_r) > 1 - \varepsilon \mid X_0 > v) > 1 - \varepsilon.
\]
Since $F_Y(\lambda) < 1 - \varepsilon$, it follows that 
\[
    \PP(Y_r > \lambda \mid X_0 > v) \geq \PP(F_Y(Y_r) > 1 - \varepsilon \mid X_0 > v) > 1 - \varepsilon.
\]
Sending $\varepsilon \to 0$ gives the claim.

We now prove \eqref{eq:1} separately for the hVAR($p$) and hNAAR($p$) models. The remainder of this proof follows closely the argument presented in Theorem~1 of \cite{BPP2024}.

\medskip

\noindent\textit{Proof for hVAR($p$) model.} Let $(\X, \Y)$ be a bivariate hVAR($p$) time series. Since $\X$ causes $\Y$, we have $\gamma^Y_r > 0$ for some $r\leq p$. We can write
\begin{align*}
    \PP(Y_r > \lambda \mid X_0 > v) 
    &= \PP \left( \gamma^Y_r X_0 + \sum_{i=1}^p \alpha_i^Y Y_{r-i} + \sum_{i=1; i\neq r}^p \gamma^Y_i X_{r-i} + \varepsilon^Y_r > \lambda 
    \Biggm\vert X_0 > v  \right) \\
    & \geq \PP \left( \gamma^Y_r v + \sum_{i=1}^p \alpha_i^Y Y_{r-i} + \sum_{i=1; i\neq r}^p \gamma^Y_i X_{r-i} + \varepsilon^Y_r > \lambda 
    \Biggm\vert X_0 > v  \right).
\end{align*}
Now using the causal representation of VAR models, we have
\begin{align*}
    X_0 
    &= \sum^{\infty}_{i=0}a^X_i \varepsilon_{-i}^X + \sum^{\infty}_{i=0}c^X_i \varepsilon_{-i}^Y, \\
    \sum_{i=1}^p \alpha_i^Y Y_{r-i} + \sum_{i=1; i\neq r}^p \gamma^Y_i X_{r-i} + \varepsilon^Y_r 
    & = \sum^{\infty}_{i=0} \tilde{a}^Y_i \varepsilon_{r-i}^X + \sum^{\infty}_{i=0} \tilde{c}^Y_i \varepsilon_{r-i}^Y,
\end{align*}
for some non-negative coefficients $a^X_i, c^X_i, \tilde{a}^Y_i, \tilde{c}^Y_i \geq 0$.
Substituting into the conditional probability, we obtain
\begin{align*}
    \PP(Y_r > \lambda \mid X_0 > v)
    &\geq \PP \left( \sum^{\infty}_{i=0} \tilde{a}^Y_i \varepsilon_{r-i}^X + \sum^{\infty}_{i=0} \tilde{c}^Y_i \varepsilon_{r-i}^Y > \lambda - \gamma^Y_r v
    \Biggm\vert \sum^{\infty}_{i=0}a^X_i \varepsilon_{-i}^X + \sum^{\infty}_{i=0}c^X_i \varepsilon_{-i}^Y > v  \right) \\
    &\geq \PP \left( \sum^{\infty}_{i=0} \tilde{a}^Y_i \varepsilon_{r-i}^X + \sum^{\infty}_{i=0} \tilde{c}^Y_i \varepsilon_{r-i}^Y > \lambda - \gamma^Y_r v \right).
\end{align*}
To justify the final inequality, observe that $a^X_i, c^X_i, \tilde{a}^Y_i, \tilde{c}^Y_i \geq 0$ implies 
\begin{align*}
    0 &\leq
    \cov \left(\1_{\{ 
    \sum^{\infty}_{i=0} \tilde{a}^Y_i \varepsilon_{r-i}^X + \sum^{\infty}_{i=0} \tilde{c}^Y_i \varepsilon_{r-i}^Y > \lambda - \gamma^Y_r v \} } ,
    \1_{ \{ 
    \sum^{\infty}_{i=0}a^X_i \varepsilon_{-i}^X + \sum^{\infty}_{i=0}c^X_i \varepsilon_{-i}^Y > v
     \} } \right)\\
    &= \PP \left( \sum^{\infty}_{i=0} \tilde{a}^Y_i \varepsilon_{r-i}^X + \sum^{\infty}_{i=0} \tilde{c}^Y_i \varepsilon_{r-i}^Y > \lambda - \gamma^Y_r v
    \ , \ \sum^{\infty}_{i=0}a^X_i \varepsilon_{-i}^X + \sum^{\infty}_{i=0}c^X_i \varepsilon_{-i}^Y > v  \right) \\
    & \qquad \ - \PP \left( \sum^{\infty}_{i=0} \tilde{a}^Y_i \varepsilon_{r-i}^X + \sum^{\infty}_{i=0} \tilde{c}^Y_i \varepsilon_{r-i}^Y > \lambda - \gamma^Y_r v
    \right)  \ \PP \left(\sum^{\infty}_{i=0}a^X_i \varepsilon_{-i}^X + \sum^{\infty}_{i=0}c^X_i \varepsilon_{-i}^Y > v  \right),
\end{align*}
and dividing both sides by $\PP \left(\sum^{\infty}_{i=0}a^X_i \varepsilon_{-i}^X + \sum^{\infty}_{i=0}c^X_i \varepsilon_{-i}^Y > v  \right)$ yields the desired inequality.
Finally, taking the limit as $v \to\infty$, we have
\begin{align*}
    \lim_{v \to \infty} \PP(Y_r > \lambda \mid X_0 > v)
    \geq \lim_{v \to \infty} \PP \left( \sum^{\infty}_{i=0} \tilde{a}^Y_i \varepsilon_{r-i}^X + \sum^{\infty}_{i=0} \tilde{c}^Y_i \varepsilon_{r-i}^Y > \lambda - \gamma^Y_r v \right) = 1,
\end{align*}
which completes the proof for the hVAR($p$) model.

\medskip

\noindent\textit{Proof for hNAAR($p$) model.} Let $(\X, \Y)$ be a bivariate hNAAR($p$) time series. Since $\X$ causes $\Y$, then $g_r$ is a non-constant function and $\lim_{x\to\infty} g_r(x) = \infty$ for some $r\leq p$. This implies that there exists $x_0 \in\R$ such that $g_r(x_0) >\lambda$ for all $\lambda\in\R$. Hence, for all $v>x_0$,
\[
    \PP \left(g_r(X_0) > \lambda \mid X_0 > v \right) = 1.
\]
Moreover, since $\varepsilon_{t}^Y$ and $g_t$ are non-negative for all $t\in \{0,1,\dots,p\}$, it follows that
\begin{align*}
    \lim_{v \to \infty} 
    \PP(Y_r > \lambda \mid X_0 > v) 
    & = \lim_{v \to \infty} \PP \left( g_0(Y_{r-1}) + g_1(X_{r-1}) + \dots + g_p(X_{r-p}) + \varepsilon^Y_r > \lambda 
    \Bigm\vert X_0 > v
    \right) \\
    &\geq \lim_{v \to \infty} \PP \left(g_r(X_0) > \lambda \mid X_0 > v \right) =1.
\end{align*}

\bigskip


\text{Claim 2:} We now prove that if $\X$ does not cause $\Y$, then $\Gamma_{\X \rightarrow h(\Y)}(p) < 1$ for all $p \in \N$.
Using the definition of the compound causal tail coefficient and the upper bound in Lemma~\ref{lemma:1}, we have
\begin{align*}
    \Gamma_{{\X} \rightarrow {h(\Y})}(p)
    &= \lim_{u\to 1^-}\E \left[ h(F_Y(Y_1), \ldots, F_Y(Y_p)) \mid F_X(X_0)>u \right] \\
    &\leq \lim_{u\to 1^-}\E \left[ \max(F_Y(Y_1), \ldots, F_Y(Y_p)) \mid F_X(X_0)>u \right].
\end{align*}
Therefore, to show that $\Gamma_{{\X} \rightarrow {h(\Y})}(p)<1 $, it suffices to prove that 
\[
    \lim_{v\to \infty}\E \left[ \max(F_Y(Y_1), \ldots, F_Y(Y_p)) \mid X_0 >v \right] <1. 
\]
This inequality follows if we are able to verify that
\begin{align}\label{eq:2}
    \lim_{v \to \infty} \PP \left(\max(Y_1, \ldots, Y_p )< \lambda \mid X_0 >v \right) > 0 ,
\end{align}
for all $\lambda\in\R$ such that $\PP(Y_0 <\lambda) >0 $.
We now prove \eqref{eq:2} separately for the hVAR($p$) and hNAAR($p$) models.

\medskip

\noindent\textit{Proof for hVAR($p$) model.} Let $(\X, \Y)$ be a bivariate hVAR($p$) time series. Since $\X$ does not cause $\Y$, the causal representation of the process takes the form
\begin{align*}
    X_t &= \sum^{\infty}_{i=0}a^X_i \varepsilon_{t-i}^X + \sum^{\infty}_{i=0}c^X_i \varepsilon_{t-i}^Y, \\
    Y_t &= \sum^{\infty}_{i=0}a^Y_i \varepsilon_{t-i}^X .
\end{align*}
It follows that
\begin{align*}
    \PP \left(\max(Y_1, \ldots, Y_p )< \lambda \mid X_0 >v \right) 
    &= \PP \left(Y_1< \lambda, \ldots, Y_p< \lambda \mid X_0 >v \right) \\
    & \geq \PP \left( \sum_{t=1}^p |Y_t| < \lambda \Bigm\vert X_0 >v \right) \\
    &= \PP \left( \sum_{t=1}^p \left| \sum^{\infty}_{i=0}a^Y_i \varepsilon_{t-i}^X \right| < \lambda 
    \Biggm\vert \sum^{\infty}_{i=0}a^X_i \varepsilon_{-i}^X + \sum^{\infty}_{i=0}c^X_i \varepsilon_{-i}^Y > v \right) \\
    & \geq \PP \left( \sum_{t=1}^p  \sum^{\infty}_{i=0}a^Y_i | \varepsilon_{t-i}^X | < \lambda 
    \Biggm\vert \sum^{\infty}_{i=0}a^X_i \varepsilon_{-i}^X + \sum^{\infty}_{i=0}c^X_i \varepsilon_{-i}^Y > v \right).
\end{align*}
Finally, let $\theta$ denote the common tail index of regular variation for the noise variables $\varepsilon_{i}^X$ and $\varepsilon_{i}^Y$. By Proposition~2 in \cite{BPP2024}, we obtain the following identity:
\begin{align*}
    \lim_{v \to \infty} & \ \PP \left( \sum_{t=1}^p  \sum^{\infty}_{i=0}a^Y_i | \varepsilon_{t-i}^X | < \lambda 
    \Biggm\vert \sum^{\infty}_{i=0}a^X_i \varepsilon_{-i}^X + \sum^{\infty}_{i=0}c^X_i \varepsilon_{-i}^Y > v \right) \\
    = & \lim_{v \to \infty} \ \PP \left( \sum_{t=1}^p  \sum^{\infty}_{i=0}a^Y_i | \varepsilon_{t-i}^X | < \lambda \right) \, 
      \frac{ \sum^{\infty}_{i=0}(c^X_i)^\theta + \sum_{i: a^X_i > a^Y_i = 0}(a^X_i)^\theta }
      { \sum^{\infty}_{i=0}(c^X_i)^\theta + \sum^{\infty}_{i=0}(a^X_i)^\theta }
      >  \,  0.
\end{align*}
The strict positivity of the final expression follows from the choice of $\lambda$ and the model assumptions in Section~\ref{sec:models}, which guarantee that both terms in the numerator and denominator are finite and non-negative.

\medskip
\noindent\textit{Proof for hNAAR($p$) model.} Let $(\X, \Y)$ be a bivariate hNAAR($p$) time series. Since $\X$ does not cause $\Y$, the model reduces to
\begin{align*}
    X_t &= f_0(X_{t-1}) + \sum_{i=1}^p f_i(Y_{t-i}) + \varepsilon^X_t, \\
    Y_t &= g_0(Y_{t-1}) + \varepsilon^Y_t.
\end{align*}
By continuity of $g_0$ and the assumption $\lim_{s\to\infty}g_0(s)=\infty$, we can choose $\lambda$ sufficiently large such that
\[
    \lambda^{\star} := \sup_{x < \lambda} g_0 (x) < \lambda \quad \text{and} \quad \PP\,(\varepsilon^Y_0 < \lambda - \lambda^{\star} ) >0.
\]
Using an iterative application of Bayes' rule, we can rewrite
\begin{align*}
    \PP \left(\max(Y_1, \ldots, Y_p )< \lambda \mid X_0 >v \right) 
    & = \PP \left(Y_1< \lambda, \ldots, Y_p< \lambda \mid X_0 >v \right) \\
    & = \prod_{i=1}^p \PP \left( Y_i < \lambda \mid Y_1 < \lambda, \ldots, Y_{i-1} < \lambda,\, X_0 > v \right).
\end{align*}
It therefore suffices to show that the limit of each factor in the product remains strictly positive as $v\to\infty$.
For each $i \in\{1,\dots, p\}$, we have
\begin{align*}
    \lim_{v \to \infty} & \PP \left( Y_i < \lambda \mid Y_1 < \lambda, \ldots, Y_{i-1} < \lambda,\, X_0 > v \right) \\
    = & \lim_{v \to \infty} \PP \left( g_0(Y_{i-1}) + \varepsilon^Y_i < \lambda \mid Y_1 < \lambda, \ldots, Y_{i-1} < \lambda,\, X_0 > v \right) \\
    \geq & \lim_{v \to \infty} \PP \left( \lambda^{\star} + \varepsilon^Y_i < \lambda \mid Y_1 < \lambda, \ldots, Y_{i-1} < \lambda,\, X_0 > v \right) \\
    = & \ \PP \left( \lambda^{\star} + \varepsilon^Y_i < \lambda \right) > 0,
\end{align*}
where the final equality follows from the independence of $\varepsilon^Y_i$ from $Y_1, \ldots, Y_{i-1}$ and $X_0$, and the final inequality holds by the choice of $\lambda$.
Hence, each conditional probability remains bounded away from zero, and \eqref{eq:2} follows. This completes the proof.
\end{proof}

\setcounter{theorem}{1}
\begin{theorem}
    Let $(\X, \Y)$ follow an hVAR($p$) model with possibly negative coefficients, satisfying the extremal causal condition. Suppose $\varepsilon_t^X$ and $\varepsilon_t^Y$ have full support on $\R$, and that $|\varepsilon_t^X|$ and $|\varepsilon_t^Y|$ are regularly varying with index $\theta$.  
    If $\X$ causes $\Y$, then $\Gamma_{|\X| \rightarrow h(|\Y|)}(p) = 1.$
    If $\X$ does not cause $\Y$, then $\Gamma_{|\X| \rightarrow h(|\Y|)}(p) < 1$ for all $p \in \N$.
\end{theorem}

\begin{proof}[Proof of Theorem 2.]
    The argument proceeds identically to the proof of Theorem~3 in \cite{BPP2024} and is therefore omitted.
\end{proof}

\setcounter{theorem}{2}
\begin{theorem}
    Let $(\X, \Y, \ZZ)$ be a stable $\mathrm{VAR}(p)$ process with i.i.d.\ regularly varying noise. Suppose $\ZZ$ is a common cause of $\X$ and $\Y$, and that neither $\X$ nor $\Y$ causes $\ZZ$. If $\X$ does not cause $\Y$, then $\Gamma_{\X \rightarrow h(\Y)}(p) < 1$ for all $p \in \N$.
\end{theorem}

\begin{proof}[Proof of Theorem 3.]
Since $\ZZ$ is a common cause of both $\X$ and $\Y$, and neither $\X$ nor $\Y$ causes $\ZZ$, and $\X$ does not cause $\Y$, the time series admits the following causal representation:
\begin{align*}
    X_t &= \sum^{\infty}_{i=0}a^X_i \varepsilon_{t-i}^X  + \sum^{\infty}_{i=0}b^X_i \varepsilon_{t-i}^Z + \sum^{\infty}_{i=0}c^X_i \varepsilon_{t-i}^Y, \\
    Y_t &= \sum^{\infty}_{i=0}a^Y_i \varepsilon_{t-i}^Y + \sum^{\infty}_{i=0}b^Y_i \varepsilon_{t-i}^Z, \\
    Z_t &= \sum^{\infty}_{i=0}b^Z_i \varepsilon_{t-i}^Z.
\end{align*}
Following the same reasoning as in the proof of Theorem~\ref{thm:1}, it suffices to show that
\[
    \lim_{v \to \infty} \PP \left( Y_p > \lambda \mid X_0 >v \right) < 1,
\]
for all $\lambda\in\R$ such that $\PP(Y_0 <\lambda) >0 $.
Substituting the causal representation of $X_0$ and $Y_p$ and applying Proposition~2 in \cite{BPP2024}, we obtain
\[
    \lim_{v \to \infty} \PP \left( \sum^{\infty}_{i=0}a^Y_i \varepsilon_{p-i}^Y + \sum^{\infty}_{i=0}b^Y_i \varepsilon_{p-i}^Z > \lambda 
    \Biggm\vert 
    \sum^{\infty}_{i=0}a^X_i \varepsilon_{-i}^X  + \sum^{\infty}_{i=0}b^X_i \varepsilon_{-i}^Z + \sum^{\infty}_{i=0}c^X_i \varepsilon_{-i}^Y >v \right) < 1.\qedhere
\]
\end{proof}

\subsection{Asymptotic results of the nonparametric estimator}\label{sec:estimator_pf}
\setcounter{theorem}{3}
\begin{theorem}
    Let (X,Y) be a stationary and ergodic bivariate time series with absolutely continuous marginal distributions supported on some neighborhood of infinity. Suppose that the compound causal tail coefficient $\Gamma_{X\to h(Y)}(p)$ is well-defined for some extremal delay $p$. Let $k=k_n$ satisfy 
    \begin{align*}
        k \to\infty \ \text{ and }\ \tfrac{k}{n} \to 0,\ \text{ as }\ n\to\infty,
    \end{align*}
    and assume that
    \begin{align}
        \frac{n}{k^{1-\xi}} \sup_{x \in\R} \left|\hat{F}_X(x) - F_X(x) \right|  & \xlongrightarrow{\PP}0, \label{rate assumption x} \\
        \frac{n}{k} \ \sup_{y \in\R} \left|\hat{F}_Y(y) - F_Y(y) \right|  & \xlongrightarrow{\PP}0, \label{rate assumption y}
    \end{align}
    for some $\xi>0$.
    Then, the nonparametric compound causal tail coefficient defined in \eqref{estimator} is consistent. That is, 
    \begin{align*}
        \hat{\Gamma}_{X\to h(Y)}(p)  \xlongrightarrow{\PP}  {\Gamma}_{X\to h(Y)}(p)
        ,\ \text{ as }\ n\to\infty.
    \end{align*}
\end{theorem}

\begin{proof}[Proof of Theorem 4.]
We begin by noting the following equivalence, which holds since $X$ is strictly stationary:
\begin{align*}
    \left\{ X_i \geq X_{(n-k+1)}\right\}
    = 
    \left\{ \hat{F}_X(X_i) > 1-\tfrac{k}{n} \right\}, 
    \ \text{ for all } i,k \in \{1, \dots, n\},
\end{align*}
where $X_{(j)}$ denotes the $j$-th order statistic of $X_1, \ldots, X_n$. Let $u := 1-\frac{k}{n}$ denote a high quantile level approaching 1, that is, $\PP(F_X(X_i) > u) = \frac{k}{n}$. Then the nonparametric estimator can equivalently be written as
\begin{align*}
    \hat{\Gamma}_{{\X} \rightarrow {h(\Y})} (p) = 
    \hat{\Gamma}_n :=
    \frac{1}{k} \sum_{i=1}^{n-p} h \left(\hat{F}_Y(Y_{i+1}), \ldots, \hat{F}_Y(Y_{i+p}) \right) 
    \cdot \1_{ \{ \hat{F}_X(X_i) > u \} }.
\end{align*}

To establish convergence in probability of this estimator to ${\Gamma}_{X\to h(Y)}(p)$, we introduce the following intermediate quantity:
\begin{align}\label{eq:intermediate}
    \tilde{\Gamma}_n = \frac{1}{\tilde{k}} \sum_{i=1}^{n-p} h \left({F}_Y(Y_{i+1}), \ldots, {F}_Y(Y_{i+p}) \right) 
    \cdot \1_{ \{ {F}_X(X_i) > u \} },
\end{align}
where $\tilde{k} := \sum_{i=1}^{n-p} \1_{ \{F_X(X_i) > u \} }$ denotes the number of exceedances above $u$ under the true distribution function of $X$. Note that \eqref{eq:intermediate} uses the true marginal distributions in both the impact function and the exceedance condition.

We will show the following two convergence claims as $n \to \infty$:
\begin{enumerate}
    \item \, $\hat{\Gamma}_n - \tilde{\Gamma}_n \xlongrightarrow{\PP} 0 $, \label{claim 1}
    \item \, $\tilde{\Gamma}_n \xlongrightarrow{\PP} \Gamma_{{\X} \rightarrow {h(\Y})}$. \label{claim 2}
\end{enumerate}
The desired consistency $\hat{\Gamma}_n \xlongrightarrow{\PP} \Gamma_{\X \rightarrow h(\Y)}(p)$ then follows directly by the triangle inequality:
\[
    \left|\hat{\Gamma}_n - \Gamma \right| \le \left|\hat{\Gamma}_n - \tilde{\Gamma}_n \right| 
    + \left|\tilde{\Gamma}_n - \Gamma \right| \xlongrightarrow{\PP} 0.
\]

To prove the first claim, we decompose the difference $\hat{\Gamma}_n - \tilde{\Gamma}_n$ into three terms:
\begin{align}
    \hat{\Gamma}_n - \tilde{\Gamma}_n
    = & \  \frac{1}{k} \sum_{i=1}^{n-p} \left\{
    h \left( \hat{F}_Y(Y_{i+1}), \dots, \hat{F}_Y(Y_{i+p})\right) 
    - 
    h \left( {F}_Y(Y_{i+1}), \dots, {F}_Y(Y_{i+p})\right) 
    \right\} \cdot 
    \1_{ \{ \hat{F}_X(X_i) > u \} }
    \tag{A}\\
    & \ + \left(\frac{1}{k} - \frac{1}{\tilde{k}} \right) \sum_{i=1}^{n-p} 
    h \left( {F}_Y(Y_{i+1}), \dots, {F}_Y(Y_{i+p})\right) 
    \cdot 
    \1_{ \{ \hat{F}_X(X_i) > u \} }
    \tag{B}\\
    & \ + \frac{1}{\tilde{k}} \sum_{i=1}^{n-p} 
    h \left( {F}_Y(Y_{i+1}), \dots, {F}_Y(Y_{i+p})\right) 
    \cdot \left\{
    \1_{\{ \hat{F}_X(X_i) > u \} } - \1_{ \{ {F}_X(X_i) > u \} }
    \right\}.
    \tag{C}
\end{align}

We begin by showing the convergence of term (A). For notational convenience, define $v_i := F_Y(Y_i)$ and $\hat{v}_i := \hat{F}_Y(Y_i)$, and let $c := 1 - e^{-\alpha}$. Then
\begin{align*}
    h(\hat{v}_1, \dots, \hat{v}_p) - h({v}_1, \dots, {v}_p) 
    &= \frac{1}{c} \left\{
    \prod_{i=1}^{p}(1-v_i c)^{w_i} - \prod_{i=1}^{p}(1-\hat{v}_i c)^{w_i}
    \right\}.
\end{align*}
Note that the impact function $h: [0,1]^p \to \R$ is Lipschitz continuous, since $c \in (0,1)$. Therefore, there exists a constant $L_c > 0$ such that
\[
    \left|h(\hat{\vv}) - h(\vv)\right| 
    \leq L_c \, \|\hat{\vv} - \vv \|_1 
    = L_c \sum_{i=1}^p |\hat{v}_i - {v}_i|
    \leq L_c\, p \ \sup_{y\in\R} \left| \hat{F}_Y(y) - F_Y(y) \right|.
\]
It then follows, under assumption \eqref{rate assumption y}, that
\[
    (A) \leq \frac{L_c\, p \, n}{k} \sup_{y\in\R} \left| \hat{F}_Y(y) - F_Y(y) \right| 
    \xlongrightarrow{\PP} 0, \ \text{ as } n\to\infty.
\]

We now show the convergence of term (B). 
Since $F_X(X_i)$ is a measurable function of a stationary and ergodic process, and the indicator $\1_{\{F_X(X_i) > u\} } \in [0,1]$ is integrable, we may apply the Birkhoff Ergodic Theorem (see, e.g., Theorem~6.2.1 in \citet{Durrett2019}):
\begin{align}\label{k tilde}
    \frac{1}{n-p} \sum_{i=1}^{n-p} \1_{ \{ F_X(X_i) > u \} } - \E \left[\1_{ \{ F_X(X_0) > u \} } \right]
    \xrightarrow{\text{a.s.}} 0
    , \ \text{ as } n\to\infty.
\end{align}
By the definition of $u$, we have that
\[
    \E \left[\1_{ \{ F_X(X_0) > u \} } \right]
    =\PP\left(F_X(X_0) > u \right) 
    =\frac{k}{n}.
\]
Hence, \eqref{k tilde} implies
\[
    \frac{\tilde{k}}{k} = \frac{\tilde{k}}{n-p} \cdot \frac{n-p}{k} 
    \xlongrightarrow{\PP}  1, \ \text{ as } n\to\infty,
\]
which is equivalent to $\tilde{k} = k + o_{\PP}(k)$.
We therefore obtain the following bound
\begin{align*}
    \left| \frac{1}{k} - \frac{1}{\tilde{k}} \right|
    = \left| \frac{\tilde{k}-k }{k \cdot \tilde{k}} \right|
    =\frac{o_{\PP}(k)}{k^2 + o_{\PP}(k^2)} = o_{\PP}\left(\frac{1}{k}\right).
\end{align*}
It follows from Proposition~\ref{prop:1} and the observation that $\sum_{i=1}^{n-p}
    \1_{ \{ \hat{F}_X(X_i) > u \} } \leq k$, that
\begin{align*}
    (\text{B}) 
    &\leq \left| \frac{1}{k} - \frac{1}{\tilde{k}} \right| \cdot 
    \sum_{i=1}^{n-p} 
    \left| h \left( {F}_Y(Y_{i+1}), \dots, {F}_Y(Y_{i+p})\right) \right|
    \cdot 
    \1_{ \{ \hat{F}_X(X_i) > u \} } \\
    & \leq \left| \frac{1}{k} - \frac{1}{\tilde{k}} \right| \cdot
    \sum_{i=1}^{n-p}
    \1_{ \{ \hat{F}_X(X_i) > u \} } \\
    & \leq \left| \frac{1}{k} - \frac{1}{\tilde{k}} \right| \cdot
    k
    = o_{\PP}(1).
\end{align*}

It remains to show that term (C) vanishes as $n\to\infty$. As before, we can bound:
\begin{align*}
     (\text{C}) 
    &\leq \frac{1}{\tilde{k}}  
    \sum_{i=1}^{n-p} 
    \left| h \left( {F}_Y(Y_{i+1}), \dots, {F}_Y(Y_{i+p})\right) \right|
    \cdot 
    \left|
    \1_{\{ \hat{F}_X(X_i) > u \} } - \1_{ \{ {F}_X(X_i) > u \} }
    \right| \\
    & \leq \frac{1}{\tilde{k}}
    \sum_{i=1}^{n-p} 
    \left|
    \1_{\{ \hat{F}_X(X_i) > u \} } - \1_{ \{ {F}_X(X_i) > u \} }
    \right| .
\end{align*}
We now bound the number of mismatched indicator terms. Observe that
\[
    \Delta_i := \left|
    \1_{\{ \hat{F}_X(X_i) > u \} } - \1_{ \{ {F}_X(X_i) > u \} } \right| = 1
\] 
only when
\[
    \left| F_X(X_i) - u \right| 
    < \left| \hat{F}_X(X_i) - F_X(X_i) \right| 
    \leq \sup_{x} \left| \hat{F}_X(x) - F_X(x) \right| .
\]
Let $S_n := \sup_{x \in \R} |\hat{F}_X(x) - F_X(x)|$. Then
\[
\sum_{i=1}^{n-p} \Delta_i \le \sum_{i=1}^{n-p} \1_{ \left\{ \left|F_X(X_i) - u \right| \le S_n \right\} }.
\]
Fix any $\varepsilon > 0$. By assumption \eqref{rate assumption x}, we can define a sequence $\delta_n := \frac{k^{1-\xi}}{n} \searrow 0$ such that
\[
    \PP\left( S_n > \delta_n\right) < \frac{\varepsilon}{2}
\]
for all sufficiently large $n$. 
On the event $\{ S_n \le \delta_n\}$, it holds that
\[
    \sum_{i=1}^{n-p} \Delta_i \le \sum_{i=1}^{n-p} \1_{ \left\{ \left|F_X(X_i) - u \right| \le \delta_n\right\} }.
\]
By the Birkhoff Ergodic Theorem , we obtain
\[
    \frac{1}{n-p} \sum_{i=1}^{n-p} \1_{\left\{ \left|F_X(X_i) - u \right| \le \delta_n\right\}  } 
    - \PP\left( \left|F_X(X_0) - u \right| \le \delta_n\right)
    \xlongrightarrow{\text{a.s.}} 0.
\]
Since $F_X(X_0) \sim \text{Uniform}[0,1]$, it follows that 
\[
    \PP\left( \left|F_X(X_0) - u \right| \le \delta_n\right) 
    = \int_{\min(0, u - \delta_n)}^{\max(1,u + \delta_n)} 1 \, dz
    \leq 2\delta_n.
\]
Therefore, with probability at least $1-\varepsilon/2$, the total number of mismatched indices satisfies
\[
    \sum_{i=1}^{n-p} \Delta_i = O(n \delta_n).
\]
We have already shown earlier that $\tilde{k} = k + o_{\PP}(k)$, so for sufficiently large $n$, with probability at least $1-\varepsilon/2$, 
\[
    {\tilde{k}} \geq \frac{k}{2}.
\]
Combining the two high-probability events using a union bound, we obtain that with probability at least $1-\varepsilon$, 
\begin{align*}
    (\text{C}) \ & 
    \leq \ \frac{1}{\tilde{k}}
    \sum_{i=1}^{n-p} \Delta_i \
    \leq O \left( \frac{n\delta_n}{k} \right).
\end{align*}
Note that with $\delta_n = \frac{k^{1-\xi}}{n}$ and $\xi>0$, we have
\[
    \frac{n\delta_n}{k} = k^{-\xi} \xlongrightarrow{} 0, \ \text{ as } n\to\infty.
\]
Then letting $\varepsilon\to 0$ gives the desired convergence (C)$\xlongrightarrow{\PP} 0$. This completes the proof of claim~\ref{claim 1}.

\medskip
To establish the second claim, we first define $\Gamma_u := \ \E \, \left[  h(F_Y(Y_1), \ldots, F_Y(Y_p)) \mid F_X(X_0)>u \right]$ and show that 
\begin{align}\label{eq:3}
    \tilde{\Gamma}_n - \Gamma_u  \xlongrightarrow{\PP} 0, \ \text{ as } n\to\infty.
\end{align}
Observe that both $\tilde{\Gamma}_n $ and $\Gamma_u$ can be written as ratios:
\begin{align*}
    \tilde{\Gamma}_n 
    &=  \frac{
    \frac{1}{n-p} \sum_{i=1}^{n-p} 
    h \left({F}_Y(Y_{i+1}), \ldots, {F}_Y(Y_{i+p}) \right) 
    \cdot \1_{ \{ {F}_X(X_i) > u \} }}
    { \frac{1}{n-p} \sum_{i=1}^{n-p} \1_{ \{F_X(X_i) > u \}}}, \\
    \Gamma_u 
    &=  \frac{ \E \, \left[  h(F_Y(Y_1), \ldots, F_Y(Y_p)) 
    \cdot\1_{ \{ {F}_X(X_0) > u \} } \right] }
    {\PP\left(F_X(X_0) > u \right)}.
\end{align*}
As established earlier in \eqref{k tilde}, the denominator of $\tilde{\Gamma}_n$ converges almost surely to the denominator of $\Gamma_u$. 
A similar argument applies to the numerator. Since $(X,Y)$ is strictly stationary and ergodic, and the function 
\[
    \phi(X_i,Y_{i+1}, \dots, Y_{i+p}):= h ({F}_Y(Y_{i+1}), \ldots, {F}_Y(Y_{i+p}) ) 
    \cdot \1_{ \{ {F}_X(X_i) > u \} }
\]
is integrable (as $h$ is bounded), the Birkhoff Ergodic Theorem yields: 
\[
    \frac{1}{n-p} \sum_{i=1}^{n-p} \phi(X_i,Y_{i+1}, \dots, Y_{i+p}) - \E \, \left[  h(F_Y(Y_1), \ldots, F_Y(Y_p)) 
    \cdot\1_{ \{ {F}_X(X_0) > u \} } \right]
    \xrightarrow{\text{a.s.}} 0 . 
\]
By the Continuous Mapping Theorem, and since $\PP\left(F_X(X_0) > u \right) = \frac{k}{n}> 0$, it follows that the convergence in \eqref{eq:3} holds.

Finally, recall that $u = 1 - \frac{k}{n} \to 1^- $ as $n\to \infty$. By assumption, the limit
\[
    \Gamma_{\X \rightarrow h(\Y)}(p) := \lim_{u \to 1^-} \E\left[ h(F_Y(Y_1), \ldots, F_Y(Y_p)) \mid F_X(X_0) > u \right]
\]
exists, so we have 
\[
    \Gamma_u  \longrightarrow \Gamma_{\X \rightarrow h(\Y)}(p), \ \text{ as } n\to\infty.
\]
Combining this with the convergence in probability $\tilde{\Gamma}_n \xlongrightarrow{\PP} \Gamma_{u} $, we conclude that as $n\to\infty$,
\[
    \tilde{\Gamma}_n \xlongrightarrow{\PP} \Gamma_{\X \rightarrow h(\Y)}(p). \qedhere
\]
\end{proof}

\medskip
In the proof of Theorem~\ref{thm:5}, the ergodicity assumptions are used to establish the convergence of terms (B) and (C) in claim~\ref{claim 1}, as well as to prove claim~\ref{claim 2}. For autoregressive processes, ergodicity is guaranteed under mild regularity conditions. In particular, a strictly stationary and causal AR($p$) process is strongly mixing and hence ergodic \citep{AP1986}.
The convergence rate assumption \eqref{rate assumption y} for the effect series $Y$ is used to show that term (A) in claim~\ref{claim 1} vanishes, while a slightly stronger rate assumption \eqref{rate assumption x} on the cause series $X$ is required to ensure that term (C) vanishes. These assumptions sharpen the standard uniform convergence in probability by imposing explicit requirements on the rate, namely:
\begin{align*}
    \sup_{x \in\R} \left|\hat{F}_X(x) - F_X(x) \right| = o_{\PP}\left( \frac{k^{1-\xi} }{n} \right) \ \text{ and } \
    \sup_{y \in\R} \left|\hat{F}_Y(y) - F_Y(y) \right| = o_{\PP}\left( \frac{k}{n} \right).
\end{align*}
Such rate conditions are not overly restrictive and are satisfied by a subset of stationary autoregressive models under certain technical conditions. For comparison, in the i.i.d.\ case, assumption \eqref{rate assumption x} holds if $k^{2(1-\xi)} / n \to \infty$ as $n \to \infty$, as implied by the Dvoretzky–Kiefer–Wolfowitz inequality.

\subsection{Asymptotic validity of the shifted MBB test}\label{sec:mbb_pf}

\begin{theorem}\label{thm:6}
Let $(\X, \Y)$ be a bivariate time series governed by either an hVAR($p$) or hNAAR($p$) model with non-negative coefficients and continuous marginal distributions. 
Suppose further that the process is $\alpha$-mixing with mixing coefficients $\alpha_n$ satisfying
\begin{align*}
    \sum_{n=1}^\infty (n+1)^7 \alpha_n^{\tau} <\infty, \ \text{ for all } \tau\in (0, \tfrac{1}{2}).
\end{align*} 
Assume that the extremal delay parameter $p$ is fixed and correctly specified, and that the block length satisfies 
\begin{align*}
  \ell = O(n^{\varepsilon}), \ \text{ for some }
  \varepsilon \in \left(0, \tfrac{1}{2}\right).
\end{align*}
Then, the moving block bootstrap procedure with time-shifting, as described in Section \ref{sec:mbb}, yields a consistent test for causal dependence in extremes,
provided that $\hat{\Gamma}_{X\to h(Y)}(p)$ is a compactly differentiable functional of the empirical measure. %
Specifically:
\begin{enumerate}
    \item Under $H_0$, the bootstrap distribution consistently approximates the sampling distribution of $\hat{\Gamma}_{X\to h(Y)}(p)$, and the bootstrap $p$-value is asymptotically uniformly distributed on $[0,1]$.
    \item Under $H_A$, the bootstrap $p$-value converges to zero in probability, so the test rejects the null with probability tending to one as $n\to\infty$.
\end{enumerate}
\end{theorem}

\begin{remark}
Instead of using the nonparametric estimator in \eqref{estimator} as the test statistic, we establish a tighter consistency proof under a slightly stronger assumption that $\hat{\Gamma}_{X\to h(Y)}(p)$ is a compactly differentiable functional. This property holds, for example, for the estimator
\begin{align}\label{eq:estimator_u}
    \hat{\Gamma}_{{\X} \rightarrow {h(\Y})} (p) = \frac{1}{k} \sum_{i=1}^{n-p} h \left(\hat{F}_Y(y_{i+1}), \ldots, \hat{F}_Y(y_{i+p}) \right) 
    \cdot \1_{ \{x_i \geq u \} },
\end{align}
where $\hat{F}_Y(y) = \frac{1}{n}\sum_{j=1}^n \1_{ \{y_j \leq y\} }$ is the empirical distribution function of $Y$, and the threshold $u$ is fixed and non-random (\textit{i.e.}, independent of the data). 
This stronger assumption circumvents the well-known challenges in proving bootstrap consistency for empirical tail processes. Extending the proof to data-dependent thresholds remains an avenue for future work. 
Nevertheless, the simulation results in Section~\ref{sec:simulation} indicate that the proposed moving block bootstrap test with time-shifting performs well in practice even when a data-dependent threshold is used.
\end{remark}

\begin{proof}[Proof of Theorem 5.]
We prove the result by adapting Theorem~2.1 of \citet{NR1994} to account for the time-shifting step and the test statistic $\hat\Gamma_n := \hat{\Gamma}_{X\to h(Y)}(p)$ defined in \eqref{eq:estimator_u}.

Under the assumption that the extremal delay correctly identifies the true causal lag $p$ in an autoregressive model, the time-shifting operation defined by $Y'_t = Y_{(t-p) \bmod n}$ removes causal dependence within the extremal delay window by breaking the temporal alignment between the cause series $\X$ and the effect series $\Y$. More generally, this procedure is valid provided that $\X$ does not Granger cause the shifted process $\Y'$ at lag order $p$. When applied prior to resampling, the shifting step ensures that the resulting bootstrap samples are, by construction, drawn from a process consistent with the null hypothesis that $\X$ does not cause $\Y$.

Regarding the statistic, we have already established in Theorem~\ref{thm:5} that $\hat{\Gamma}_n$ is consistent. It follows from Theorem~\ref{thm:1} that, under the null hypothesis, $\hat{\Gamma}_n \xlongrightarrow{\PP} \gamma < 1$, and under the alternative, $\hat{\Gamma}_n \xlongrightarrow{\PP} 1$.
Furthermore, each term $\hat{F}_Y(y_{i+j})$ in $\hat{\Gamma}_n$ is a step function of the empirical measure, and the impact function $h$ is continuously differentiable. Therefore, by the chain rule, the statistic $\hat{\Gamma}_n$ is a smooth (Hadamard differentiable) functional of the empirical process, allowing us to invoke the block bootstrap consistency results from \citet{NR1994}.

We first establish validity under the null.
Let $\hat{\Gamma}_n^j$, for $j = 1, \dots, b$, denote the $j$-th bootstrap replicate computed on resampled blocks from the shifted series $(X_t, Y'_t)$. Theorem~2.1 of \citet{NR1994} states that for smooth statistical functionals, the block bootstrap distribution converges weakly in probability to the same limit as the distribution of the test statistic under the null. Therefore, under $H_0$,
\[
    \hat{\Gamma}_n^j
    \xlongequal{d} 
    \hat{\Gamma}_{\X \rightarrow h(\Y')}(p)
    \xlongequal{d} 
    \hat{\Gamma}_{\X \rightarrow h(\Y)}(p) ,
\]
where $\xlongequal{d}$ denotes asymptotic equivalence in distribution as $n \to \infty$.
Hence, the one-sided bootstrap $p$-value satisfies
\[
    p_n := \frac{1}{b} \sum_{j=1}^b \1_{ \left\{ \hat{\Gamma}_n^j \geq \hat{\Gamma}_n \right\} }
    \xlongrightarrow{d} \text{Uniform}[0,1], 
    \ \text{ under } H_0.
\]

We now establish consistency under the alternative. 
Under $H_A$, where $\X$ causes $\Y$,
\[
    \hat{\Gamma}_{\X \rightarrow h(\Y)}(p) \xlongrightarrow{\PP} 1, \quad 
    \hat{\Gamma}_{\X \rightarrow h(\Y')}(p) \xlongrightarrow{\PP} \gamma < 1.
\]
Thus, the observed statistic converges in probability to $1$, while the bootstrap replicates converge to a smaller value $\gamma < 1$, as guaranteed by Theorem~2.1 of \citet{NR1994}:
\[
    \hat\Gamma_n \xlongrightarrow{\PP} 1, \quad
    \hat\Gamma_n^j \xlongrightarrow{\PP} \gamma < 1 .
\]
This separation arises because the time-shifting step removes causal dependence, thereby enforcing the null structure.
It follows that $\PP\left( \hat\Gamma_n^j \geq \hat\Gamma_n \right) \to 0$, and thus,
\[
    p_n := \frac{1}{b} \sum_{j=1}^b \1_{ \left\{ \hat{\Gamma}_n^j \geq \hat{\Gamma}_n \right\} }
    \xlongrightarrow{\PP} 0, \ \text{ under } H_A.
\]
Therefore, the test rejects the null hypothesis with probability tending to one.

In conclusion, the moving block bootstrap procedure with time-shifting yields a valid level-$\alpha$ test under the null and is consistent under the alternative.
\end{proof}

The mixing condition required in Theorem~\ref{thm:6} follows exactly that imposed by \citet{NR1994} for establishing block bootstrap consistency. This condition is automatically satisfied by any stable VAR($p$) process, which is known to be geometrically $\beta$-mixing \citep{Mokkadem1988}. 
In contrast, NAAR($p$) processes require additional assumptions to ensure geometric mixing.
Specifically, contractive conditions must be imposed on the component functions $f_i$ and $g_i$, for $i = 0, \dots, p$, along with regularity assumptions on the innovations, such as the existence of finite moments of sufficiently high order.
A sufficient condition for contractivity is that each $f_i$ and $g_i$ is Lipschitz continuous with constants $L_{f,i}, L_{g,i} > 0$ such that, for all $x, x' \in \R$,
\[
    \left|f_i(x) - f_i(x') \right| \leq L_{f,i} \left|x - x' \right|, \quad \left|g_i(x) - g_i(x') \right| \leq L_{g,i} \left|x - x' \right|, 
\]
and the sum of Lipschitz constants across all lags satisfies
\[
    \sum_{i=0}^p (L_{f,i} + L_{g,i}) < 1.
\]
See, for example, Section~2.4.2 of \citet{Doukhan1994} for further details.
The specifications for Models \ref{model:7} and \ref{model:9} in Section~\ref{sec:mc_setup} satisfy the geometric mixing assumption. However, the heavy-tailed noise employed in our simulations does not possess high-order moments. This condition would be met if less extreme noise were used, such as Pareto distribution with a larger shape parameter or Student’s $t$ distribution with higher degrees of freedom. Nevertheless, the proposed hypothesis testing procedure is expected to perform well even when the moment condition is violated.

\section{Monte Carlo simulations}\label{sec:setup}
\subsection{Setup}\label{sec:mc_setup}
This section details the model specifications used to study the finite-sample behavior of $\hat{\Gamma}_{{\X} \rightarrow {h(\Y})}$ and to evaluate the performance of our bootstrap hypothesis test in Section~\ref{sec:simulation}.
We start with some basic VAR(3) models. Model \ref{model:1} is a baseline random noise model, where there is no causality between $X$ and $Y$. In Model \ref{model:2}, $X$ is an AR(1) process that has a straightforward causal impact on $Y$ with a time delay of 3; in Model \ref{model:3}, this single lag-3 causal impact of $X$ on $Y$ is complicated by the autoregressive structures present in both the cause and the effect series. A causal loop where both $X$ causes $Y$ and $Y$ causes $X$ is considered in Model \ref{model:4}.

\begin{model}\label{model:1}
Independent noise model: 
\begin{align*}
    X_t &= \epsilon^X_t , \\
    Y_t &= \epsilon^Y_t .
\end{align*}
\end{model}

\begin{model}\label{model:2}
Single-lag model: 
\begin{align*}
    X_t &= 0.5 X_{t-1} + \epsilon^X_t , \\
    Y_t &= 0.5 X_{t-3} + \epsilon^Y_t .
\end{align*}
\end{model}

\begin{model}\label{model:3}
Single-lag autoregressive-effect model: 
\begin{align*}
    X_t &= 0.5 X_{t-1} + \epsilon^X_t , \\
    Y_t &= 0.5 Y_{t-1} + 0.5 X_{t-3} + \epsilon^Y_t .
\end{align*}
\end{model}

\begin{model}\label{model:4}
Single-lag bidirectional-causality model: 
\begin{align*}
    X_t &= 0.25 X_{t-1} + 0.5 Y_{t-3} + \epsilon^X_t , \\
    Y_t &= 0.25 Y_{t-1} + 0.5 X_{t-3} + \epsilon^Y_t .
\end{align*}
\end{model}

Models \ref{model:5} and \ref{model:6} below extend single-lag VAR models \ref{model:2} and \ref{model:3} to include cross-dependence in multiple lags to reflect the compounding effect in extremes.

\begin{model}\label{model:5}
Multi-lag model: 
\begin{align*}
    X_t &= 0.5 X_{t-1} + \epsilon^X_t , \\
    Y_t &= 0.25 X_{t-1} + 0.25 X_{t-2} + 0.25 X_{t-3} + \epsilon^Y_t .
\end{align*}
\end{model}
\noindent 

\begin{figure}[htbp]
  \centering

  \begin{minipage}{0.33\textwidth}
    \centering
    \includegraphics[width=\linewidth]{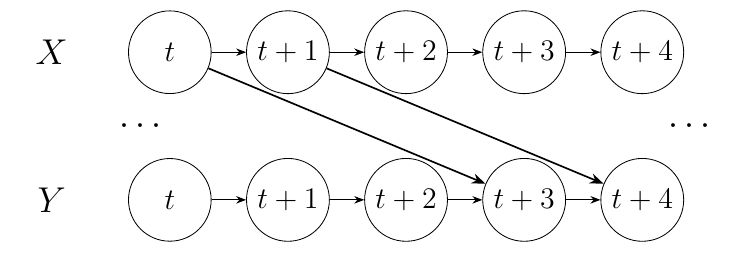}
    {\footnotesize Model \ref{model:3}}
  \end{minipage}\hfill
  \begin{minipage}{0.33\textwidth}
    \centering
    \includegraphics[width=\linewidth]{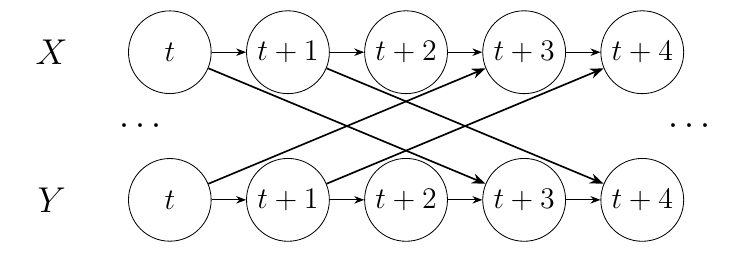}
    {\footnotesize Model \ref{model:4}}
  \end{minipage}\hfill
  \begin{minipage}{0.33\textwidth}
    \centering
    \includegraphics[width=\linewidth]{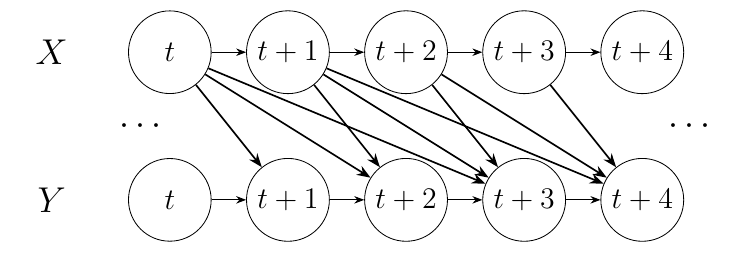}
    {\footnotesize Model \ref{model:6}}
  \end{minipage}

  \caption{\footnotesize Directed acyclic graph representation of the causal configurations between two time series.}
  \label{fig:DAG}
\end{figure}

\begin{model}\label{model:6}
Multi-lag autoregressive-effect model: 
\begin{align*}
    X_t &= 0.5 X_{t-1} + \epsilon^X_t , \\
    Y_t &= 0.5 Y_{t-1} + 0.25 X_{t-1} + 0.25 X_{t-2} + 0.25 X_{t-3} + \epsilon^Y_t .
\end{align*}
\end{model}
\noindent A visual comparison of Models \ref{model:3}, \ref{model:4}, and \ref{model:6} is provided using directed acyclic graphs in Figure \ref{fig:DAG}.

Model \ref{model:7} introduces nonlinearity into the single-lag VAR model \ref{model:3}. $X$ has a nonlinear causal impact on $Y$ only when $X$ lies in the upper tail of its distribution. This captures a causal relationship that is specifically tied to extreme events. 
\begin{model}\label{model:7}
Thresholded NAAR(3) model: 
\begin{align*}
    X_t &= 0.5 X_{t-1} + \epsilon^X_t , \\
    Y_t &= 0.5 Y_{t-1} + (X_{t-3})^\frac{3}{4}\1_{ \{ X_{t-3}>u_X \} } +  \epsilon^Y_t ,
\end{align*}
where $u_X$ denotes a high threshold that depends on the distribution of the noise variable $\epsilon^X_t$.
\end{model}

Lastly, to test the robustness of our proposed causal discovery method in the presence of hidden confounders, Models \ref{model:8} and \ref{model:9} add a common cause variable $Z$ to Models \ref{model:2} and \ref{model:7}, respectively. We assume $\ZZ$ is an AR(1) process that has the same type of noise distribution as $\X$ and $\Y$, and the tails of $\epsilon^Z_t$ are no heavier than those of $\epsilon^X_t$ and $\epsilon^Y_t$. 

\begin{model}\label{model:8}
VAR(3) model with a confounding time series $\ZZ$: 
\begin{align*}
    X_t &= 0.5 X_{t-1} + 0.5 Z_{t-2} + \epsilon^X_t , \\
    Y_t &= 0.5 X_{t-3} + 0.5 Z_{t-1} + \epsilon^Y_t , \\
    Z_t &= 0.5 Z_{t-1} + \epsilon^Z_t . 
\end{align*}
\end{model}

\begin{model}\label{model:9}
Thresholded NAAR(3) model with a confounding time series $\ZZ$: 
\begin{align*}
    X_t &= 0.5 X_{t-1} + 0.5 Z_{t-2} + \epsilon^X_t , \\
    Y_t &= 0.5 Y_{t-1} + 0.5 Z_{t-1} + (X_{t-3})^\frac{3}{4}\1_{ \{ X_{t-3}>u_X \} } + \epsilon^Y_t , \\
    Z_t &= 0.5 Z_{t-1} + \epsilon^Z_t ,
\end{align*}
where $u_X$ denotes a high threshold that depends on the distribution of the noise variable $\epsilon^X_t$.
\end{model}

For each model, we considered three noise distributions for $\epsilon^X_t$, $\epsilon^Y_t$, and $\epsilon^Z_t$: Student’s $t$, standard Pareto, and Poisson$(3)$. Unless otherwise specified, the degrees of freedom (df) were set to $10$ for $\epsilon^X_t$ and $\epsilon^Z_t$, and $2$ for $\epsilon^Y_t$ in the $t$ case. These distributions represent a range of tail behaviors.
Pareto and Student’s $t$ with small df are heavy-tailed, exhibiting polynomial tail decay: Pareto(1,1) has infinite mean and is prone to extreme values, whereas the Student’s $t$ distribution with $df = 2$ has finite mean but infinite variance, with tails lighter than Pareto yet still heavier than exponential families. In contrast, the Poisson distribution is light-tailed, with exponentially decaying tail probabilities that make extreme values rare. Note that Poisson noise violates the regular variation assumptions of Section \ref{sec:prelim}.

Unless otherwise stated, each time series had length $n=1000$, with the largest $k=\sqrt{n}$ values treated as extremes. This choice of $k$, following \citet{BPP2024}, is pragmatic but potentially suboptimal (see Section~\ref{sec:threshold}). For each simulated series under a given model and noise type, we estimated the observed causal tail coefficient $\hat{\Gamma}_{{\X} \rightarrow {h(\Y})}^{obs}$ and conducted a block bootstrap test. 
The extremal delay was assumed to be known and fixed at the oracle value $p=3$ across all models and methods. This corresponds to the largest cross-series causal lag specified by the data-generating processes and was also used in the non-causal directions to ensure comparability.
A sensitivity analysis of the choice of $p$ is presented later in Section~\ref{sec:sensi_lag}.
The block length was set to $\ell=n^{1/3}$. Its selection involves a bias--variance trade-off: larger blocks better preserve serial dependence but reduce the effective number of blocks available for resampling.
We generated $b=100$ bootstrap samples, each yielding an estimate $\hat{\Gamma}_{{\X} \rightarrow {h(\Y})}^{(j)}$, $j=1,\dots,b$, and computed a one-sided $p$-value as the proportion of bootstrap estimates no smaller than $\hat{\Gamma}_{{\X} \rightarrow {h(\Y})}^{obs}$. This procedure was repeated 100 times to obtain Monte Carlo distributions of both the coefficients and the $p$-values.

\subsection{Results}\label{sec:mc_results}
\begin{figure}[htbp]
    \centering
    \begin{minipage}{0.4\textwidth}
        \centering
        \includegraphics[width=\linewidth]{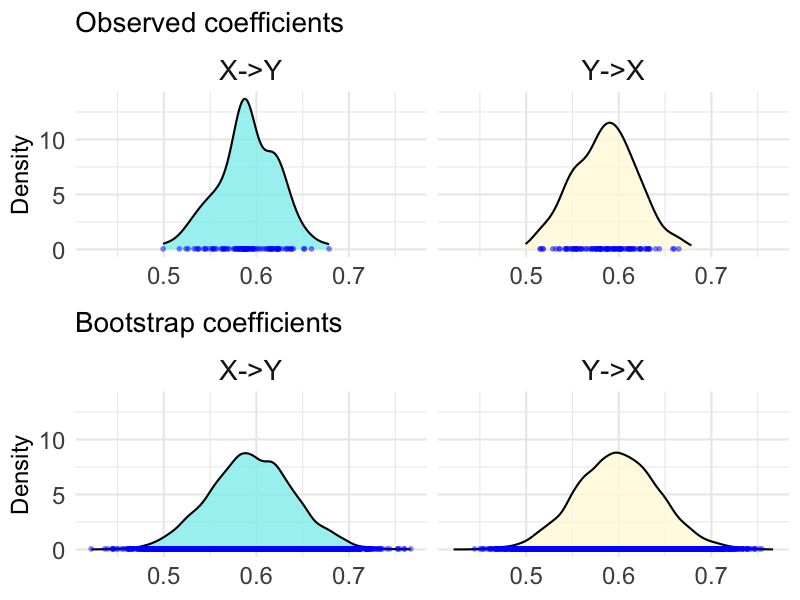}
    \end{minipage}%
    \hspace{0cm}
    \begin{minipage}{0.4\textwidth}
        \centering
        \includegraphics[width=\linewidth]{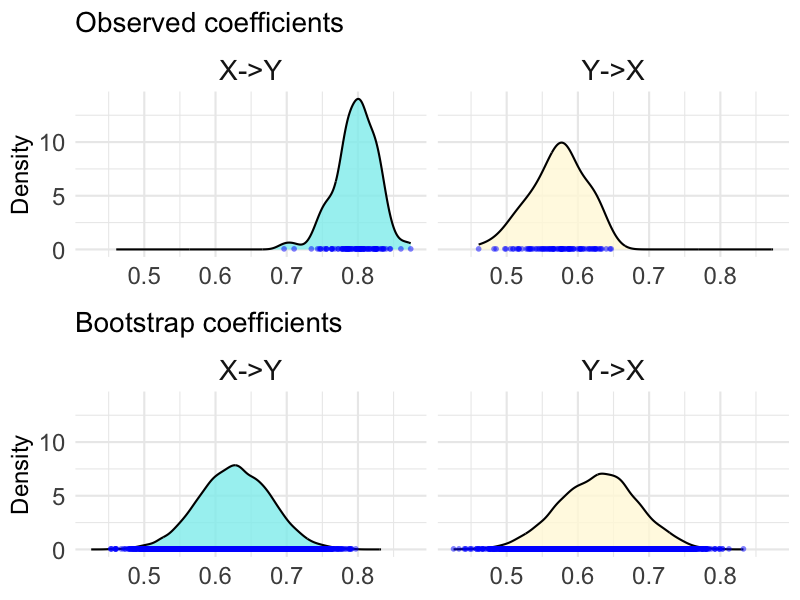}
    \end{minipage}

    \vspace{0cm}

    \begin{minipage}{0.4\textwidth}
        \centering
        \includegraphics[width=\linewidth]{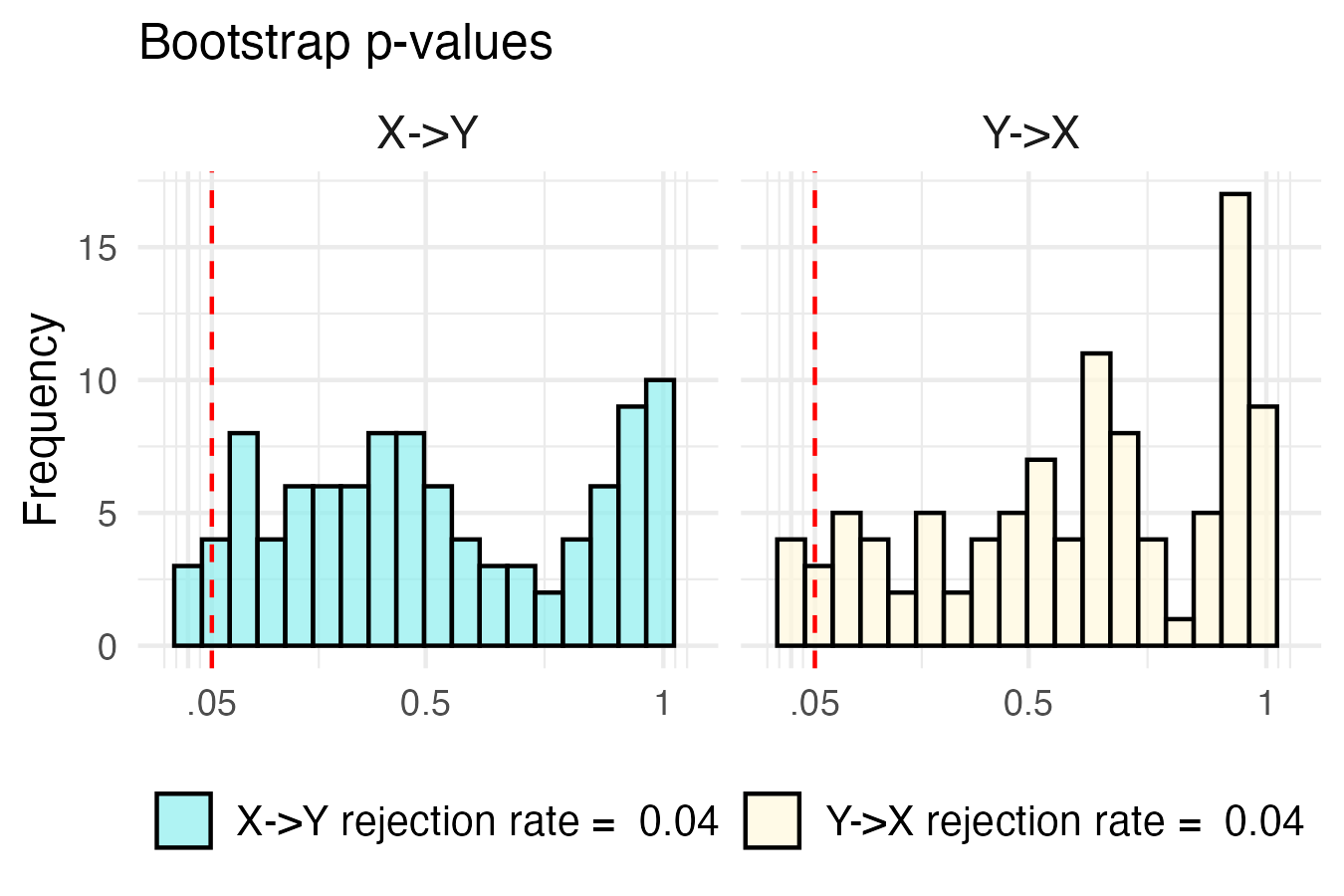}
        {Model \ref{model:1}: independent noise}
    \end{minipage}%
    \hspace{0cm}
    \begin{minipage}{0.4\textwidth}
        \centering
        \includegraphics[width=\linewidth]{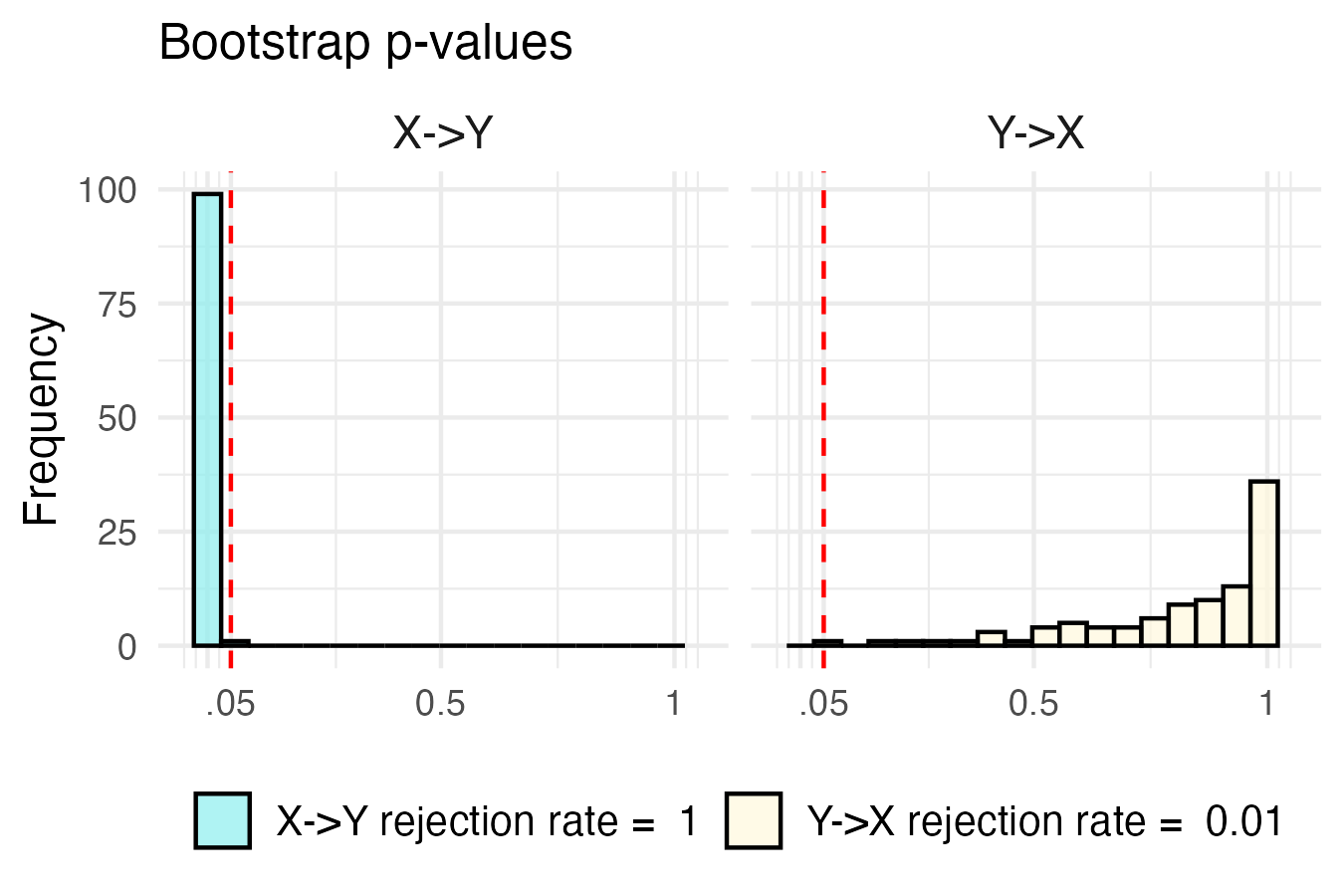}
        {Model \ref{model:5}: multi-lag VAR}
    \end{minipage}

    \vspace{0.8cm}
    
    \begin{minipage}{0.4\textwidth}
        \centering
        \includegraphics[width=\linewidth]{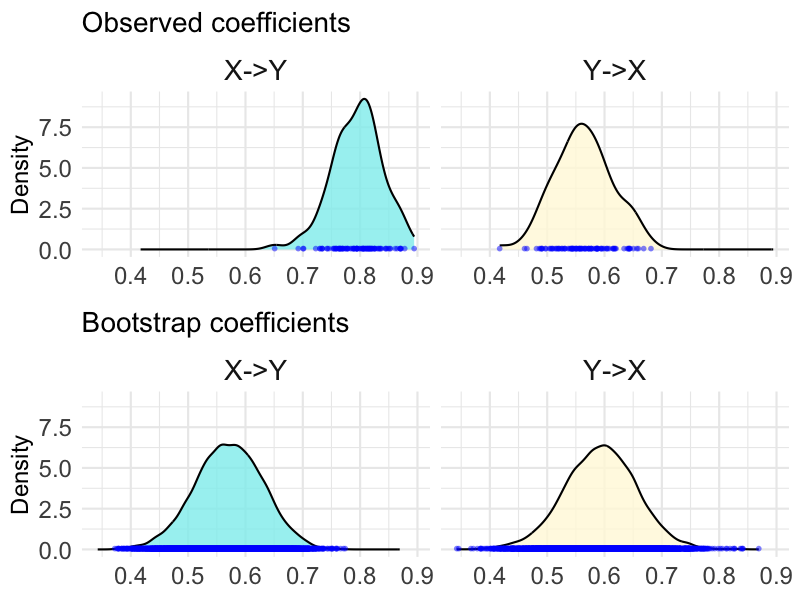}
    \end{minipage}%
    \hspace{0cm}
    \begin{minipage}{0.4\textwidth}
        \centering
        \includegraphics[width=\linewidth]{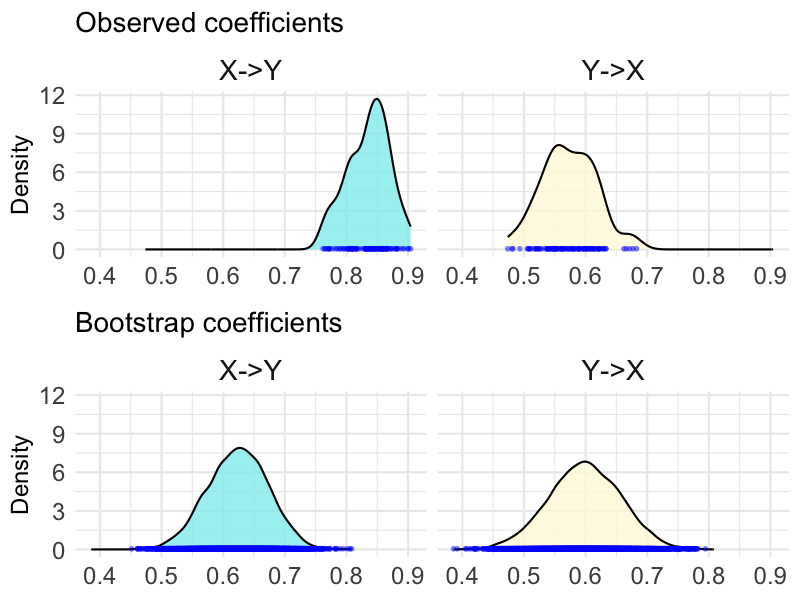}
    \end{minipage}

    \vspace{0cm}

    \begin{minipage}{0.4\textwidth}
        \centering
        \includegraphics[width=\linewidth]{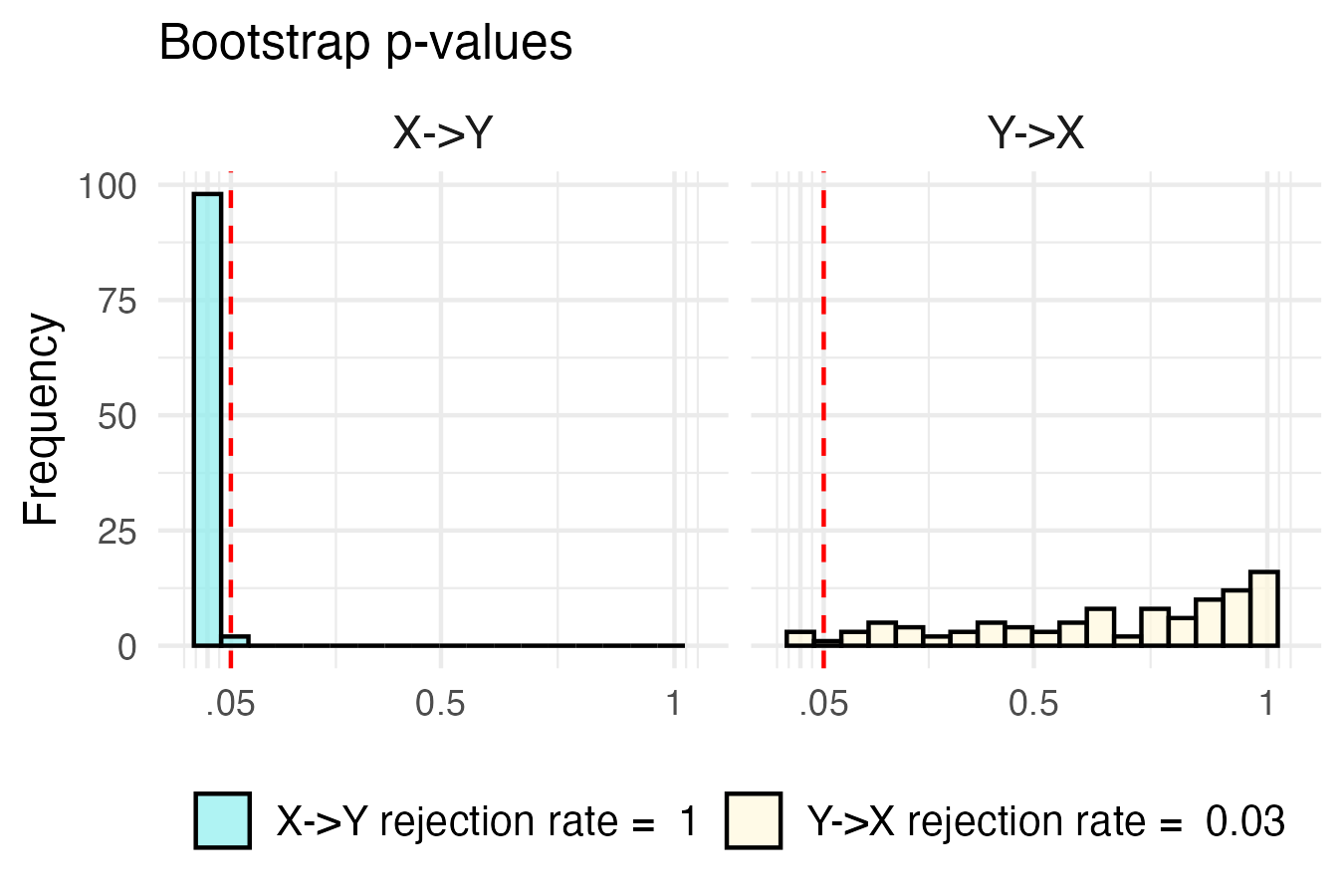}
        {Model \ref{model:7}: thresholded NAAR}
    \end{minipage}%
    \hspace{0cm}
    \begin{minipage}{0.4\textwidth}
        \centering
        \includegraphics[width=\linewidth]{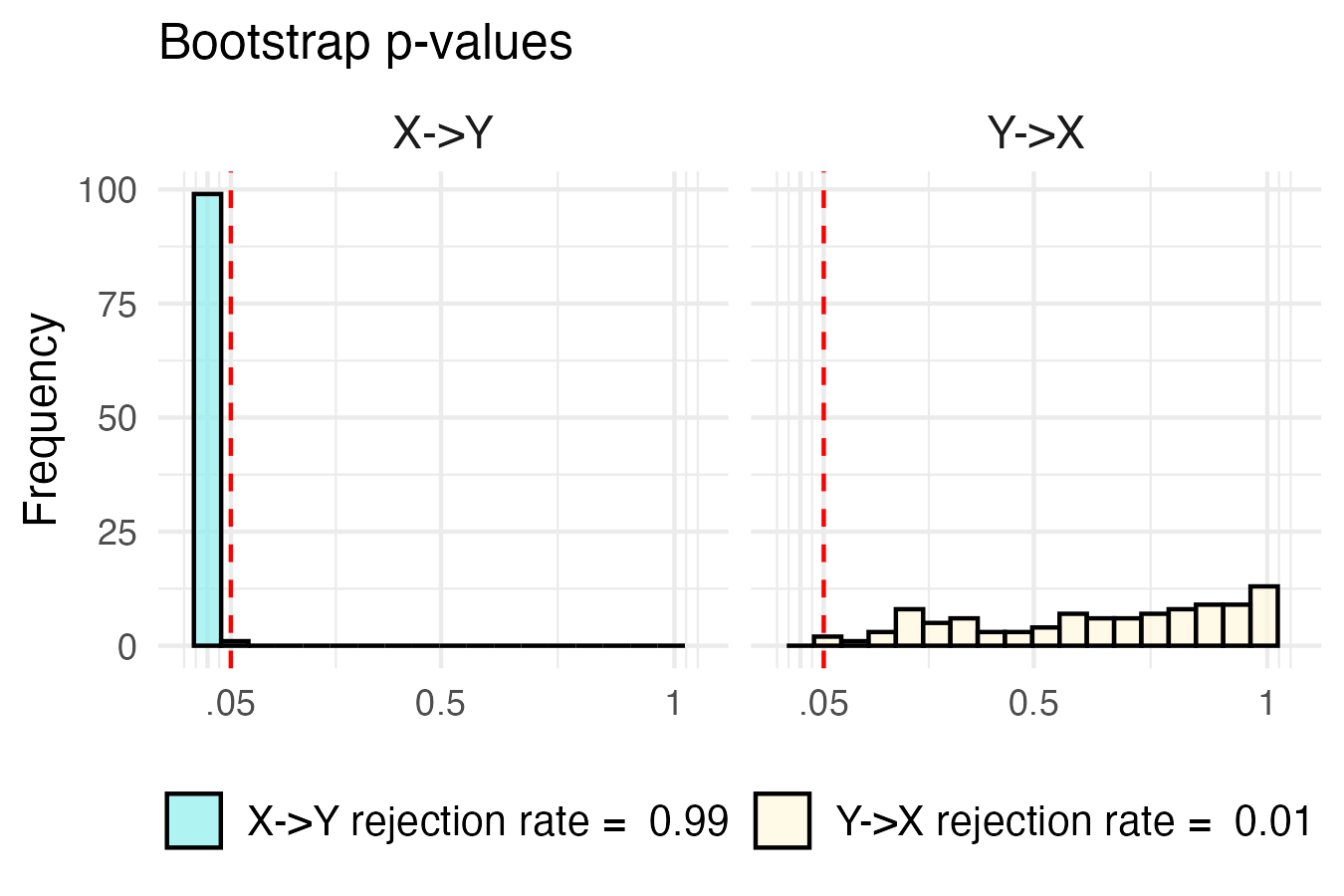}
        {Model \ref{model:9}: NAAR with confounder}
    \end{minipage}

    \caption{\footnotesize Repeated simulation results of bootstrap hypothesis test on pair of time series with Student's $t$ noise.  
    Each density plot displays the kernel density estimate of observed or bootstrapped causal tail coefficient values, aggregated over 100 time series realizations. Each accompanying histogram shows the distribution of bootstrap $p$-values across the same 100 realizations. Blue corresponds to the true causal direction $X \rightarrow Y$, and yellow to the reverse direction $Y \rightarrow X$.}
    \label{fig:results}
\end{figure}

Figure~\ref{fig:results} shows selected Monte Carlo simulation results. Histograms (bottom row) give $p$-value distributions from 100 independent simulations of the same model; kernel densities show observed (top row) and bootstrap (middle row) compound CTCs aggregated across simulations. Blue denotes the true causal direction $X \to Y$ and yellow the non-causal direction $Y \to X$ for Models~\ref{model:5}, \ref{model:7}, and \ref{model:9}.
Simulation results for additional models with varying noise types are provided in Section~\ref{sec:compare}, with corresponding Tables~\ref{tab:t}, \ref{tab:pareto}, and \ref{tab:poisson}.

The distributions of observed and bootstrapped CTCs vary across models. In the independent noise model, densities for $X \to Y$ and $Y \to X$ are nearly identical and centered at small values. By contrast, in Models~\ref{model:5}, \ref{model:7}, and \ref{model:9}, the bootstrapped coefficients not only diverge between directions but also deviate from the independent case. Within each model, bootstrapped coefficients are typically smaller than the observed coefficients in the true causal direction. These patterns suggest that the time-shifting step before block bootstrapping effectively removes direct causality while preserving dependence from feedback structures.

The $p$-values from repeated simulations show that the block bootstrap test reliably detects true causal effects while controlling Type I errors. In Model~\ref{model:1}, false positives occur at about $5\%$, matching the test’s significance level. Models~\ref{model:5}, \ref{model:7}, and \ref{model:9} yield results consistent with their model specifications: for the true causal direction $X \to Y$, nearly all tests reject the null, whereas rejection rates for $Y \to X$ remain very low. These findings align with the distributional contrasts between observed and bootstrapped CTCs: in the true direction, observed coefficients are larger and nearly disjoint from the bootstrap distribution, whereas in the reverse direction they overlap and the observed values lie slightly to the left.

Including a confounder in Model~\ref{model:9} shifts the distributions of observed and bootstrapped coefficients in both directions relative to Model~\ref{model:7}. In particular, the observed coefficients shift to the right, aligning with the intuition that a common cause induces spurious causality between $X$ and $Y$. Nonetheless, the distribution of bootstrap $p$-values remains largely unchanged, suggesting that our causal discovery framework is robust to hidden confounders.

Beyond its strong accuracy in simulations, our method is computationally efficient. Estimating a compound CTC with \texttt{DEoptim} has complexity $O(n \log n + \sqrt{n} p^2)$, and the one-sided bootstrap test $O\left(n^{4/3} + b(n \log n + \sqrt{n} p^2)\right)$, where $n$ is the time series length, $p$ the extremal delay, and $b$ the number of bootstrap samples. For $n=1000$, $p=3$, and $b=100$, bidirectional testing takes about 0.6s on a 2021 MacBook Pro (M1 Pro, 16GB RAM).

\section{Sensitivity analysis}\label{sec:sensitivity}
In this section, we present simulation results showing that both the estimated compound causal tail coefficients and the bootstrap hypothesis test in Section~\ref{sec:mc_results} remain robust across a broad range of parameter settings. 

\subsection{Impact shape parameter}\label{sec:shape}

\begin{figure}[htbp]
    \centering
    \includegraphics[width=0.9\linewidth]{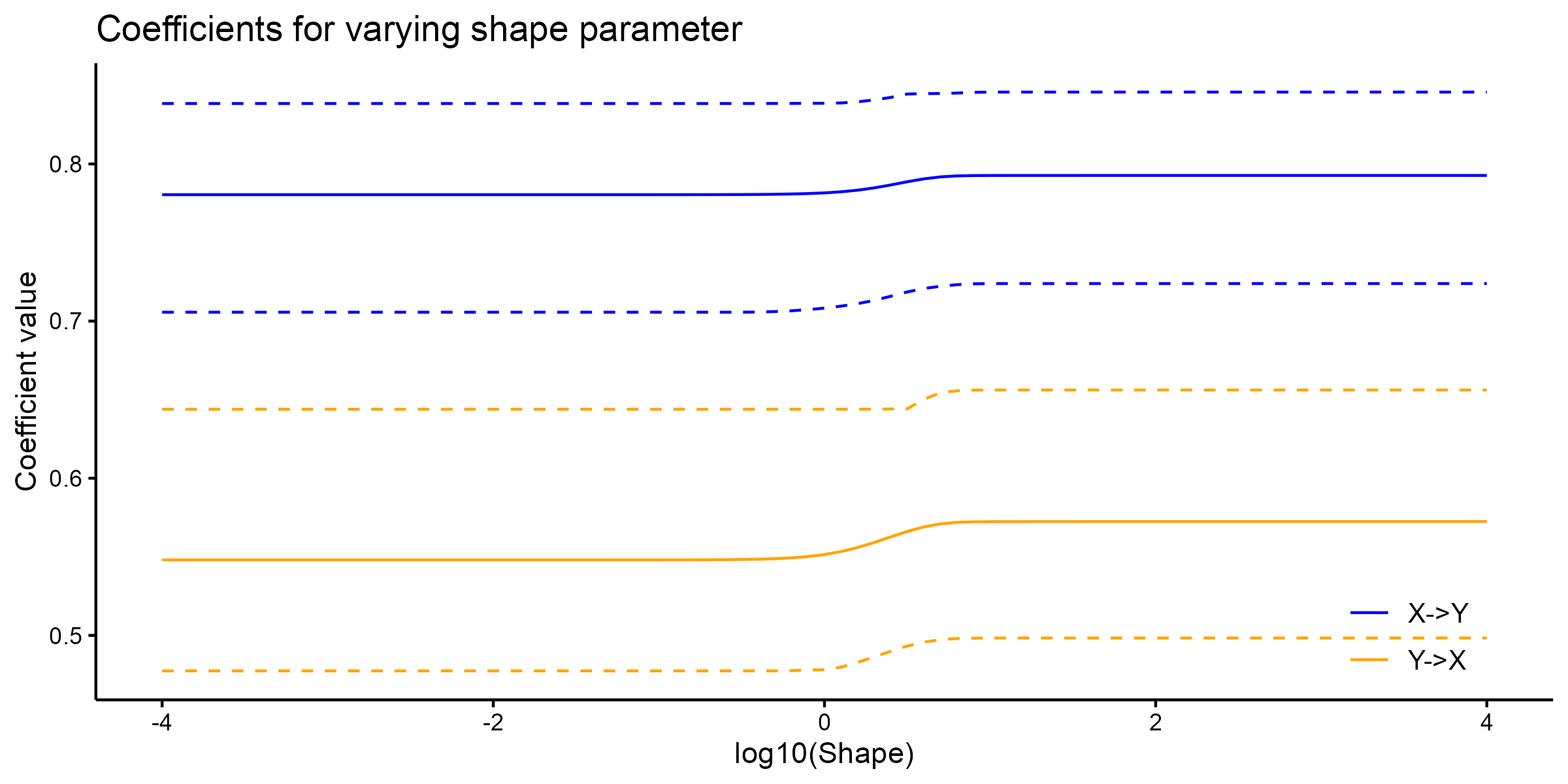}
    \caption{\footnotesize Behavior of the estimators $\hat{\Gamma}_{{\X} \rightarrow {h(\Y)}}(3)$ (blue) and $\hat{\Gamma}_{{\Y} \rightarrow {h(\X)}}(3)$ (yellow) across varying values of the shape parameter $\alpha$ in the impact function. 
    Solid lines indicate the mean coefficient estimates across 100 realizations, while dotted lines denote the empirical 5\% and 95\% pointwise quantiles.
    }
    \label{fig:shape}
\end{figure}

We begin with the impact shape parameter $\alpha$ in the impact function \eqref{impactfunction}, using Model~\ref{model:6}, the multi-lag autoregressive-effect model, to illustrate how the compound causal tail coefficient $\hat{\Gamma}_{{\X} \rightarrow {h(\Y)}}(p)$ varies with $\alpha$.
Similar to the approach in Section \ref{sec:mc_setup}, we generated time series of length $n=1000$ from Model~\ref{model:6} with Student's $t$ noise and computed $\hat{\Gamma}_{{\X} \rightarrow {h(\Y)}}(p)$ and $\hat{\Gamma}_{{\Y} \rightarrow {h(\X)}}(p)$ using $k=\sqrt{n}$ extreme values and the true extremal delay $p=3$, across a range of impact shape parameter values. Specifically, $\alpha$ was varied over a log-scale grid from $10^{-4}$ to $10^{4}$ in steps of $0.1$ on the exponent. This procedure was repeated 100 times, and Figure~\ref{fig:shape} displays the mean along with the 5\% and 95\% empirical quantiles of the resulting estimates.

The coefficient values in both the correct and incorrect directions remain largely flat as the impact shape parameter $\alpha$ varies, with a slight increase near $\sqrt{10}$. Following this rise, the 90\% empirical confidence intervals narrow slightly, while the separation between the two directions remains evident.
Recall that the impact shape parameter $\alpha$ governs the type of cancellation effect among lagged variables in the impact function, with $\alpha\to \infty$ corresponding to a nil cancellation and $\alpha\to 0$ to a linear cancellation. Although these cancellation effects have only a marginal impact on the causal tail coefficient value, which is also the objective function for optimizing the weight vector $\vw$, they lead to notable differences in the resulting weight distributions, as discussed in Section \ref{sec:weight_distr}.

\begin{figure}[htbp]
    \centering
    {\footnotesize $\alpha=10^4$} \\
    \begin{minipage}{0.4\textwidth}
        \centering
        \includegraphics[width=\linewidth]{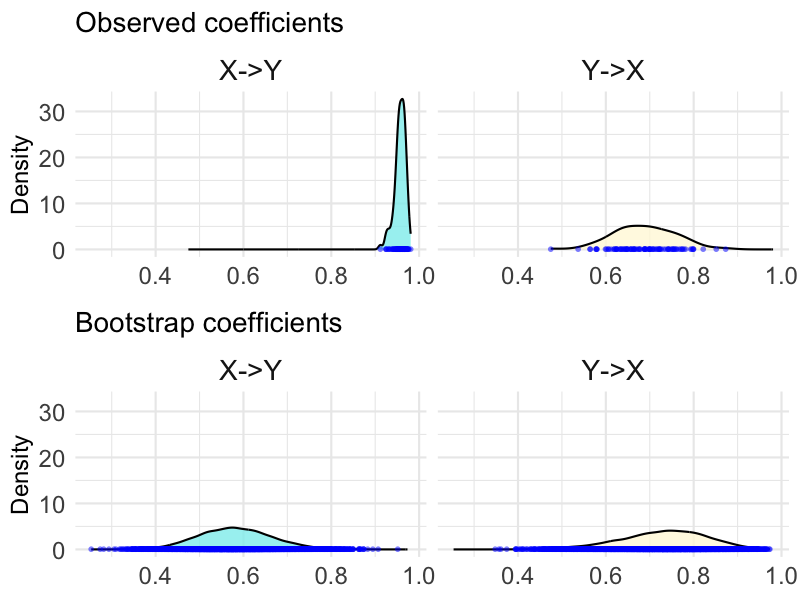}
    \end{minipage}%
    \hspace{0cm}
    \begin{minipage}{0.4\textwidth}
        \centering
        \includegraphics[width=\linewidth]{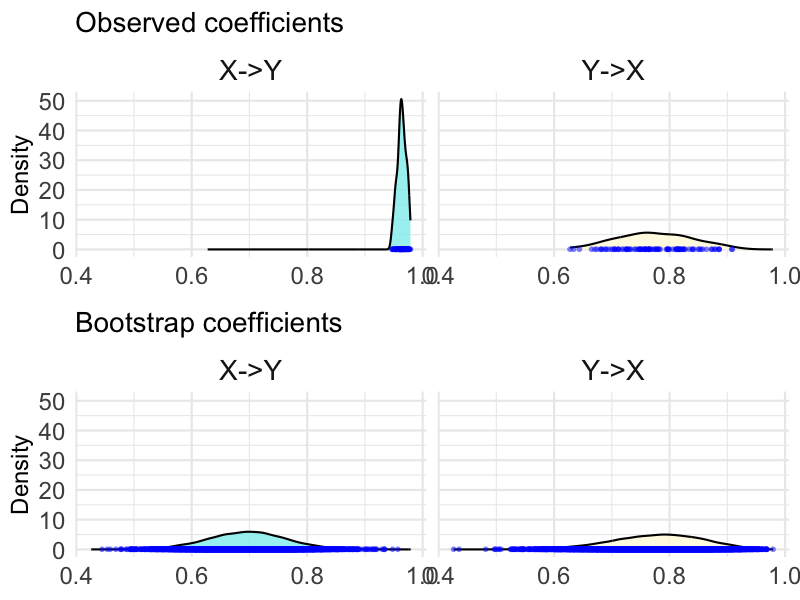}
    \end{minipage}

    \vspace{0cm}
    {\footnotesize $\alpha=1$} \\
    \begin{minipage}{0.4\textwidth}
        \centering
        \includegraphics[width=\linewidth]{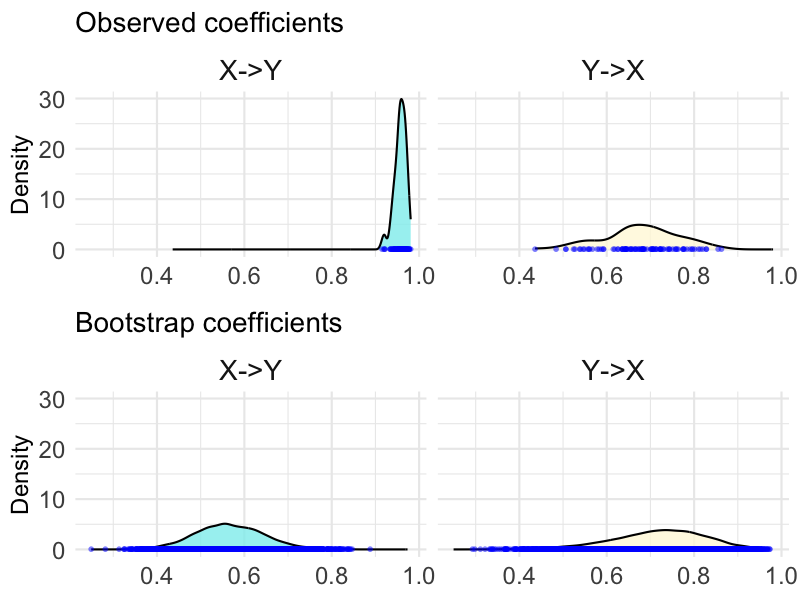}
    \end{minipage}%
    \hspace{0cm}
    \begin{minipage}{0.4\textwidth}
        \centering
        \includegraphics[width=\linewidth]{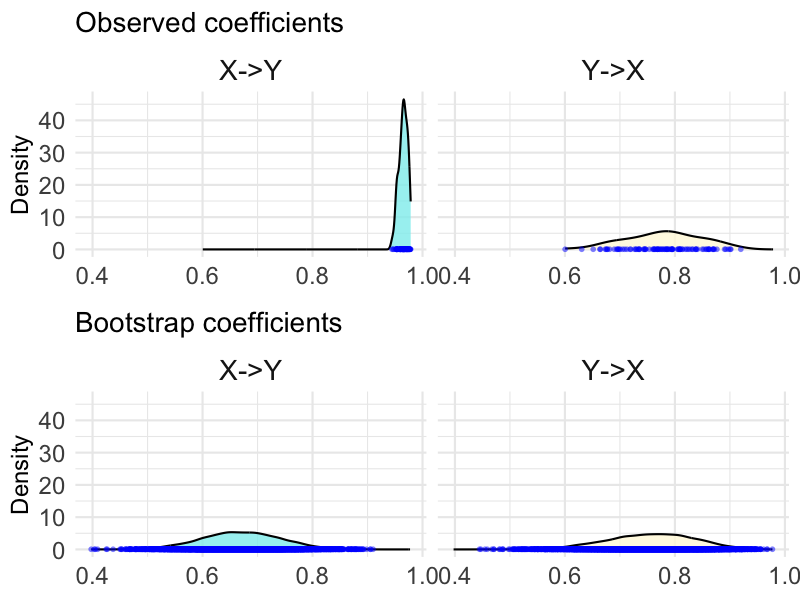}
    \end{minipage}

    \vspace{0cm}
    {\footnotesize $\alpha=10^{-4}$} \\
    \begin{minipage}{0.4\textwidth}
        \centering
        \includegraphics[width=\linewidth]{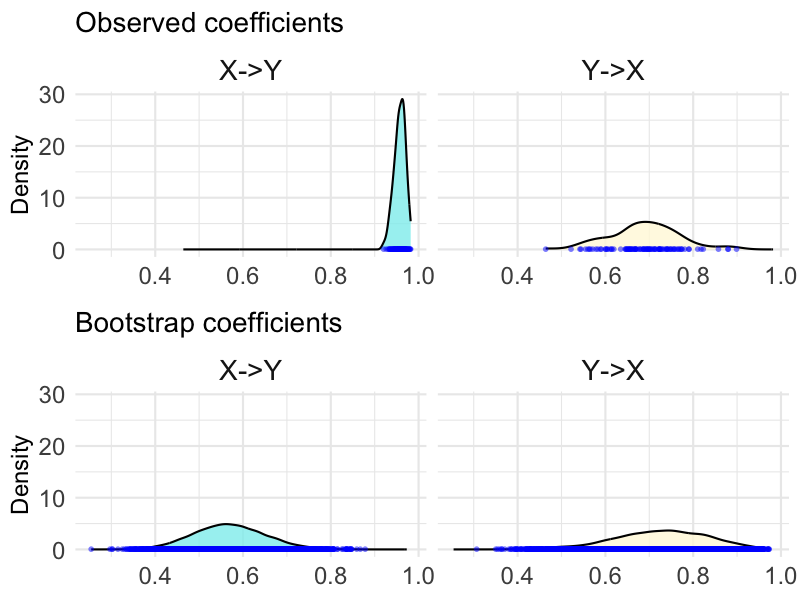} \\
        \small Model \ref{model:3}: single-lag VAR
    \end{minipage}%
    \hspace{0cm}
    \begin{minipage}{0.4\textwidth}
        \centering
        \includegraphics[width=\linewidth]{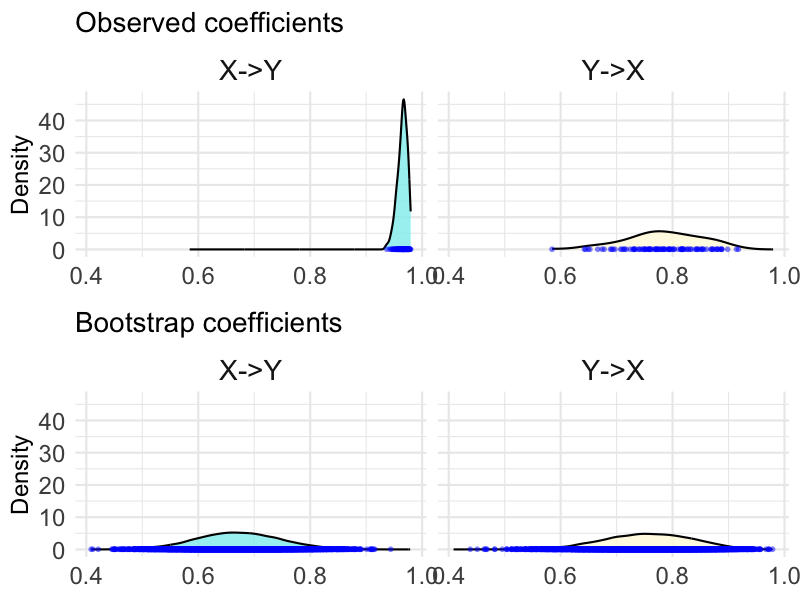} \\
        \small Model \ref{model:5}: multi-lag VAR
    \end{minipage}

    \caption{\footnotesize Comparison of Monte Carlo distributions of observed and bootstrapped causal tail coefficients across different shape parameter $\alpha$ for Models \ref{model:3} and \ref{model:5}.
    Each density plot displays the kernel density estimate of observed or bootstrapped causal tail coefficient values, aggregated over 100 time series realizations. Blue corresponds to the true causal direction $X \rightarrow Y$, and yellow to the reverse direction $Y \rightarrow X$.}
    \label{fig:shape2a}
\end{figure}

\begin{figure}[htbp]
    \centering
    {\footnotesize $\alpha=10^4$} \\
    \begin{minipage}{0.4\textwidth}
        \centering
        \includegraphics[width=\linewidth]{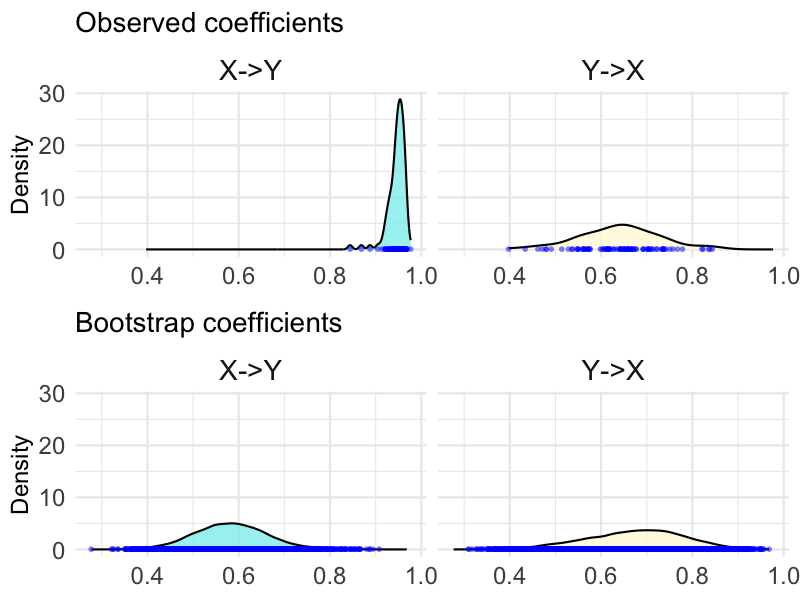}
    \end{minipage}%
    \hspace{0cm}
    \begin{minipage}{0.4\textwidth}
        \centering
        \includegraphics[width=\linewidth]{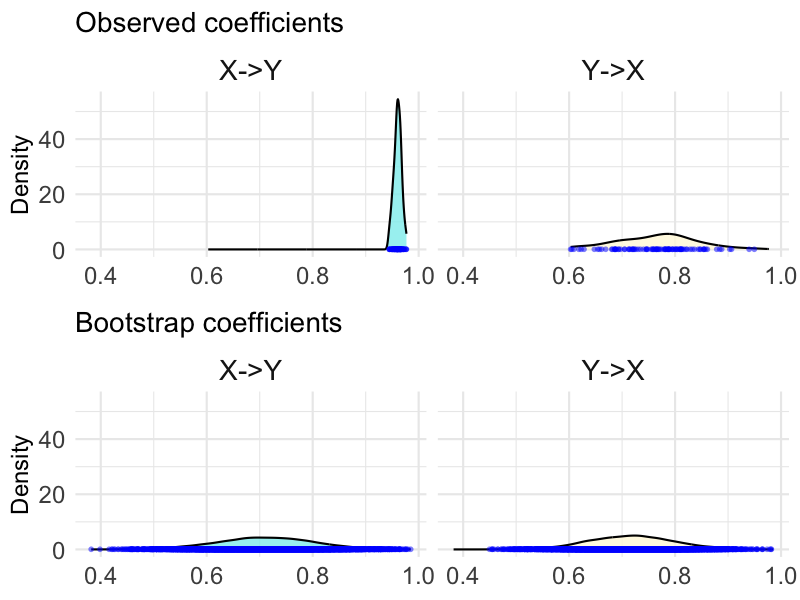}
    \end{minipage}

    \vspace{0cm}
    {\footnotesize $\alpha=1$} \\
    \begin{minipage}{0.4\textwidth}
        \centering
        \includegraphics[width=\linewidth]{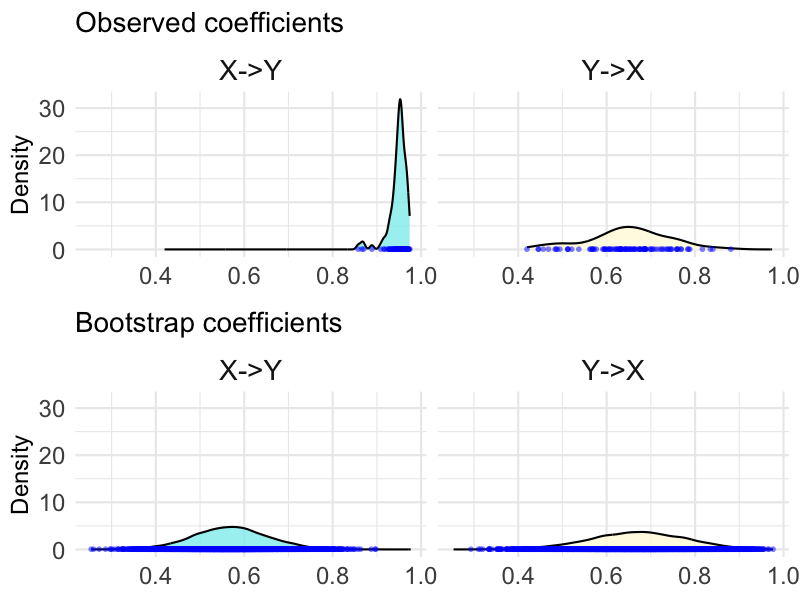}
    \end{minipage}%
    \hspace{0cm}
    \begin{minipage}{0.4\textwidth}
        \centering
        \includegraphics[width=\linewidth]{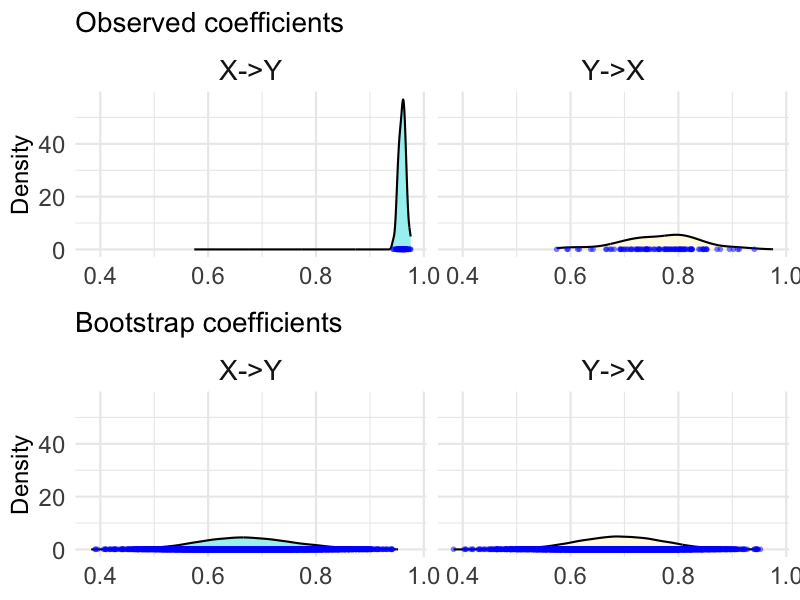}
    \end{minipage}

    \vspace{0cm}
    {\footnotesize $\alpha=10^{-4}$} \\
    \begin{minipage}{0.4\textwidth}
        \centering
        \includegraphics[width=\linewidth]{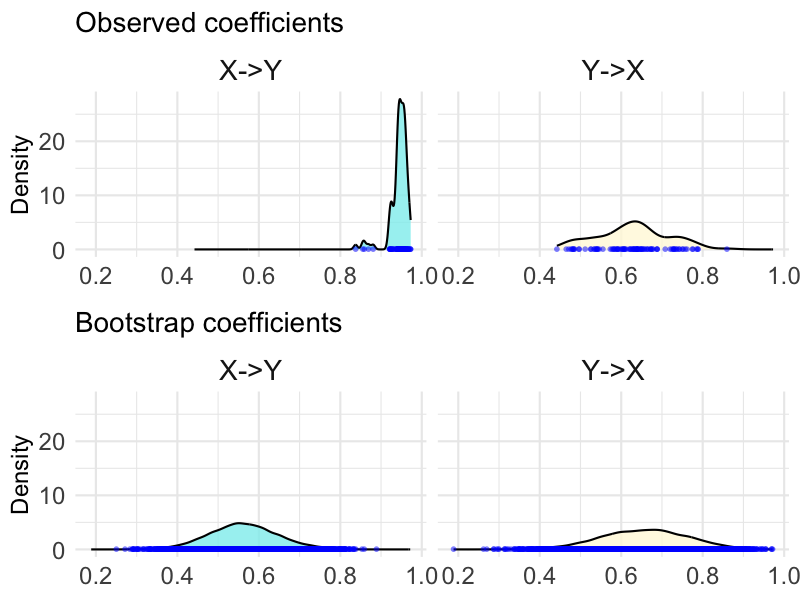} \\
        \small Model \ref{model:7}: thresholded NAAR
    \end{minipage}%
    \hspace{0cm}
    \begin{minipage}{0.4\textwidth}
        \centering
        \includegraphics[width=\linewidth]{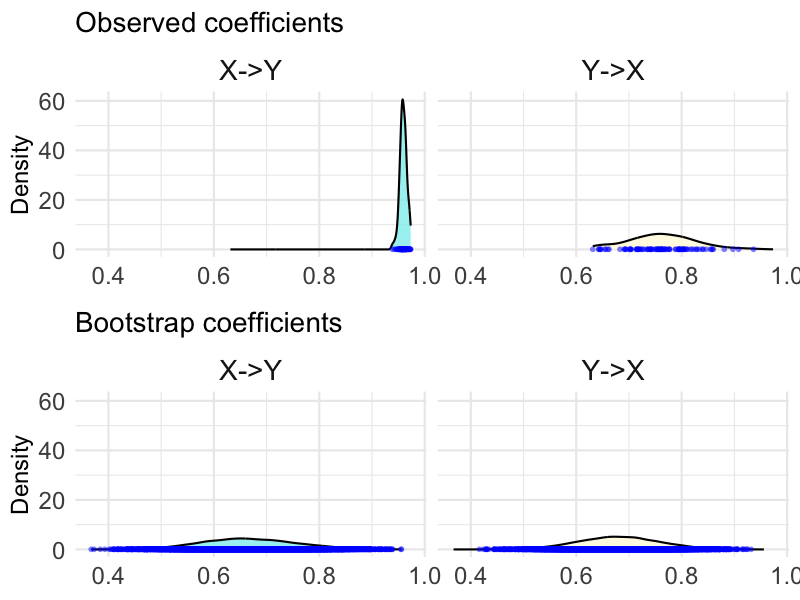} \\
        \small Model \ref{model:9}: NAAR with confounder
    \end{minipage}

    \caption{\footnotesize Comparison of Monte Carlo distributions of observed and bootstrapped causal tail coefficients across different shape parameter $\alpha$ for Models \ref{model:7} and \ref{model:9}.
    Each density plot displays the kernel density estimate of observed or bootstrapped causal tail coefficient values, aggregated over 100 time series realizations. Blue corresponds to the true causal direction $X \rightarrow Y$, and yellow to the reverse direction $Y \rightarrow X$.}
    \label{fig:shape2b}
\end{figure}

We further compared the Monte Carlo distributions of the observed and bootstrapped coefficients in both directions for three representative impact shape parameter values: $\alpha = 10^4$, $10^0$, and $10^{-4}$. 
Figures \ref{fig:shape2a} and \ref{fig:shape2b} show results for four time series models. For each model, we simulated 100 time series of length $n = 1000$ with Pareto$(1,1)$ noise and generated 100 bootstrap samples per series using the time-shifting procedure described in Section~\ref{sec:mbb}. 
The shape and the range of observed and bootstrapped coefficient distributions remain highly consistent across all three $\alpha$ values. This suggests that the impact shape parameter has little effect on the $p$-values or rejection rates of the proposed bootstrap tests.

\subsection{Extreme threshold}\label{sec:threshold}
Selecting an appropriate threshold is a well-known challenge in extreme value theory. In our setting, this involves choosing the threshold parameter $k$, which specifies the number of upper-order statistics treated as extremes in the estimation of the causal tail coefficient via \eqref{estimator}. The choice of $k$ involves a bias–variance trade-off: smaller values of $k$ produce estimates that more closely align with the limiting definition in \eqref{CCTC}, thereby reducing the bias; conversely, larger values of $k$ incorporate more data into the estimation, resulting in greater stability and lower variance. 

\begin{figure}[htbp]
    \centering
    \includegraphics[width=0.9\linewidth]{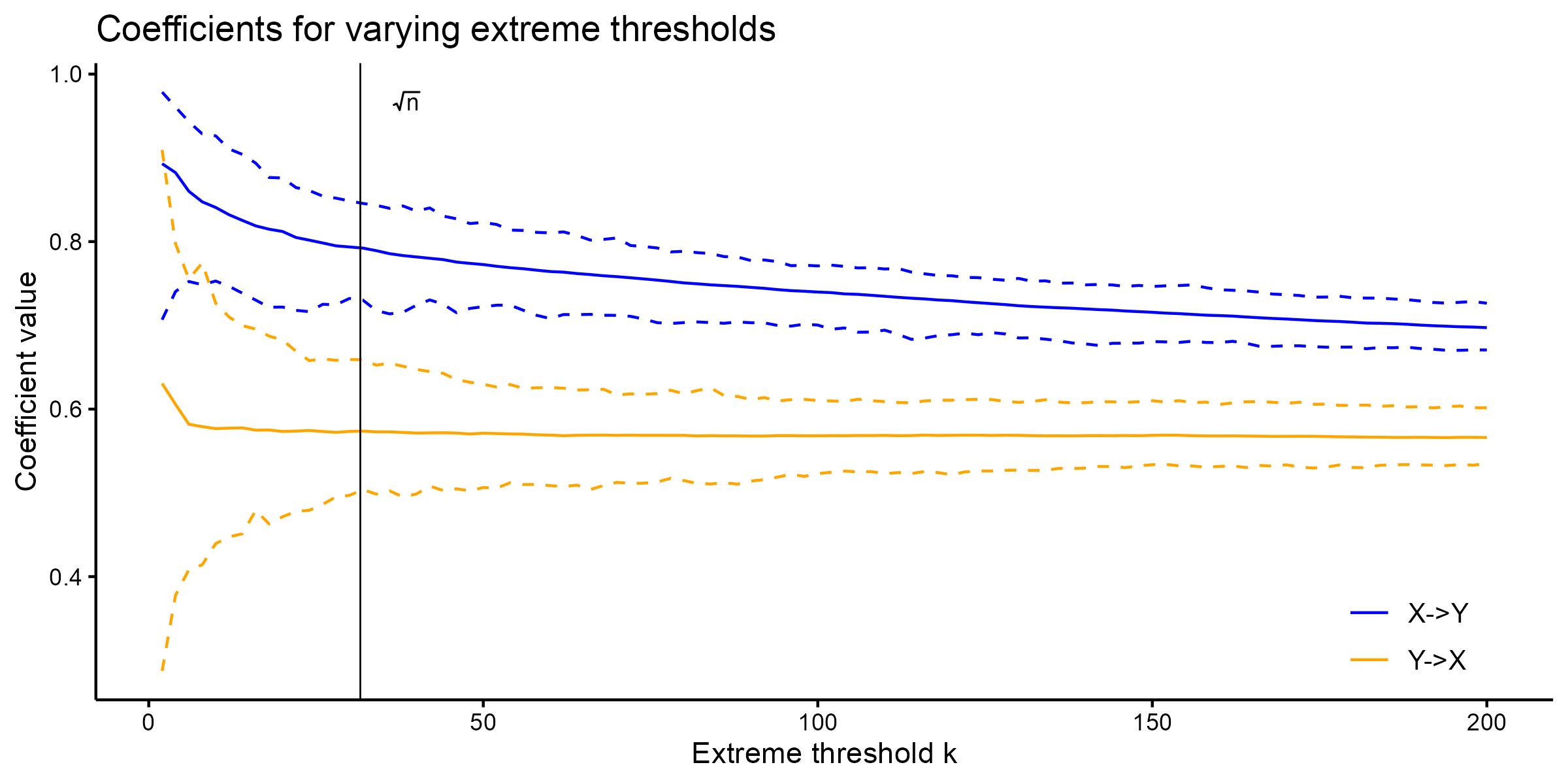}
    \caption{\footnotesize Behavior of the estimators $\hat{\Gamma}_{{\X} \rightarrow {h(\Y)}}(3)$ (blue) and $\hat{\Gamma}_{{\Y} \rightarrow {h(\X)}}(3)$ (yellow) as a function of the threshold parameter $k$, which determines the number of extremes used in estimating the compound causal tail coefficient. 
    Solid lines indicate the mean coefficient estimates across 100 realizations, while dotted lines denote the empirical 5\% and 95\% pointwise quantiles.
    }
    \label{fig:threshold}
\end{figure}

To illustrate the impact of extreme threshold selection on the compound causal tail coefficient, we again used Model \ref{model:6} and followed a procedure similar to that in Section \ref{sec:shape}. In this case, we fixed the impact shape parameter at $\alpha = 10^4$ and varied the threshold parameter $k$ over a range of integer values from $2$ to $200$. Figure \ref{fig:threshold} displays the mean along with the 5\% and 95\% empirical quantiles of the coefficient estimators $\hat{\Gamma}_{{\X} \rightarrow {h(\Y)}}(3)$ and $\hat{\Gamma}_{{\Y} \rightarrow {h(\X)}}(3)$.

The observed patterns align with theoretical expectations. The variances of both $\hat{\Gamma}_{{\X} \rightarrow {h(\Y)}}(3)$ and $\hat{\Gamma}_{{\Y} \rightarrow {h(\X)}}(3)$ decrease with increasing $k$, though the variance associated with the incorrect direction ($\Y \rightarrow \X$) is substantially larger at smaller $k$ values. 
The estimates in the true causal direction, $\hat{\Gamma}_{{\X} \rightarrow {h(\Y)}}(3)$, exhibit a continued decline in $k$, suggesting increasing bias, whereas $\hat{\Gamma}_{{\Y} \rightarrow {h(\X)}}(3)$ shows an initial decrease but stabilizes as $k$ becomes large.
Visual inspection of this model and others suggests that $k = \sqrt{n}$ offers a reasonable balance between bias and variance, although it may not be optimal.

\subsection{Tail index}\label{sec:sensi_tail}

The simulation study in Section~\ref{sec:simulation} considers several innovation distributions representing qualitatively different tail regimes. However, comparisons across distributional families do not isolate the effect of tail heaviness. We therefore examine the sensitivity of the compound causal tail coefficient to the tail index within the Student's $t$ family. A Student's $t$ distribution with $\nu$ degrees of freedom is regularly varying with tail index $\nu$, so smaller values of $\nu$ correspond to heavier tails.

\begin{figure}[htbp]
    \centering
    \includegraphics[width=0.9\linewidth]{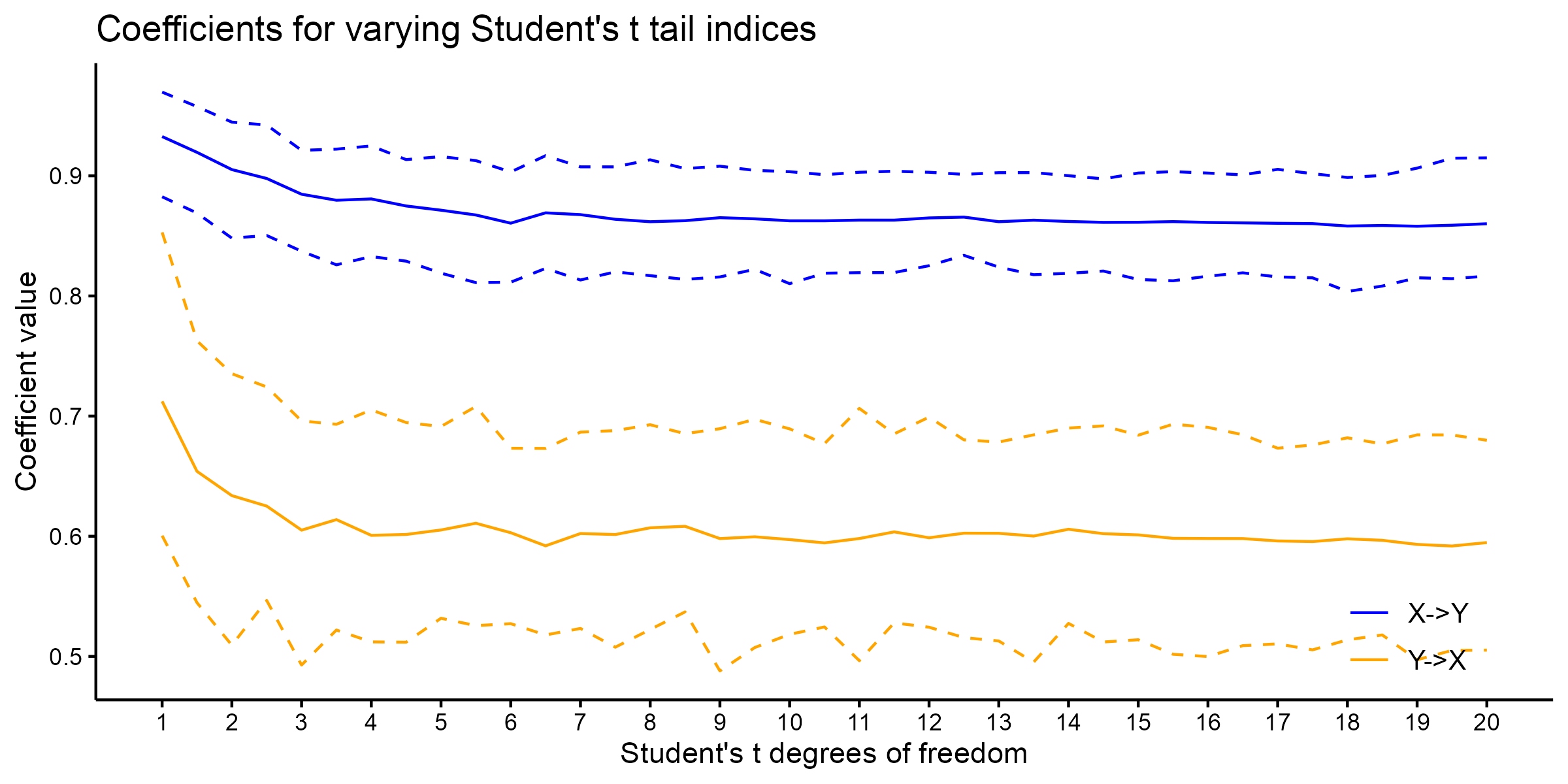}
    \caption{\footnotesize
    Behavior of the estimators
    $\hat{\Gamma}_{{\X}\rightarrow {h(\Y)}}(3)$ (blue) and
    $\hat{\Gamma}_{{\Y}\rightarrow {h(\X)}}(3)$ (yellow)
    as a function of the degrees-of-freedom parameter $\nu$ of the Student's $t$ innovations. Smaller values of $\nu$ correspond to heavier tails.
    Solid lines indicate the mean coefficient estimates across 100 realizations, while dotted lines denote the empirical 5\% and 95\% pointwise quantiles.
    }
    \label{fig:tail_index}
\end{figure}

Following Sections~\ref{sec:shape} and~\ref{sec:threshold}, we used Model~\ref{model:6} and generated both innovation series from Student's $t$ distributions with a common degrees-of-freedom parameter $\nu$. We varied $\nu$ from $1$ to $20$ in increments of $0.5$, while fixing the time series length at $n=1000$, the impact shape parameter at $\alpha=10^4$, the threshold at $k=\lfloor\sqrt{n}\rfloor$, and the extremal delay at $p=3$. Figure~\ref{fig:tail_index} displays the mean and the 5\% and 95\% empirical quantiles of the coefficient estimators $\hat{\Gamma}_{{\X}\rightarrow {h(\Y)}}(3)$ and $\hat{\Gamma}_{{\Y}\rightarrow {h(\X)}}(3)$ across 100 realizations.

Both $\hat{\Gamma}_{{\X} \rightarrow {h(\Y)}}(3)$ and $\hat{\Gamma}_{{\Y} \rightarrow {h(\X)}}(3)$ initially decrease as $\nu$ increases, with the most pronounced changes occurring between $\nu=1$ and $\nu=3$. For larger values of $\nu$, both estimators stabilize, with $\hat{\Gamma}_{{\X} \rightarrow {h(\Y)}}(3)$ remaining close to $0.86$ and $\hat{\Gamma}_{{\Y} \rightarrow {h(\X)}}(3)$ remaining close to $0.60$. Their empirical variability is broadly stable across the range of tail indices considered. Most importantly, the estimators in the two directions remain clearly separated throughout the entire range, with no overlap between their empirical 5\%--95\% quantile bands. These results suggest that, although extremely heavy-tailed innovations affect the magnitudes of the estimated compound causal tail coefficients, the directional separation is robust to substantial variation in tail heaviness within the Student's $t$ family.

\subsection{Extremal delay parameter}\label{sec:sensi_lag}

Thus far, our simulation studies have assumed that the extremal delay parameter is known from the cross-lag structure of the underlying time series models. 
In this section, we relax this assumption and consider the case where the extremal delay is unknown. We examine the behavior of $\hat{\Gamma}_{{\X} \rightarrow {h(\Y)}}(p)$ as the extremal delay $p$ varies and present empirical results on the estimation of $p$.

\begin{figure}[htbp]
    \centering
    \includegraphics[width=0.9\linewidth]{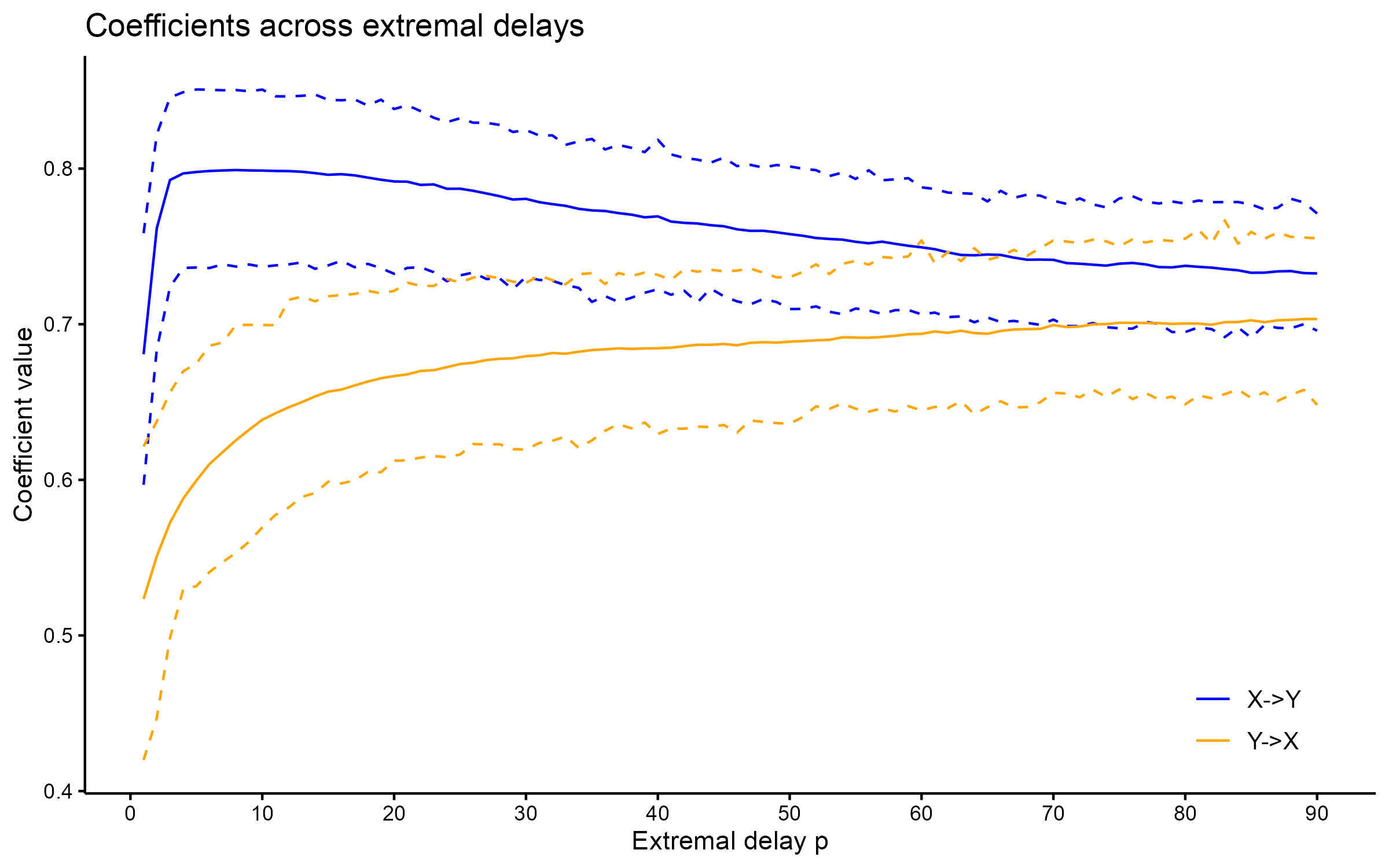}
    \caption{\footnotesize Behavior of the estimators $\hat{\Gamma}_{{\X} \rightarrow {h(\Y)}}(p)$ (blue) and $\hat{\Gamma}_{{\Y} \rightarrow {h(\X)}}(p)$ (yellow) across varying values of the extremal delay $p$. 
    Solid lines indicate the mean coefficient estimates across 100 realizations, while dotted lines denote the empirical 5\% and 95\% pointwise quantiles.
    }
    \label{fig:lag}
\end{figure}

To demonstrate the behavior of the compound causal tail coefficient under varying choices of the extremal delay $p$, we again considered Model \ref{model:6} and adopted a procedure similar to that of Section \ref{sec:shape}. The impact shape parameter was fixed at $\alpha = 10^4$, the extreme threshold was set to $k=\sqrt{n}$, while the extremal delay $p$ varied over integer values from 1 to 90. Figure~\ref{fig:lag} shows the mean and the 5\% and 95\% empirical quantiles of the estimators $\hat{\Gamma}_{{\X} \rightarrow {h(\Y)}}(p)$ and $\hat{\Gamma}_{{\Y} \rightarrow {h(\X)}}(p)$. 

Increasing the number of lags in estimating the causal tail coefficient does not necessarily yield larger values. In the correct causal direction, $\hat{\Gamma}_{{\X} \rightarrow {h(\Y)}}(p)$ rises quickly until the true extremal delay $p=3$, then declines as additional lags are included. In contrast, the reverse coefficient $\hat{\Gamma}_{{\Y} \rightarrow {h(\X)}}(p)$ grows slowly and eventually converges to the value of $\hat{\Gamma}_{{\X} \rightarrow {h(\Y)}}(p)$. In this model, the optimal $p$ coincides with the cross-lag order of the underlying bivariate time series, a pattern observed across many VAR models. Moderate over-specification of $p$ is tolerated, but excessive lags do not improve estimation and may introduce negative bias.

\begin{figure}[htbp]
    \centering

    \begin{minipage}[b]{0.32\textwidth}
        \centering
        {\footnotesize Model 1}\\
        \includegraphics[width=\textwidth]{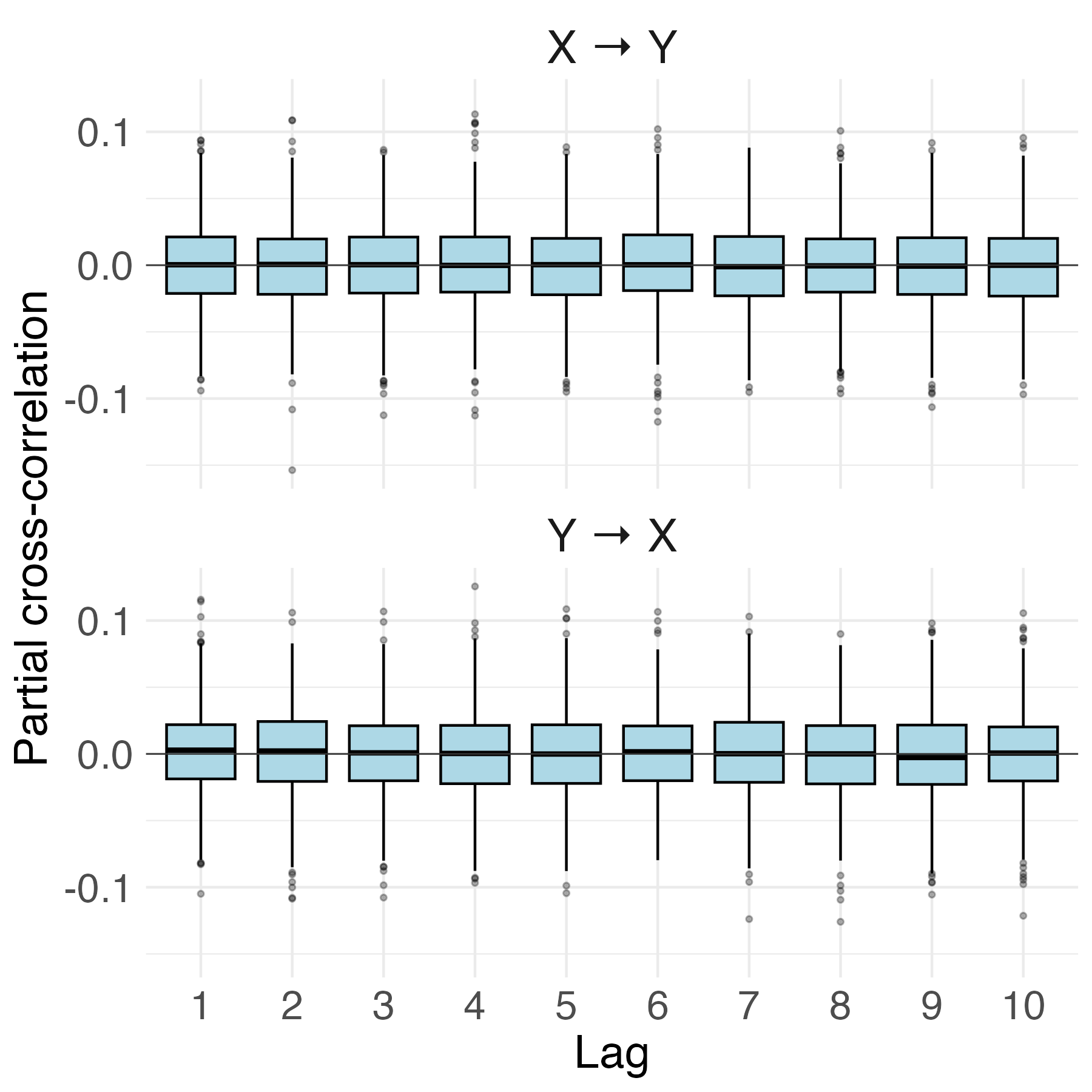}
    \end{minipage}\hfill
    \begin{minipage}[b]{0.32\textwidth}
        \centering
        {\footnotesize Model 2}\\
        \includegraphics[width=\textwidth]{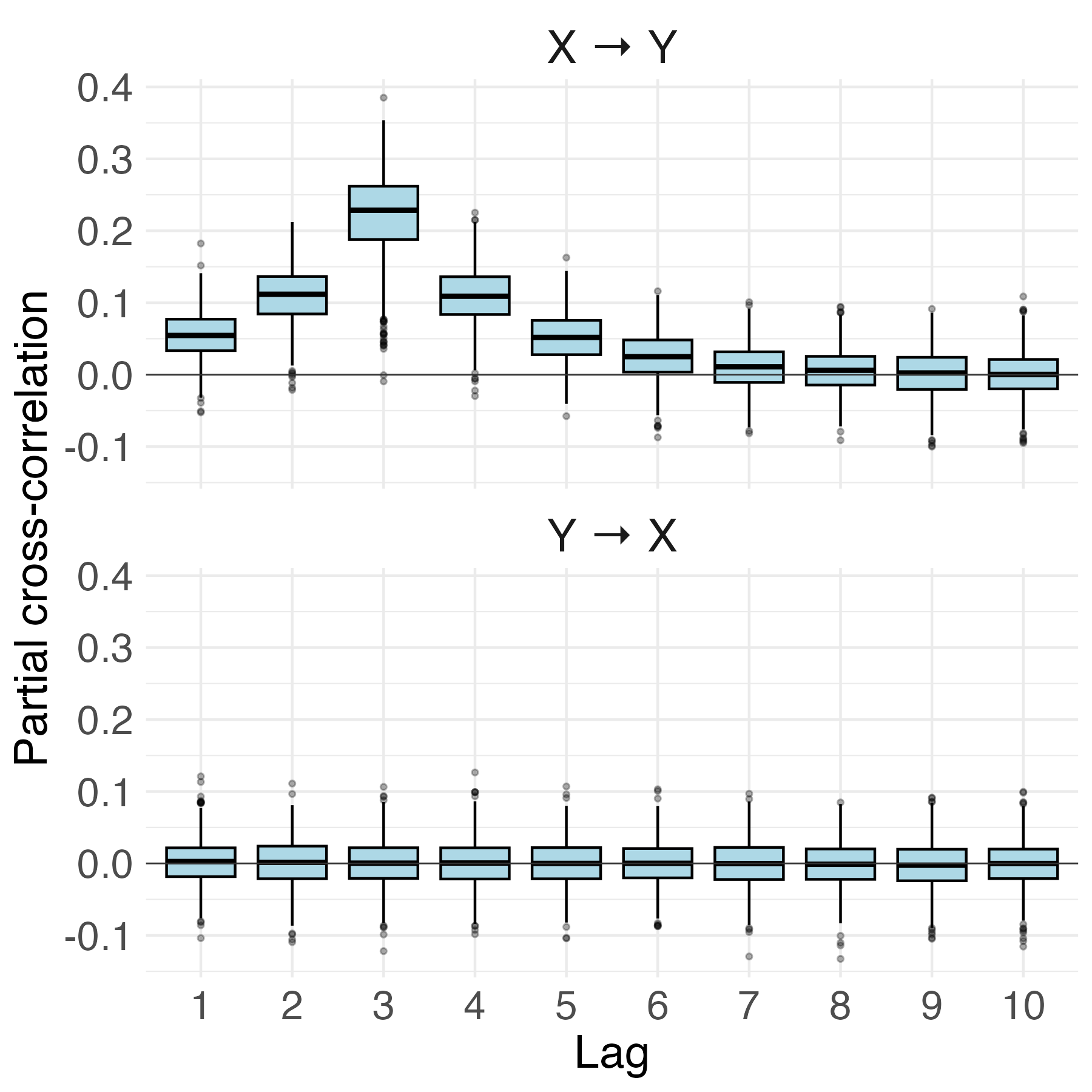}
    \end{minipage}\hfill
    \begin{minipage}[b]{0.32\textwidth}
        \centering
        {\footnotesize Model 3}\\
        \includegraphics[width=\textwidth]{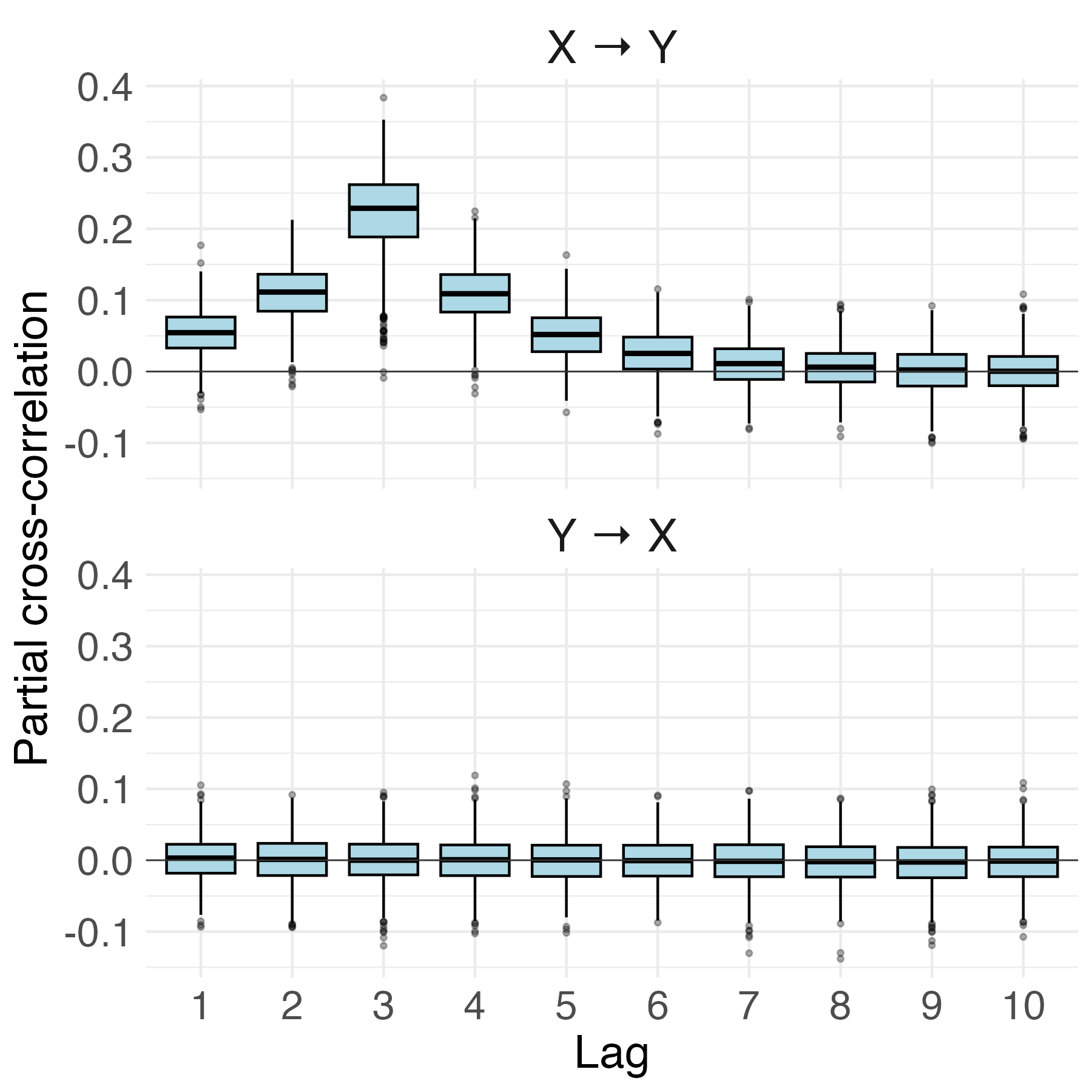}
    \end{minipage}

    \vspace{0.6em}

    \begin{minipage}[b]{0.32\textwidth}
        \centering
        {\footnotesize Model 4}\\
        \includegraphics[width=\textwidth]{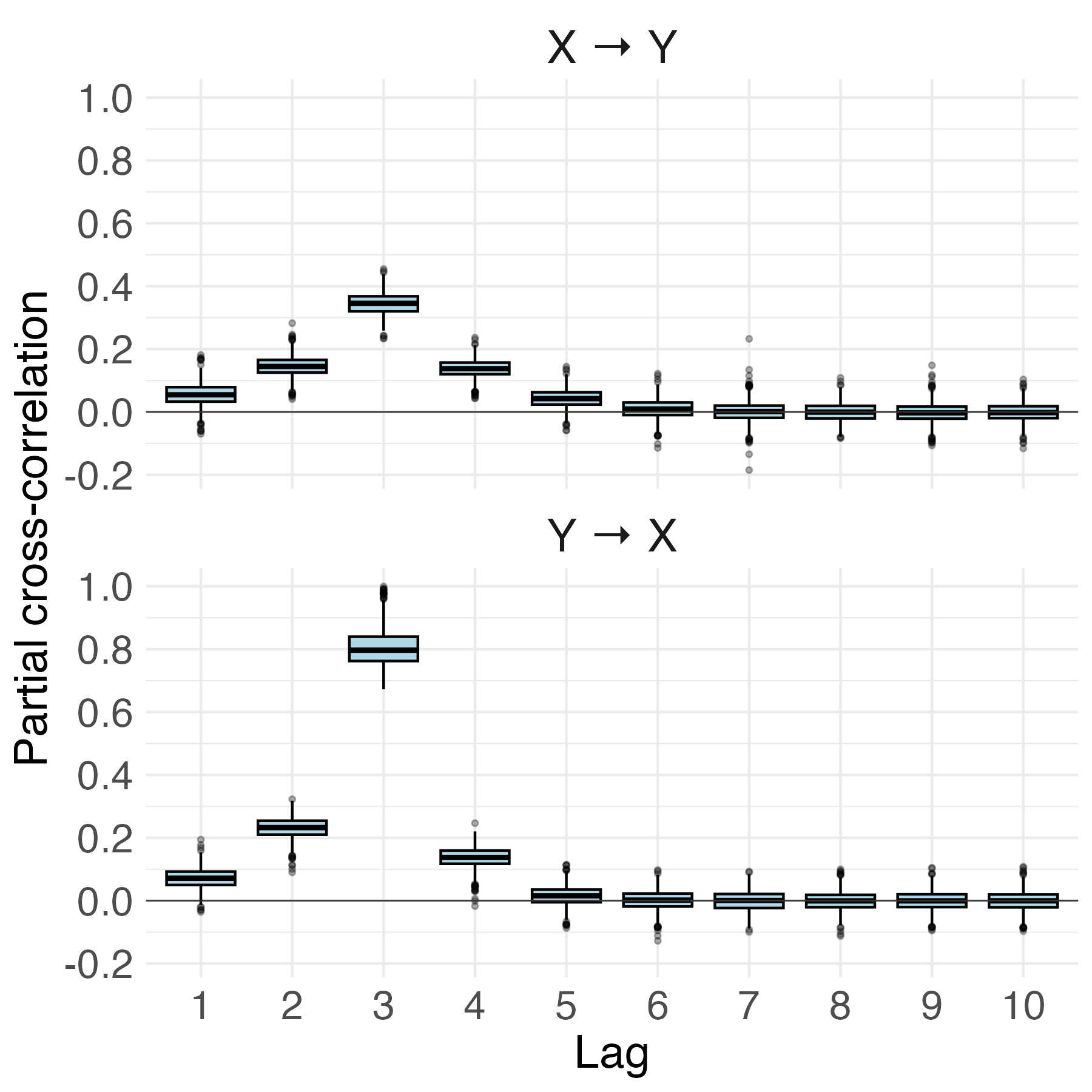}
    \end{minipage}\hfill
    \begin{minipage}[b]{0.32\textwidth}
        \centering
        {\footnotesize Model 5}\\
        \includegraphics[width=\textwidth]{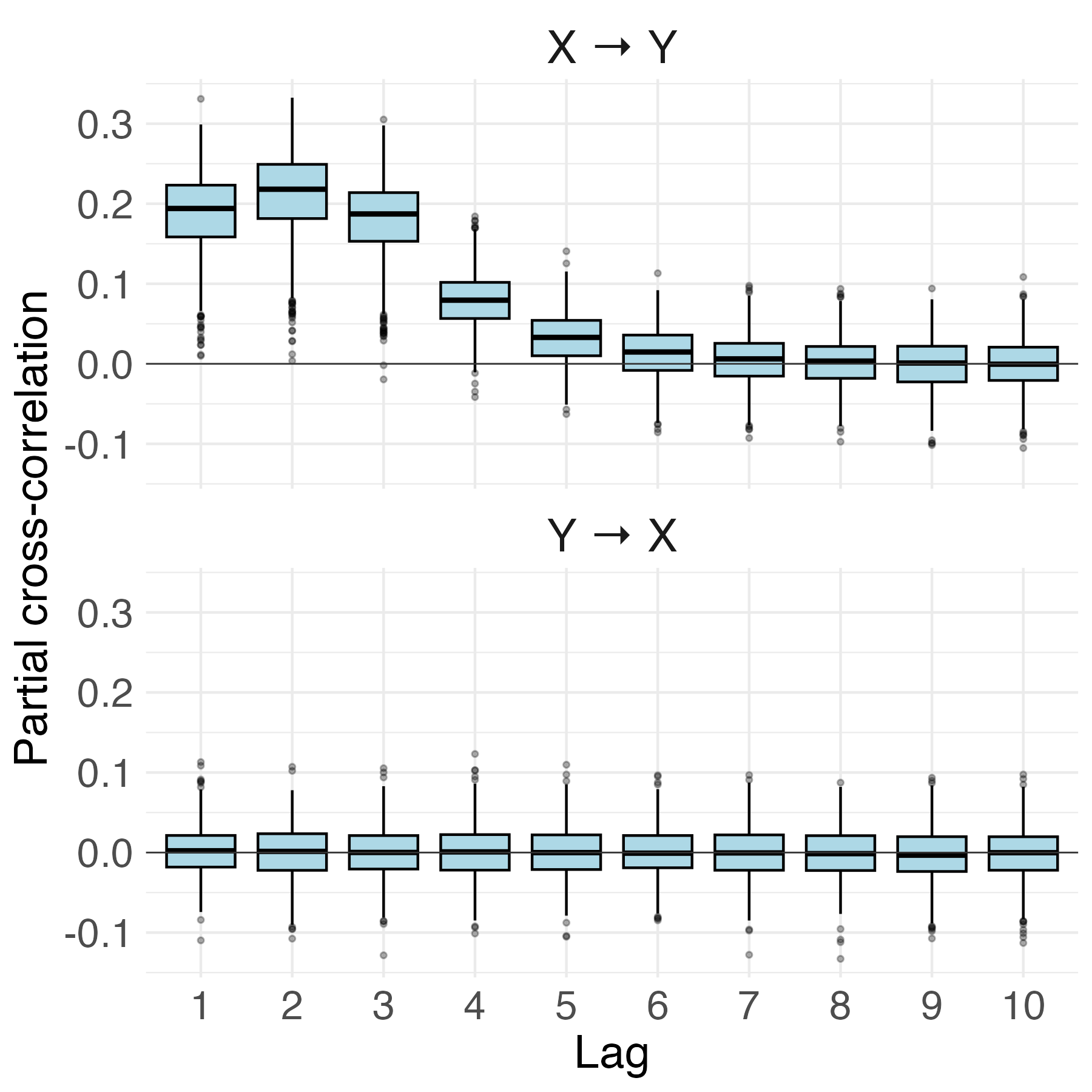}
    \end{minipage}\hfill
    \begin{minipage}[b]{0.32\textwidth}
        \centering
        {\footnotesize Model 6}\\
        \includegraphics[width=\textwidth]{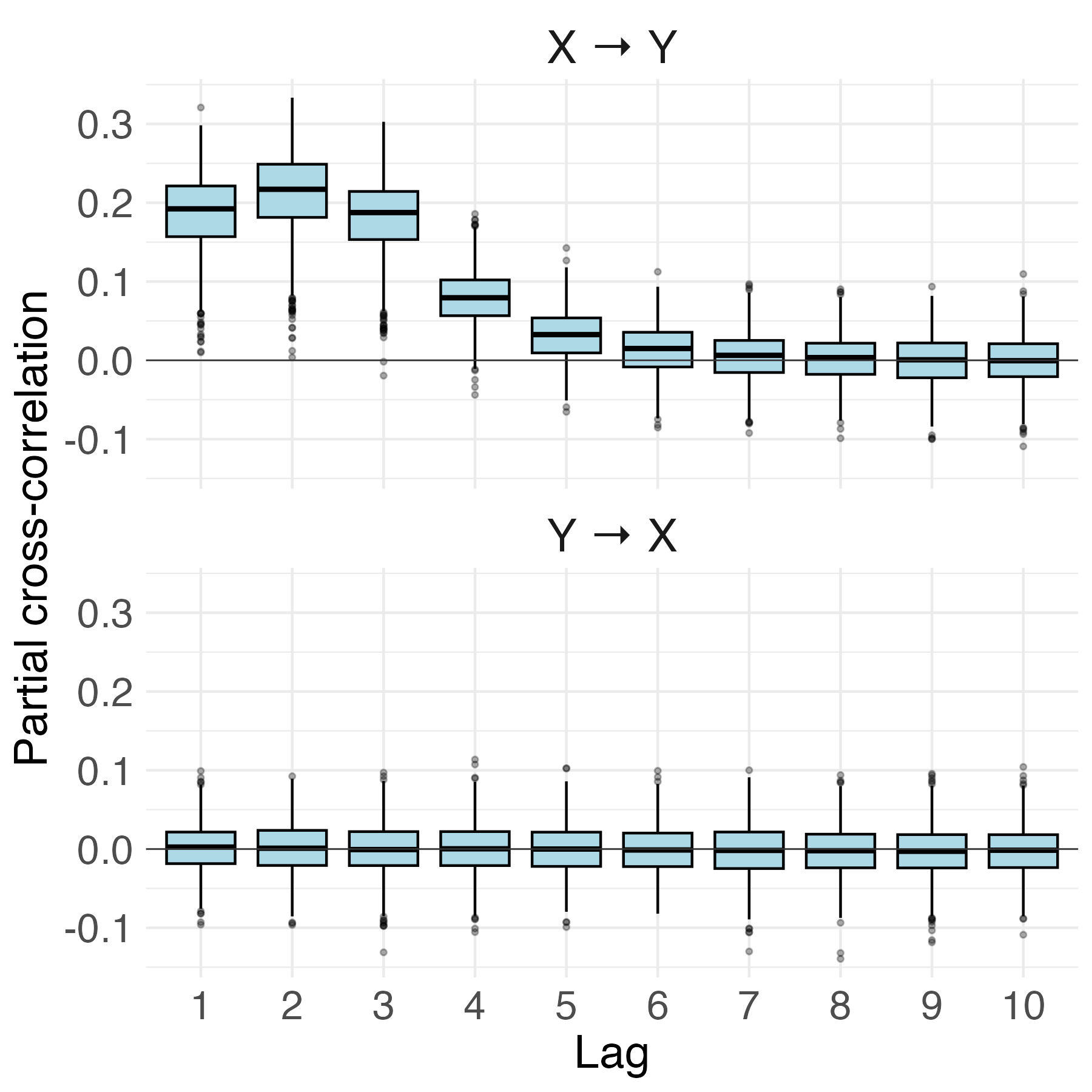}
    \end{minipage}

    \vspace{0.6em}

    \begin{minipage}[b]{0.32\textwidth}
        \centering
        {\footnotesize Model 7}\\
        \includegraphics[width=\textwidth]{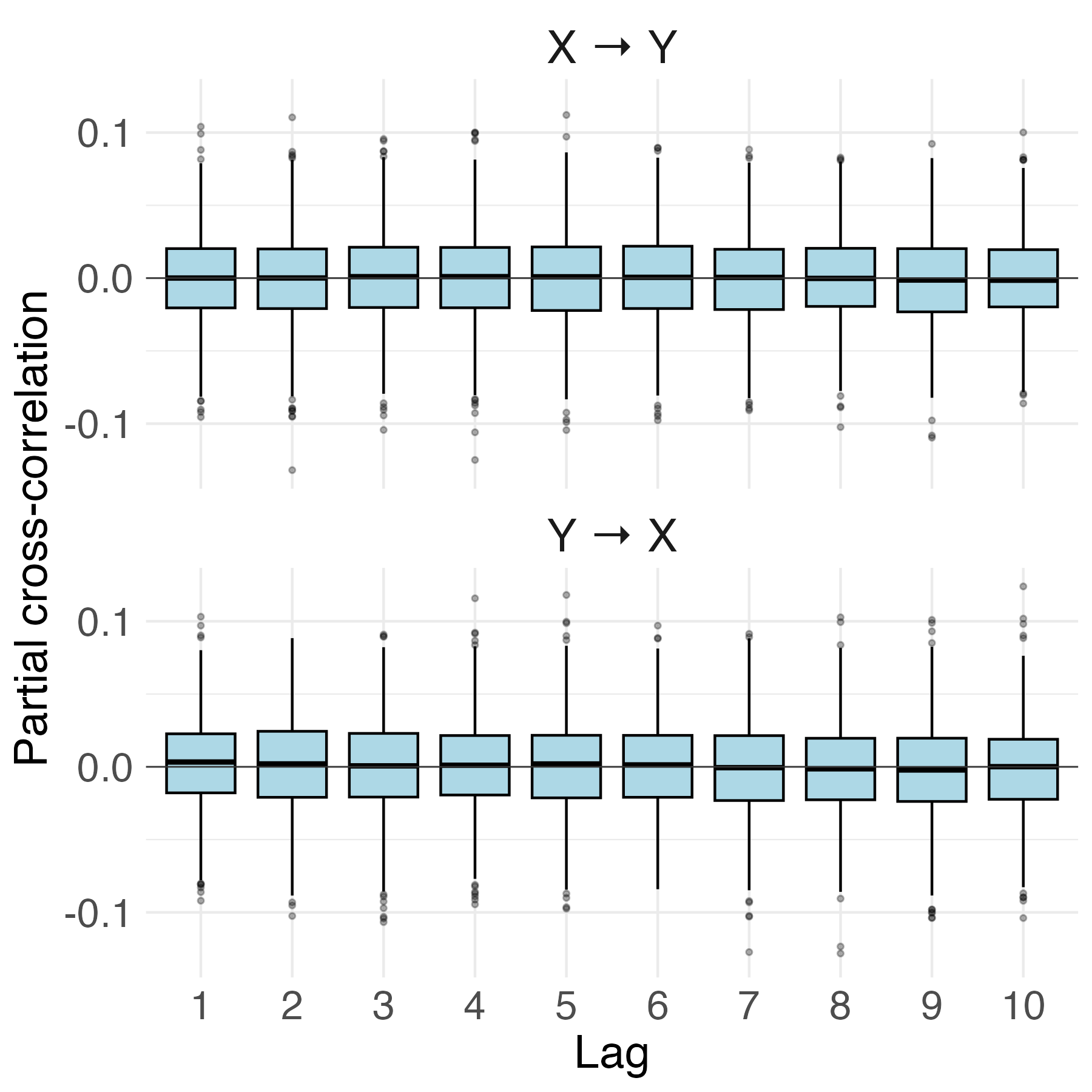}
    \end{minipage}\hfill
    \begin{minipage}[b]{0.32\textwidth}
        \centering
        {\footnotesize Model 8}\\
        \includegraphics[width=\textwidth]{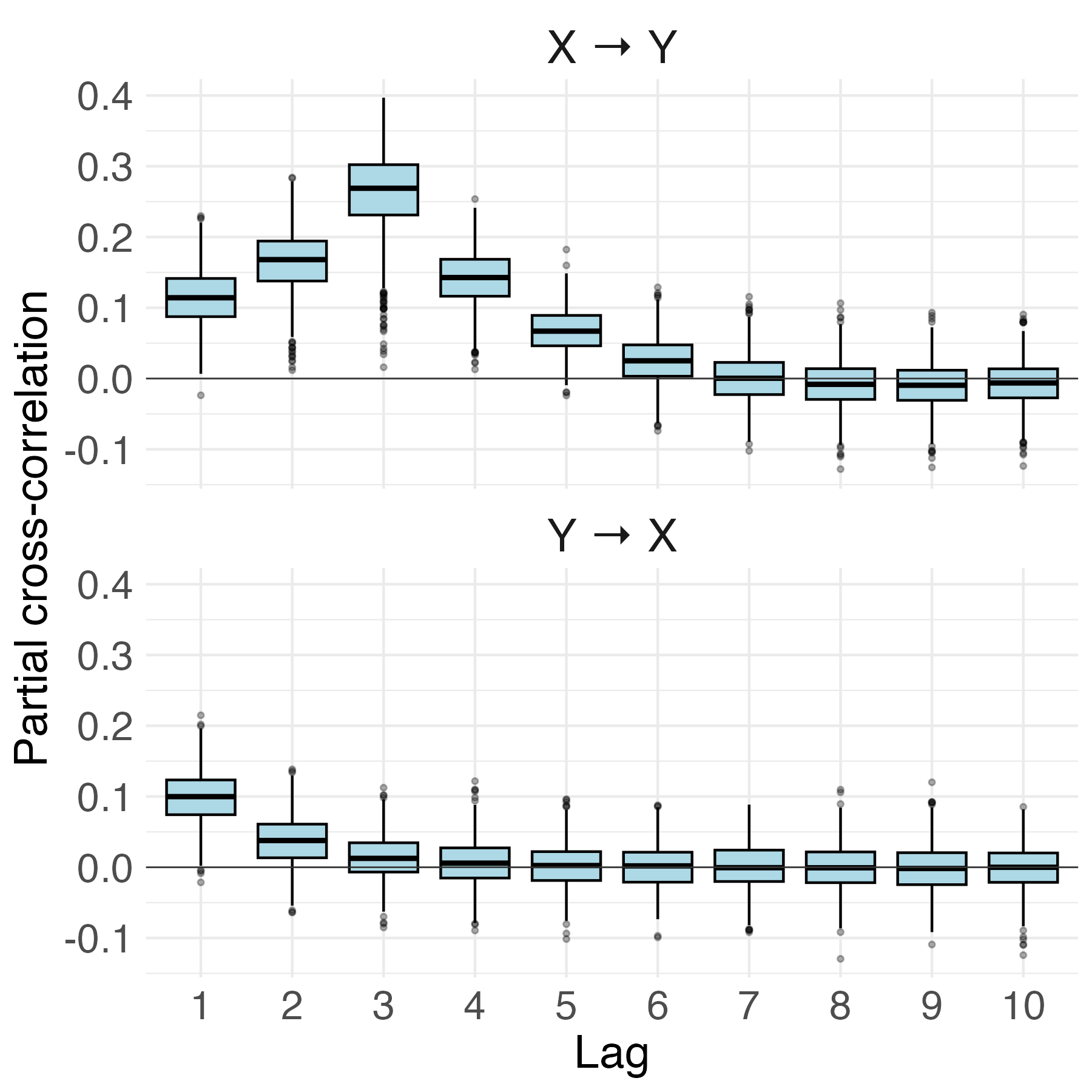}
    \end{minipage}\hfill
    \begin{minipage}[b]{0.32\textwidth}
        \centering
        {\footnotesize Model 9}\\
        \includegraphics[width=\textwidth]{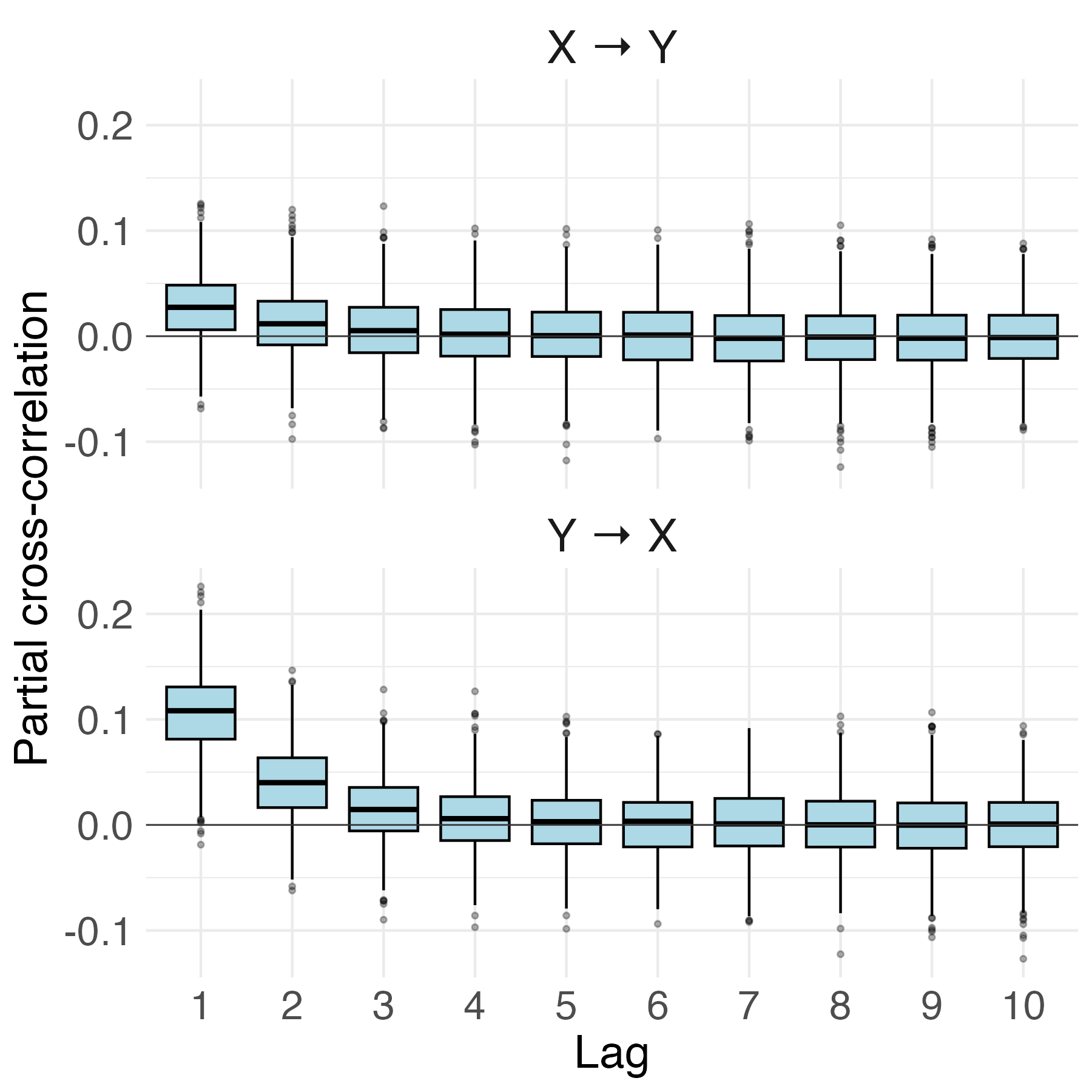}
    \end{minipage}

    \caption{\footnotesize Estimated partial cross-correlation functions across lags. 
    Time series models are repeatedly generated with Student's $t$ noise.
    Each boxplot summarizes results over 1000 realizations, showing the interquartile range (box), median (horizontal line), and outliers beyond 1.5 times the interquartile range (dots).}
    \label{fig:pccf_student_models}
\end{figure}

\begin{figure}[htbp]
    \centering

    \begin{minipage}[b]{0.32\textwidth}
        \centering
        {\footnotesize Model 1}\\
        \includegraphics[width=\textwidth]{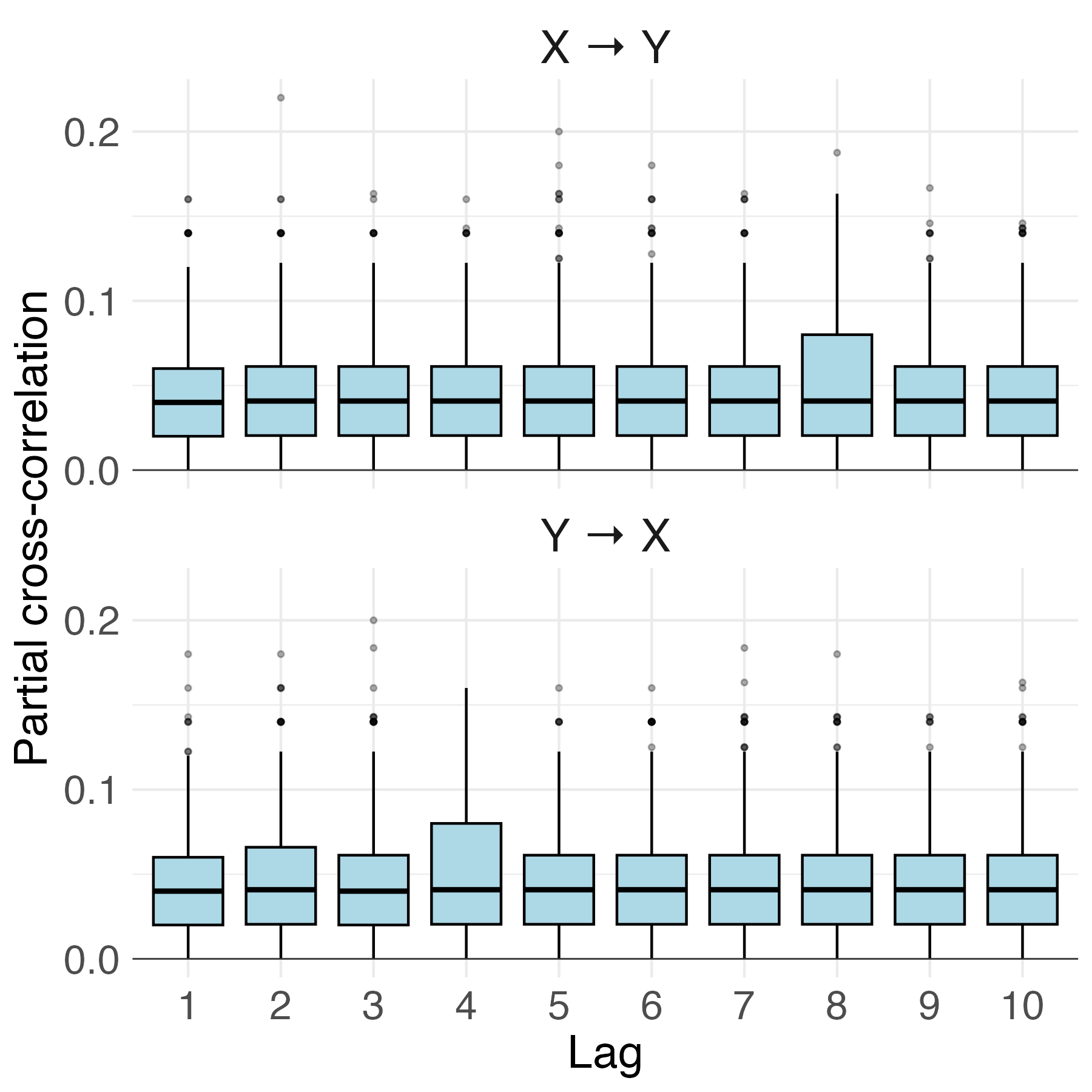}
    \end{minipage}\hfill
    \begin{minipage}[b]{0.32\textwidth}
        \centering
        {\footnotesize Model 2}\\
        \includegraphics[width=\textwidth]{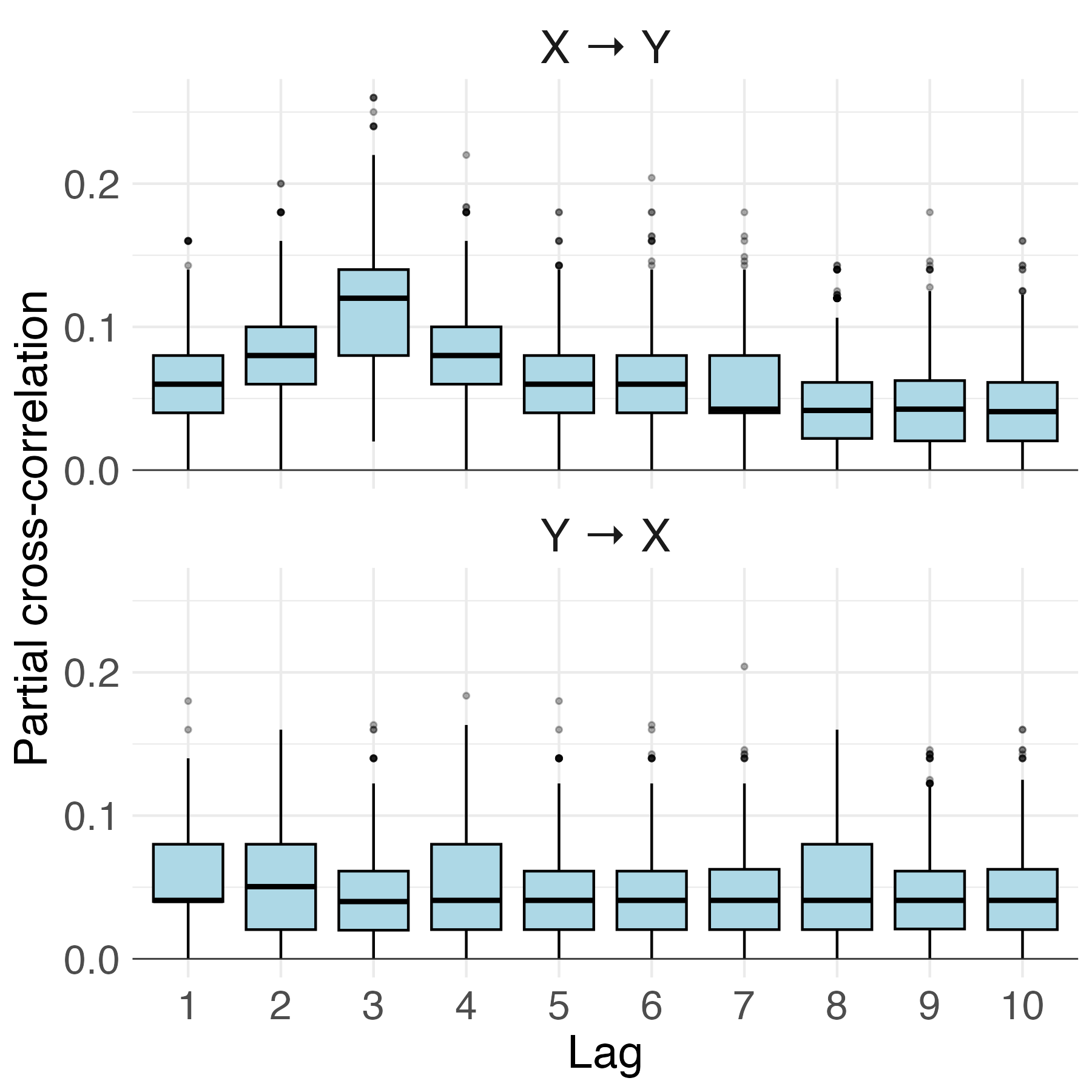}
    \end{minipage}\hfill
    \begin{minipage}[b]{0.32\textwidth}
        \centering
        {\footnotesize Model 3}\\
        \includegraphics[width=\textwidth]{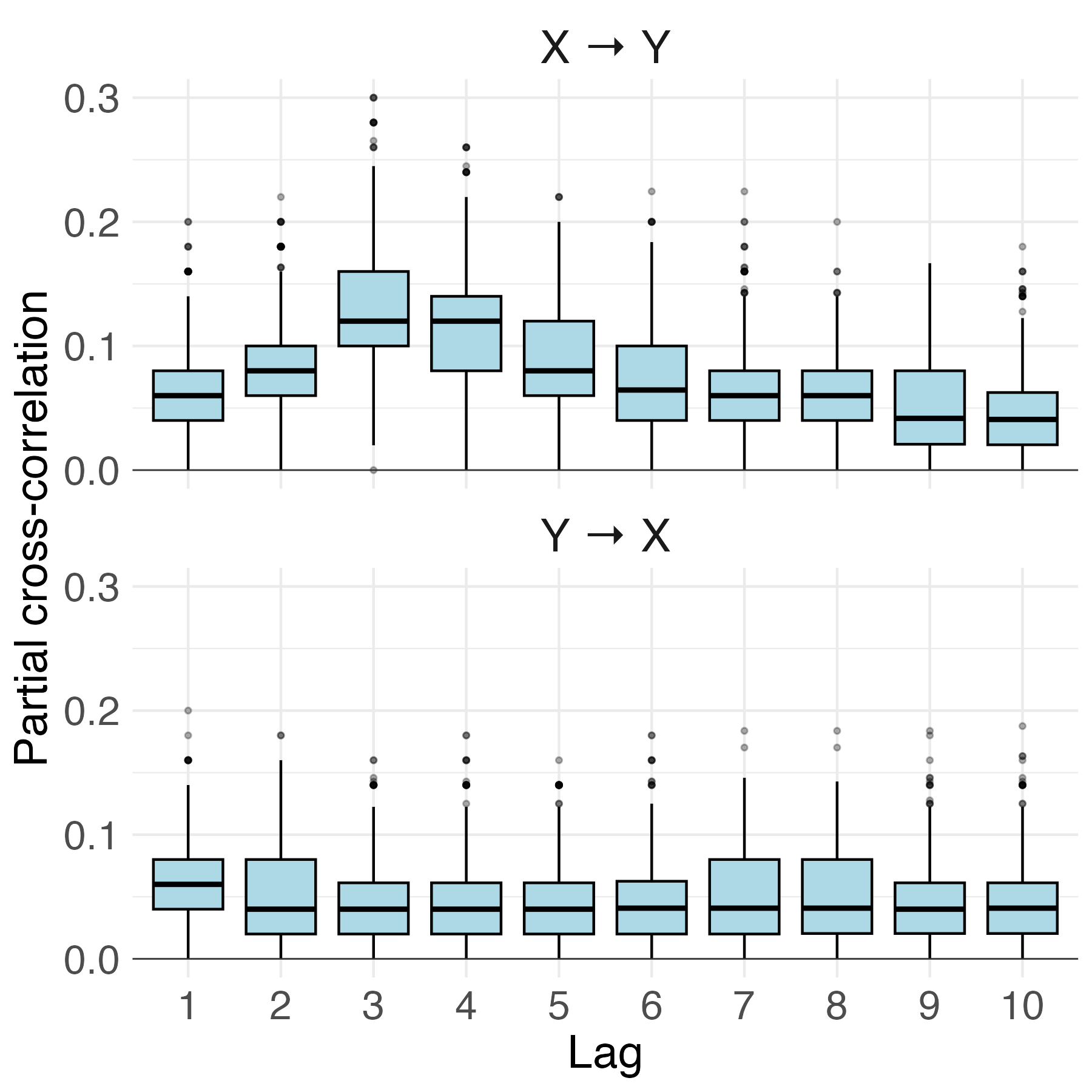}
    \end{minipage}

    \vspace{0.6em}

    \begin{minipage}[b]{0.32\textwidth}
        \centering
        {\footnotesize Model 4}\\
        \includegraphics[width=\textwidth]{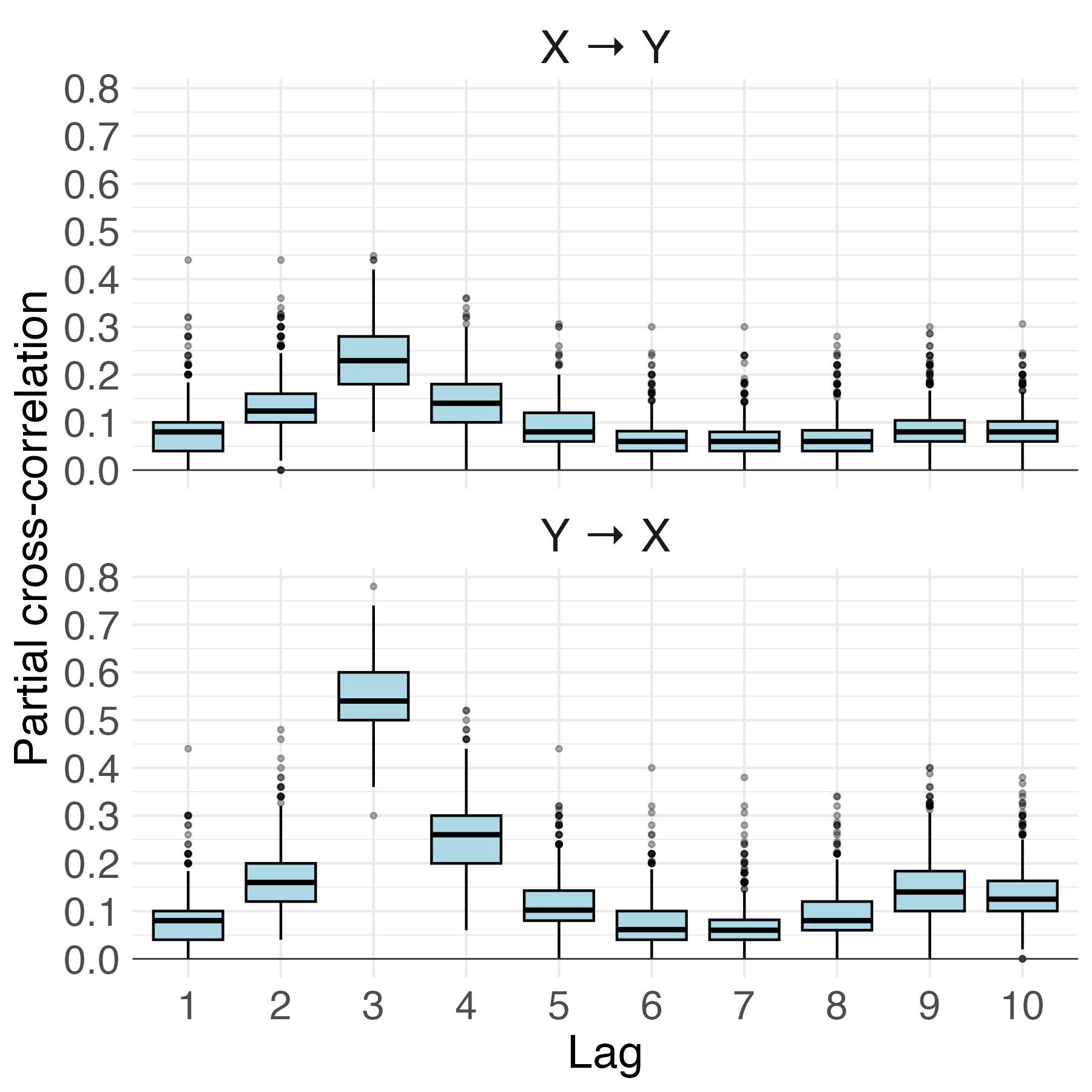}
    \end{minipage}\hfill
    \begin{minipage}[b]{0.32\textwidth}
        \centering
        {\footnotesize Model 5}\\
        \includegraphics[width=\textwidth]{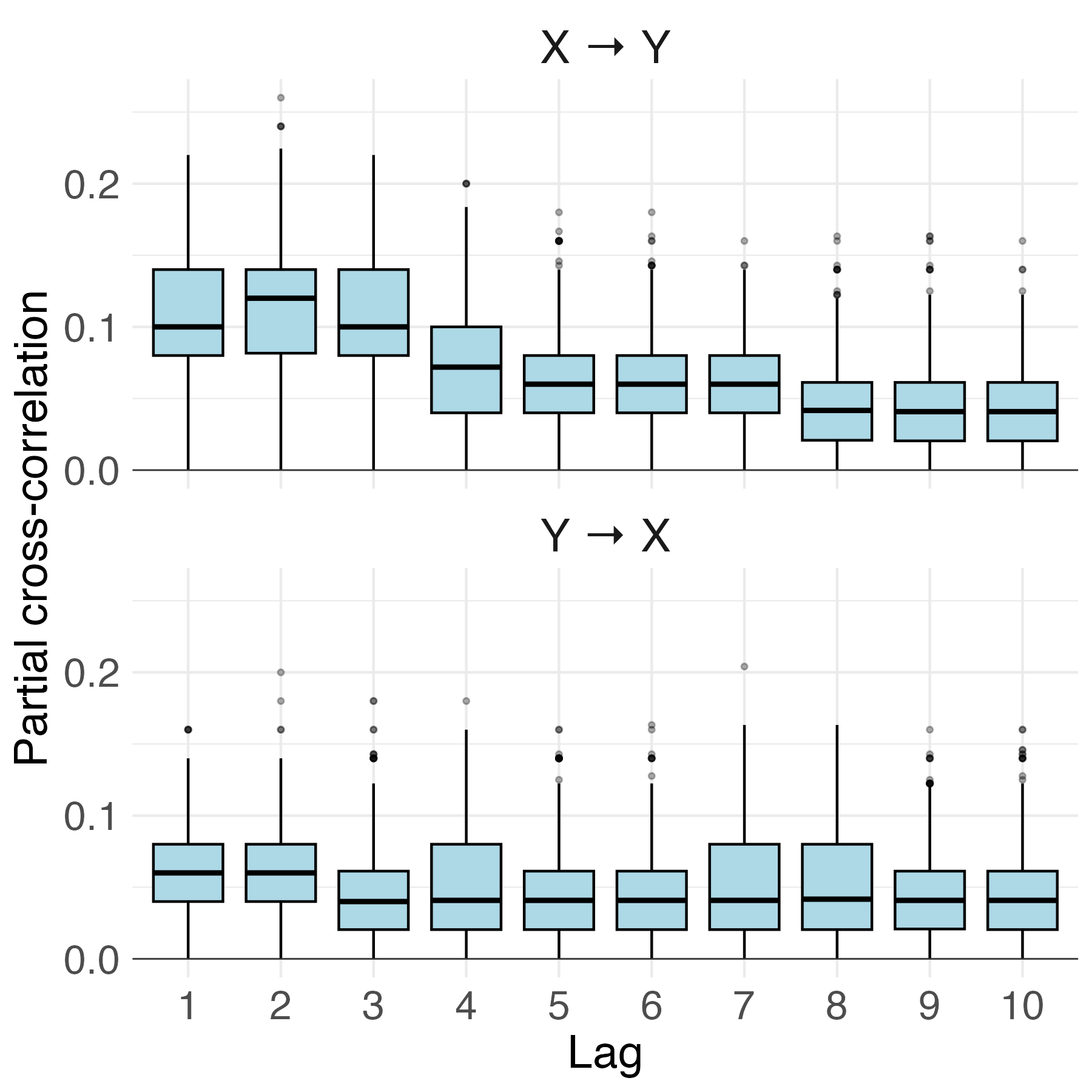}
    \end{minipage}\hfill
    \begin{minipage}[b]{0.32\textwidth}
        \centering
        {\footnotesize Model 6}\\
        \includegraphics[width=\textwidth]{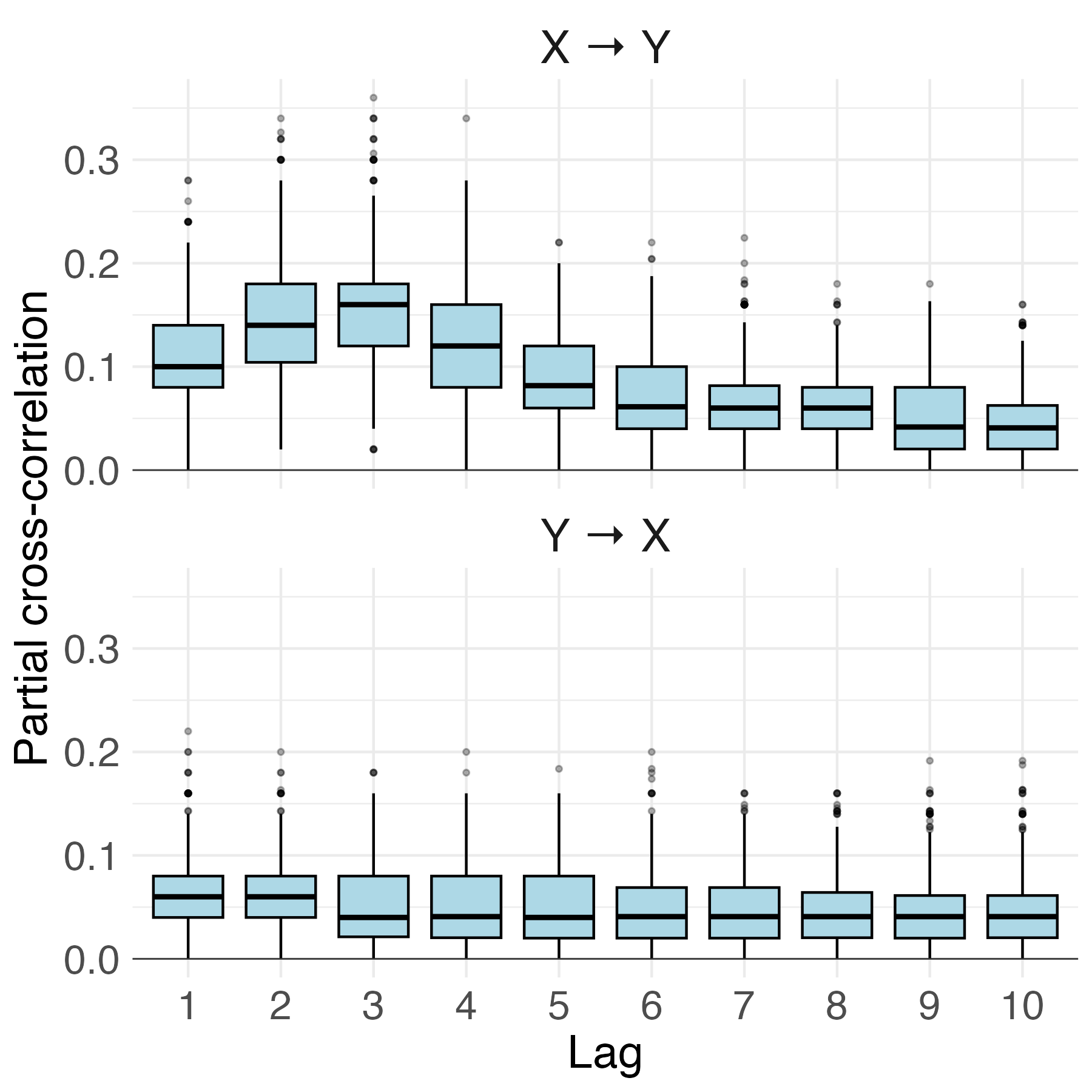}
    \end{minipage}

    \vspace{0.6em}

    \begin{minipage}[b]{0.32\textwidth}
        \centering
        {\footnotesize Model 7}\\
        \includegraphics[width=\textwidth]{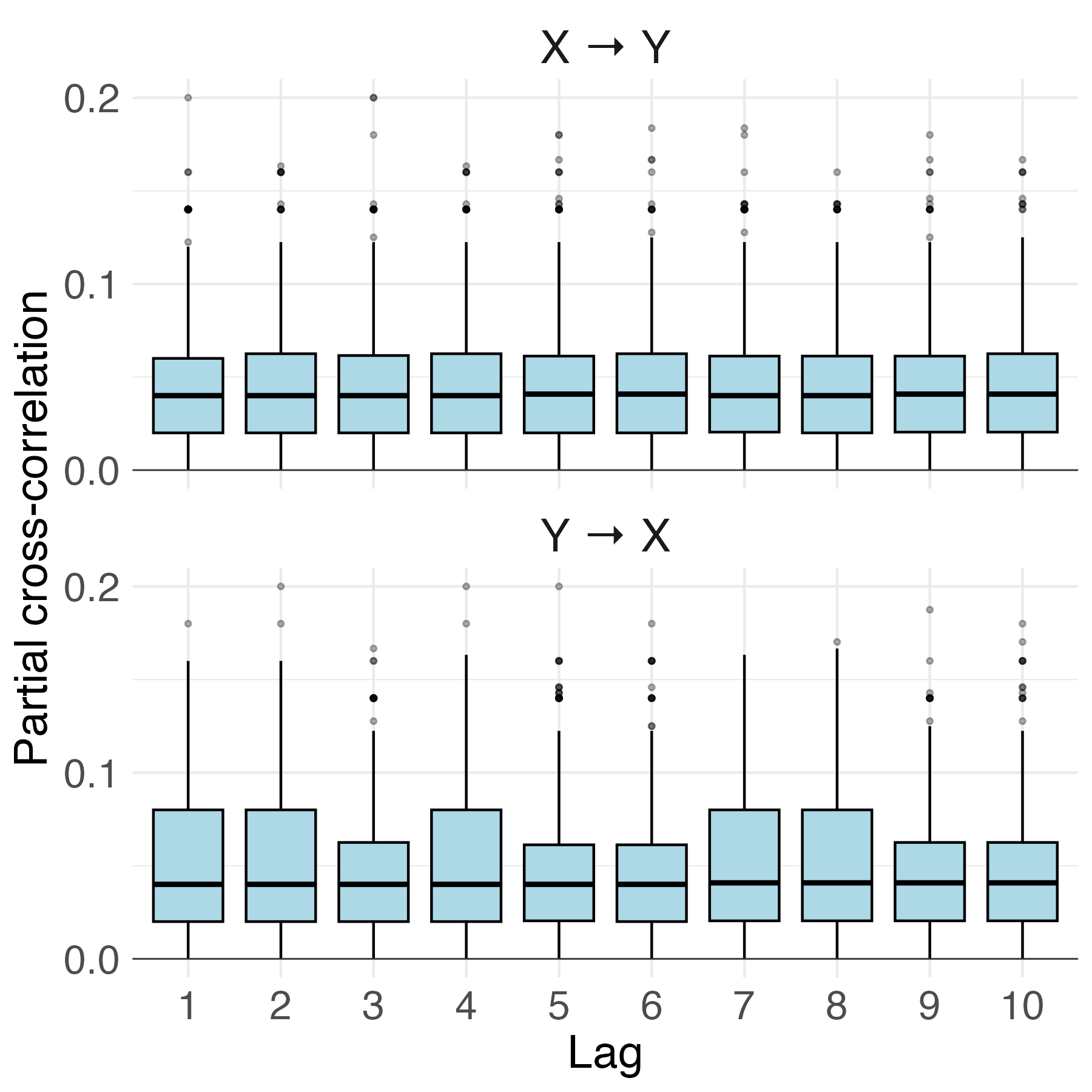}
    \end{minipage}\hfill
    \begin{minipage}[b]{0.32\textwidth}
        \centering
        {\footnotesize Model 8}\\
        \includegraphics[width=\textwidth]{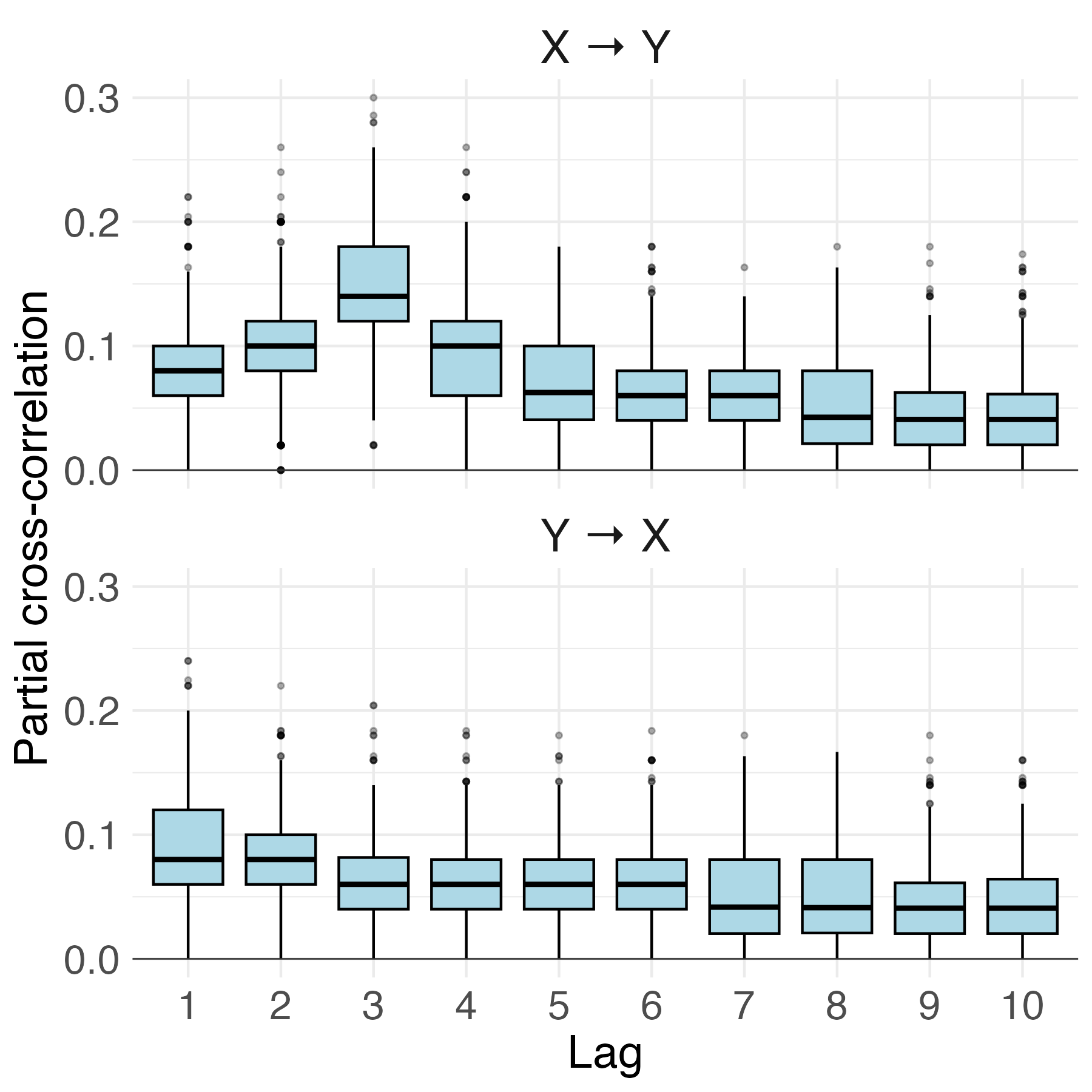}
    \end{minipage}\hfill
    \begin{minipage}[b]{0.32\textwidth}
        \centering
        {\footnotesize Model 9}\\
        \includegraphics[width=\textwidth]{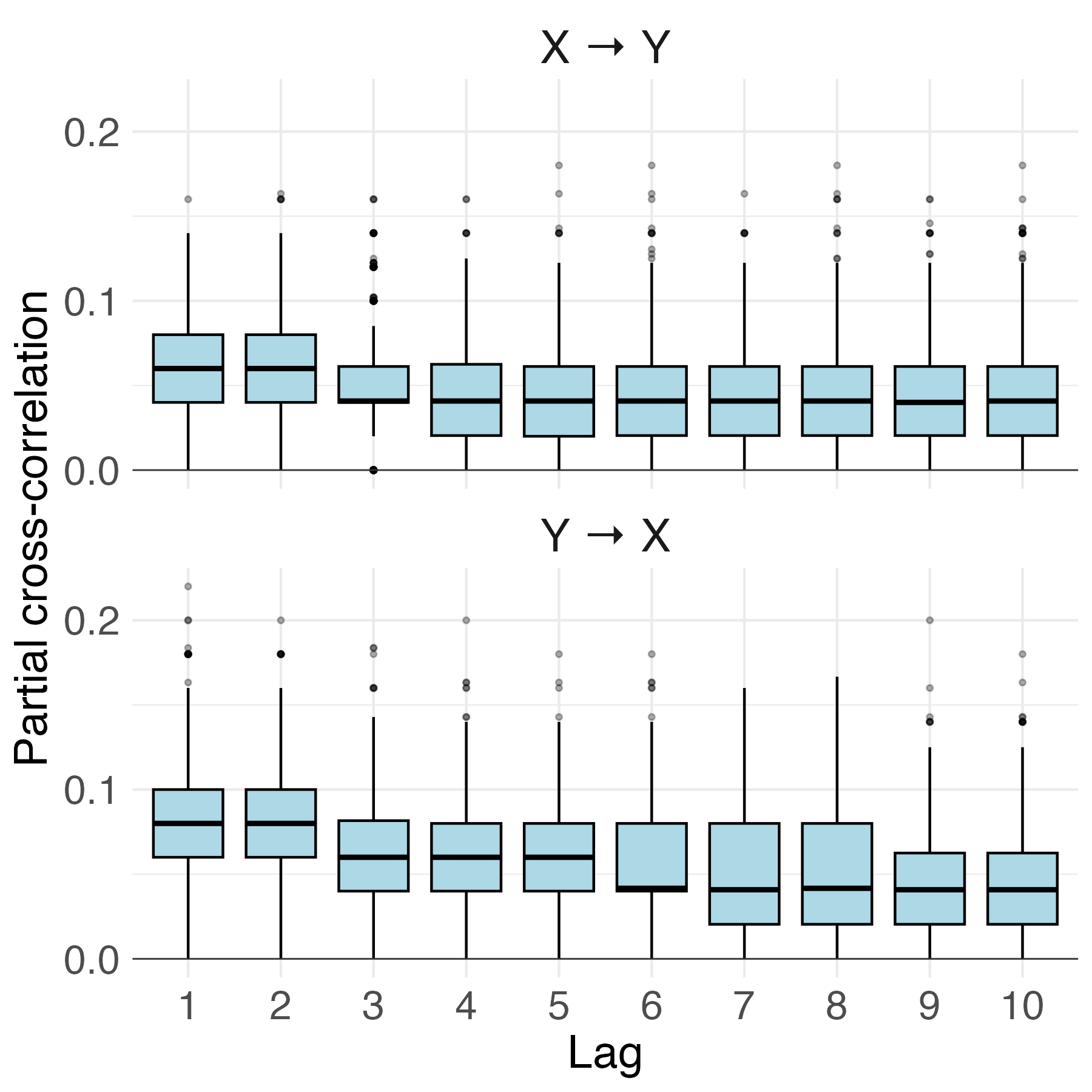}
    \end{minipage}

    \caption{\footnotesize Estimated cross-extremograms across lags. 
    Time series models are repeatedly generated with Student's $t$ noise.
    Each boxplot summarizes results over 1000 realizations, showing the interquartile range (box), median (horizontal line), and outliers beyond 1.5 times the interquartile range (dots).}
    \label{fig:extremogram_student_models}
\end{figure}

We now present simulation results for estimating the extremal delay parameter in the compound causal tail coefficients, using the two approaches outlined in Section \ref{sec:pccf}. 
For each model presented in Section \ref{sec:mc_setup}, we simulated time series of length $n=1000$ with Student's $t$ noise and estimated both the partial cross-correlation function and the cross-extremogram for ten lag values. This procedure was repeated 1000 times. 
Figures~\ref{fig:pccf_student_models} and \ref{fig:extremogram_student_models} display boxplots of the resulting estimates. Each boxplot shows the interquartile range (box), median (horizontal line), and outliers beyond $1.5$ times the interquartile range (dots).

For VAR models \ref{model:1}-\ref{model:6}, and \ref{model:8}, the PCCF reliably detects the cross-lag structure, with $\hat{\phi}_{XY}(3)$ typically above $0.15$. In the single-lag models \ref{model:2}–\ref{model:4}, the gradual decay of $\hat{\phi}_{XY}(\tau)$ around lag 3 reflects their autoregressive structure, in contrast to the consistently high PCCF values in the multi-lag models~\ref{model:5}–\ref{model:6}. For all non-causal directions in these models, the estimated partial cross-correlations remain near zero. 
However, the PCCF fails in the thresholded models~\ref{model:7} and \ref{model:9}, where causality is confined to the tail. This limitation is expected, since the PCCF relies on linear regressions over the full distribution, making tail-specific dependencies too subtle to detect. The issue is mitigated when the models are simulated with standard Pareto noise, where extremes are more frequent and pronounced. 
Overall, the PCCF offers a reasonable approach for selecting the extremal delay parameter, particularly when the underlying data exhibit a VAR-like structure.

The cross-extremogram yields comparable but slightly weaker performance than the PCCF across all nine models with Student’s $t$ noise. For VAR models, the estimated extremogram $\hat{\rho}_{XY}(\tau)$ attains slightly higher values at the true causal lags than at non-causal lags, but the difference is marginal since the extremogram, defined via conditional probabilities, is inherently non-negative. Estimation also shows wider interquartile ranges than the PCCF, reflecting higher variance from relying only on upper-tail observations.
Surprisingly, the extremogram does not outperform the PCCF in the thresholded models~\ref{model:7} and \ref{model:9}, where no causal lags are detected (Figure~\ref{fig:extremogram_student_models}). Its performance improves considerably, however, under standard Pareto noise.

In practice, selecting the extremal delay parameter is best approached by combining multiple criteria. When the number of time series is small, as in the space-weather application discussed in Section \ref{sec:spacew}, it is computationally feasible to compute and visualize the compound causal tail coefficients $\hat{\Gamma}_{{\X} \rightarrow {h(\Y)}}(p)$ over a range of lag values. In high dimensional settings, such as those involving network data, one may either estimate a separate extremal delay parameter for each pair of time series or assign a common parameter to groups of similar series. It is also advisable to estimate both the PCCF and the extremogram, examine their decay patterns as the lag increases, and select the extremal delay based on visual inspection or a predefined thresholding rule, as outlined in Section \ref{sec:pccf}. Finally, domain knowledge should be incorporated whenever available, as the selected extremal delay parameter should align with the maximum expected time delay in the underlying physical system.

\subsection{Weight distribution} \label{sec:weight_distr}

\begin{figure}[htbp]
    \centering
    \includegraphics[width=0.9\textwidth]
    {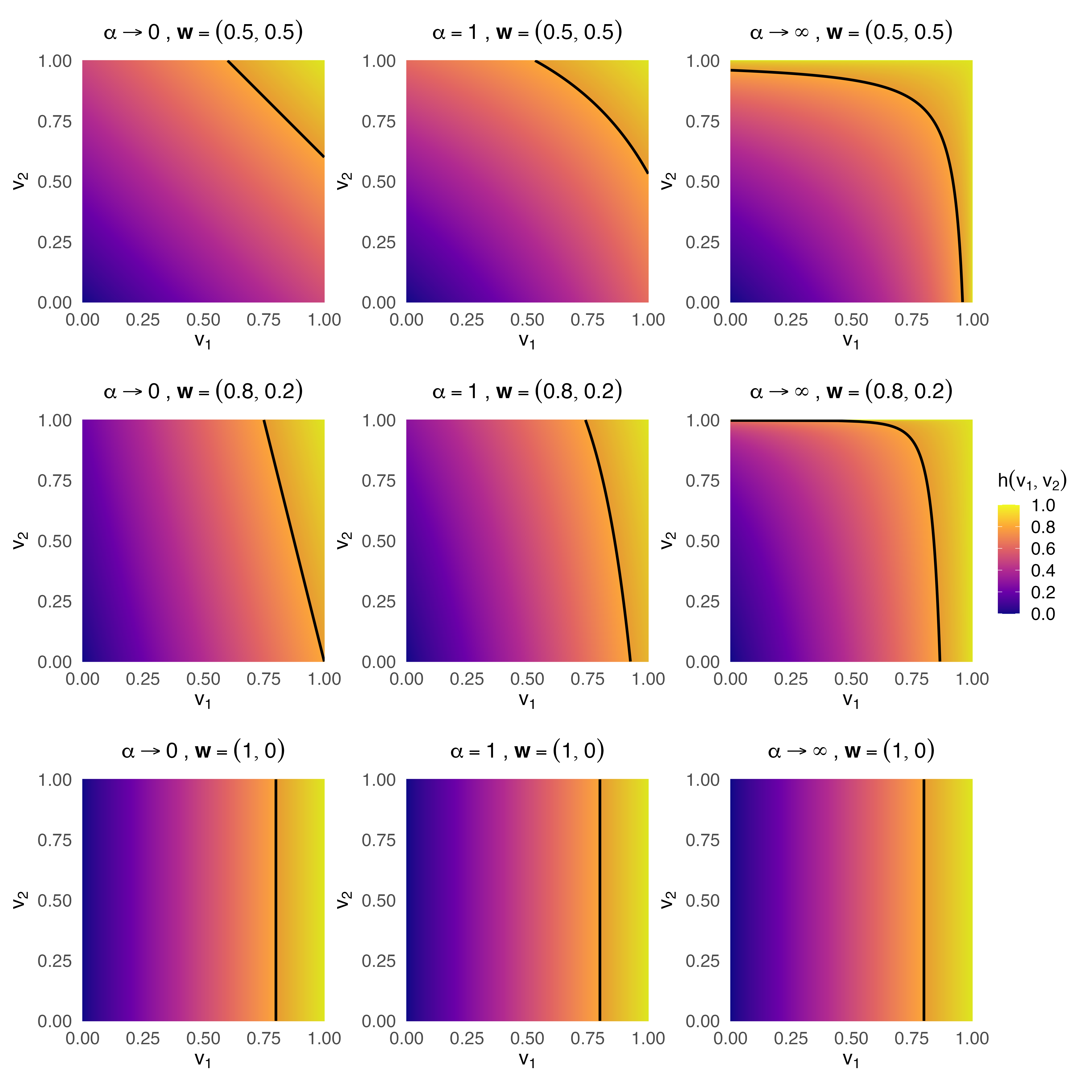}
    \caption{\footnotesize
    Expanded geometry of the two component impact function. Columns correspond to $\alpha\to0$, $\alpha=1$, and $\alpha\to\infty$, while rows correspond to $\vw=(0.5,0.5)$, $\vw=(0.8,0.2)$, and $\vw=(1,0)$. Colors indicate the values of $h(v_1,v_2;\vw,\alpha)$, and the black contour marks the boundary of $\mathcal{E}_{0.8}(\vw,\alpha)$.}
    \label{fig:impact_geometry_full}
\end{figure}

Figure~\ref{fig:impact_geometry} in Section~\ref{sec:impact} illustrates selected combinations of the impact shape parameter $\alpha$ and the weight vector $\vw$. For completeness, Figure~\ref{fig:impact_geometry_full} provides an expanded $3\times3$ comparison covering all combinations of $\alpha\to0$, $\alpha=1$, and $\alpha\to\infty$ with $\vw=(0.5,0.5)$, $\vw=(0.8,0.2)$, and $\vw=(1,0)$. 
Across the first two rows, increasing $\alpha$ produces progressively more curved contours, while unequal weights shift the impact toward the more heavily weighted component. In the final row, $\vw=(1,0)$ gives $h(v_1,v_2;(1,0),\alpha)=v_1$, so the impact and its contour are unchanged across all values of $\alpha$.

\subsubsection{Weight optimization}
We next provide the practical details of the weight optimization procedure introduced in Section~\ref{sec:weight_opt}. The optimization is carried out using the differential evolution algorithm, a global optimization method from the R package \texttt{DEoptim} \citep{MAGWC2011}, or its updated version \texttt{RcppDE} \citep{Eddelbuettel2025}. The number of candidate solutions is set to ten times the dimension $p$ of the weight vector.
Recall that the weights are constrained to be non-negative and normalized to sum to one. 
To incorporate these constraints into the optimization, we re-parameterize the weights using the softmax function:
\begin{align*}
    w_i = \frac{e^{\tilde{w}_i}}{\sum_{j=1}^p e^{\tilde{w}_j}}.
\end{align*}
The vector $\tilde{\vw} = (\tilde{w}_1, \dots, \tilde{w}_p)$ lies in the full Euclidean space $\R^p$, allowing us to perform an unconstrained optimization of the causal tail coefficient over $\tilde{\vw}$. The final weight parameters ${w}_1, \dots, {w}_p$ are then recovered by applying a simplex transformation to $\tilde{\vw}$.
This simplex-based approach is preferable to direct optimization over $\vw$ followed by \textit{post hoc} normalization, as it ensures the constraints are satisfied inherently throughout the optimization process. Given that the softmax transformation yields weights in the interval $(0,1)$, future work may explore incorporating sparsity constraints to identify the most relevant weights within $[0,1]$.

\begin{figure}[htbp]
  \centering

  \begin{minipage}[b]{0.88\linewidth}
    \raggedright \footnotesize Model \ref{model:2}: $Y_t = 0.5 X_{t-3} + \epsilon^Y_t$ \\
    \centering
    \includegraphics[width=\linewidth]{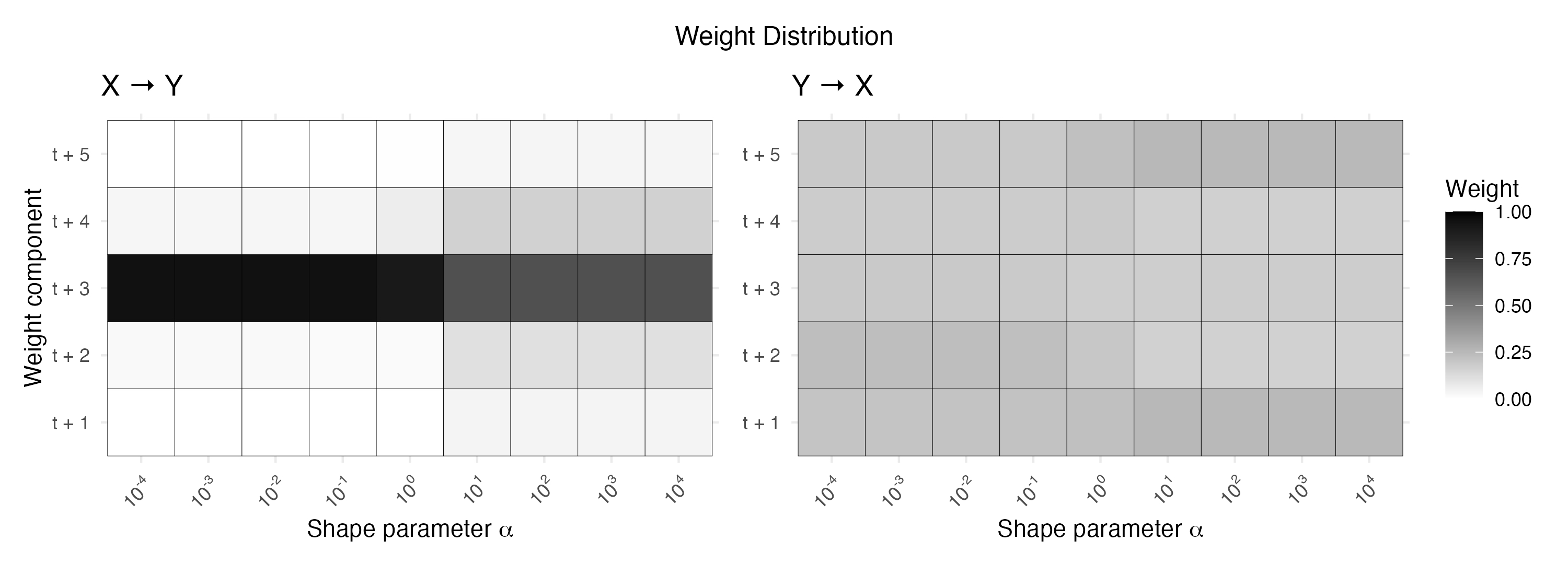}
  \end{minipage}

  \vspace{1em}

  \begin{minipage}[b]{0.88\linewidth}
    \raggedright \footnotesize Model \ref{model:3}: $Y_t = 0.5 Y_{t-1} + 0.5 X_{t-3} + \epsilon^Y_t$ \\
    \centering
    \includegraphics[width=\linewidth]{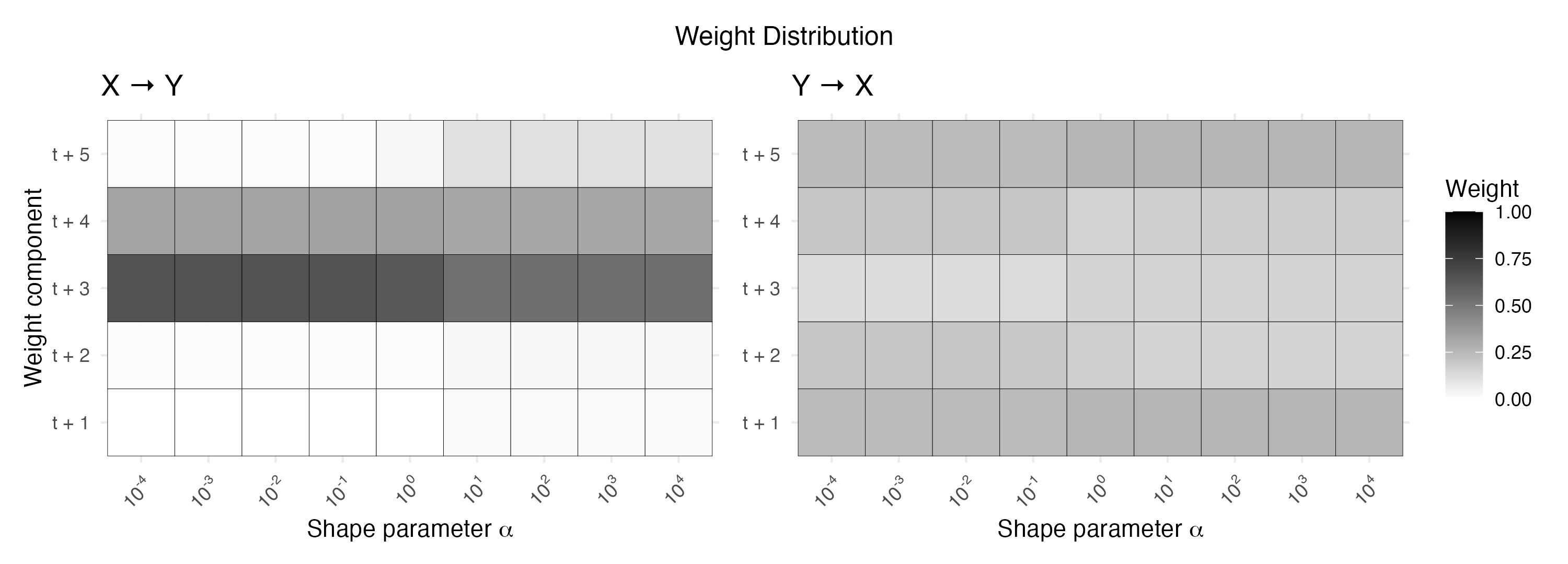}
  \end{minipage}

  \vspace{1em}

  \begin{minipage}[b]{0.88\linewidth}
    \raggedright \footnotesize Model \ref{model:6}: $Y_t = 0.5 Y_{t-1} + 0.25 X_{t-1} + 0.25 X_{t-2} + 0.25 X_{t-3} + \epsilon^Y_t$ \\
    \centering
    \includegraphics[width=\linewidth]{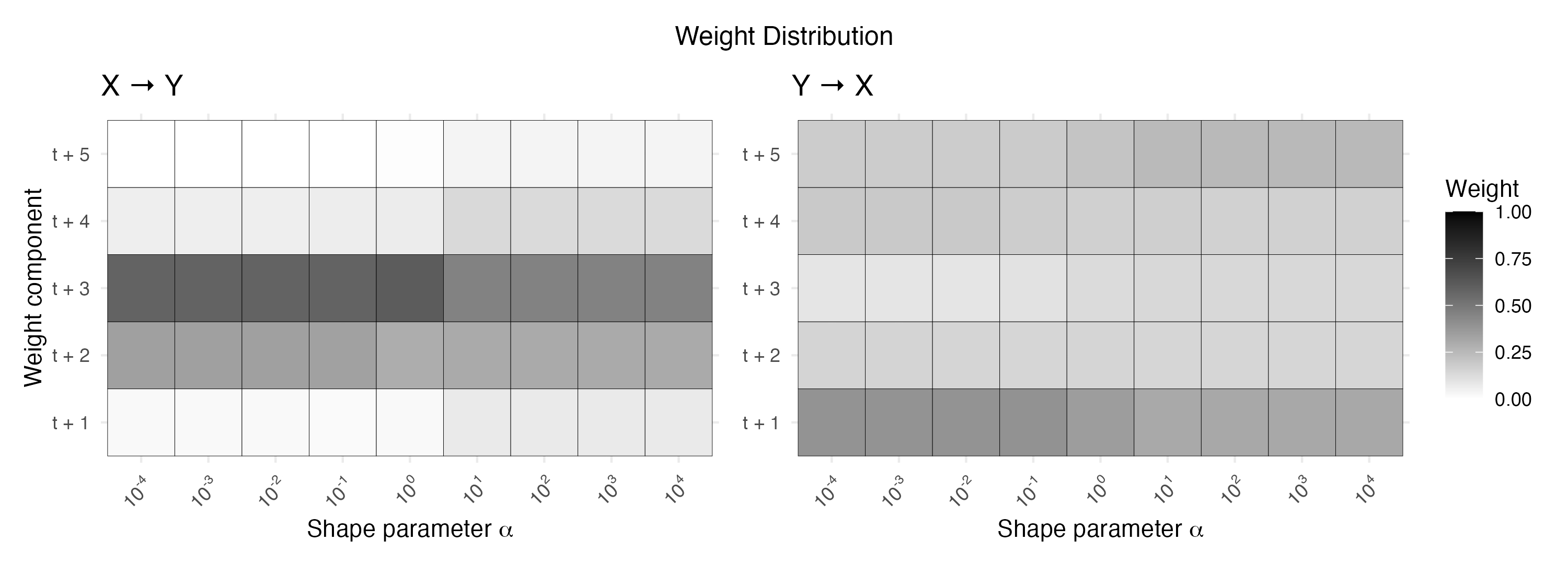}
  \end{minipage}

  \caption{\footnotesize Heatmap of unit weight allocation across lagged components for varying shape parameter values. 
  Weights are obtained by maximizing the compound causal tail coefficient, and each cell in the heatmap represents the mean weight across 100 realizations.}
  \label{fig:weights}
\end{figure}

The remainder of this section examines the distribution of optimized weights from Section \ref{sec:weight_opt} for various values of the impact shape parameter $\alpha$ in the impact function, across different time series models. Simulations followed the same design as in Section \ref{sec:shape}, with the exception that the extremal delay parameter was set to $p = 5$. This slightly misspecified choice of $p$ allows us to better assess whether the resulting weight distribution aligns with the true causal lag. We focused on Models \ref{model:2}, \ref{model:3}, and \ref{model:6} with Student's $t$ noise to illustrate how differing cross-lag structures affects the resulting weight distributions.
Results, averaged over $100$ repetitions, are visualized using heatmaps in Figure \ref{fig:weights}. Each column of the heatmap represents the distribution of a unit weight across components of the weight vector $\vw = (w_1, \dots, w_5)$ for a given impact shape parameter value. The columns correspond to different values of $\alpha$, ranging over a logarithmic grid from $10^{-4}$ to $10^{4}$ in integer steps on the exponent.

The weight distribution obtained under the optimization scheme in Section \ref{sec:weight_opt} aligns with the true cross-lag structure of the underlying time series models. In the single-lag model \ref{model:2}, weights are concentrated on the lag-3 component $w_3$ in the correct causal direction $X \to Y$. By comparison, Model \ref{model:3} assigns additional weight to $w_4$. This is consistent with its autoregressive structure in the effect series $\Y$, where part of the causal influence at lag 3 propagates to lag 4. In a similar vein, the multi-lag model \ref{model:6} exhibits weight distribution across lags 1 to 4, with the highest concentration at lag 3. This pattern again reflects causal propagation due to its autoregressive structure.
In the reverse, non-causal direction $Y \to X$, the weights in all three models are distributed more uniformly across lag components.
Therefore, the optimized weight parameters effectively reflect the significance of individual lagged variables in a compound extreme event.

The impact shape parameter $\alpha$ affects the dispersion of the weight distribution: larger values result in weights being more spread out across lag components, whereas smaller values concentrate the weights on fewer lags. As $\alpha \to 0$, the optimization tends to assign all weight to a single lag within each simulation. However, because the selected lag may differ across simulations of the same model, the averaged outcomes in Figure \ref{fig:weights} show shaded grid cells at multiple lag components.
Note that the effect of $\alpha$ on the weight distribution is consistent across different time series models. Hence, the optimal choice of $\alpha$ for approximating the underlying ``true" weight distribution is inherently model-dependent.

\subsection{Weight specification}\label{sec:weight_spec}

Although the results in Section~\ref{sec:weight_distr} indicate that the optimized weights generally reflect the underlying cross-lag structure, the optimization procedure does not guarantee that Assumption~1 is satisfied. We therefore examine the sensitivity of the compound CTC and bootstrap test to alternative weight specifications.
To this end, we conduct an additional simulation study under Models~\ref{model:2} and~\ref{model:5}. These two settings represent complementary cases: Model~\ref{model:2} contains a single causal lag, whereas Model~\ref{model:5} contains causal effects distributed across all three lags and is therefore closer to the intended compound extreme setting. For each model, we compare the optimized weight vector used in the main simulation study with several fixed benchmark choices. 
In Model~\ref{model:2}, where the true causal effect occurs at lag 3, the oracle vector is $\vw=(0,0,1)$. We then consider three deliberately misspecified choices: a partially misspecified vector $\vw=(1/2,0,1/2)$, which assigns some weight to a non-causal lag; a severely misspecified vector $\vw=(1,0,0)$, which places all weight on a non-causal lag; and a uniform misspecified vector $\vw=(1/3,1/3,1/3)$, which dilutes the true causal lag with irrelevant lags. 
In Model~\ref{model:5}, where all three lags contribute causally, the oracle vector is taken to be the uniform vector $\vw=(1/3,1/3,1/3)$. To examine the effect of under-representing the compound lag structure, we compare this oracle choice with single-lag specifications $\vw=(1,0,0)$, $\vw=(0,1,0)$, and $\vw=(0,0,1)$, together with the optimized weights.

Following the setup in Section~\ref{sec:mc_setup}, for each model and weight specification, we generated time series of length $n=1000$ with Student's $t$ noise. We computed the compound CTC in both directions using $k=\sqrt{n}$ extreme observations, the true extremal delay $p=3$, and the impact shape parameter $\alpha=10^4$. We then applied the MBB hypothesis test in each direction by shifting the corresponding effect series forward by $s=p=3$ time units and generating $b=100$ bootstrap samples with block length $\ell=n^{1/3}$. The one-sided test rejected the null hypothesis when the resulting bootstrap $p$-value was at most $0.05$. This procedure was repeated over 100 independent realizations. For each specification, we report the mean compound CTC and the empirical rejection rate of the moving block bootstrap test across the 100 realizations in both directions. 
This comparison allows us to distinguish the behavior of the proposed optimized weights from the idealized oracle case and to quantify the finite sample impact of violating, or only partially representing, the theoretical weight condition in Assumption~\ref{weight assumption}.

\begin{table}[htbp]
\centering
\caption{\footnotesize Sensitivity of the compound CTC and bootstrap test to weight specification. Entries report the mean compound CTC and the percentage of rejections at the 5\% significance level across 100 independent repetitions.}
\label{tab:weight_sensitivity}
\begin{tabular}{llcccc}
\toprule
& & \multicolumn{2}{c}{Mean compound CTC}
& \multicolumn{2}{c}{Rejection rate} \\
\cmidrule(lr){3-4}\cmidrule(lr){5-6}
Model
& Weights $\vw$
& $\X \rightarrow \Y$
& $\Y \rightarrow \X$
& $\X \rightarrow \Y$
& $\Y \rightarrow \X$ \\
\midrule
2
& Optimized
& 0.773
& 0.565
& 99\%
& 2\% \\

& $(0,0,1)$
& 0.759
& 0.502
& 100\%
& 10\% \\

& $(1/2,0,1/2)$
& 0.730
& 0.550
& 97\%
& 2\% \\

& $(1/3,1/3,1/3)$
& 0.731
& 0.554
& 98\%
& 3\% \\

& $(1,0,0)$
& 0.572
& 0.513
& 35\%
& 1\% \\

\midrule
5
& Optimized
& 0.797
& 0.579
& 99\%
& 0\% \\

& $(1/3,1/3,1/3)$
& 0.792
& 0.567
& 99\%
& 0\% \\

& $(1,0,0)$
& 0.726
& 0.533
& 95\%
& 0\% \\

& $(0,1,0)$
& 0.741
& 0.519
& 98\%
& 1\% \\

& $(0,0,1)$
& 0.723
& 0.509
& 95\%
& 6\% \\
\bottomrule
\end{tabular}
\end{table}

The results in Table~\ref{tab:weight_sensitivity} indicate that our proposed methodology is generally robust to moderate weight misspecification, but may lose power when the chosen weights poorly represent the true causal lag structure. The optimized weights perform similarly to the oracle choices in both models, yielding comparable mean compound CTCs and rejection rates. 
In Model~\ref{model:2}, partial and uniform misspecification lead to only modest reductions in the estimated coefficient and test power, whereas placing all weight on the non-causal first lag substantially weakens directional separation and reduces the rejection rate in the true causal direction to $35\%$. 
In Model~\ref{model:5}, the method remains robust when only a single causal lag is used, although the estimated compound CTC is lower than under the optimized or oracle weights. 
Across all settings, rejection rates in the reverse direction are generally low, suggesting that weight misspecification primarily reduces power rather than producing spurious causal detection.

\subsection{Shift parameter in block bootstrap}\label{sec:sensi_shift}
So far, we have considered parameters that directly affect the value of the compound causal tail coefficient. We now turn to the bootstrap hypothesis test, which evaluates whether the coefficient is statistically significant to support the existence of a causal relationship. The test uses a shifting technique that shifts one time series by the extremal delay $p$ (Section~\ref{sec:mbb}). We now perform a sensitivity analysis to assess how the results change when the shift is $s$ time units, where $s$ is not necessarily equal to $p$. We refer to $s$ as the shift parameter.

\begin{figure}[htbp]
    \centering
    \includegraphics[width=0.9\textwidth]{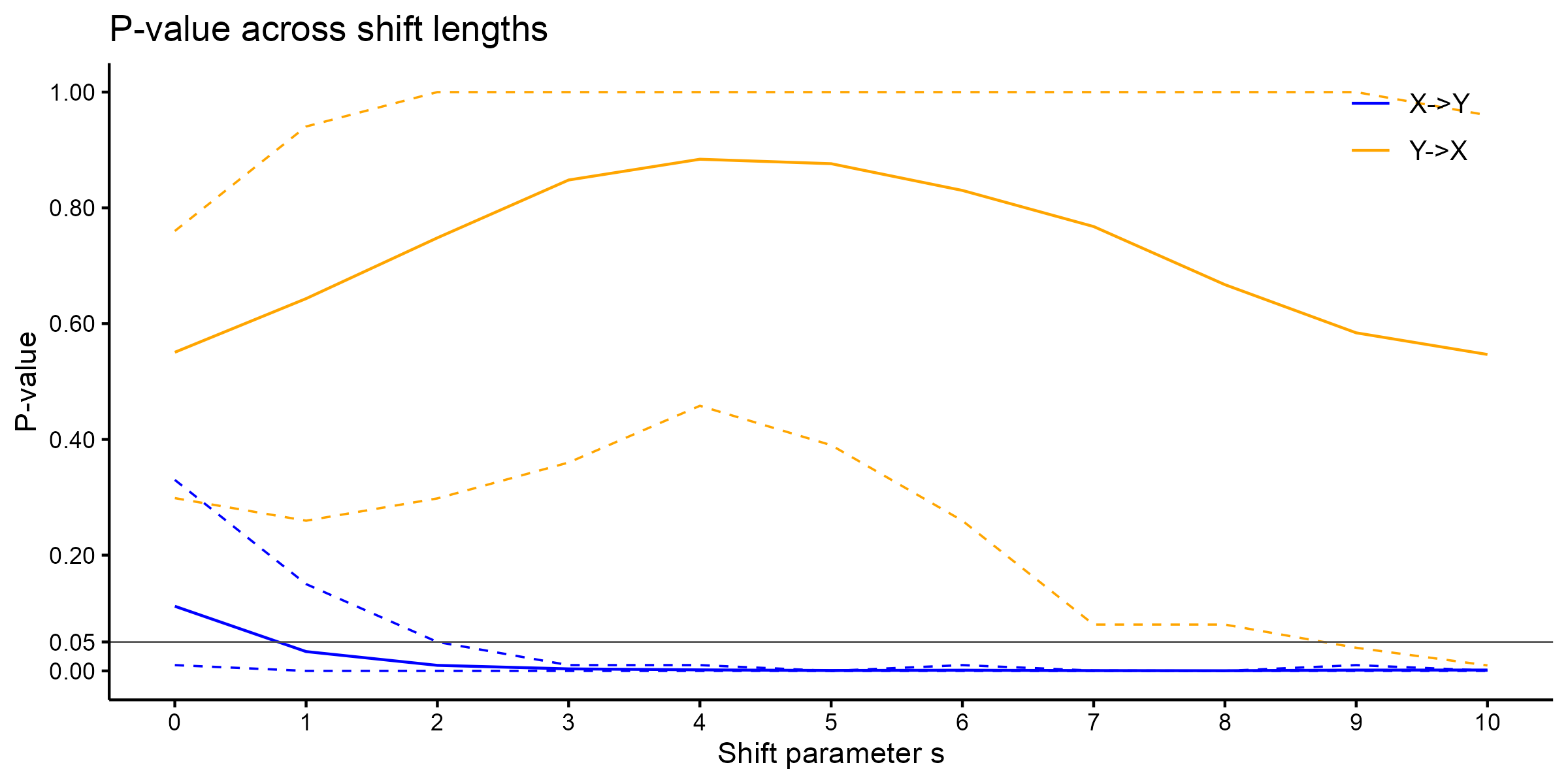}
    \includegraphics[width=0.9\textwidth]{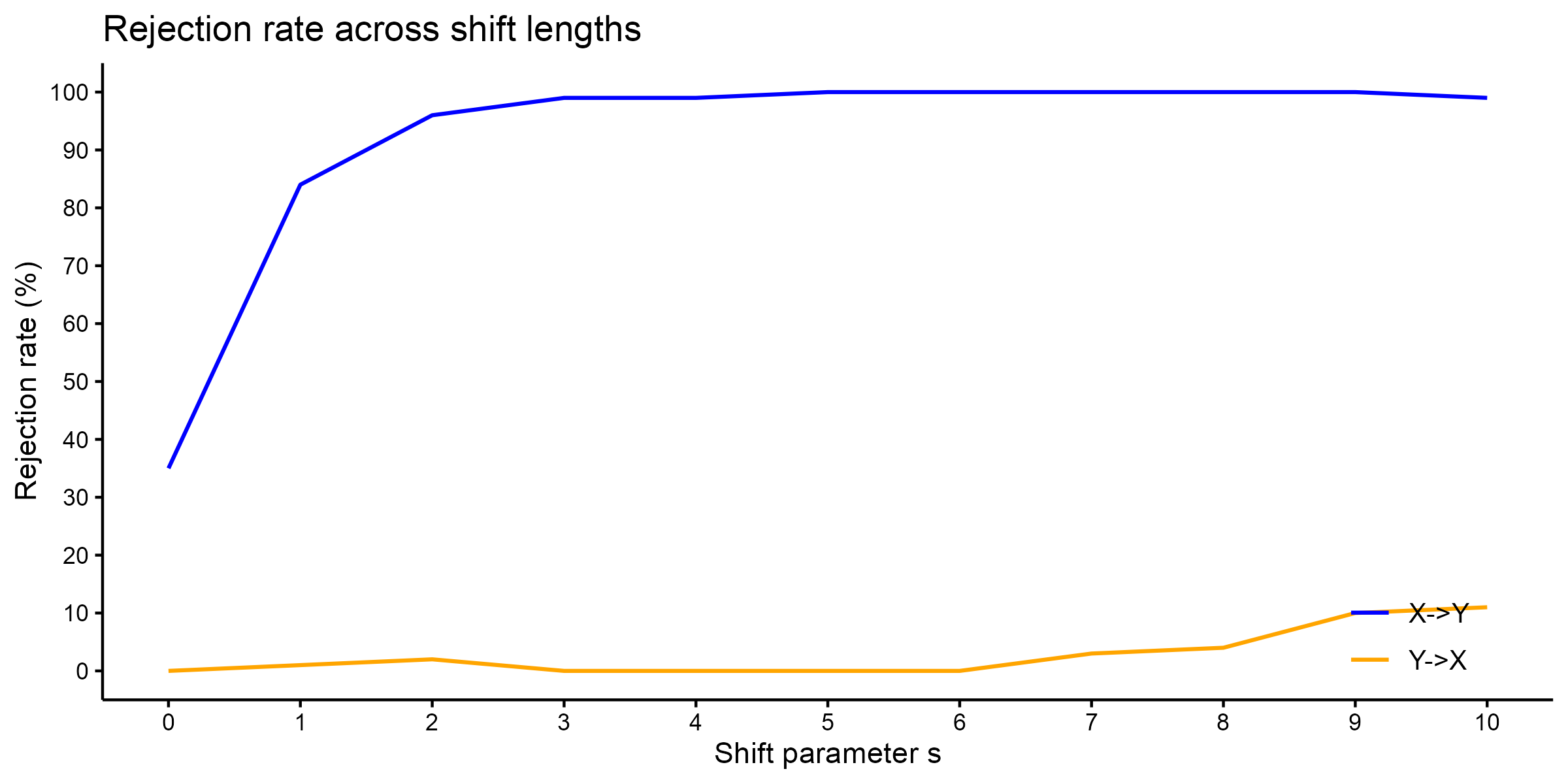}
    \caption{\footnotesize Behavior of $p$-values and rejection rates across varying shift lengths $s$ in the block bootstrap hypothesis test.
    In the upper panel, solid lines show the mean bootstrap $p$-values across 100 realizations, with dotted lines indicating the empirical 5\% and 95\% pointwise quantiles. The lower panel displays the proportion of rejections at the 5\% significance level across 100 realizations. Blue corresponds to the true causal direction $X \rightarrow Y$, and yellow to the reverse direction $Y \rightarrow X$.}
    \label{fig:shift}
\end{figure}

We performed Monte Carlo simulations using Model \ref{model:6}, the multi-lag autoregressive-effect model, to examine how the shift parameter affects the performance of the proposed bootstrap test in Section \ref{sec:mbb}. Following the setup in Section \ref{sec:mc_setup}, we generated a time series of length $n=1000$ with Student's $t$ noise and computed the observed compound causal tail coefficients for both directions using $k=\sqrt{n}$ extreme values, the true extremal delay $p=3$, and a impact shape parameter $\alpha=10^4$. An MBB hypothesis test was then conducted: we shifted the second time series forward in time by a given $s$, generated $b=100$ bootstrap samples with block length $\ell = n^{1/3}$, estimated bootstrapped coefficients, and calculated the one-sided $p$-value. This procedure was repeated 100 times, and the number of rejections (i.e., $p$-value $\leq 0.05$) was recorded for each shift parameter value $s$. We considered $s$ from 0 to 10, as shift values beyond the block length in the MBB are unnecessary. 
The first panel in Figure \ref{fig:shift} shows the mean along with the 5\% and 95\% empirical quantiles of the calculated $p$-values and the second panel presents the rejection rate from 100 repeated simulations across different shift values.

For the true causal direction $X \to Y$, $p$-values rapidly decrease below $0.05$ as the shift parameter becomes non-trivial and converge to zero at the true extremal delay $p = 3$. Accordingly, the rejection rate approaches 100\% once the shift parameter equals 3. These patterns persist up to the optimal block length $\ell = n^{1/3} = 10$. 
In contrast, for the reverse direction, $p$-values stay well above the 5\% significance level until $s$ reaches twice the true extremal delay. While the rejection rate stays near 0 for $s \leq 6$, the variance of the $p$-values increases substantially after $s = 4$. As such, an optimal choice for $s$ lies within the range of 3 to 6, with values closer to 3 being preferable.

The changes in the behavior of $p$-values and rejection rate between the true extremal delay and twice that value are not specific to this time series model. Similar patterns are observed across other models discussed in Section \ref{sec:mc_setup}, although the precise optimal choice of $s$, in terms of minimizing bias and variance, differs across settings. Based on empirical findings, setting the shift parameter equal to the extremal delay parameter, $s = p$, appears to be a reasonable choice, especially given that the true extremal delay may be unknown.

\subsection{Adjustment for confounding variables}\label{sec:sensi_confounding}

Results for Models \ref{model:8} and \ref{model:9} in Section \ref{sec:compare} have demonstrated that the proposed block bootstrap hypothesis test, combined with the compound causal tail coefficient, effectively detects extremal causality between pairs of time series even in the presence of a confounding time series $\ZZ$. In this section, we discuss how the conditional version of the compound causal tail coefficient, as defined in \eqref{CCTC_C} can further help distinguish true structural causality from spurious causality induced by confounders.

To incorporate the conditioning set $\ZZ_{\past(p)} = {Z_1, \dots,Z_{p-1}, Z_p}$ - as specified in the definition of conditional compound causal tail coefficient in \eqref{CCTC_C} - into the empirical estimator, we modify the nonparametric estimator of the unconditional coefficient in \eqref{estimator} and propose the following \textit{tail-adjusted filtering estimator} for the conditional coefficient:
\begin{align*}
    \resizebox{\textwidth}{!}{$
    \hat{\Gamma}_{{\X} \rightarrow {h(\Y}) \mid \ZZ} (p) = \frac{1}{k_c} \sum_{i=1, \ldots, n-p} h \left( \hat{F}_Y(y_i), \ldots, \hat{F}_Y(y_{i+p}) \right) 
    \cdot \1_{ \{x_{i+1} \geq x_{(n-k+1)} \} }
    \cdot \1_{ \left\{ \bigcap_{j=1}^{\min(p,i)} \{z_{i-j} < z_{(n-k+1)} \} \right\} }
    $,}
\end{align*}
where 
$
k_c := \left| \left\{ i \in \{1, \dots, n - p\} : x_i \geq x_{(n - k + 1)} \ \wedge \ \forall j \in \{1, \dots, \min(p,i) \},\ z_{i-j} < z_{(n - k + 1)} \right\} \right| \leq k
$ helps compute the empirical average of the impact function over the conditioning set. 
This adjustment accounts for the confounding time series $\ZZ$ by conditioning on all its values within the extremal delay, $Z_{t-1},\dots, Z_{t-p}$, being non-extreme. By doing so, we eliminate scenarios where an extreme value in the confounding series $\ZZ$ causes a compound extreme event in the effect series $\Y$, thereby ensuring all remaining extremes in $\Y$ are attributable to extreme values of $\X$.
Note that the contemporaneous value $Z_t$ is excluded for consistency, as immediate causal effects are not considered in this study.

\begin{figure}[htbp]
    \centering

    \begin{minipage}[b]{0.48\textwidth}
        \centering
        \includegraphics[width=\linewidth]{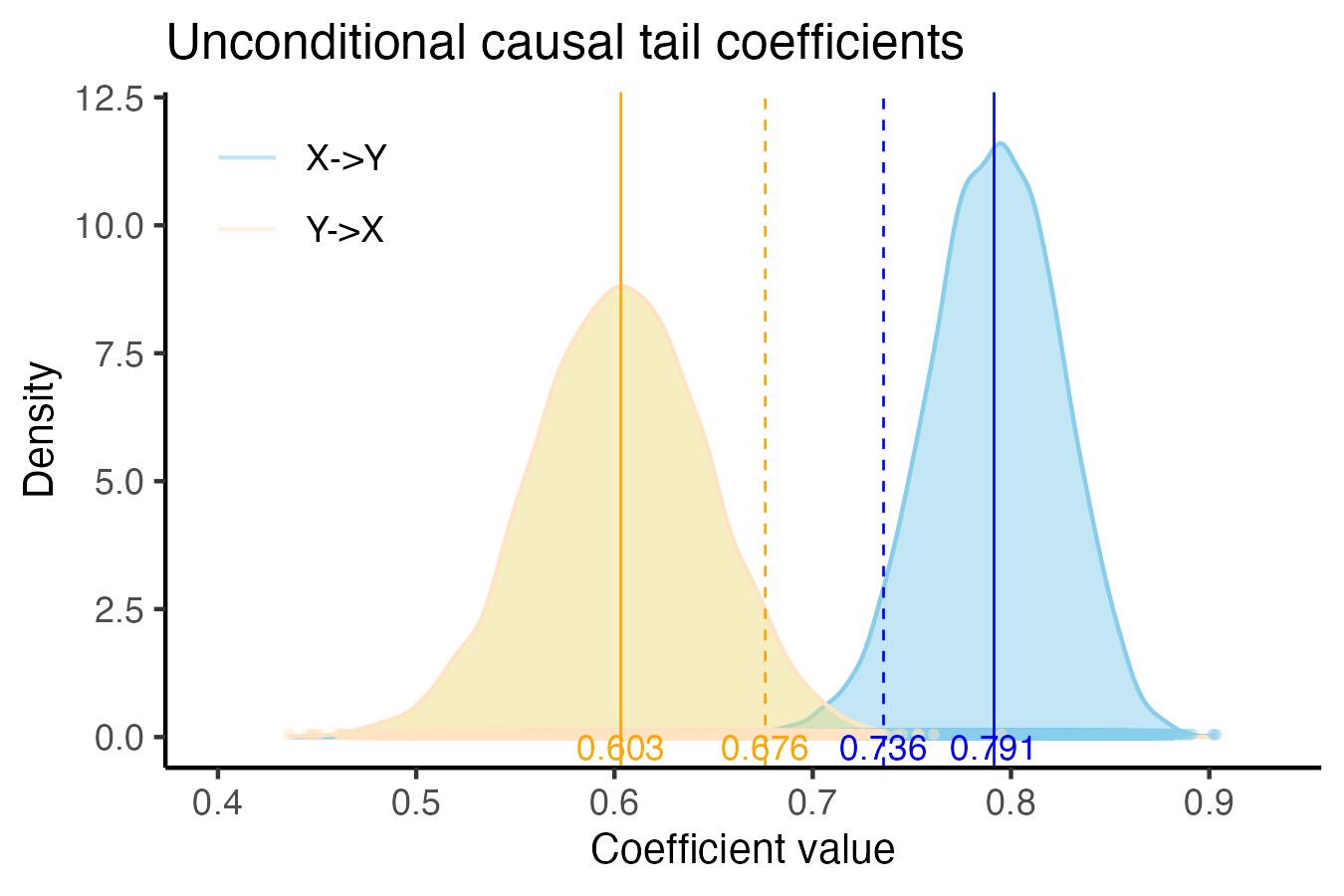}
    \end{minipage}
    \hfill
    \begin{minipage}[b]{0.48\textwidth}
        \centering
        \includegraphics[width=\linewidth]{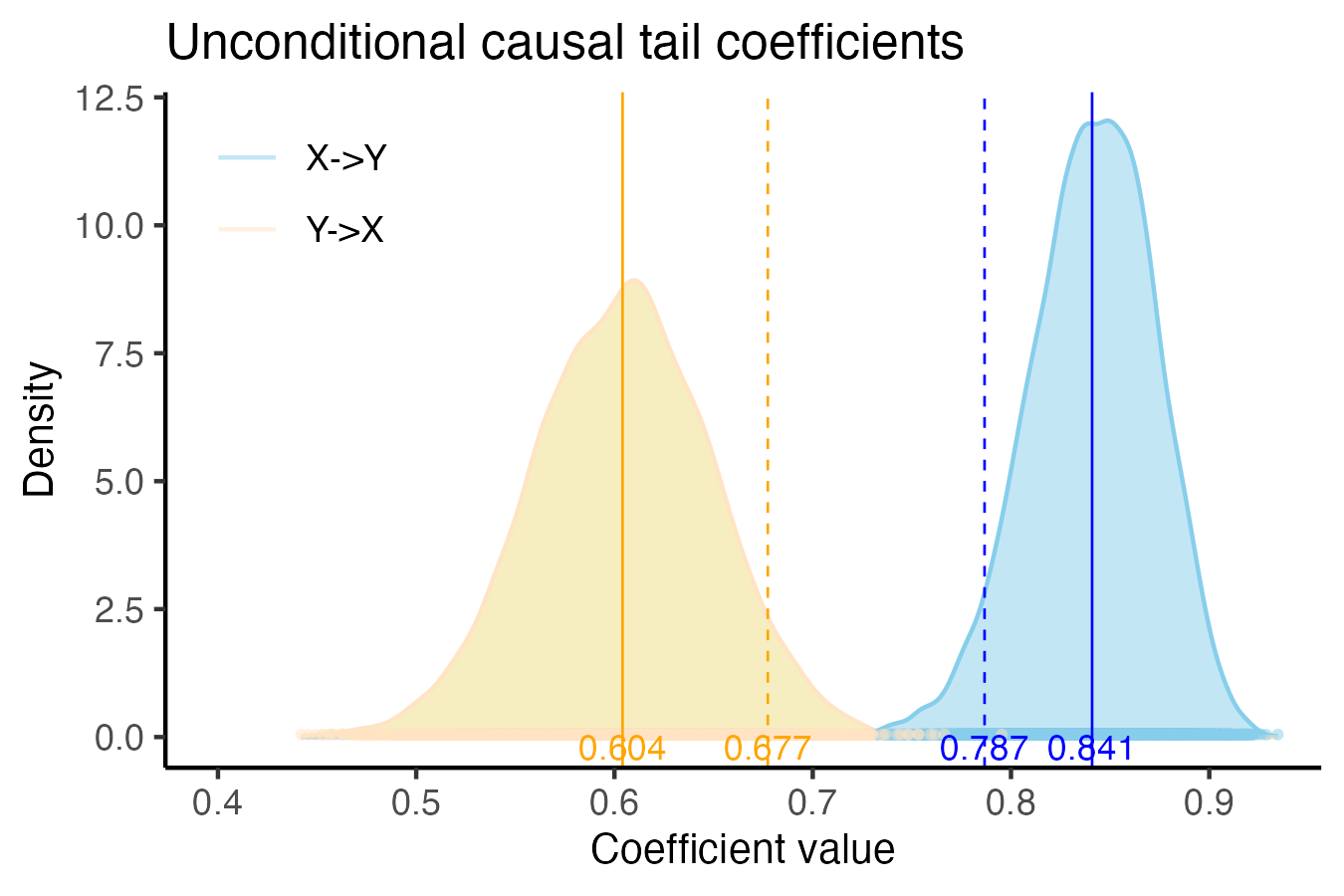}
    \end{minipage}

    \vskip\baselineskip

    \begin{minipage}[b]{0.48\textwidth}
        \centering
        \includegraphics[width=\linewidth]{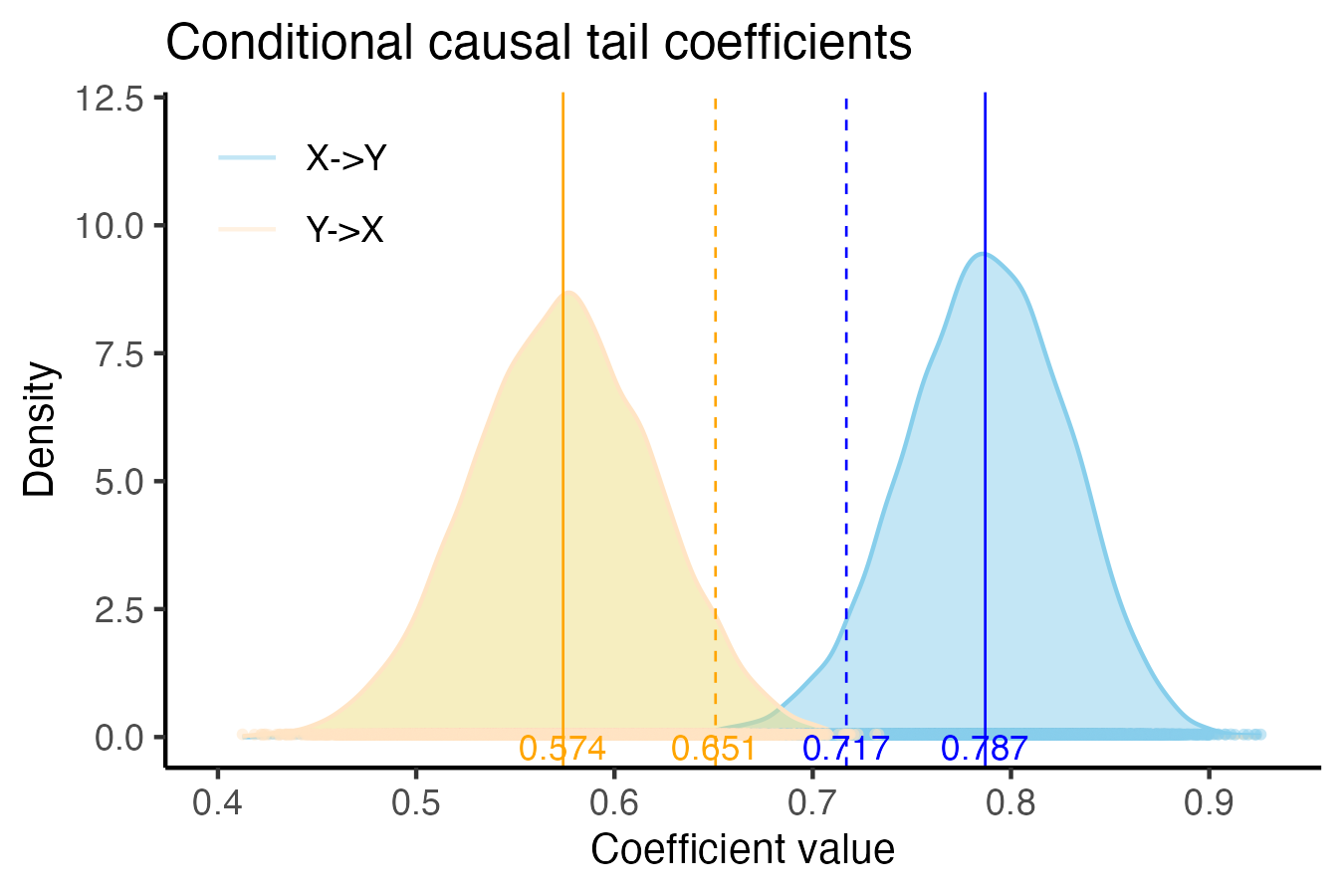}
        {\footnotesize (a) Model \ref{model:8}: VAR(3)}
    \end{minipage}
    \hfill
    \begin{minipage}[b]{0.48\textwidth}
        \centering
        \includegraphics[width=\linewidth]{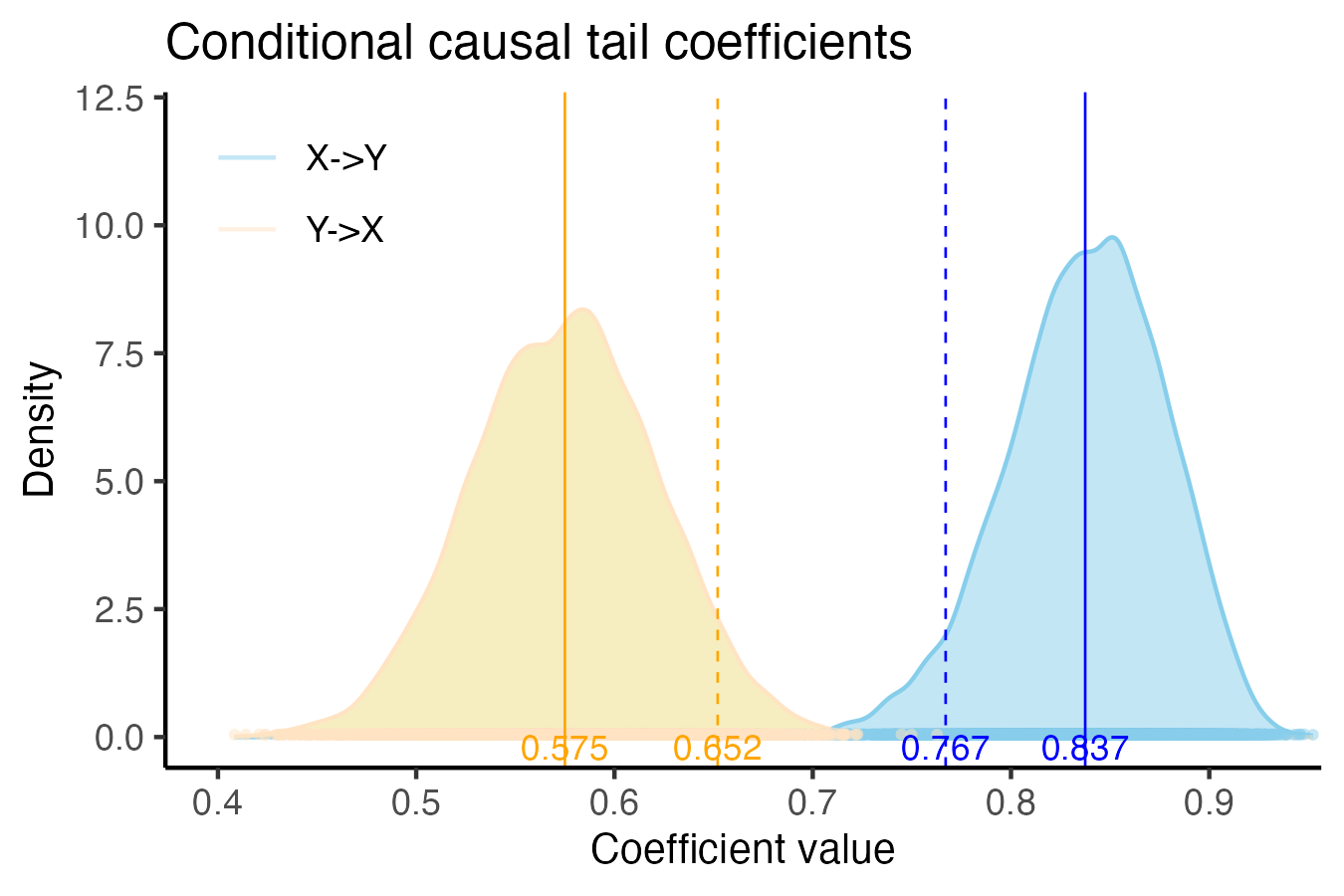}
        {\footnotesize (b) Model \ref{model:9}: thresholded NAAR(3)}
    \end{minipage}

    \caption{\footnotesize Comparison of Monte Carlo distributions for the unconditional and the conditional compound CTCs.
    Each density plot displays the kernel density estimate of unconditional or conditional coefficient values, aggregated over 1000 time series realizations. Solid vertical lines indicate the mean of each distribution, and dashed lines mark the 5th or 95th percentiles. Blue corresponds to the true causal direction $X \rightarrow Y$, and yellow to the reverse direction $Y \rightarrow X$.}
    \label{fig:confounder}
\end{figure}

Using nonparametric estimators for both unconditional and conditional versions of the compound causal tail coefficient, we followed the Monte Carlo simulation procedure outlined in Section~\ref{sec:mc_setup}. We repeatedly generated $10^4$ time series from Models \ref{model:8} and \ref{model:9} with Student's $t$ noise. For each realization, we computed the estimated compound causal tail coefficient using a representative impact shape parameter $\alpha = 10^4$. 
The resulting coefficient density distributions are shown in Figure~\ref{fig:confounder}, where blue curves correspond to the true causal direction $X \to Y$, and yellow curves to the reverse, non-causal direction $Y \to X$. Solid vertical lines indicate means of the distributions, while dashed lines denote either the 5th or 95th percentile, depending on the direction of interest in each distribution.

Conditioning on the confounder $\ZZ$ shifts the distribution of the compound causal tail coefficient leftward in both directions, with a stronger reduction in the non-causal direction $Y \to X$. This suggests that the conditional estimator effectively mitigates spurious causal effects from confounding.
The distributions of conditional coefficients also show greater spread in both directions, as using non-extreme values of $\ZZ$ reduces the effective sample size and thereby increases variance. This dispersion is more pronounced in the true causal direction, where the conditional estimates exhibit lower kurtosis compared to the unconditional case.
Meanwhile, the 95th percentile of $Y \to X$ coefficients and the 5th percentile of $X \to Y$ coefficients become more separated, suggesting a reduced overlap between the two coefficient distributions. This increased separation makes it easier to distinguish causal from non-causal directions through the asymmetry of the two coefficients.

\section{Multivariate extension}\label{sec:mv_ap}
\setcounter{model}{0}
\renewcommand{\themodel}{M\arabic{model}}
\subsection{Setup}\label{sec:mvsetup}

In this section, we investigate the behavior of the multivariate extension of the compound causal tail coefficient, ${\Gamma}_{{\X} \rightarrow {h(\bm{Y})}}$, as defined in Section \ref{sec:mv}, and assess the performance of the corresponding bootstrap hypothesis test through Monte Carlo simulations.
We extend the time series models \ref{model:1}–\ref{model:6} from Section~\ref{sec:mc_setup} by introducing an additional effect variable $\Y^2$, while preserving the original specifications of the cause variable $\X$ and the effect variable $\Y \equiv Y^1$. The full effect series is now denoted by $\bm{Y} = (\Y^1, \Y^2)$, and the extended models are detailed below.

\begin{model}\label{model:1mv}
Independent noise model: 
\begin{align*}
    X_t &= \epsilon^X_t , \\
    Y^1_t &= \epsilon^{Y^1}_t , \\
    Y^2_t &= \epsilon^{Y^2}_t .
\end{align*}
\end{model}

\begin{model}\label{model:2mv}
Single-lag model: 
\begin{align*}
    X_t &= 0.5 X_{t-1} + \epsilon^X_t , \\
    Y^1_t &= 0.5 X_{t-3} + \epsilon^{Y^1}_t , \\
    Y^2_t &= 0.5 X_{t-3} + \epsilon^{Y^2}_t .
\end{align*}
\end{model}

\begin{model}\label{model:3mv}
Single-lag autoregressive-effect model: 
\begin{align*}
    X_t &= 0.5 X_{t-1} + \epsilon^X_t , \\
    Y^1_t &= 0.5 Y^1_{t-1} + 0.5 X_{t-3} + \epsilon^{Y^1}_t , \\
    Y^2_t &= 0.5 Y^2_{t-1} + 0.5 X_{t-3} + \epsilon^{Y^2}_t .
\end{align*}
\end{model}

\begin{model}\label{model:4mv}
Single-lag bidirectional model: 
\begin{align*}
    X_t &= 0.25 X_{t-1} + 0.5 Y^1_{t-3} + \epsilon^X_t , \\
    Y^1_t &= 0.5 X_{t-3} + 0.25 Y^1_{t-1} + \epsilon^{Y^1}_t , \\
    Y^2_t &= 0.5 X_{t-3} + 0.25 Y^1_{t-3} + \epsilon^{Y^2}_t .
\end{align*}
\end{model}

\begin{model}\label{model:5mv}
Multi-lag model: 
\begin{align*}
    X_t &= 0.5 X_{t-1} + \epsilon^X_t , \\
    Y^1_t &= 0.25 X_{t-1} + 0.25 X_{t-2} + 0.25 X_{t-3} + \epsilon^{Y^1}_t , \\
    Y^2_t &= 0.5 X_{t-3} + \epsilon^{Y^2}_t .
\end{align*}
\end{model}
\noindent 

\begin{model}\label{model:6mv}
Multi-lag autoregressive-effect model: 
\begin{align*}
    X_t &= 0.5 X_{t-1} + \epsilon^X_t , \\
    Y^1_t &= 0.5 Y^1_{t-1} + 0.25 X_{t-1} + 0.25 X_{t-2} + 0.25 X_{t-3} + \epsilon^{Y^1}_t , \\
    Y^2_t &= 0.5 Y^2_{t-1} + 0.25 X_{t-1} + 0.25 X_{t-2} + 0.25 X_{t-3} + \epsilon^{Y^2}_t .
\end{align*}
\end{model}

In Models \ref{model:1mv}–\ref{model:3mv}, \ref{model:5mv}, and \ref{model:6mv}, the ground truth is that $\X$ has a compound causal effect on the bivariate series $(\Y^1, \Y^2)$, whereas $\Y^1$ has no such effect on $(\X, \Y^2)$. In contrast, Model \ref{model:2mv} is bidirectional, with $\Y^1$ also having a compound causal effect on $(\X, \Y^2)$.
For each of the models described above, we examined three different distributions for the noise variables $\epsilon^X_t$, $\epsilon^Y_t$, and $\epsilon^Z_t$: Student's $t$, standard Pareto, and Poisson with rate 3. Student's $t$ noise is generated with $df= 2$ for $\Y^1$ and $df = 10$ for $\X$ and $\Y^2$. The multivariate compound causal tail coefficient is estimated using a nonparametric estimator analogous to \eqref{estimator}, defined by
\begin{align*}
    \hat{\Gamma}_{{\X} \rightarrow {h(\Y^1, \Y^2})} (p) = \frac{1}{k} \sum_{i=1}^{n-p} h 
    \left(\hat{F}_{Y^1}(y^1_{i+1}), \ldots, \hat{F}_{Y^1}(y^1_{i+p}), 
    \hat{F}_{Y^2}(y^2_{i+1}), \ldots, \hat{F}_{Y^2}(y^2_{i+p}) \right) 
    \cdot \1_{ \{x_i \geq x_{(n-k+1)} \} }.
\end{align*}
The rest of the simulation procedure follows exactly as described in Section \ref{sec:mc_setup}.

\subsection{Results}\label{sec:mv_results}
We now present Monte Carlo simulation results for the multivariate extension of our causal inference method, applying the proposed bootstrap hypothesis test to detect extremal causality from a univariate cause series to a bivariate effect series. Specifically, we compare the causal direction $\X \to (\Y^1, \Y^2)$ with the non-causal direction $\Y^1 \to (\X, \Y^2)$.

\begin{figure}[htbp]
    \centering
    \begin{minipage}{0.4\textwidth}
        \centering
        \includegraphics[width=\linewidth]{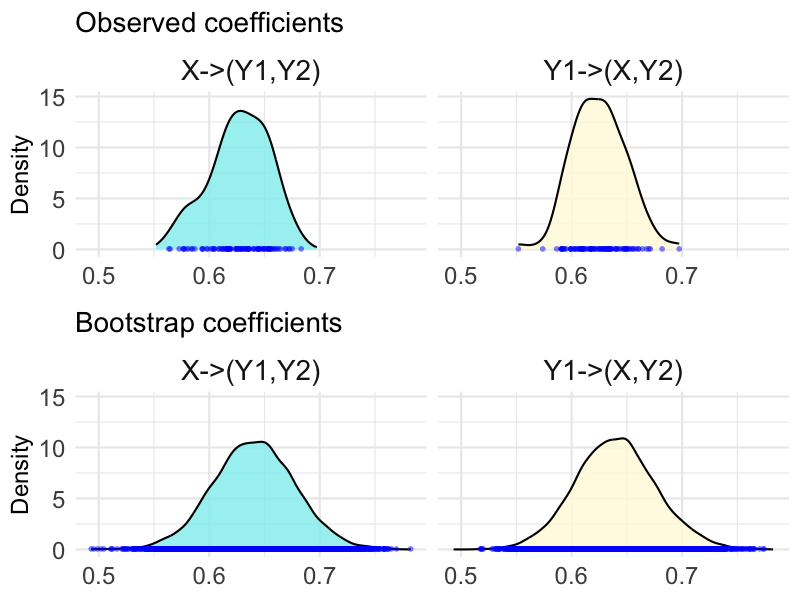}
    \end{minipage}%
    \hspace{0cm}
    \begin{minipage}{0.4\textwidth}
        \centering
        \includegraphics[width=\linewidth]{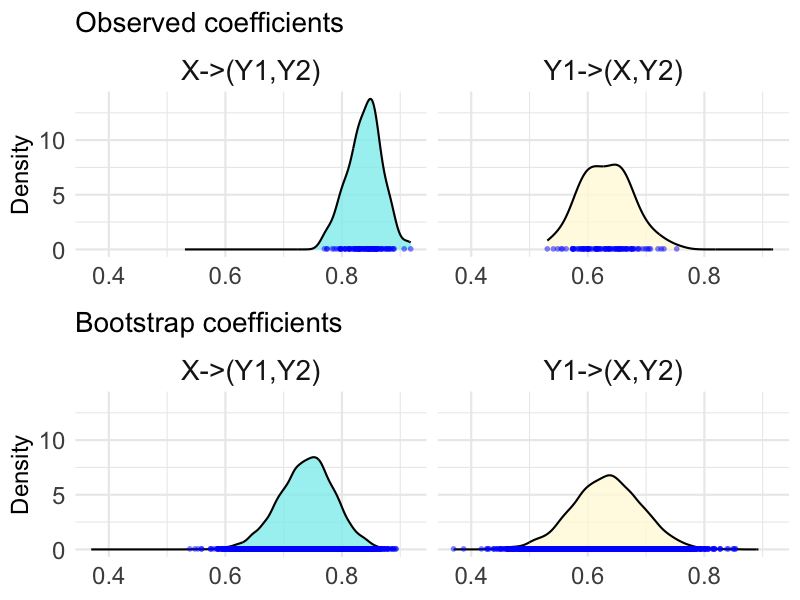}
    \end{minipage}

    \vspace{0cm}

    \begin{minipage}{0.4\textwidth}
        \centering
        \includegraphics[width=\linewidth]{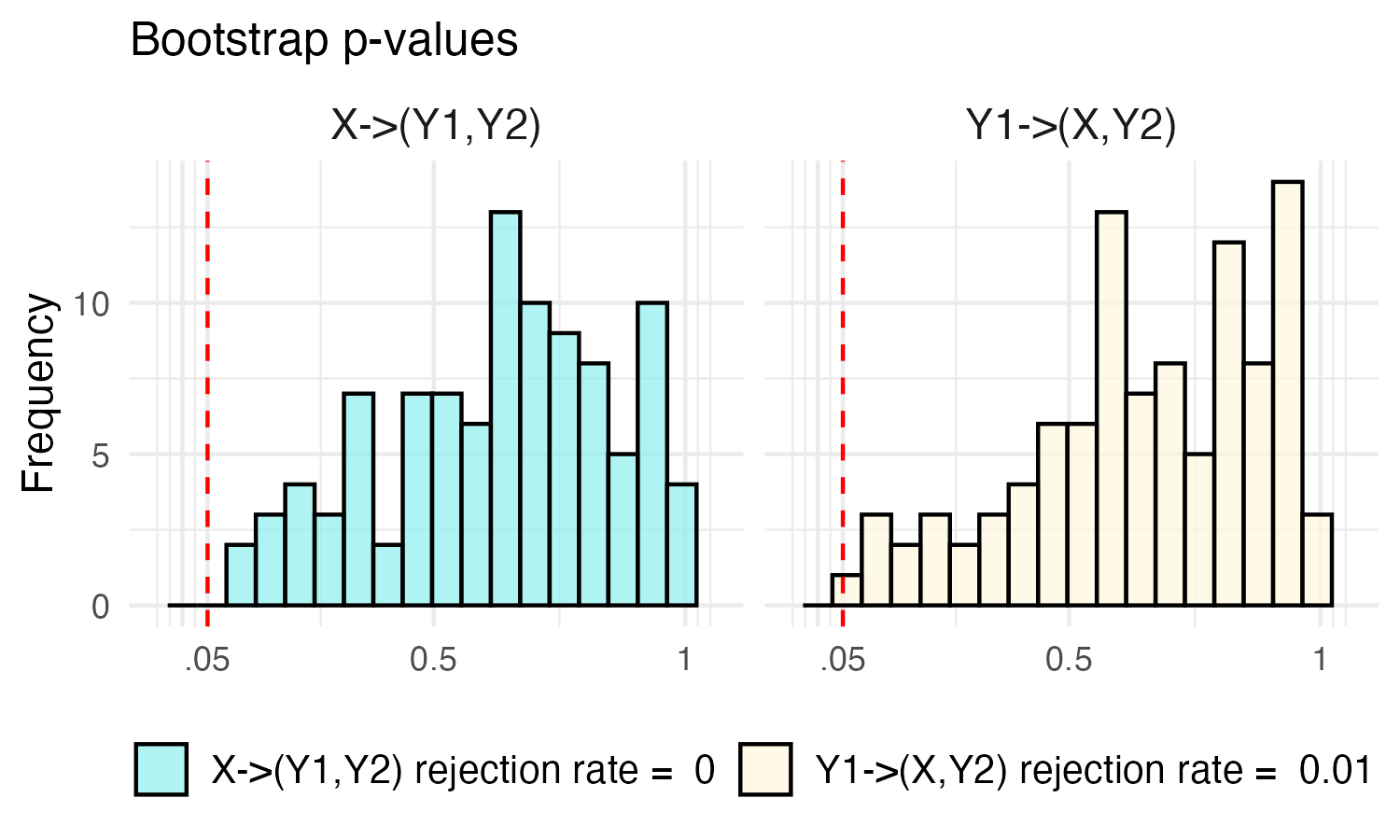}
        {Model \ref{model:1mv}: independent noise}
    \end{minipage}%
    \hspace{0cm}
    \begin{minipage}{0.4\textwidth}
        \centering
        \includegraphics[width=\linewidth]{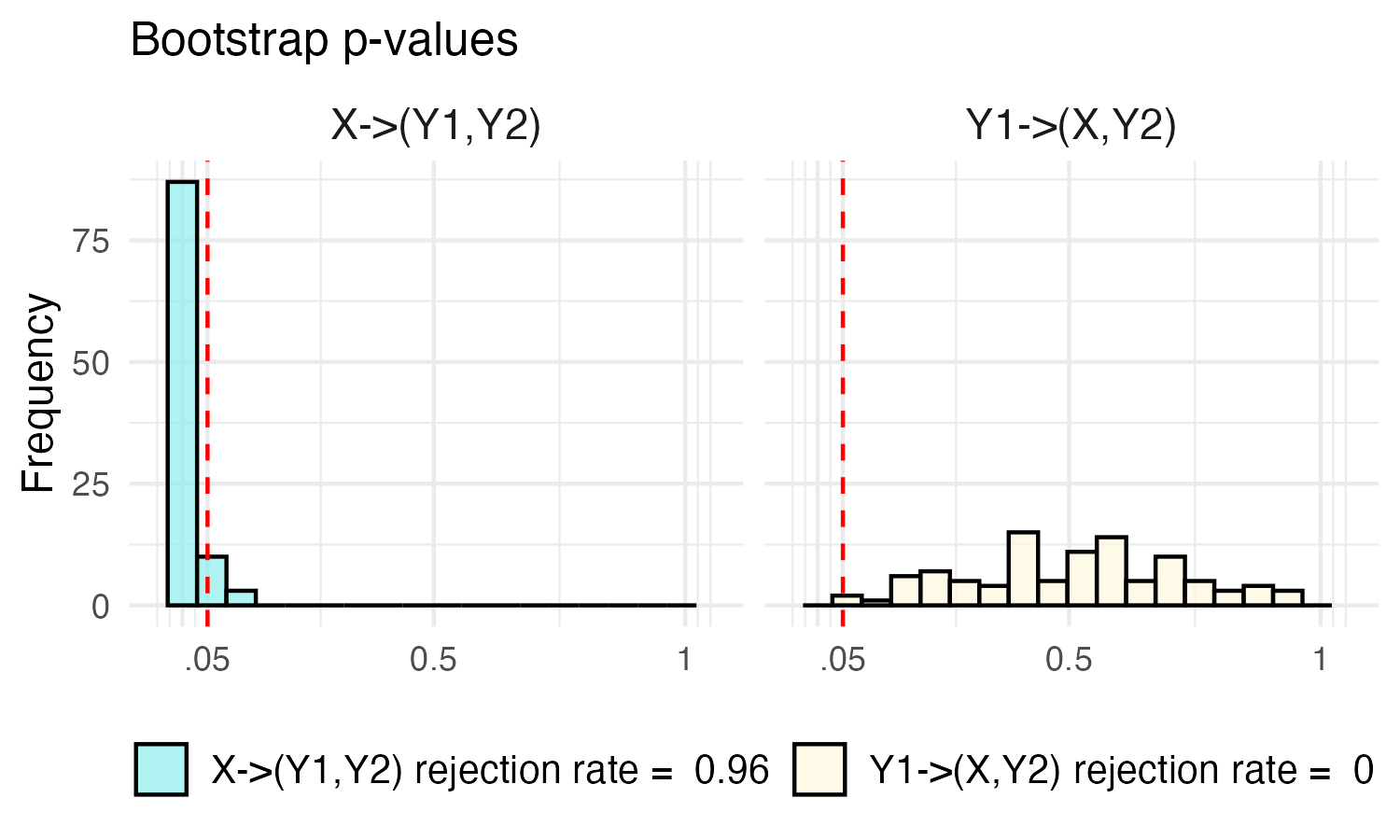}
        {Model \ref{model:3mv}: single-lag}
    \end{minipage}

    \vspace{0.8cm}
    
    \begin{minipage}{0.4\textwidth}
        \centering
        \includegraphics[width=\linewidth]{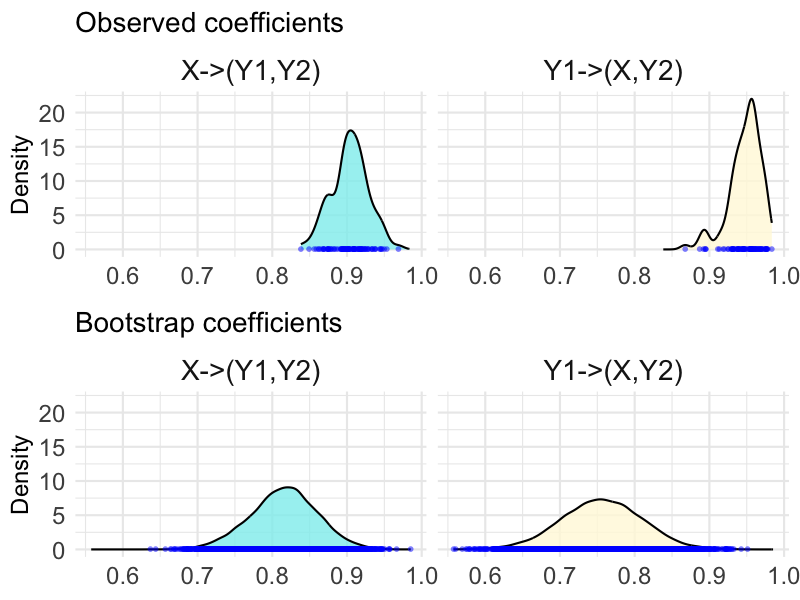}
    \end{minipage}%
    \hspace{0cm}
    \begin{minipage}{0.4\textwidth}
        \centering
        \includegraphics[width=\linewidth]{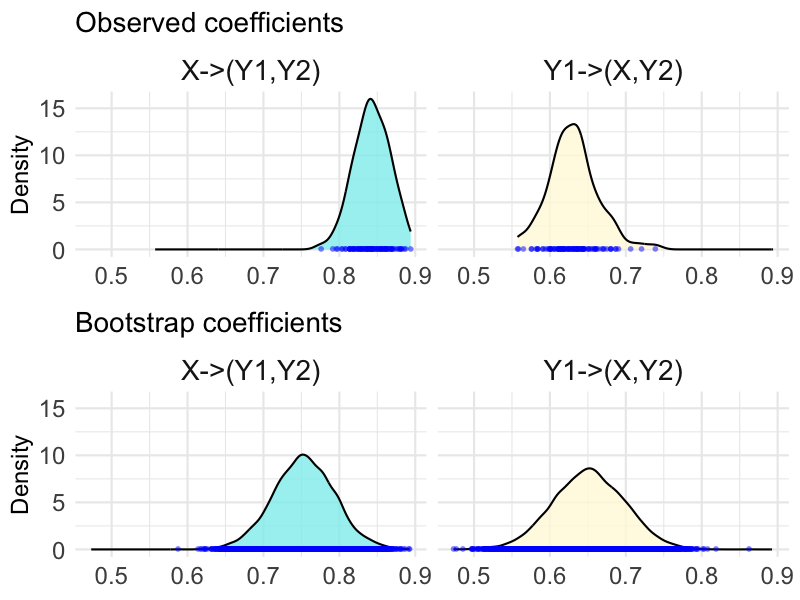}
    \end{minipage}

    \vspace{0cm}

    \begin{minipage}{0.4\textwidth}
        \centering
        \includegraphics[width=\linewidth]{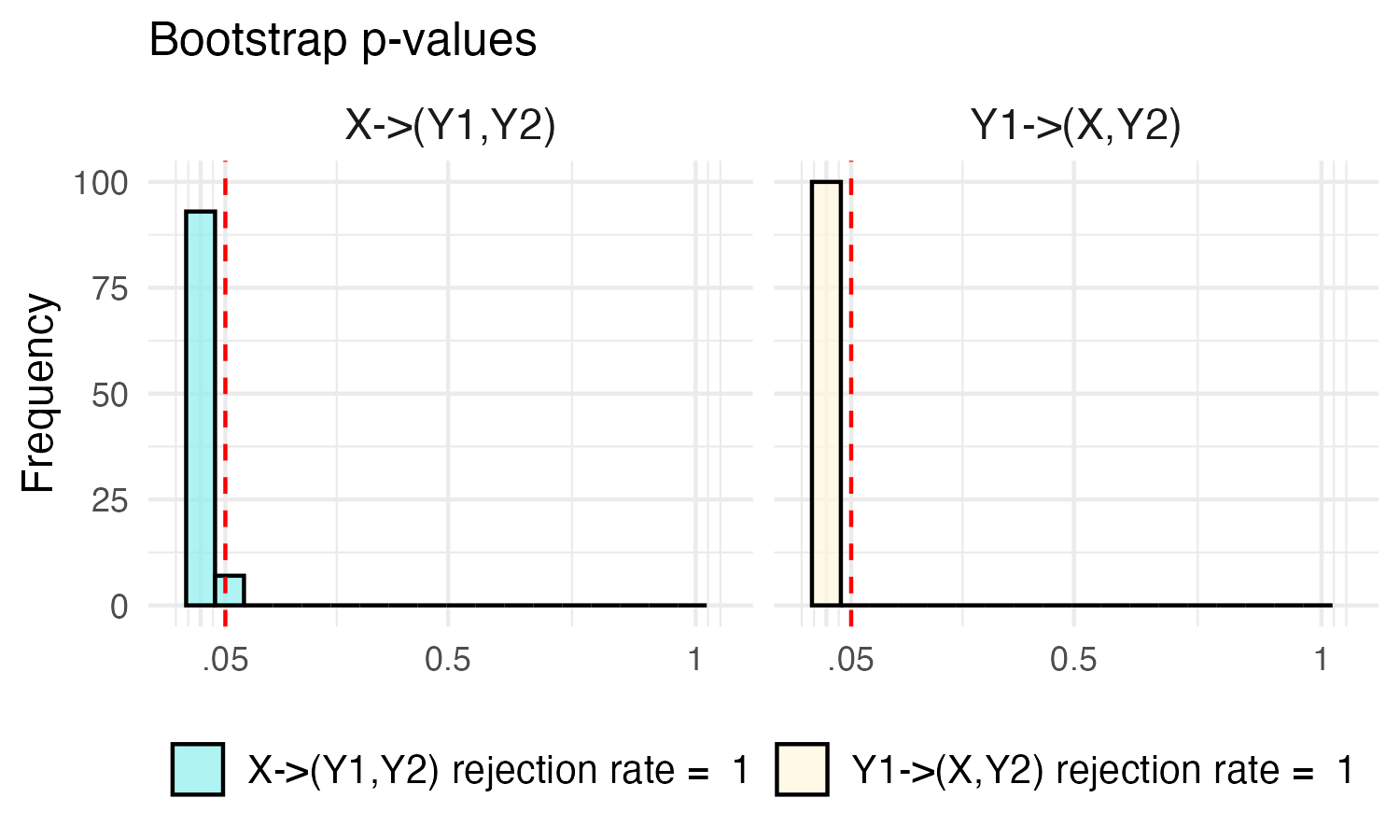}
        {Model \ref{model:4mv}: bidirectional}
    \end{minipage}%
    \hspace{0cm}
    \begin{minipage}{0.4\textwidth}
        \centering
        \includegraphics[width=\linewidth]{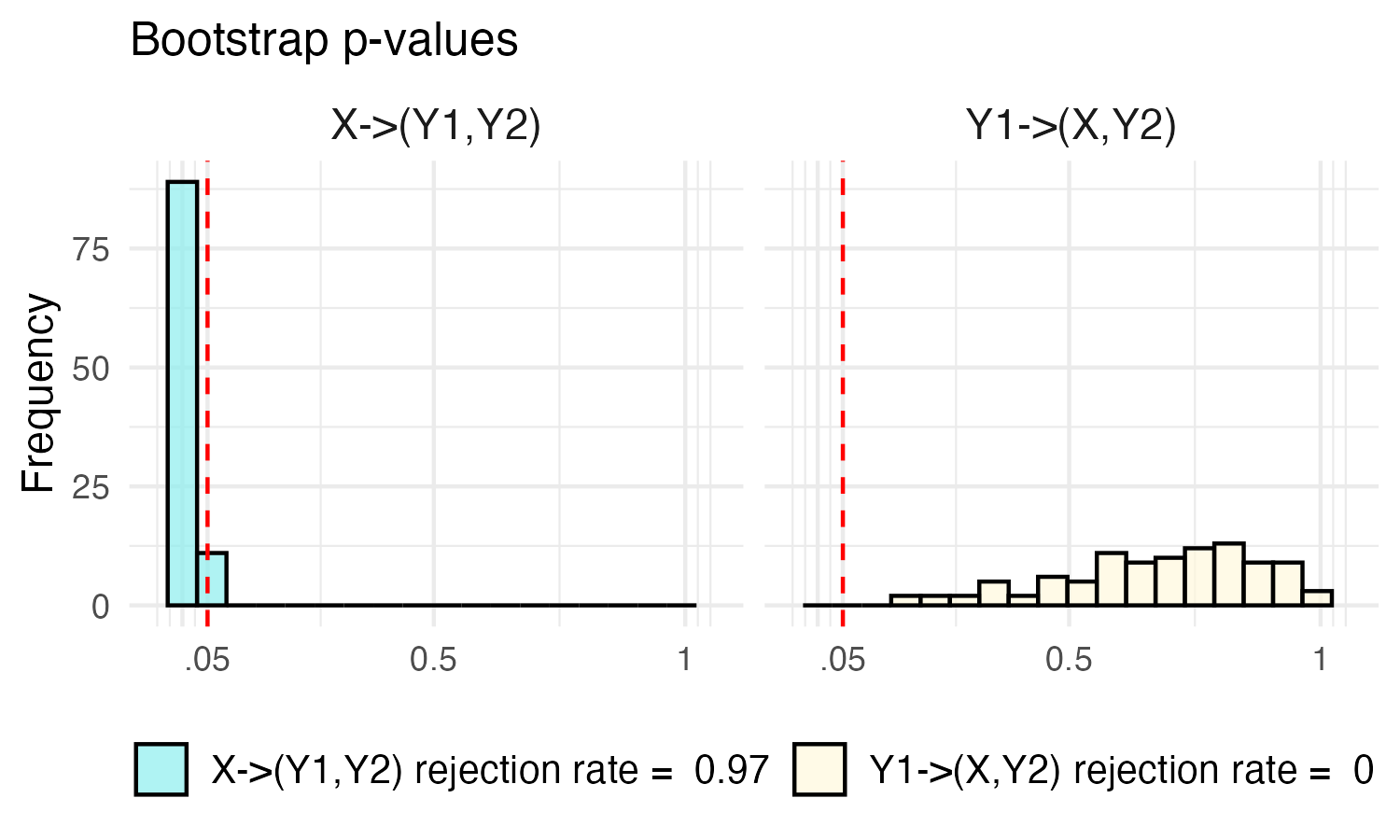}
        {Model \ref{model:5mv}: multi-lag}
    \end{minipage}

    \caption{\footnotesize Repeated simulation results of bootstrap hypothesis test on a three-dimensional time series with Student's $t$ noise. 
    Each density plot displays the kernel density estimate of observed or bootstrapped values of the multivariate compound causal tail coefficient, aggregated over 100 time series realizations. Each accompanying histogram shows the distribution of bootstrap $p$-values across the same 100 realizations. Blue corresponds to the true causal direction $X \to (Y^1,Y^2)$, and yellow to the non-causal direction $Y^1 \to (X,Y^2)$.}
    \label{fig:mv}
\end{figure}

Table \ref{tab:mv} reports the percentage of correct causal inferences for each model and noise configuration described in Section \ref{sec:mvsetup}.
Figure \ref{fig:mv} shows Monte Carlo distributions for four selected models under Student’s $t$ noise. Each histogram (bottom row) shows the empirical distribution of bootstrap $p$-values computed from 100 independent simulations of the same time series model. The corresponding kernel density plots display the distributions of the observed (top row) and bootstrapped (middle row) values of the compound causal tail coefficient $\hat{\Gamma}_{{\X} \rightarrow {h(\Y^1, \Y^2})}(3)$ and $\hat{\Gamma}_{{\Y^1} \rightarrow {h(\X, \Y^2})}(3)$, aggregated over all bootstrap samples and repeated simulations. In Models \ref{model:1mv}, \ref{model:3mv}, and \ref{model:5mv}, the true causal direction $\X \to (\Y^1, \Y^2)$ is shown in blue, while the non-causal direction $\Y^1 \to (\X, \Y^2)$ is shown in yellow.

\begin{table}[htbp]
\caption{\footnotesize Performance of the multivariate compound causal tail coefficient with bootstrap method across six time series models under three noise distributions. Each percentage is based on 100 repetitions and represents the proportion of correct causal inferences.}
\label{tab:mv}
\centering
\begin{tabular}{c|cc|cc|cc}
\toprule
 & \multicolumn{2}{c|}{Student's $t$} & \multicolumn{2}{c|}{Pareto(1,1)} & \multicolumn{2}{c}{Poisson(3)} \\
\cmidrule(lr){2-3} \cmidrule(lr){4-5} \cmidrule(lr){6-7}
Model & $X$→$(Y^1,Y^2)$ & $Y^1$→$(X,Y^2)$ & $X$→$(Y^1,Y^2)$ & $Y^1$→$(X,Y^2)$ & $X$→$(Y^1,Y^2)$ & $Y^1$→$(X,Y^2)$ \\
\midrule
\ref{model:1mv} & 100\% & 99\% & 99\% & 99\% & 99\% & 99\% \\
\ref{model:2mv} & 98\% & 100\% & 100\% & 100\% & 100\% & 100\% \\
\ref{model:3mv} & 96\% & 100\% & 100\% & 99\% & 100\% & 99\% \\
\ref{model:4mv} & 100\% & 100\% & 100\% & 100\% & 100\% & 100\% \\
\ref{model:5mv} & 97\% & 100\% & 100\% & 99\% & 100\% & 100\% \\
\ref{model:6mv} & 87\% & 100\% & 100\% & 99\% & 100\% & 100\% \\
\bottomrule
\end{tabular}
\end{table}

The distributional patterns of the observed and bootstrapped causal tail coefficients in Figure~\ref{fig:mv} closely resemble those in the bivariate setting shown in Figure \ref{fig:results}. In the baseline random noise model \ref{model:1mv}, the density plots for the two directions, $X \to (Y^1, Y^2)$ and $Y^1 \to (X,Y^2)$, are nearly identical, consistent with the absence of any underlying causal structure. In contrast, for Models~\ref{model:3mv}, \ref{model:4mv}, and \ref{model:5mv}, the bootstrapped coefficients are systematically smaller than the observed coefficients in the true causal direction, whereas in the non-causal direction, the two remain indistinguishable.

Compared to their bivariate counterparts in Figure \ref{fig:results} of Section \ref{sec:simulation} (e.g., Model \ref{model:1mv} vs.~\ref{model:1}, and \ref{model:5mv} vs.~\ref{model:5}), all density plots in the multivariate setting are centered at slightly larger coefficient values. This upward shift reflects the effect of weight optimization in the impact function, which can only increase the compound causal tail coefficient as additional variables are included. Intuitively, adding additional variables into the compound event increases the chance of capturing further associations between the cause and the effect, thereby inflating the estimated causal strength.

Importantly, this upward shift affects both the observed and bootstrapped coefficients symmetrically, and therefore does not compromise the validity of the test. The bootstrap $p$-values confirm that the block bootstrap procedure reliably detects the true causal directions while controlling the Type I error rate. In particular, the histograms for Model \ref{model:1mv} show that false positives occur at a rate consistent with the nominal significance level; meanwhile, Models \ref{model:3mv}, \ref{model:4mv}, and \ref{model:5mv} yield results that align with their respective ground truth causal structures.
The strong performance of the block bootstrap test based on the multivariate compound causal tail coefficient is further supported by its high accuracy across all scenarios reported in Table~\ref{tab:mv}. Taken together, these findings demonstrate that our proposed methodology generalizes well to high-dimensional settings, achieving good power and size across a wide range of models.

\section{Space-weather application}\label{sec:25yr}

\begin{figure}[htb]
    \centering
    \includegraphics[width=0.8\linewidth]{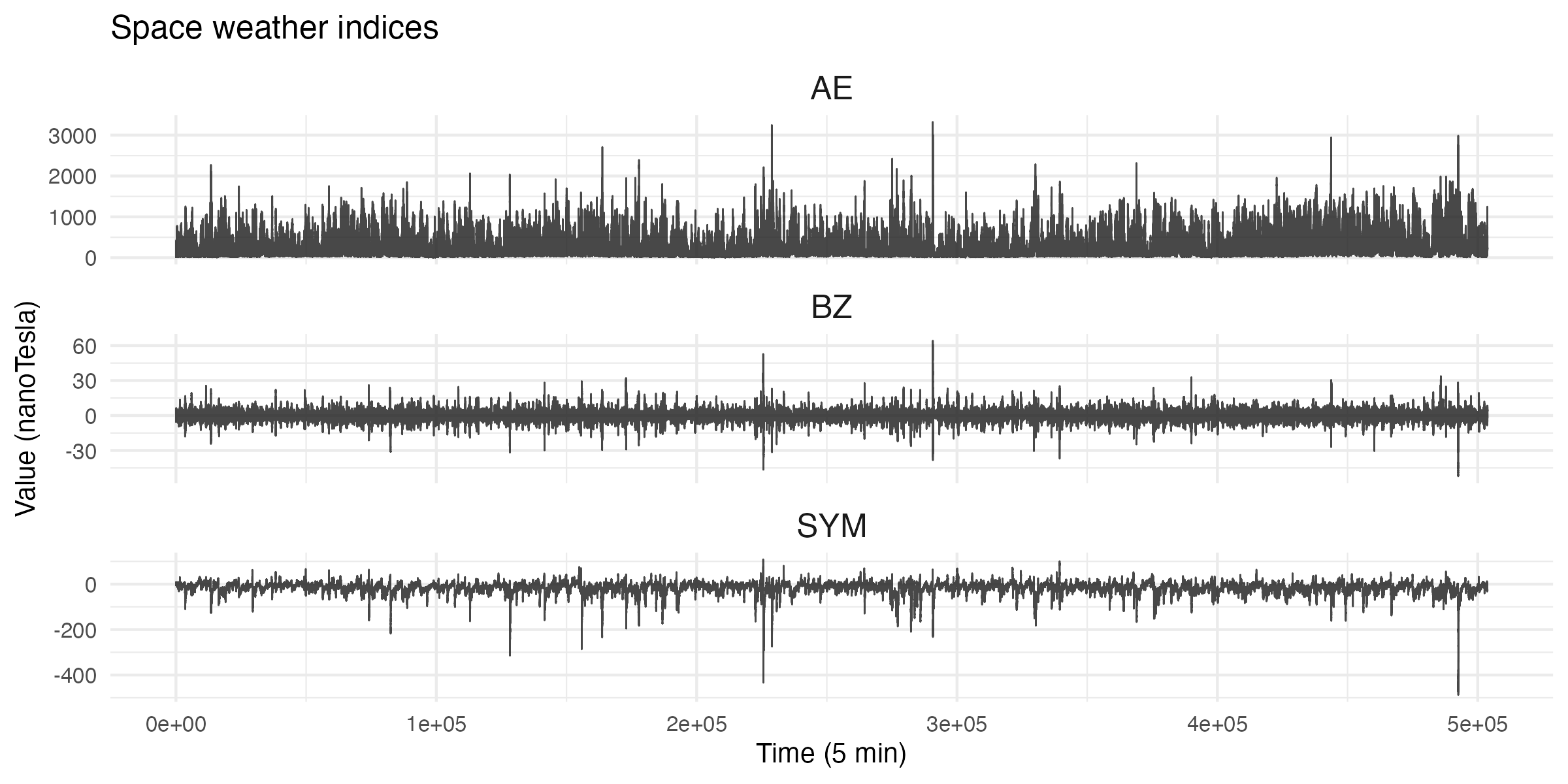}
    \caption{\footnotesize Space-weather time series.
    AE represents the substorm index; BZ measures a vertical component of the interplanetary magnetic field; SYM is the magnetic storm index. All variables were recorded by NASA at 5-minute intervals during the period 1999-2003,  reported in nanoteslas (nT).}
    \label{fig:spacew_ts}
\end{figure}

In this section, we report supplementary results for the space-weather application. A visual overview of the data used in Section~\ref{sec:spacew} is provided in Figure \ref{fig:spacew_ts}. Based on the observed characteristics, we adopted the approach of \cite{BPP2024} to analyze extreme events characterized by significantly low SYM values, significantly high AE values, and significantly low BZ values. Specifically, we considered the transformed variables -SYM, +AE, -BZ and focused on their upper-tail behavior.

In addition, we verified the key assumptions of stationarity and regular variation for the time series under study. The augmented Dickey-Fuller test indicates that all three series are stationary. However, this test is known to produce overly optimistic assessments of stationarity, and its results should be interpreted with caution \citep{Dickey2005}. To assess regular variation, we estimated the upper tail indices using the \texttt{HTailIndex} function from the \texttt{ExtremeRisks} package in R \citep{PS2020}, with the threshold parameter set to $k = 500$. Confidence intervals were constructed using the normal approximation described in \citet[page 1288]{Drees2000}. The estimated tail indices and corresponding 95\% confidence intervals are as follows: AE, $0.14 \; (0.10, 0.20)$; BZ, $0.28 \; (0.11, 0.47)$; and SYM, $0.25 \; (0.03, 0.48)$. 
These results suggest that all three series exhibit regular variation, and the observed overlap among the confidence intervals provides empirical support for the common tail index assumption used in our theoretical development.

\begin{figure}[htbp]
  \centering

  \begin{minipage}[b]{\textwidth}
    \centering
    \includegraphics[width=0.95\textwidth]{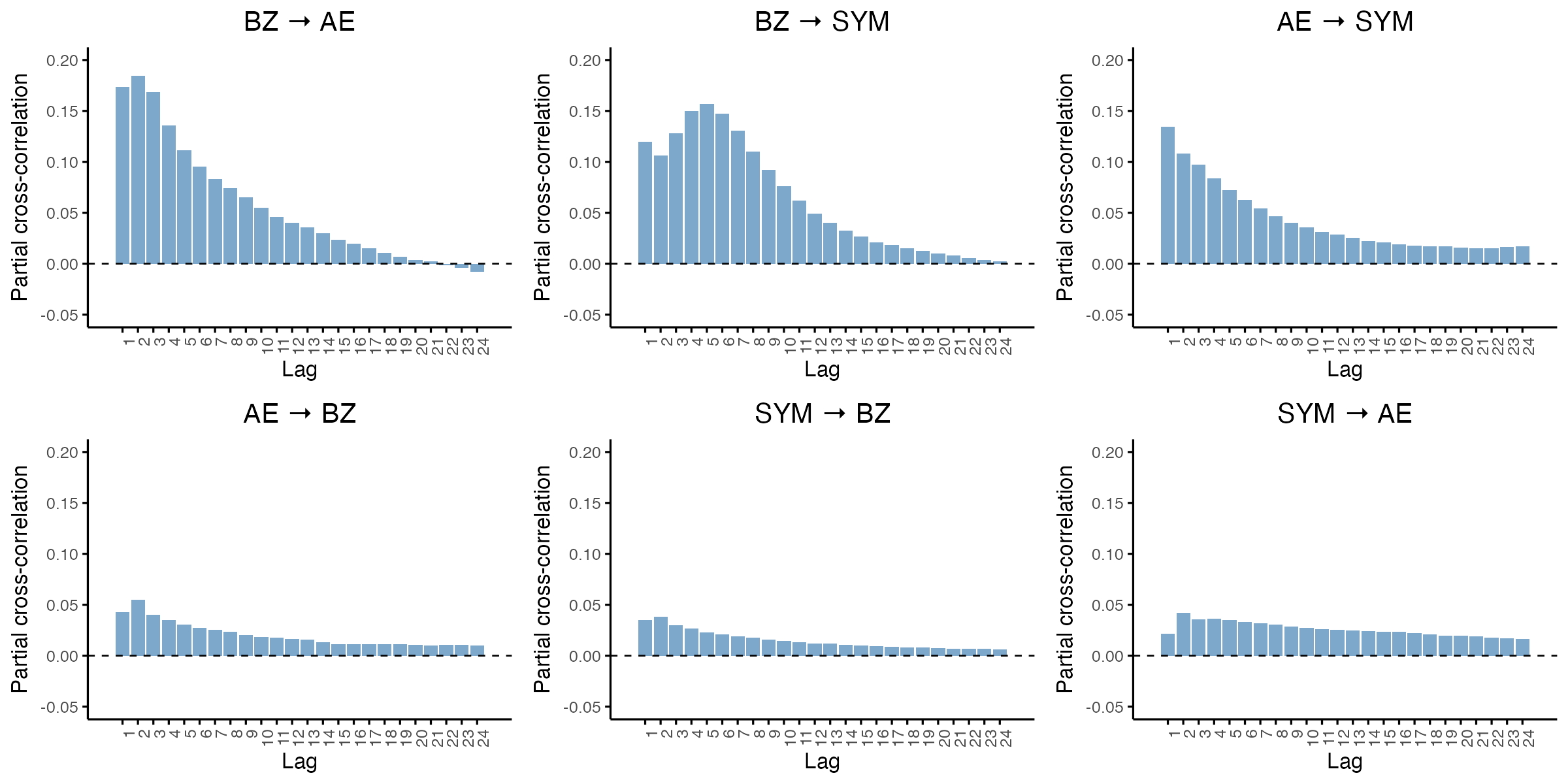}
    \par\vspace{0.3em}
    {\footnotesize (a) Partial cross-correlation over 24 lags}
  \end{minipage}

  \vskip 1em

  \begin{minipage}[b]{\textwidth}
    \centering
    \includegraphics[width=0.95\textwidth]{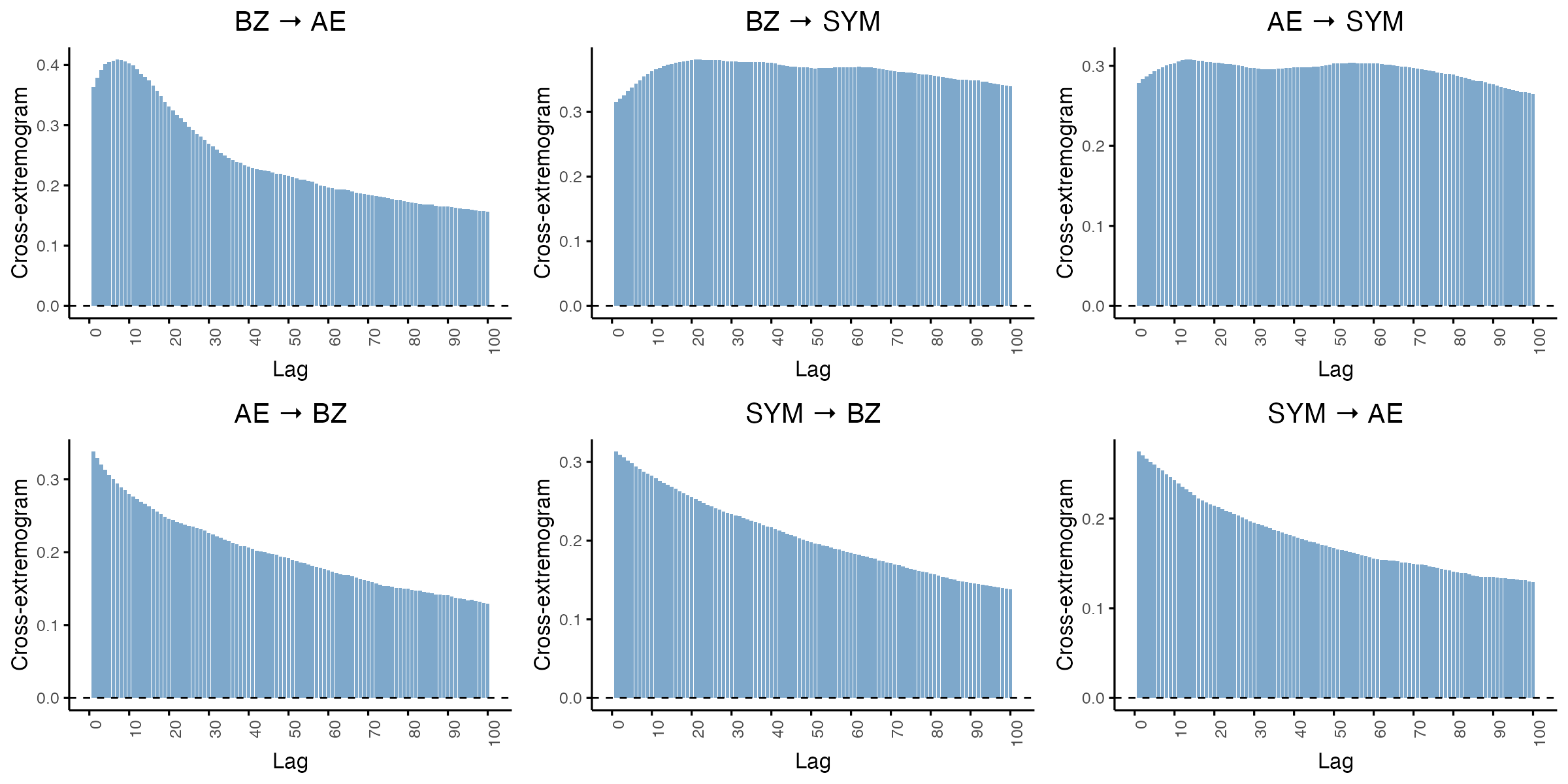}
    \par\vspace{0.3em}
    {\footnotesize (b) Cross-extremogram over 100 lags}
  \end{minipage}

  \caption{\footnotesize Pairwise cross-dependencies between space-weather variables.}
  \label{fig:spacew_pccf}
\end{figure}

We provide insights on the selection of the extremal delay parameter based on the two methods introduced in Section \ref{sec:pccf}. Figure \ref{fig:spacew_pccf} displays the estimated PCCFs alongside the estimated cross-extremograms for each pair of variables. The PCCFs exhibit a much more rapid decay to zero compared to the cross-extremogram. Using a threshold of $\bar{c} = 0.1$, the extremal delay $p$ is selected as 5 for the pair (BZ, AE), 8 for (BZ, SYM), and 2 for (AE, SYM). In contrast, the cross-extremogram remains above the threshold $\bar{c} = 0.15$ even beyond 100 lags, which exceeds the range of extremal delays that are physically meaningful. These findings suggest that the PCCF provides a more reliable basis for selecting the extremal delay parameter in the context of this space-weather dataset.

\subsection{Extended horizon analysis (1995-2019)}\label{sec:extended_horizon}

\begin{figure}[htbp]
    \centering
    \includegraphics[width=0.8\linewidth]{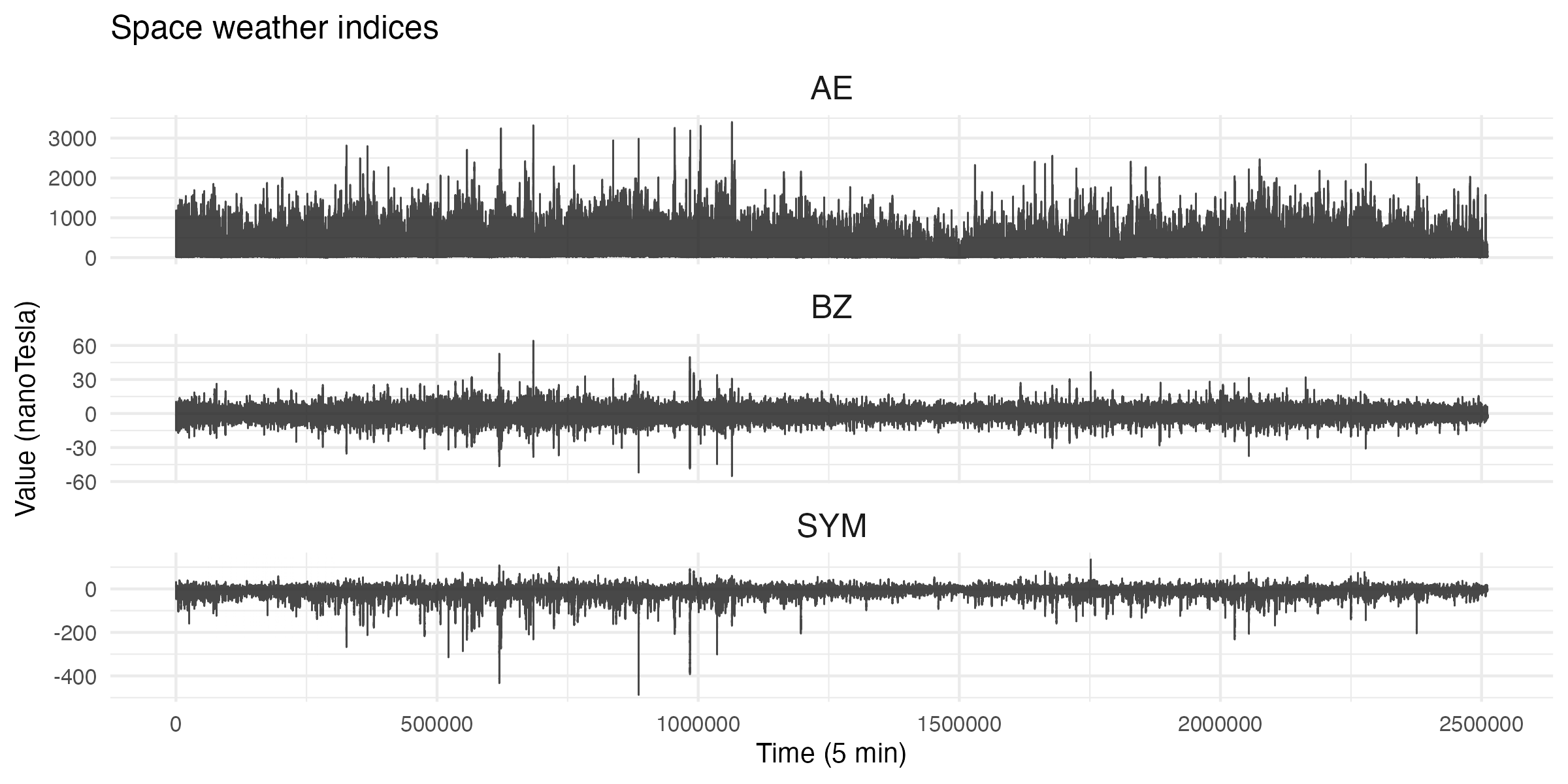}
    \caption{\footnotesize Space-weather time series.
    AE represents the substorm index; BZ measures a vertical component of the interplanetary magnetic field; SYM is the magnetic storm index. All variables were recorded by NASA at 5-minute intervals during the period 1995–2019,  reported in nanoteslas (nT).}
    \label{fig:spacew_ts_all}
\end{figure}

In this section, we present the empirical results for the estimated causal tail coefficients and corresponding bootstrap hypothesis tests using the space-weather dataset over an extended time period. The analysis covers the 25-year span from January 1, 1995, to December 31, 2019, which includes both Solar Cycles 23 and 24. The same variables introduced in Section \ref{sec:spacew} are used, and a visual summary of the extended time series is provided in Figure \ref{fig:spacew_ts_all}.

\begin{figure}[htbp]
    \centering
    \includegraphics[width=0.8\linewidth]{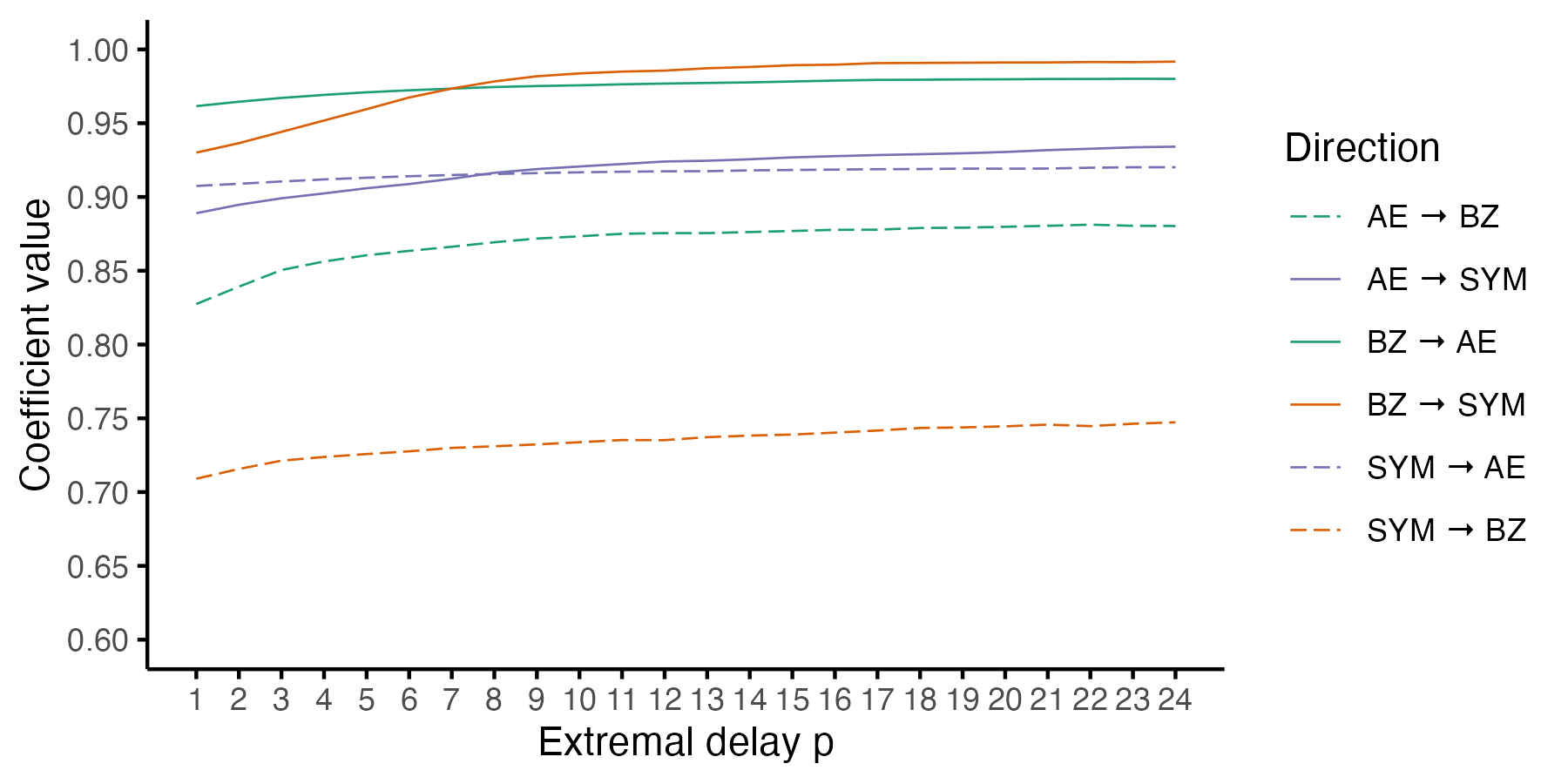}
    \caption{\footnotesize Compound causal tail coefficient values for all directed variable pairs across extremal delays $p \in [1, 24]$. 
    }
    \label{fig:spacew_coef_all}
\end{figure}

\begin{figure}[htbp]
  \centering

  \begin{minipage}[b]{\textwidth}
    \centering
    \includegraphics[width=0.95\textwidth]{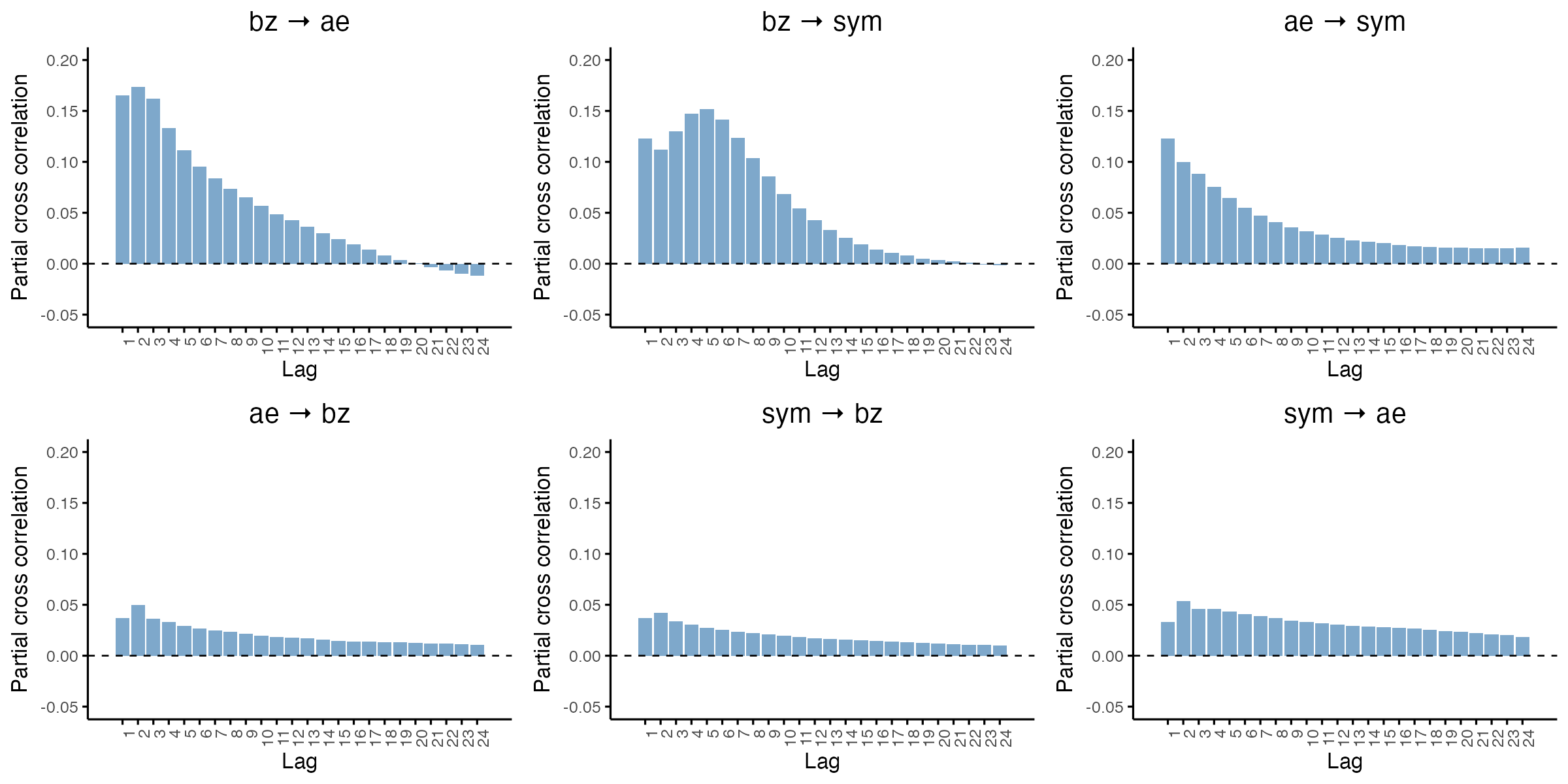}
    \par\vspace{0.3em}
    {\footnotesize (a) Partial cross-correlation over 24 lags}
  \end{minipage}

  \vskip 1em

  \begin{minipage}[b]{\textwidth}
    \centering
    \includegraphics[width=0.95\textwidth]{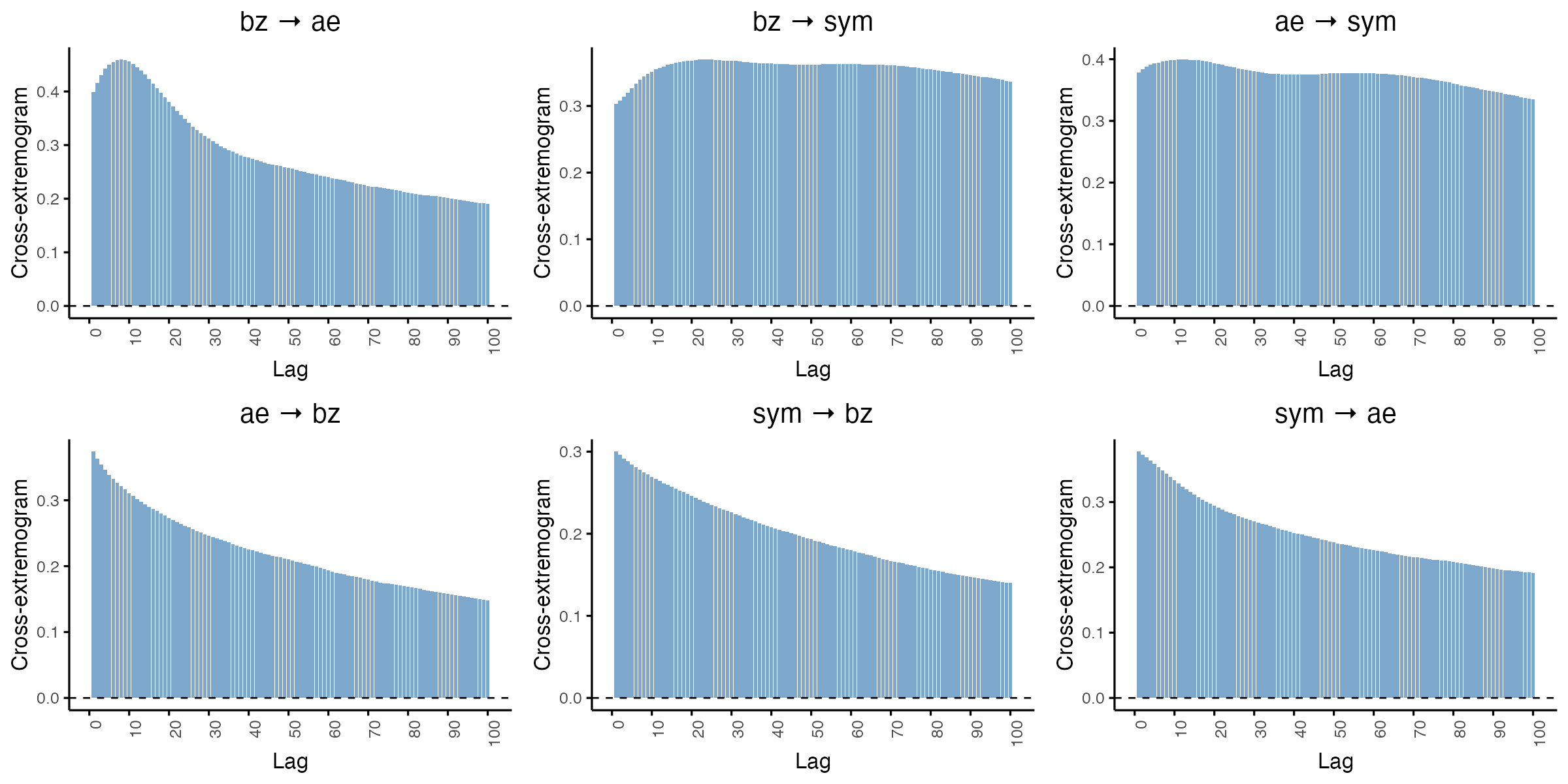}
    \par\vspace{0.3em}
    {\footnotesize (b) Cross-extremogram over 100 lags}
  \end{minipage}

  \caption{\footnotesize Pairwise cross-dependencies between space-weather variables.}
  \label{fig:spacew_pccf_all}
\end{figure}

We applied the same methodology used in the preceding analysis. Stationarity was verified using the augmented Dickey–Fuller test, while the assumption of regular variation was assessed using the \texttt{HTailIndex} function. The estimated tail indices and their corresponding 95\% confidence intervals are as follows: AE, $0.14 \; (0.10, 0.18)$; BZ, $0.27 \; (0.12, 0.41)$; and SYM, $0.22 \; (0.07, 0.38)$. Figure \ref{fig:spacew_coef_all} displays the estimated compound causal tail coefficients across a range of extremal delay parameter $p \in [1, 24]$. To guide the selection of $p$, Figure \ref{fig:spacew_pccf_all} presents the estimated PCCFs alongside the estimated cross-extremograms for each pair of variables. Table \ref{tab:spacew_all} reports the estimated compound causal tail coefficients and the corresponding bootstrap $p$-values, with extremal delay parameter $p$ selected based on three different threshold levels applied to the PCCF.

\begin{table}[htbp]
    \centering
    \caption{\footnotesize Estimated compound causal tail coefficients, with bootstrap $p$-values reported in parentheses, across varying PCCF thresholds used to select extremal delay parameter $p$.
    }
    \label{tab:spacew_all}
    \begin{tabular}{l|c c c}
    \toprule
    \multicolumn{1}{c|}{} & $\bar{c}=0.15$ & $\bar{c}=0.10$ & $\bar{c}=0.05$ \\
    \midrule
    AE $\rightarrow$ BZ  & 0.851 (1.00) & 0.861 (1.00) & 0.874 (1.00) \\
    AE $\rightarrow$ SYM & 0.889 (0.35) & 0.889 (0.30) & 0.910 (0.02) \\
    BZ $\rightarrow$ AE  & 0.967 (0.01) & 0.971 (0.00) & 0.976 (0.00) \\
    BZ $\rightarrow$ SYM & 0.960 (0.01) & 0.978 (0.00) & 0.985 (0.00) \\
    SYM $\rightarrow$ BZ & 0.726 (0.69) & 0.731 (0.81) & 0.736 (0.87) \\
    SYM $\rightarrow$ AE & 0.908 (0.42) & 0.908 (0.55) & 0.914 (0.71) \\
    \bottomrule
    \end{tabular}
\end{table}

The estimated compound causal tail coefficients and corresponding bootstrap $p$-values closely align with those reported in Section \ref{sec:results}, reinforcing the robustness of the findings. As before, the coefficients for the directed pairs BZ$\to$SYM (solid orange curve) and BZ$\to$AE (solid green curve) remain substantially larger than those for the other pairs, while the coefficients for AE$\to$SYM and SYM$\to$AE (purple curves) are more similar than previously observed, suggesting limited directional asymmetry. Consistent with these patterns, the bootstrap hypothesis tests continue to indicate strong causal effects in the directions BZ$\to$SYM and BZ$\to$AE, with corresponding $p$-values decreased further relative to the five-year analysis. In contrast, although the bootstrap test for AE$\to$SYM rejects the null hypothesis when the PCCF threshold is set to 0.05, it fails to do so at a threshold of 0.1 or 0.15, suggesting a weaker causal relationship. Collectively, these results reinforce our earlier conclusion that BZ acts as a common driver of both SYM and AE, with AE serving as a potential secondary driver of SYM. Thus, the conclusions drawn in Section \ref{sec:results} remain highly robust.


\clearpage


\begin{thebibliography}{42}
\providecommand{\natexlab}[1]{#1}
\providecommand{\url}[1]{\texttt{#1}}
\expandafter\ifx\csname urlstyle\endcsname\relax
  \providecommand{\doi}[1]{doi: #1}\else
  \providecommand{\doi}{doi: \begingroup \urlstyle{rm}\Url}\fi

\bibitem[Athreya and Pantula(1986)]{AP1986}
K.~B. Athreya and S.~G. Pantula.
\newblock A note on strong mixing of {ARMA} processes.
\newblock \emph{Statistics \& Probability Letters}, 4\penalty0 (4):\penalty0 187--190, June 1986.

\bibitem[Baumann and McCloskey(2021)]{BM2021}
C.~Baumann and A.~E. McCloskey.
\newblock Timing of the solar wind propagation delay between {L}1 and {E}arth based on machine learning.
\newblock \emph{Journal of Space Weather and Space Climate}, 11:\penalty0 1--13, 2021.
\newblock \doi{10.1051/swsc/2020073}.

\bibitem[Bevacqua et~al.(2017)Bevacqua, Maraun, Haff, Widmann, and Vrac]{BMHWV2017}
E.~Bevacqua, D.~Maraun, I.~H. Haff, M.~Widmann, and M.~Vrac.
\newblock Multivariate statistical modelling of compound events via pair-copula constructions: analysis of floods in {R}avenna ({I}taly).
\newblock \emph{Hydrology and Earth System Sciences}, 21:\penalty0 2701–2723, 2017.
\newblock \doi{10.5194/hess-21-2701-2017}.

\bibitem[Bevacqua et~al.(2021)Bevacqua, De Michele, Manning, Couasnon, Ribeiro, Ramos, Vignotto, Bastos, Blesić, Durante, Hillier, Oliveira, Pinto, Ragno, Rivoire, Saunders, van der Wiel, Wu, Zhang, and Zscheischler]{BMM2021}
E.~Bevacqua, C.~De Michele, C.~Manning, A.~Couasnon, A.~F.~S. Ribeiro, A.~M. Ramos, E.~Vignotto, A.~Bastos, S.~Blesić, F.~Durante, J.~Hillier, S.~C. Oliveira, J.~G. Pinto, E.~Ragno, P.~Rivoire, K.~Saunders, K.~van der Wiel, W.~Wu, T.~Zhang, and J.~Zscheischler.
\newblock Guidelines for studying diverse types of compound weather and climate events.
\newblock \emph{Earth's Future}, 9(11):\penalty0 e2021EF002340, 2021.
\newblock \doi{10.1029/2021EF002340}.

\bibitem[Blalock(1982)]{Blalock1982}
H.~M. Blalock.
\newblock \emph{Conceptualization and Measurement in the Social Sciences}.
\newblock SAGE Publications, 1982.

\bibitem[Bodik et~al.(2024)Bodik, Paluš, and Pawlas]{BPP2024}
J.~Bodik, M.~Paluš, and Z.~Pawlas.
\newblock Causality in extremes of time series.
\newblock \emph{Extremes}, 27:\penalty0 67--121, 2024.
\newblock \doi{10.1007/s10687-023-00479-5}.

\bibitem[Courgeau and Veraart(2021)]{CV2021}
V.~Courgeau and A.~E.~D. Veraart.
\newblock Extreme event propagation using counterfactual theory and vine copulas.
\newblock \emph{arXiv Preprint 2106.13564}, 2021.
\newblock \doi{10.48550/arXiv.2106.13564}.

\bibitem[Davis and Mikosch(2009)]{DM2009}
R.~A. Davis and T.~Mikosch.
\newblock The extremogram: A correlogram for extreme events.
\newblock \emph{The Annals of Statistics}, 37\penalty0 (5B):\penalty0 2046--2082, 2009.
\newblock \doi{10.1214/08-AOS620}.

\bibitem[Dickey(2005)]{Dickey2005}
D.~A. Dickey.
\newblock Stationarity issues in time series models.
\newblock \emph{SAS Users Group International Conference Proceedings, Paper 192-30}, 2005.
\newblock URL \url{https://citeseerx.ist.psu.edu/document?doi=a37e31ed2969afce603128b4c871b1f1e332bf6f}.

\bibitem[Doukhan(1994)]{Doukhan1994}
P.~Doukhan.
\newblock \emph{Mixing: Properties and Examples}, volume~85 of \emph{Lecture Notes in Statistics}.
\newblock Springer, New York, 1994.
\newblock ISBN 978-0-387-94214-8.

\bibitem[Drees(2000)]{Drees2000}
H.~Drees.
\newblock Weighted approximations of tail processes for $\phi$-mixing random variables.
\newblock \emph{Annals of Applied Probability}, 10:\penalty0 1274--1301, 2000.
\newblock \doi{10.1214/aoap/1019487518}.

\bibitem[Durrett(2019)]{Durrett2019}
R.~Durrett.
\newblock \emph{Probability: Theory and Examples}.
\newblock Cambridge University Press, 5 edition, 2019.

\bibitem[Eddelbuettel(2025)]{Eddelbuettel2025}
D.~Eddelbuettel.
\newblock {RcppDE}: Global optimization by differential evolution in {C++}.
\newblock \url{https://CRAN.R-project.org/package=RcppDE}, 2025.
\newblock R package version 0.1.8.

\bibitem[Embrechts et~al.(1997)Embrechts, Kluppelberg, and Mikosch]{EKM1997}
P.~Embrechts, C.~Kluppelberg, and T.~Mikosch.
\newblock \emph{Modelling Extremal Events: For Insurance and Finance}.
\newblock Springer, Berlin, Heidelberg, 1997.

\bibitem[Engelke and Hitz(2020)]{EH2020}
S.~Engelke and A.~S. Hitz.
\newblock Graphical models for extremes.
\newblock \emph{Journal of the Royal Statistical Society: Series B (Statistical Methodology)}, 82:\penalty0 871--932, 2020.
\newblock \doi{10.1111/rssb.12399}.

\bibitem[Engelke and Ivanovs(2021)]{EI2021}
S.~Engelke and J.~Ivanovs.
\newblock Sparse structures for multivariate extremes.
\newblock \emph{Annual Review of Statistics and Its Application}, 8:\penalty0 241--270, 2021.
\newblock \doi{10.1146/annurev-statistics-042720-124255}.

\bibitem[Fayers and Hand(2002)]{FH2002}
P.~M. Fayers and D.~J. Hand.
\newblock Causal variables, indicator variables and measurement scales: an example from quality of life.
\newblock \emph{Journal of the Royal Statistical Society: Series A (Statistics in Society)}, 165\penalty0 (2):\penalty0 233--253, 2002.
\newblock \doi{10.1111/1467-985x.02020}.

\bibitem[Firth(2002)]{Firth2002}
D.~Firth.
\newblock Discussion of ‘{C}ausal variables, indicator variables and measurement scales: an example from quality of life’ by {Fayers and Hand}.
\newblock \emph{Journal of the Royal Statistical Society: Series A (Statistics in Society)}, 165\penalty0 (2):\penalty0 257, 2002.
\newblock \doi{10.1111/1467-985X.t01-1-02020}.

\bibitem[Gnecco et~al.(2021)Gnecco, Meinshausen, Peters, and Engelke]{GMPE2021}
N.~Gnecco, N.~Meinshausen, J.~Peters, and S.~Engelke.
\newblock Causal discovery in heavy-tailed models.
\newblock \emph{The Annals of Statistics}, 49\penalty0 (3):\penalty0 1755--1778, 2021.
\newblock \doi{10.1214/20-aos2021}.

\bibitem[Granger(1969)]{Granger1969}
C.~Granger.
\newblock Investigating causal relations by econometric models and cross-spectral methods.
\newblock \emph{Econometrica}, 37\penalty0 (3):\penalty0 424--438, 1969.
\newblock \doi{10.2307/1912791}.

\bibitem[Granger(1980)]{Granger1980}
C.~Granger.
\newblock Testing for causality: a personal viewpoint.
\newblock \emph{Journal of Economic Dynamics and Control}, 2:\penalty0 329--352, 1980.

\bibitem[Hafner and Herwartz(2008)]{HH2008}
C.~M. Hafner and H.~Herwartz.
\newblock Testing for causality in variance using multivariate {GARCH} models.
\newblock \emph{Annals of Economics and Statistics}, 89:\penalty0 215--241, 2008.
\newblock \doi{10.2307/27715168}.

\bibitem[Hannart and Naveau(2018)]{HN2018}
A.~Hannart and P.~Naveau.
\newblock Probabilities of causation of climate changes.
\newblock \emph{Journal of Econometrics}, 31\penalty0 (14):\penalty0 5507--5524, 2018.
\newblock \doi{10.1175/JCLI-D-17-0304.1}.

\bibitem[Hong et~al.(2009)Hong, Liu, and Wang]{HLW2009}
Y.~Hong, Y.~Liu, and S.~Wang.
\newblock Granger causality in risk and detection of extreme risk spillover between financial markets.
\newblock \emph{Journal of Econometrics}, 150\penalty0 (2):\penalty0 271--287, 2009.
\newblock \doi{10.1016/j.jeconom.2008.12.013}.

\bibitem[Kamide et~al.(1998)Kamide, Baumjohann, Daglis, Gonzalez, Grande, Joselyn, McPherron, Phillips, Reeves, Rostoker, et~al.]{KBDG1998}
Y.~Kamide, W.~Baumjohann, I.~A. Daglis, W.~D. Gonzalez, M.~Grande, J.~A. Joselyn, R.~L. McPherron, J.~L. Phillips, E.~G.~D. Reeves, G.~Rostoker, et~al.
\newblock Current understanding of magnetic storms: Storm-substorm relationships.
\newblock \emph{Journal of Geophysical Research: Space Physics}, 103\penalty0 (A8):\penalty0 17705--17728, 1998.
\newblock \doi{10.1029/98JA01426}.

\bibitem[Kiriliouk and Naveau(2020)]{KN2020}
A.~Kiriliouk and P.~Naveau.
\newblock Climate extreme event attribution using multivariate peaks-over-thresholds modeling and counterfactual theory.
\newblock \emph{The Annals of Applied Statistics}, 14\penalty0 (3), 2020.
\newblock \doi{10.1214/20-aoas1355}.

\bibitem[Kuersteiner(2010)]{Kuersteiner2010}
G.~M. Kuersteiner.
\newblock {Granger-Sims} causality.
\newblock In S.~N. Durlauf and L.~E. Blume, editors, \emph{Macroeconometrics and Time Series Analysis}, The New Palgrave Economics Collection. Palgrave Macmillan, London, 2010.
\newblock \doi{10.1057/9780230280830_14}.

\bibitem[Kunsch(1989)]{Kunsch1989}
H.~R. Kunsch.
\newblock The jackknife and the bootstrap for general stationary observations.
\newblock \emph{The Annals of Statistics}, 17\penalty0 (3):\penalty0 1217–1241, 1989.
\newblock \doi{10.1214/aos/1176347265}.

\bibitem[Lahiri(1999)]{Lahiri1999}
S.~N. Lahiri.
\newblock Theoretical comparisons of block bootstrap methods.
\newblock \emph{The Annals of Statistics}, 27\penalty0 (1):\penalty0 386--404, 1999.
\newblock \doi{10.1214/aos/1018031117}.

\bibitem[Liu and Singh(1992)]{LS1992}
R.~Liu and K.~Singh.
\newblock Moving blocks jackknife and bootstrap capture weak dependence.
\newblock In R.~LePage and L.~Billard, editors, \emph{Exploring the Limits of Bootstrap}. Wiley, New York, 1992.

\bibitem[Manshour et~al.(2021)Manshour, Balasis, Consolini, Papadimitriou, and Paluš]{MBCPP2021}
P.~Manshour, G.~Balasis, G.~Consolini, C.~Papadimitriou, and M.~Paluš.
\newblock Causality and information transfer between the solar wind and the magnetosphere--ionosphere system.
\newblock \emph{Entropy}, 23\penalty0 (4):\penalty0 390, 2021.
\newblock \doi{10.3390/e23040390}.

\bibitem[Mokkadem(1988)]{Mokkadem1988}
A.~Mokkadem.
\newblock Mixing properties of {ARMA} processes.
\newblock \emph{Stochastic Processes and their Applications}, 29\penalty0 (2):\penalty0 309--315, 1988.
\newblock \doi{10.1016/0304-4149(88)90040-2}.

\bibitem[Mullen et~al.(2011)Mullen, Ardia, Gil, Windover, and Cline]{MAGWC2011}
K.~M. Mullen, D.~Ardia, D.~L. Gil, D.~Windover, and J.~Cline.
\newblock {DE}optim: An {R} package for global optimization by differential evolution.
\newblock \emph{Journal of Statistical Software}, 40\penalty0 (6):\penalty0 1--26, 2011.
\newblock \doi{10.18637/jss.v040.i06}.

\bibitem[Naik-Nimbalkar and Rajarshi(1994)]{NR1994}
U.~V. Naik-Nimbalkar and M.~B. Rajarshi.
\newblock Validity of blockwise bootstrap for empirical processes with stationary observations.
\newblock \emph{The Annals of Statistics}, 22\penalty0 (2):\penalty0 980--994, 1994.
\newblock \doi{10.1214/aos/1176325519}.

\bibitem[Noble et~al.(2006)Noble, Wright, Smith, and Dibben]{NWSD2006}
M.~Noble, G.~Wright, G.~Smith, and C.~Dibben.
\newblock Measuring multiple deprivation at the small-area level.
\newblock \emph{Environment and Planning A: Economy and Space}, 38\penalty0 (1):\penalty0 169--185, 2006.
\newblock \doi{10.1068/a37168}.

\bibitem[Padoan and Stupfler(2020)]{PS2020}
S.~A. Padoan and G.~Stupfler.
\newblock Extreme expectile estimation for heavy-tailed time series.
\newblock \emph{arXiv preprint arXiv:2004.04078}, 2020.
\newblock \doi{10.48550/arXiv.2004.04078}.
\newblock URL \url{https://arxiv.org/abs/2004.04078}.

\bibitem[Pasche et~al.(2023)Pasche, Chavez-Demoulin, and Davison]{PCDD2023}
O.~C. Pasche, V.~Chavez-Demoulin, and A.~C. Davison.
\newblock Causal modelling of heavy-tailed variables and confounders with application to river flow.
\newblock \emph{Extremes}, 26:\penalty0 573--594, 2023.
\newblock \doi{10.1007/s10687-022-00456-4}.

\bibitem[Pickands(1975)]{Pickands1975}
J.~Pickands.
\newblock Statistical inference using extreme order statistics.
\newblock \emph{Annals of Statistics}, 3:\penalty0 119--131, 1975.
\newblock \doi{10.1214/aos/1176343003}.

\bibitem[Reichenbach(1956)]{Reichenbach1956}
H.~Reichenbach.
\newblock \emph{The Direction of Time}.
\newblock University of California Press, Berkeley, CA, 1956.
\newblock Reprinted by Dover Publications, 1971. ISBN 0-486-40926-0.

\bibitem[Resnick(1987)]{Resnick1987}
S.~I. Resnick.
\newblock \emph{Extreme Values, Regular Variation, and Point Processes}.
\newblock Springer, New York, 1987.

\bibitem[Sharma et~al.(2003)Sharma, Baker, Grande, Kamide, Lakhina, McPherron, Reeves, Rostoker, Vondrak, and Zelenyiio]{SBGK2003}
A.~S. Sharma, D.~N. Baker, M.~Grande, Y.~Kamide, G.~S. Lakhina, R.~M. McPherron, G.~D. Reeves, G.~Rostoker, R.~Vondrak, and L.~Zelenyiio.
\newblock The storm-substorm relationship: Current understanding and outlook.
\newblock In \emph{Disturbances in Geospace: The Storm-Substorm Relationship}, pages 1--14. American Geophysical Union (AGU), Washington, DC, 2003.

\bibitem[Vybostokova et~al.(2020)Vybostokova, Revallo, Bochníček, and Minarovjech]{VRBM2020}
O.~Vybostokova, M.~Revallo, J.~Bochníček, and M.~Minarovjech.
\newblock Magnetospheric response to solar wind driving.
\newblock In \emph{Proceedings of the 2020 Workshop on the Dynamics of the Sun and Space Weather (WDS 2020)}, pages 29--35. Charles University, Prague, 2020.
\newblock URL \url{https://www.mff.cuni.cz/veda/konference/wds/proc/pdf20/WDS20_04_f1_Vybostokova.pdf}.

\bibitem[Zscheischler et~al.(2018)Zscheischler, Westra, van~den Hurk, Seneviratne, Ward, Pitman, AghaKouchak, Bresch, Leonard, Wahl, and Zhang]{Zscheischler2018}
J.~Zscheischler, S.~Westra, B.~J. J.~M. van~den Hurk, S.~I. Seneviratne, P.~J. Ward, A.~Pitman, A.~AghaKouchak, D.~N. Bresch, M.~Leonard, T.~Wahl, and X.~Zhang.
\newblock Future climate risk from compound events.
\newblock \emph{Nature Climate Change}, 8\penalty0 (6):\penalty0 469--477, June 2018.
\newblock \doi{10.1038/s41558-018-0156-3}.

\bibitem[Zscheischler et~al.(2020)Zscheischler, Martius, Westra, Bevacqua, Raymond, Horton, van~den Hurk, AghaKouchak, Jézéquel, Mahecha, Maraun, Ramos, Ridder, Thiery, and Vignotto]{Zscheischler2020}
J.~Zscheischler, O.~Martius, S.~Westra, E.~Bevacqua, C.~Raymond, R.~M. Horton, B.~van~den Hurk, A.~AghaKouchak, A.~Jézéquel, M.~D. Mahecha, D.~Maraun, A.~M. Ramos, N.~N. Ridder, W.~Thiery, and E.~Vignotto.
\newblock A typology of compound weather and climate events.
\newblock \emph{Nature Reviews Earth \& Environment}, 1\penalty0 (7):\penalty0 333--347, July 2020.
\newblock \doi{10.1038/s43017-020-0060-z}.

\end{thebibliography}
\end{document}